# Կոմբինատորային ալգորիթմներ և ալգորիթմների վերլուծություն

Վահան Վ. Մկրտչյան

## Դասախոսություն 1: Դասընթացի նպատակը և հիմնական խնդիրները

Չնայած այն բանին, որ ալգորիթմի գաղափարը լայնորեն օգտագործվել է մաթեմատիկայում իր ողջ պատմության ընթացքում, 1936 թվականը կարելի է համարել ալգորիթմի ծննդյան տարեթիվը, քանի որ հենց այդ թվականին էր, որ Թյուրինգի, Պոստի, Գյոդելի, Չյորչի աշխատանքներում տրվեց ալգորիթմի գաղափարի հստակ, մաթեմատիկական սահմանում: Ոչ շատ ուշ ստացվեցին առաջին "հիասթափությունները", արդյունքներ, որոնք ցույց էին տալիս կոնկրետ խնդիրների օրինակներ, որոնց լուծման համար հնարավոր չէ առաջարկել ալգորիթմ: Այդ արդյունքներից ամենատպավորիչը և շատ հետաքրքիր ոչ մաթեմատիկական հետևանքներ ունեցողը, հավանաբար, Ալֆրեդ Տարսկու կողմից՝ 1936 թվականին ապացուցված թեորեմն է.

**Թեորեմ** [1]: Գոյություն չունի ալգորիթմ, որը թույլ կտա ֆորմալ թվաբանության ցանկացած բանաձևի համար պարզել, թե այն տեսության թեորեմ է, թե՞ ոչ:

Նշենք, որ առարկայի շրջանակներում մենք կուսումնասիրենք խնդիրներ, որոնք միշտ ալգորիթմորեն լուծելի են: Ավելի կոնկրետ, մեր կողմից դիտարկվող խնդիրները կունենան հետևյալ տեսքը.

Տրված են $E=\{e_1,\ldots,e_n\}$ էլեմենտների բազմությունը և $c:E\to R$ գնային ֆունկցիան: Պահանջվում է գտնել $e\in E$ տարր, այնպես որ $c(e)\to\min$ կամ $c(e)\to\max$:

Պարզ է, որ նմանատիպ խնդիրն ի սկզբանե ալգորիթմորեն լուծելի է. բավական է հերթով դիտարկել $e_1,\ldots,e_n$ −ը և ընտրել այն տարրն, որն օպտիմիզացնում է $c$ −ի արժեքը: Այս եղանակը մասնագիտական գրականության մեջ հայտնի է հատարկման մեթոդ (brute force/exhaustive search, метод перебора):

Չնայած մեթոդի (ալգորիթմի) պարզ ձևակերպմանը, պետք է նշել, որ այս եղանակի գործնական իրականացումը կապված է որոշակի դժվարությունների հետ: Բանը նրանում է, որ սովորաբար էլեմենտների բազմությունը, որի վրա տրված է գնային ֆունկցիան, կարող է լինել շատ մեծ (այսինքն՝ էքսպոնենցիալ, ինչպես օրինակ, գրաֆի զուգակցումների բազմության, անկախ բազմությունների դեպքում), և մուտքային տվյալների չափի մեծացմանը զուգընթաց, էլեմենտների բազմության հատարկումը կարող է պահանջել այնքան գործողություն, ինչքան ասենք տիեզերքի տարիքն է գրված վայրկյաններով:

Եվ հենց սրանով է բացատրվում այն պարզ հանգամանքը, որ շատ ու շատ կիրառական խնդիրների լուծման համար էֆեկտիվ ալգորիթմների նախագծման ընթացքում, որոնք սկիզբ են առնում մարդկային գործունեության տարբեր ոլորտներում, ներգրավված են բազմաթիվ պրոֆեսիոնալ մաթեմատիկոսներ, որոնց հիմնական խնդիրը, կոպիտ ասած, “խորը թեորեմների” (deep theorems) ապացույցն է: Շնորհիվ հենց այս թեորեմների է, որ հաջողվում է շատ խնդիրների համար առաջարկել էֆեկտիվ (բազմանդամային) ալգորիթմներ, որոնք խուսափում են բոլոր էլեմենտների հատարկումից: Այս տեսանկյունից հատկանշական է գծային ծրագրավորման խնդրի համար 1980-ին Լեոնիդ Խաչյանի կողմից առաջարկված ալգորիթմը [2] կամ առաջին բազմանդամային բարդություն ունեցող ալգորիթմը, որը թույլ է տալիս պարզել տրված թվի պարզ լինելը [10]:

Դասընթացի շրջանակներում կսահմանվեն ալգորիթմի բարդության գաղափարը և կուսումնասիրվեն խնդիրներ, որոնց համար կնշվեն այդ խնդիրները լուծող լավագույն ալգորիթմները և նրանց բարդությունը: Այնուհետև, կուսումնասիրվեն թվաբանական, երկրաչափական, գրաֆներին առնչվող մի շարք խնդիրներ, որոնց համար կառաջարկվեն էֆեկտիվ ալգորիթմներ: Նախապես նշենք, որ այդ ալգորիթմները կարող են լավագույնը չլինել:

Կսահմանվեն նաև խնդիրների $P$ և $NP$ դասերը, որոնցից առաջինը ($P$-ն) կպարունակի վերը նշված (“հեշտ լուծելի”) խնդիրները: Ցույց կտրվի, որ $P \subseteq NP$ և կդիտարկվեն խնդիրներ, որոնք պատկանում են $NP$ դասին, և որոնց $P$ դասին պատկանելու հարցը բաց է մինչ այսօր: Ցույց կտրվեն, որ այս խնդիրները համարժեք են և հանդիսանում են $NP$ դասի ամենադժվար խնդիրները, այսինքն եթե նշված խնդիրներից որևէ մեկը պատկանում է $NP$ դասին, ապա $P = NP$:

Այնուհետև, կառաջարկվեն էֆեկտիվ ալգորիթմներ, որոնք կլուծեն այս խնդիրները ինչ-որ ճշտությամբ (մոտավոր ալգորիթմներ): Կուսումնասիրվեն այս ալգորիթմների մոտարկման գործակիցները, կնշվեն այս ալգորիթմների և մատրոիդների կապը, կբերվեն խնդիրների օրինակներ, որոնց մոտավոր լուծման հարցը համարժեք է ճիշտ լուծմանը (խնդիրներ, որոնք թույլ չեն տալիս մոտարկում, inapproximability results):

Նշենք, որ 2000 թվականին $P/NP$ պրոբլեմի (ճիշտ է արդյոք, որ $P = NP$) լուծման համար առաջարկվել է 1.000.000 դոլլար: [3,9] աշխատաքներում կարելի է ծանոթանալ խնդրի ձևակերպմանը, կարևորությանը: Կարելի է նշել նաև Ա. Վիգդերսոնի զեկույցի տեքստը 2006 թվականի օգոստոսին Մադրիդում կայացած մաթեմատիկոսների միջազգային կոնգրեսում [4]: Ս. Սմեյլը իր "Mathematical problems for the next century" աշխատաքում [5] $P/NP$ պրոբլեմը համարում է նվեր ինֆորմատիկայից (computer science) մաթեմատիկային:

Մեր ժամանակի ականավոր մաթեմատիկոսներ Լ. Ֆորտնոուն և Լ. Տրեվիսանը ունեն weblog-եր, համապատասխանաբար, Computational Complexity և In-Theory անուններով, որոնք կարելի է ընթերցել, գրանցվել և ստանալ պրոբլեմին վերաբերվող նորությունները email–ի վրա [6,7]:

Հաշվի առնելով խնդրի կարևորությունը, Internet-ում կարելի է գտնել $P/NP$ պրոբլեմի բազմաթիվ "լուծումներ", որոնց անընդհատ թարմացվող ցուցակը կարելի է գտնել Վոեջինջերի կայքում [8]:

## Գրականություն

Գտնված սխալների, առաջարկությունների, ինչպես նաև դասախոսություն-ներն e-mail-ով ստանալու համար կարող էք դիմել vahanmkrtchyan2002@yahoo.com հասցեով:

Կոմբինատորային ալգորիթմներ և ալգորիթմների վերլուծություն
Վահան Վ. Մկրտչյան

Դասախոսություն 2: Որոնման ալգորիթմներ և նրանց ներկայացումը ծառերի միջոցով: Ալգորիթմի բարդություն: Կարգավոր բազմության տարրի որոնում:

Մեր մոտակա նպատակը որոշակի տիպի խնդիրներ լուծող ալգորիթմների համար որոնման ծառի գաղափարի ներմուծումն է: Այս գաղափարը մեզ թույլ կտա սահմանել այդպիսի ալգորիթմների բարդության գաղափարը: Այնուհետև, կդիտարկենք խնդիրներ, որոնց համար կկառուցվեն այդ խնդիրները լուծող լավագույն ալգորիթմները:

Նկատենք, որ տրված ալգորիթմի՝ լավագույնը լինելու ապացույցը բաղկացած է երկու քայլից.

- այն բանի ապացույցից, որ տրված ալգորիթմն իրոք լուծում է խնդիրը,
- այն բանի ապացույցից, որ խնդիրը լուծող և ավելի արագ աշխատող ալգորիթմ գոյություն չունի:

Երկրորդ կետն իրականացնելու ժամանակ կուսումնասիրվեն պարզագույն եղանակներ, որոնք թույլ կտան ստանալ ստորին գնահատականներ տրված խնդիրը լուծող ցանկացած ալգորիթմի քայլերի քանակի համար:

Որոնման ալգորիթմի համապատասխան ծառի գաղափարը պարզաբանելու համար դիտարկենք հետևյալ խնդիրը.

տրված են $n$ գնդիկներ՝ համարակալված $1,\ldots,n$ թվերով: Հայտնի է, որ նրանցից մեկը ռադիոակտիվ է (մեզ հայտնի չէ, թե որը): Ունենք սարք, որը մեկ ստուգումով պարզում է ռադիոակտիվ գնդիկի առկայությունը մեր կողմից ընտրած գնդիկների ենթաբազմության մեջ: Պահանջվում է հնարավորին չափ քիչ ստուգումների միջոցով գտնել ռադիոակտիվ գնդիկը:

Նկատենք, որ գոյություն ունեն այս խնդիրները լուծող բազմաթիվ ալգորիթմներ: Դիտարկենք նրանցից երկուսը.

Ալգորիթմ 1: Հերթով դիտարկել $1,\dots,n$ գնդիկները մինչև ռադիոակտիվ գնդիկի գտնելը:

Ալգորիթմ 2:
Քայլ 1: $I := \{1,\dots,n\}$;
Քայլ 2: $I$ -ն բաժանել երկու համարյա հավասար մասերի, այսինքն

$$I = I_1 \cup I_2 ,$$
$$I_1 \cap I_2 = \varnothing ,$$
$$\left\| I_1 \right| - \left| I_2 \right\| \le 1 :$$

Քայլ 3: Ստուգել $I_1$-ում ռադիոակտիվ գնդիկի առկայությունը;
Քայլ 4: Եթե $I_1$-ում կա ռադիոակտիվ գնդիկ, ապա $I := I_1$, հակառակ դեպքում՝ $I := I_2$;
Քայլ 5: Եթե $|I| = 1$, ապա ավարտել ալգորիթմի աշխատանքը, հակառակ դեպքում՝ անցնել Քայլ 2-ի կատարմանը:

Նախ նկատենք, որ այս երկու ալգորիթմներն էլ լուծում են խնդիրը, այսինքն գտնում են ռադիոակտիվ գնդիկը:
Ալգորիթմին համապատասխան որոնման ծառի գաղափարը պարզաբանելու համար դիտարկենք մասնավոր դեպք, երբ $n = 4$:
Դիտարկենք $2$–ծառ, որի արմատին համապատասխանեցված է ալգորիթմի առաջին քայլում ստուգվող ենթաբազմությունը: Եթե ծառի որևէ գագաթին համապատասխանեցված է $M$ ենթաբազմությունը, ապա

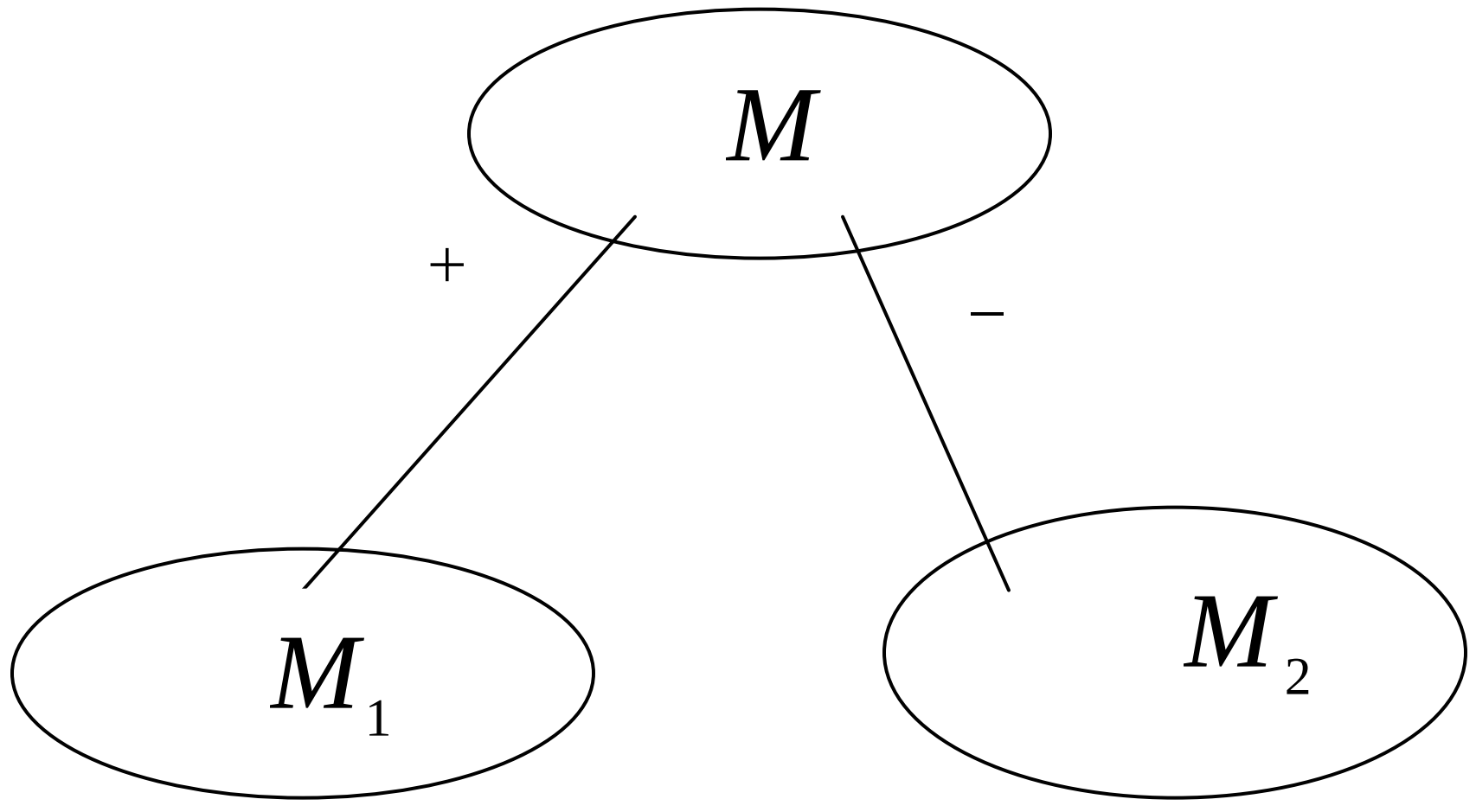

նկար 1

նրան հաջորդող երկու գագաթներին համապատասխանեցնենք գնդիկների $M_1$ և $M_2$ ենթաբազմությունները, որոնք ստուգվում են $M$ -ում ռադիոակտիվ գնդիկի առկայության կամ բացակայության դեպքում (նկար 1): Ալգորիթմի աշխատանքի ավարտին համապատասխանող գագաթները ծառի տերևներն են:

Դժվար չէ համոզվել, որ Ալգորիթմներ 1 և 2-ին համապատասխանող ծառերը կլինեն

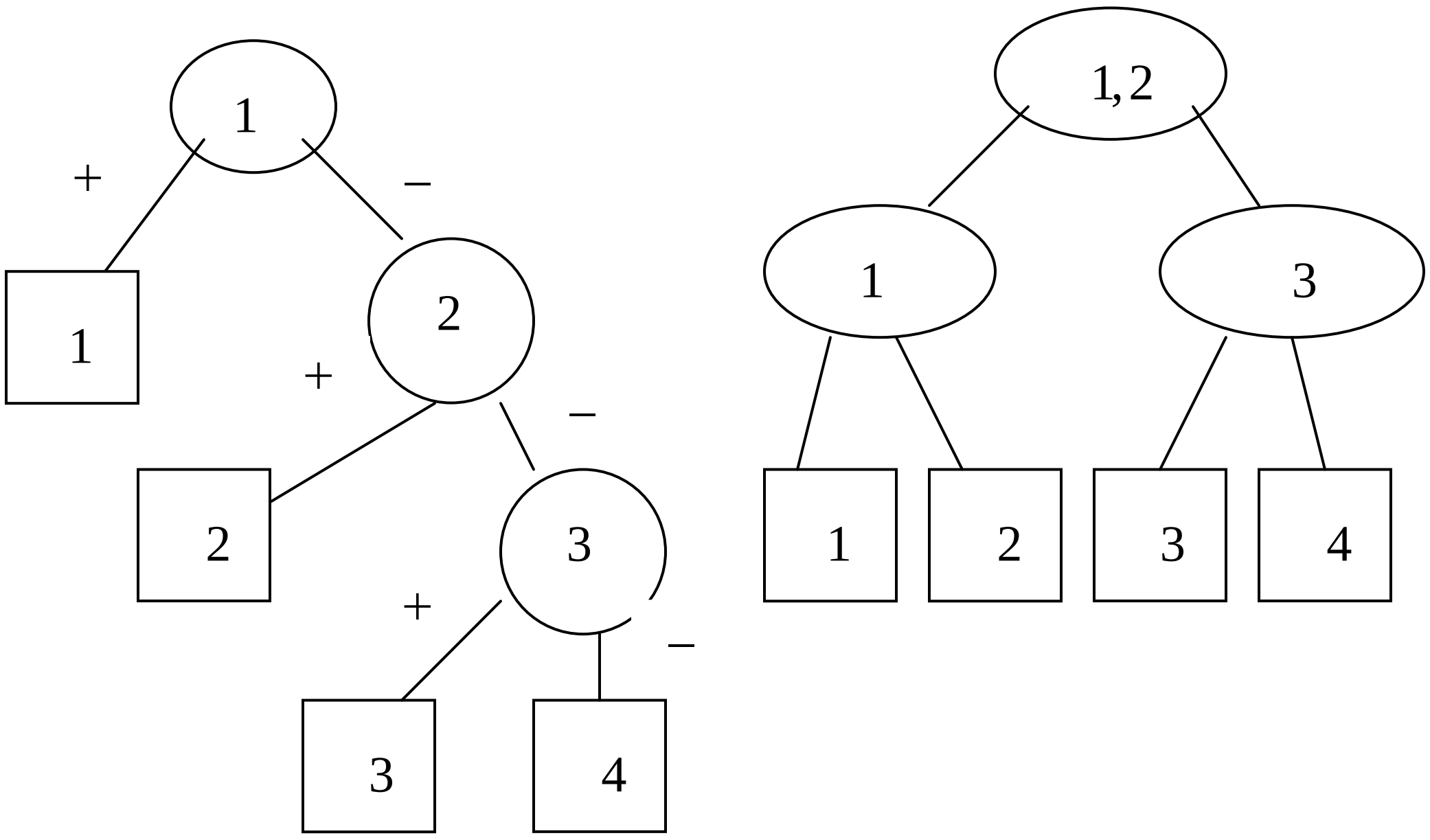

նկար 2

Նկար 2-ից երևում է, որ տերևի բարձրությունը (այդ գագաթի՝ արմատին միացնող շղթայի երկարությունը) ալգորիթմի կատարած ստուգումների քանակն է, երբ ռադիոակտիվ է այդ գագաթին համապատասխանող գնդիկը:

Նկատենք, որ եթե 1գնդիկը ռադիոակտիվ է, ապա Ալգորիթմ 1-ը ավելի շուտ է պարզում նրա ռադիոակտիվ լինելը (1 քայլ), քան Ալգորիթմ 2-ը (2 քայլ): Հակառակը, եթե ռադիոակտիվ է 4 գնդիկը, ապա Ալգորիթմ 2-ը ավելի շուտ է պարզում նրա ռադիոակտիվ լինելը (2 քայլ), քան Ալգորիթմ 1-ը (3 քայլ):

Դասընթացի շրջանակներում ալգորիթմի բարդություն ասելով կհասկանանք *վատագույն* դեպքում ստուգումների քանակը (նկատենք, որ սա համապատասխանում է ալգորիթմի որոնման ծառի ամենաերկար ճյուղի երկարությանը): Նշենք, որ սա ալգորիթմի բարդության սահմանման միակ հնարավոր եղանակը չէ: Ներկայումս, լայնորեն օգտագործվում է, այսպես կոչված միջին բարդության գաղափարը, որն իրենից ներկայացնում է հնարավոր դեպքերի ստուգումների քանակի գումարի հարաբերությունը դեպքերի քանակին: Օրինակ՝ Ալգորիթմ 1-ի միջին բարդությունը$=\dfrac{1+2+3+3}{4}=2.25$, այնինչ՝ Ալգորիթմ 2-ի միջին բարդությունն է $\dfrac{2+2+2+2}{4}=2$:

Նշենք նաև, որ ալգորիթմի բարդության մեր սահմանումից հետևում է, որ Ալգորիթմ 1-ի բարդությունը 3-է, իսկ Ալգորիթմ 2-ինը՝ 2: Հետևաբար, ըստ մեր սահմանման, Ալգորիթմ 2-ը "գերադասելի է" Ալգորիթմ 1-ից:

Ստորև, կդիտարկվեն որոշակի կոմբինատոր խնդիրներ և կառաջարկվեն այդ խնդիրները լուծող լավագույն ալգորիթմներ:

**Խնդիր 1**: Գտնել ռադիոակտիվ զնդիկը որոշելու այնպիսի ալգորիթմ, որի բարդությունը ամենափոքրն է (այսինքն վատագույն դեպքում ստուգումների քանակը հնարավորինս փոքր է):

Նկատենք, որ խնդիր 1-ը լուծող լավագույն ալգորիթմ միշտ գոյություն ունի: Իրոք, այդպիսի ալգորիթմի գոյությունը հետևում է այն բանից, որ Ալգորիթմ 1-ը խնդիր 1-ը լուծում է ոչ ավել քան $n-1$ ստուգմամբ, իսկ $n-1$-ից փոքր կամ հավասար բարձրություն ունեցող ծառերի քանակը վերջավոր է:

$t_1(n)$-ով նշանակենք այդ ալգորիթմի բարդությունը: Դիտարկենք նաև հետևյալ խնդիրը

**Խնդիր 2**: Տրված է $A=\{a_1,...,a_n\}$ բազմությունը և $x\in A$ մեզ անհայտ տարրը: Թույլատրվում է $x$–ի մասին տալ "այո" կամ "ոչ" պատասխան ունեցող հարցեր և ստանալ պատասխաններ: Պահանջվում է նվազագույն թվով հարցերի միջոցով պարզել, թե $a_1,...,a_n$-ից որ մեկն է $x$ տարրը:

Նկատենք, որ այս խնդիրը նույնպես ունի լավագույն ալգորիթմ ( ալգորիթմին համապատասխանեցնել ծառ՝ գագաթներում գրելով հարցերը): $t_2(n)$-ով նշանակենք այդ ալգորիթմի բարդությունը:

**Թեորեմ 1**: $t_2(n)\le t_1(n)$

**Ապացույց**: Դիցուք $A$-ն ռադիոակտիվ զնդիկը որոշող լավագույն ալգորիթմ է, $T_A$-ն՝ նրան համապատասխան որոնման ծառը, որի բարձրությունն է $t_1(n)$–ը: Վերցնենք $T_A$-ի որևէ գագաթ, որին համապատասխանեցված է $M$ բազմությունը: Այդ գագաթին համապատասխանեցնենք $(x\in\{a_i / i\in M\})$ հարցը (ճիշտ է արդյոք, որ $x$ տարրը պատկանում է $\{a_i / i\in M\}$ բազմությանը):

Նկատենք, որ ստացանք $t_1(n)$ բարձրությամբ ծառ, որին համապատասխանում է **Խնդիր 2**-ը լուծող ալգորիթմ, հետևաբար՝ $t_2(n) \le t_1(n)$:

**Թեորեմ 2:** $\lceil \log_2 n \rceil \le t_2(n)$

**Ապացույց:** Դիցուք $A$-ն խնդիր 2-ը լուծող լավագույն ալգորիթմ է, $T_A$-ն՝ նրան համապատասխան որոնման ծառը, որի բարձրությունն է $t_2(n)$–ը: Նկատենք, որ $T_A$-ն պետք է ունենա առնվազն $n$ տերև, որոնց համապատասխանում են $x = a_1, \ldots, x = a_n$ դեպքերը: Քանի որ $T_A$-ն $2$–ծառ է, ապա

$$2^{t_2(n)} \ge T_A\text{-ի տերևների քանակից} \ge n$$

կամ

$$t_2(n) \ge \log_2 n \text{, և հետևաբար՝ } t_2(n) \ge \lceil \log_2 n \rceil :$$

**Թեորեմ 3** (Հիմնական): $t_1(n) = t_2(n) = \lceil \log_2 n \rceil$

**Ապացույց:** Նկատենք, որ Թեորեմներ 1 և 2-ից հետևում է, որ բավական է ցույց տալ, որ $t_1(n) \le \lceil \log_2 n \rceil$, ինչի համար բավական է կառուցել ռադիոակտիվ գնդիկը որոշող ալգորիթմ, որի ստուգումների քանակը $\le \lceil \log_2 n \rceil$:

Դիտարկենք վերը նշված Ալգորիթմ 2-ը, որի բարդությունը նշանակենք $t(n)$–ով: Նկատենք, որ $t(1) = 0$, $t(1) \le t(2) \le t(3) \le \ldots$ և $t(n) \le 1 + t(\lceil \frac{n}{2} \rceil)$: Ցույց տանք, որ $t(n) \le \lceil \log_2 n \rceil$: Ապացույցը կատարենք ինդուկցիայով:

$n = 1$: $t(1) = 0 = \lceil \log_2 1 \rceil$

Ենթադրենք, որ $t(i) \le \lceil \log_2 i \rceil$ $i = 1, \ldots, n-1$ ($n \ge 2$): Ունենք՝

$$t(n) \le 1 + t(\left\lceil \frac{n}{2} \right\rceil) \le 1 + \left\lceil \log_2 \left\lceil \frac{n}{2} \right\rceil \right\rceil = \left\lceil \log_2 2 \left\lceil \frac{n}{2} \right\rceil \right\rceil = \lceil \log_2 n \rceil$$

(դիտարկել $n = 2k$ և $n = 2k + 1$ դեպքերը):

## Գրականություն

Գտնված սխալների, առաջարկությունների, ինչպես նաև դասախոսություն-ներն e-mail-ով ստանալու համար կարող էք դիմել vahanmkrtchyan2002@yahoo.com հասցեով:

Կոմբինատորային ալգորիթմներ և ալգորիթմների վերլուծություն
Վահան Վ. Մկրտչյան

## Դասախոսություն 3: Կեղծ մետաղադրամի որոնում:

Դիցուք ունենք $n$ մետաղադրամ, և հայտնի է, որ նրանցից ամենաշատը մեկը կեղծ է (այն կարող է լինել ծանր կամ թեթև մյուսներից): Լրացուցիչ տրված է մեկ իսկական մետաղադրամ: Ունենք նաև նժարավոր կշեռք, որի միջոցով կարող ենք համեմատել մեր կողմից ընտրված մետաղադրամների երկու խմբերի քաշերը: Պահանջվում է հնարավորինս փոքր թվով կշռումների միջոցով գտնել կեղծ մետաղադրամը (եթե այդպիսին կա) և իմանալ՝ այն թեթև, թե ծանր է մյուսներից:

Կասկածելի մետաղադրամները համարակալենք $1,\ldots,n$ թվերով, իսկ իսկականին վերագրենք $0$ համարը: $i$-րդ մետաղադրամի ծանր կամ թեթև լինելը համապատասխանաբար, նշանակենք $i+$ և $i-$: Թող $S=\{0,1,\ldots,n\}$: Հնարավոր է $2n+1$ դեպք՝ $1+,\ldots,n+,1-,\ldots,n-,$ և $0$, որտեղ՝ $0$-ն այն դեպքն է, երբ բոլոր մետաղադրամները իսկական են:

Նախ նկատենք, որ քանի որ յուրաքանչյուր $S_1:S_2$ բաղդադում ($S_1,S_2\subseteq S$) ունի երեք հնարավոր արդյունք, ապա կեղծ մետաղադրամի խնդիրը լուծող ցանկացած ալգորիթմի կարելի է համապատասխանեցնել $3$-ծառ:

Իրոք, դիցուք ունենք կեղծ մետաղադրամի խնդիրը լուծող որևէ ալգորիթմ: Դիտարկենք $3$-ծառ, որի արմատին համապատասխանեցված է ալգորիթմի առաջին կշռումը:

Եթե ծառի որևէ որևէ գագաթին համապատասխանեցված է $S_1:S_2$ բաղդադումը, ապա նրանից դուրս եկնող երեք կողերին համապատասխանեցնենք $>,<$ և $=$ նշանները (նկար 2.1):

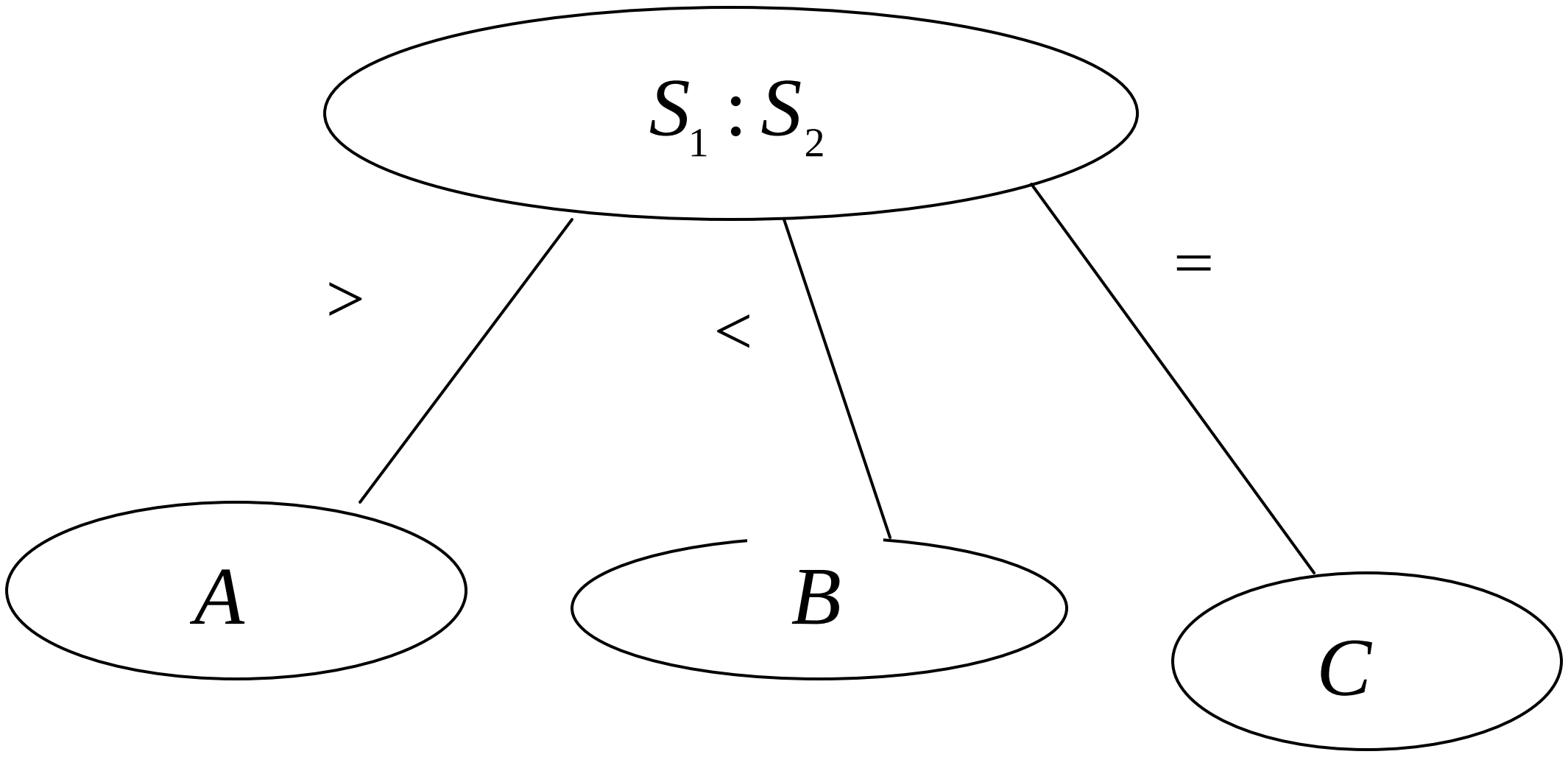

նկար 2.1

$A$ գագաթին համապատասխանեցնենք այն կշռումը, որը կատարվում է, երբ $S_1$–ի քաշը մեծ է $S_2$–ի քաշից: $B$ գագաթին համապատասխանեցնենք այն կշռումը, որը կատարվում է, երբ $S_1$–ի քաշը փոքր է $S_2$–ի քաշից: $C$ գագաթին համապատասխանեցնենք այն կշռումը, որը կատարվում է, երբ $S_1$–ի քաշը հավասար է $S_2$–ի քաշին:

Ալգորիթմի ավարտին համապատասխանող գագաթները համարենք ծառի տերևներ:

$\tau(n)$-ով նշանակենք $\{1,...,n\}$ կասկածելի մետաղադրամների բազմությունից կեղծ մետաղադրամը գտնող լավագույն ալգորիթմի բարդությունը, երբ հայտնի է, որ կա ամենաշատը մեկ կեղծ մետաղադրամ:

**Թեորեմ**: $\tau(n)=\lceil \log_3(2n+1)\rceil$

**Ապացույց**: Քանի որ կեղծ մետաղադրամը գտնող լավագույն ալգորիթմին համապատասխանող ծառը $3$-ծառ է, ապա նրա տերևների քանակը կարող է լինել ամենաշատը $3^{\tau(n)}$: Մյուս կողմից, պարզ է, որ այդ ծառի տերևների քանակն առնվազն $2n+1$-է, հետևաբար՝ $3^{\tau(n)}\geq 2n+1$ կամ $\tau(n)\geq\lceil\log_3(2n+1)\rceil$:

Ցույց տանք, որ $\tau(n)\leq\lceil\log_3(2n+1)\rceil$: Նկատենք, որ $2n+1\leq 3^{\lceil\log_3(2n+1)\rceil}$, հետևաբար պնդման ապացույցն ավարտելու համար բավական է ապացուցել, որ երբ $2n+1\leq 3^l$, ապա $\tau(n)\leq l$, այսինքն՝ գոյություն ունի կեղծ մետաղադրամի խնդիրը լուծող ալգորիթմ, որը կատարում է ոչ ավել քան $l$ կշռում:

Նախ դիտարկենք մասնավոր դեպք, երբ $2n+1=3^l$, այսինքն՝ երբ $n=\frac{3^l-1}{2}=K_l$: Նկատենք, որ $K_1=1, K_2=4$ և $K_l=3K_{l-1}+1$, երբ $l\geq 2$:

Ստորև կտրվի $K_l$ մետաղադրամներից $l$ կշռումների միջոցով կեղծը գտնելու ընդհանուր նկարագիրը: Ալգորիթմի աշխատանքի ժամանակ կհանդիպեն հետևյալ երեք դեպքերը:

**Առաջին դեպք**: Մնացել է կասկածելի $K_j$ $1\leq j\leq l$ մետաղադրամ:
Անհրաժեշտ է $j$ կշռումների միջոցով գտնել կեղծ մետաղադրամը, եթե այն կա (ունենք նաև գոնե մեկ իսկական մետաղադրամ):

Նկատենք որ սկզբնական պահին բավարարվում է հենց առաջին դեպքի պայմանները:

**Երկրորդ դեպք**: Կեղծ մետաղադրամ կա. այն կամ ծանր է և $K_j+1$՝ ծանրության մեջ կասկածվող մետաղադրամների մեջ է, կամ էլ թեթև է և մեկն է $K_j$՝ թեթևության մեջ կասկածվող մետաղադրամներից, $1\leq j\leq l$:
Անհրաժեշտ է $j$ կշռումների միջոցով գտնել կեղծ մետաղադրամը:

**Երրորդ դեպք**: Կեղծ մետաղադրամ կա. այն կամ թեթև է և $K_j+1$՝ թեթևության մեջ կասկածվող մետաղադրամների մեջ է, կամ էլ ծանր է և մեկն է $K_j$՝ ծանրության մեջ կասկածվող մետաղադրամներից, $1\leq j\leq l$:
Անհրաժեշտ է $j$ կշռումների միջոցով գտնել կեղծ մետաղադրամը:

Ստորև կնկարագրվի ալգորիթմի վարքը դեպքերից յուրաքանչյուրում:

**Կշռման եղանակը առաջին դեպքում**: Կշեռքի $A$ նժարին դնում ենք $K_{j-1}+1$ կասկածելի մետաղադրամ, իսկ $B$ նժարին՝ $K_{j-1}$ կասկածելի և մեկ իսկական մետաղադրամ: Չի օգտագործվում $K_{j-1}$ կասկածելի մետաղադրամ:

*Կշռման արդյունքի վերլուծությունը*:

ա) համեմատվող կշռաքարերի քաշերը հավասար են՝ $A=B$, և հետևաբար, բոլոր բաղադրվող մետաղադրամները իսկական են: Խնդրի լուծումն ավարտելու համար պետք է $j-1$ կշռումների միջոցով չօգտագործված $K_{j-1}$ մետաղադրամներից գտնել կեղծը եթե այն կա:

Երբ $j-1>1$ ապա հանգեցինք նույն առաջին դեպքին, իսկ $j-1=1$ դեպքում ունենք $K_1=1$ կասկածելի մետաղադրամ և պետք է մեկ կշռումով որոշել թե այն ինչպիսին է: Լուծումն ակնհայտ է:

բ) $A>B$: Կեղծ մետաղադրամ կա. Այն կամ ծանր է և գտնվում է $A$ նժարի վրա (մեկն է $K_{j-1}+1$ մետաղադրամներից, որոնք կասկածվում են ծանրության մեջ), կամ էլ թեթև է և գտնվում է $B$ նժարի վրա (մեկն է $K_{j-1}$ մետաղադրամներից, որոնք կասկածվում են թեթևության մեջ): Խնդրի լուծումն ավարտելու համար պետք է $j-1$ կշռումների միջոցով գտնել այն:

Երբ $j-1>1$ ապա հանգեցինք երկրորդ դեպքին, իսկ $j-1=1$ դեպքում ունենք երկու՝ ծանրության մեջ կասկածվող մետաղադրամներ, և մեկ թեթևության մեջ կասկածվող մետաղադրամ: Պահանջվում է մեկ կշռման միջոցով ավարտել խնդրի լուծումը: Լուծումն ակնհայտ է. բավական է համեմատել ծանրության մեջ կասկածվող մետաղադրամները:

գ) $A<B$: Կեղծ մետաղադրամ կա. Այն կամ թեթև է և գտնվում է $A$ նժարի վրա (մեկն է $K_{j-1}+1$ մետաղադրամներից, որոնք կասկածվում են թեթևության մեջ), կամ էլ ծանր է և գտնվում է $B$ նժարի վրա (մեկն է $K_{j-1}$ մետաղադրամներից, որոնք կասկածվում են ծանրության մեջ): Խնդրի լուծումն ավարտելու համար պետք է $j-1$ կշռումների միջոցով գտնել այն:

Երբ $j-1>1$ ապա հանգեցինք երրորդ դեպքին, իսկ $j-1=1$ դեպքում ունենք երկու՝ թեթևության մեջ կասկածվող մետաղադրամներ, և մեկ ծանրության մեջ կասկածվող մետաղադրամ: Պահանջվում է մեկ կշռման միջոցով ավարտել խնդրի լուծումը: Լուծումն ակնհայտ է. բավական է համեմատել թեթևության մեջ կասկածվող մետաղադրամները:

**Կշռման եղանակը երկրորդ դեպքում**: Կշեռքի $A,B$ նժարներից յուրաքանչյուրին դնում ենք $K_{j-1}+1$ ծանրության մեջ և $K_{j-1}$ թեթևության մեջ կասկածվող մետաղադրամ: Չի օգտագործվում ծանրության մեջ կասկածվող $K_{j-1}$ և թեթևության մեջ կասկածվող $K_{j-1}+1$ մետաղադրամ:

*Կշռման արդյունքի վերլուծությունը*:

ա) համեմատվող մետաղադրամի քաշերը հավասար են և հետևաբար, նրանք իսկական են: Չօգտագործված ծանրության մեջ կասկածվող $K_{j-1}$ և թեթևության մեջ կասկածվող $K_{j-1}+1$ մետաղադրամներից մեկը կեղծ է: Խնդրի լուծումն ավարտելու համար անհրաժեշտ է $j-1$ կշռումների միջոցով գտնել այն:

$j-1=1$ դեպքն ակնհայտ է, իսկ $j-1>1$ դեպքում հանգեցինք երրորդ դեպքին:

բ) $A>B$: Կեղծ մետաղադրամը համեմատվողների մեջ է: Այն կամ ծանր է, և $A$ նժարի վրա գտնվող $K_{j-1}+1$ ծանրության մեջ կասկածվող մետաղադրամներից մեկն է. կամ էլ թեթև է, և $B$ նժարի $K_{j-1}$ թեթևության մեջ կասկածվող մետաղադրամներից մեկն է: Խնդրի լուծումն ավարտելու համար պետք է $j-1$ կշռումների միջոցով գտնել այն:

Նկատենք, որ կրկին ստացվեց նույն երկրորդ դեպքը $j-1$-ի համար:

գ) $A<B$: նորից բերվում է երկրորդ դեպքին:

**Կշռման եղանակը երրորդ դեպքում**: Կշեռքի $A,B$ նժարներից յուրաքանչյուրին դնում ենք $K_{j-1}+1$ թեթևության մեջ և $K_{j-1}$ ծանրության մեջ կասկածվող մետաղադրամ: Չի օգտագործվում թեթևության մեջ կասկածվող $K_{j-1}$ և ծանրության մեջ կասկածվող $K_{j-1}+1$ մետաղադրամ:

*Կշռման արդյունքի վերլուծությունը*:

կատարվում է երկրորդ դեպքի նման՝ փոխելով "ծանրության" և "թեթևության" բառերի տեղերը:

$K_{l-1}<n<K_l$ դեպքում առանձնացնենք $K_{l-1}$ մետաղադրամ, իսկ մնացածը բաժանենք երկու հավասար մասի (եթե մնացածը կենտ թիվ է կազմում, ապա ավելացնենք նախապես տրված իսկականը) և համեմատենք: Եթե հավասար են, ապա ըստ վերը նշվածի, մնացածից կեղծը (եթե կա) կարելի է պարզել $l-1$ քայլով, իսկ եթե հավասար չեն, ապա կեղծը $n-K_{l-1}<K_l-K_{l-1}=2K_{l-1}+1=3^{l-1}$-ի մեջ է: Ցույց տանք, որ $l-1$ կշռումով կարող ենք գտնել կեղծը: Ապացուցենք ավելի ընդհանուր փաստ. դիցուք ունենք $x+y\leq 3^k$ մետաղադրամ, որոնցից մեկը կեղծ է: Հայտնի է, որ նրանցից $x$-ը կասկածվում են թեթևության մեջ, իսկ մնացած $y$-ը՝ ծանրության մեջ: Ապացուցենք, որ կեղծը կարելի է գտնել $k$ քայլում: Ապացույցը կատարենք ինդուկցիայով ըստ $k$-ի:

$k=1$ դեպքում պնդումն ակնհայտ է: Ենթադրենք, որ այն ճիշտ է $<k$ դեպքում, և փորձենք ապացուցել $k$-ի համար: Կշեռքի յուրաքանչյուր

նժարին դնենք $x'$ մետաղադրամ թեթևության մեջ կասկածվող $x$ մետաղադրամներից, և ծանրության մեջ կասկածվող $y'$ մետաղադրամներից, և $x', y'$ թվերն ընտրենք այնպես, որ

$$\begin{cases} x'+y' \le 3^{k-1} \\ x-2x'+y-2y' \le 3^{k-1} \end{cases}$$

կամ, որ նույնն է՝

$$\frac{x+y-3^{k-1}}{2} \le x'+y' \le 3^{k-1}:$$

Եթե նժարները հավասար են, ապա կեղծը մնացած $x-2x'+y-2y' \le 3^{k-1}$ մետաղադրամների մեջ է, հետևաբար ըստ ինդուկցիոն ենթադրության այն կարելի է գտնել $k-1$ կշռման միջոցով:
Եթե նժարները հավասար չեն, ապա այն նժարներից մեկի վրա գտնվող $x'$ թեթևության մեջ կասկածվողների մեջ է, կամ մյուսի վրա գտնվող $y'$ ծանրության մեջ կասկածվողների: Քանի որ $x'+y' \le 3^{k-1}$, ապա կրկին ըստ ինդուկցիոն ենթադրության այն կարելի է գտնել $k-1$ կշռման միջոցով:
Թեորեմն ապացուցված է:

Ցույց տանք, որ մեկ իսկական մետաղադրամ ունենալը անհրաժեշտ է $l$ կշռումների միջոցով խնդիրը լուծելու համար: Իրոք, ենթադրենք տրված չէ իսկական մետաղադրամը, և դիտարկենք մասնավոր դեպք, երբ $n = \frac{3^l - 1}{2} = K_l$: Ենթադրենք, որ գոյություն ունի $l$ համեմատում կատարող ալգորիթմ, որը գտնում է կեղծ մետաղադրամը, եթե այն կա: Դիցուք այդ ալգորիթմը առաջին քայլում համեմատում է $x$ թվով կշռաքարեր:
Եթե

նժարները հավասար են, ապա ալգորիթմը $l-1$ կշռումների միջոցով կարողանում է մնացած $K_l - 2x$ մետաղադրամներից գտնել կեղծը, եթե այն կա, հետևաբար՝ $K_l - 2x \le K_{l-1}$, կամ՝ $2x \ge K_l - K_{l-1} = 2K_{l-1} + 1 = 3^{l-1}$, նժարները հավասար չեն, ապա կեղծը նժարների վրա գտնվող $2x$ մետաղադրամների մեջ է, հետևաբար՝ $2x \le 3^{l-1}$ (ծառը պետք է ունենա առնվազն $2x$ տերև),

հետևաբար՝ $2x = 3^{l-1}$: Հակասություն:

# Գրականություն

Գտնված սխալների, առաջարկությունների, ինչպես նաև դասախոսություն-ներն e-mail-ով ստանալու համար կարող էք դիմել vahanmkrtchyan2002@yahoo.com հասցեով:

Կոմբինատորային ալգորիթմներ և ալգորիթմների վերլուծություն
Վահան Վ. Մկրտչյան

Դասախոսություն 4: Բազմությունների հավասարության ստուգում: Երկընթաց հաջորդականության ամենամեծ տարրի որոնում:

**Երկընթաց հաջորդականության ամենամեծ տարրի որոնում**: Դրական թվերի $a_1, a_2, \ldots, a_m, \ldots$ հաջորդականությունն անվանենք երկընթաց, եթե գոյություն ունի $k$ բնական թիվ այնպես, որ $0 < a_1 < a_2 < \ldots < a_k$ և $a_k > a_{k+1} > a_{k+2} > \ldots$ :

$Z_n$-ով նշանակենք $n$ անդամ պարունակող երկընթաց հաջորդականությունների բազմությունը: Դիտարկենք հետևյալ խնդիրը.
տրված է մեզ անհայտ $\alpha = (x_1, \ldots, x_n) \in Z_n$ երկընթաց հաջորդականությունը: Թույլատրվում է տալ հետևյալ տիպի հարց՝ << Որն է հաջորդականության $i$ –րդ անդամի արժեքը>> ($1 \le i \le n$) և ստանալ պատասխան: Պահանջվում է նվազագույն թվով հարցերի միջոցով որոշել $\alpha$ հաջորդականության ամենամեծ անդամը:

**Դիտողություն**: Քանի որ հաջորդականության անդամները կարող են լինել ցանկացած դրական թվեր, ապա այս դեպքում հնարավոր չէ խնդիրը լուծող ալգորիթմները պատկերել $2$-ծառերի, $3$-ծառերի, $4$-ծառերի ,…միջոցով:
**Դիտողություն**: Երկընթաց հաջորդականության ամենամեծ տարրը որոշող ալգորիթմ գոյություն ունի:

$\lambda_k$-ով նշանակենք ամենամեծ ամբողջ թիվը, որը բավարարում է հետևյալ պայմանին. ցանկացած $\alpha = (x_1, \ldots, x_{\lambda_k}) \in Z_{\lambda_k}$ երկընթաց հաջորդականության ամենամեծ անդամը կարելի է որոշել $k$ հարցերի միջոցով:

**Դիտողություն**: Ցանկացած $k$ թվի համար $\lambda_k$-ն գոյություն ունի և $\lambda_1 = 1$, $\lambda_2 = 2$:

**Թեորեմ 1**: Ցանկացած $k \geq 3$ թվի համար տեղի ունի հետևյալ անհավասարությունը.

$$\lambda_k \leq \lambda_{k-1} + \lambda_{k-2} + 1:$$

**Ապացույց**: Դիցուք $k \geq 3$ և Ա-ն երկընթաց հաջորդականության ամենամեծ տարրը որոշող լավագույն ալգորիթմ է: Ենթադրենք, որ $Z_{\lambda_k}$ բազմությանը պատկանող $\alpha$ հաջորդականության համար այդ ալգորիթմի առաջին և երկրորդ հարցերը որոշում են նրա $i_1$-րդ և $i_2$-րդ անդամի արժեքները ($i_1 < i_2$):

Նկատենք, որ $i_1 - 1 \leq \lambda_{k-2}$: Իրոք, եթե $i_1 - 1 > \lambda_{k-2}$, ապա Ա ալգորիթմը չի կարող մնացած $k-2$ քայլերի միջոցով որոշել $\alpha = (x_1, ..., x_{\lambda_k}) \in Z_{\lambda_k}$ հաջորդականության ամենամեծ անդամը, երբ այն պատկանում է $\{x_1, ..., x_{i_1 - 1}\}$ բազմությանը:

Մյուս կողմից, եթե $\alpha = (x_1, ..., x_{\lambda_k}) \in Z_{\lambda_k}$ հաջորդականության ամենամեծ անդամը պատկանում է $\{x_{i_1+1}, ..., x_{\lambda_k}\}$ բազմությանը, ապա $\lambda_k - i_1 \leq \lambda_{k-1}$, քանի որ մեկ հարցի պատասխան՝ $x_{i_2}$-ի արժեքն արդեն ունենք, և լրացուցիչ հարցերի քանակը $k-2$-ից շատ լինել չի կարող:

Գումարելով ստացված երկու անհավասարությունները, կստանանք՝

$$\lambda_k - 1 \leq \lambda_{k-1} + \lambda_{k-2}:$$

Թեորեմն ապացուցված է:

Նշանակենք $\Phi_0 = 1$, $\Phi_1 = 1$ և $k \geq 2$ համար $\Phi_k = \Phi_{k-1} + \Phi_{k-2}$: Ինչպես գիտենք այս հաջորդականությունն իրենից ներկայացնում է հանրահայտ Ֆիբոնաչիի հաջորդականությունը, որի $k$-րդ անդամը՝ $\Phi_k$-ն որոշվում է հետևյալ բանաձևով՝

$$\Phi_k = \frac{1}{\sqrt{5}}\left(\left(\frac{1}{2} + \frac{\sqrt{5}}{2}\right)^{k+1} - \left(\frac{1}{2} - \frac{\sqrt{5}}{2}\right)^{k+1}\right)$$

(բանաձևը կարելի է դուրս բերել՝ լուծելով վերոհիշյալ երկրորդ կարգի անրադարձ առնչությունը):

**Թեորեմ 2**: Ցանկացած $k$ թվի համար տեղի ունի հետևյալ անհավասարությունը.

$$\lambda_k \leq \Phi_{k+1} - 1:$$

**Ապացույց**: $k = 1, 2$ դեպքերում պնդումն ակնհայտ է: Իրոք՝

$$1=\lambda_1\le\Phi_2-1=2-1,$$
$$2=\lambda_2\le\Phi_3-1=3-1:$$

Ենթադրելով, որ պնդումը ճիշտ է $i<k$ համար, ապացուցենք այն $i=k$ դեպքում: Թեորեմներ 1,2-ից հետևում է, որ

$$\lambda_k\le\lambda_{k-1}+\lambda_{k-2}+1\le\Phi_k-1+\Phi_{k-1}-1+1=\Phi_{k+1}-1:$$

Պնդումն ապացուցված է:

**Թեորեմ 3**: Եթե հայտնի է $\alpha=(x_1,...,x_{\Phi_k-1})\in Z_{\Phi_k-1}$, $k\ge 3$ հաջորդականության $x_{\Phi_{k-1}}$ անդամի արժեքն, ապա այդ հաջորդականության ամենամեծ անդամը կարելի է գտնել $k-2$ հարցերի միջոցով:

**Ապացույց**: $k=3$ դեպքում պնդումը հետևյալն է. տրված է $x_1,x_2$ երկընթաց հաջորդականությունը, հայտնի է $x_2$-ի արժեքը և պետք է մեկ հարցի միջոցով որոշել հաջորդականության ամենամեծ անդամի արժեքը: Ակնհայտ է, որ իմանալով $x_1$-ի արժեքը, կիմանանք նաև այս հաջորդականության ամենամեծ անդամի արժեքը:

Ենթադրենք, որ պնդումը ճիշտ է $k=p$ դեպքում, և փորձենք ապացուցել այն $k=p+1$ դեպքում:

Դիցուք ունենք $x_1,...,x_{\Phi_p},...,x_{\Phi_{p+1}-1}$ երկընթաց հաջորդականության $x_{\Phi_p}$ անդամի արժեքը: Մեկ հարցի միջոցով որոշենք $x_{\Phi_{p-1}}$–ի արժեքը:

Քննարկենք հետևյալ երկու հնարավոր դեպքերը.

ա) $x_{\Phi_{p-1}}>x_{\Phi_p}$: Պարզ է, որ հաջորդականության ամենամեծ անդամը պետք է փնտրել $x_1,...,x_{\Phi_p-1}$ անդամների մեջ (ընդ որում հայտնի է $x_{\Phi_{p-1}}$–ի արժեքը): Համաձայն ինդուկցիոն ենթադրության դա կարելի է անել $p-2$ հարցերի միջոցով:

բ) $x_{\Phi_{p-1}}\le x_{\Phi_p}$: Այս դեպքում հաջորդականության ամենամեծ անդամը պետք է փնտրել $x_{\Phi_{p-1}+1},x_{\Phi_{p-1}+2},...,x_{\Phi_{p+1}-1}$ ենթահաջորդականության անդամների մեջ: Նկատենք, որ եթե դիտարկենք այս հաջորդականության շուռ տված հաջորդականությունը՝ $y_1=x_{\Phi_{p+1}-1},....,y_{t-1}=x_{\Phi_{p-1}+2},y_t=x_{\Phi_{p-1}+1}$-ն, ապա այն կլինի երկընթաց, $t=\Phi_{p+1}-\Phi_{p-1}-1=\Phi_p-1$ և հայտնի կլինի նրա $y_{\Phi_{p-1}}=x_{\Phi_p}$ անդամի արժեքը: Համաձայն ինդուկցիոն ենթադրության այս հաջորդականության ամենամեծ անդամը կարելի է գտնել $p-2$ հարցերի միջոցով:

$t(n)$-ով նշանակենք $n$ անդամ պարունակող երկընթաց հաջորդականության մեջ ամենամեծ անդամի արժեքը գտնող լավագույն ալգորիթմի բարդությունը:

**Թեորեմ 4**: Ցանկացած $n \geq 3$ համար, եթե $\Phi_k \leq n < \Phi_{k+1}$ ապա $t(n) = k$:
**Ապացույց**: Օգտվելով Թեորեմ 2-ից, կստանանք՝ $n \geq \Phi_k > \lambda_{k-1}$, հետևաբար՝ $t(n) \geq k$:

Քանի որ $n < \Phi_{k+1}$ ապա, պնդման ապացույցն ավարտելու համար բավական է ցույց տալ, որ եթե $n = \Phi_{k+1} - 1$, ապա $Z_n$ բազմության ցանկացած հաջորդականության ամենամեծ անդամի արժեքը կարելի է գտնել $k$ հարցերի միջոցով:

Դիցուք ունենք $x_1, ..., x_{\Phi_{k+1}-1}$ երկընթաց հաջորդականությունը: Առաջին երկու հարցերի միջոցով պարզենք $x_{\Phi_{k-1}}$ և $x_{\Phi_k}$ անդամների արժեքները:
Եթե $x_{\Phi_{k-1}} > x_{\Phi_k}$, ապա հաջորդականության ամենամեծ անդամը գտնվում է $x_1, ..., x_{\Phi_k - 1}$ ենթահաջորդականության անդամների մեջ: Ավելին, այս հաջորդականությունը բավարարում է Թեորեմ 3-ի պայմաններին, հետևաբար՝ $k-2$ հարցերի միջոցով կարելի է գտնել նրա ամենամեծ անդամի արժեքը:
Եթե $x_{\Phi_{k-1}} \leq x_{\Phi_k}$, ապա հաջորդականության ամենամեծ անդամը գտնվում է $x_{\Phi_{k-1}+1}, x_{\Phi_{k-1}+2}, ..., x_{\Phi_{k+1}-1}$ ենթահաջորդականության անդամների մեջ: Նկատենք, որ այս հաջորդականության շրջումից ստացված հաջորդականությունը բավարարում է Թեորեմ 3-ի պայմաններին, հետևաբար՝ $k-2$ հարցերի միջոցով կարելի է գտնել նրա ամենամեծ անդամի արժեքը:

Այսպիսով, երկու դեպքում էլ $k$ հարցերի միջոցով կարելի է գտնել հաջորդականության ամենամեծ անդամի արժեքը: Թեորեմն ապացուցված է:

**Բազմությունների հավասարության ստուգում**: Տրված են $A = \{a_1, ..., a_n\}$ և $B = \{b_1, ..., b_n\}$ բազմությունները: Թույլատրվում է ցանկացած $a_i \in A$ և $b_j \in B$ համար տալ $a_i = b_j$ և ստանալ պատասխանը: Պահանջվում է իմանալ $A = B$ թե ոչ, տալով հնարավորին չափ քիչ թվով հարցեր:

**Թեորեմ**: Բազմությունների հավասարությունը ստուգող լավագույն ալգորիթմի բարդությունը $\frac{n(n+1)}{2}$ է:

Ապացույց: Նախ նկարագրենք ալգորիթմ, որը $\frac{n(n+1)}{2}$ քայլում լուծում է խնդիրը:

Քայլ 1: Որպես $A$ բազմության հերթական տարր ընտրել $a_1$-ը:
Քայլ 2: Դիտարկել $A$ բազմության հերթական տարր և այն հերթով համեմատել $B$ բազմության տարրերի հետ: Եթե այն չկա, ապա ավարտել ալգորիթմի աշխատանքը $A \neq B$ պատասխանով, հակառակ դեպքում՝ $B$ բազմությունից հեռացնել այդ տարրը:
Քայլ 3: Եթե $A$ բազմության հերթական տարր $a_n$-է, ապա ավարտել ալգորիթմի աշխատանքը $A = B$ պատասխանով, հակառակ դեպքում որպես $A$ բազմության հերթական տարր ընտրել հաջորդը և անցնել Քայլ 2-ի կատարմանը:

Առաջարկված ալգորիթմում հարցերի առավելագույն քանակ կօգտագործվի, երբ $A$ բազմության յուրաքանչյուր տարրին հավասար տարրը $B$ բազմությունից գտնվում է վերջին հնարավոր քայլում: Հետևաբար, այդ ալգորիթմի բարդությունը կլինի

$$n + (n-1) + ... + 2 + 1 = \frac{n(n+1)}{2}:$$

Թեորեմի ապացույցն ավարտելու համար բավական է բազմությունների հավասարությունը ստուգող ցանկացած ալգորիթմում ցույց տալ ճյուղ, որի երկարությունը առնվազն $\frac{n(n+1)}{2}$ է:

Մենք կվարվենք հետևյալ կերպ. այս և հետագա որոշ խնդիրների համար կառաջարկենք որոշակի կանոններ, որոնք թույլ են տալիս խնդիրը լուծող ցանկացած ալգորիթմին համապատասխան ծառից ընտրել ճյուղ: Նշենք, որ *այս կանոնների ընտրությունը կախված կլինի խնդիրներից, և ոչ թե խնդիրը լուծող ալգորիթմներից*: Ընդունված է ալգորիթմի համապատասխան ծառում ճյուղն ընտրելու այս կանոններին անվանել գուշակ:

Դիտարկենք $n \times n$ կարգի աղյուսակ, որի տողերին համապատասխանում են $a_1,...,a_n$ տարրերը, իսկ սյուներին՝ $b_1,...,b_n$: Աղյուսակի $i$–րդ տողի և $j$-րդ սյան համապատասխան վանդակում կգրենք $a_i = b_j$ հարցին գուշակի թելադրած պատասխանը:

Աղյուսակի $n$ վանդակներ կանվանենք անկախ, եթե նրանք գտնվում են տարբեր տողեր, տարբեր սյուներում:

Աղյուսակի վանդակը կհամարենք թույլատրելի, եթե այդ վանդակում “ոչ” չի գրված: Ալգորիթմի աշխատանքի սկզբում աղյուսակի վանդակները դատարկ են, և հետևաբար՝ կլինեն թույլատրելի:

**Գուշակի (ճյուղն ընտրող կանոնների) նկարագիրը**: Հերթական $a_i = b_j$ հարցին պատասխանել “ոչ” և աղյուսակի համապատասխան վանդակում գրել “ոչ”, եթե դրանից հետո աղյուսակում կարելի է նշել $n$ անկախ թույլատրելի վանդակներ, հակառակ դեպքում՝ $a_i = b_j$ հարցին պատասխանել “այո” և աղյուսակի համապատասխան վանդակում գրել “այո”:

Նկատենք, որ այս ձևով սահմանված գուշակը (ճյուղն ընտրող կանոնների բազմությունը) $A$ և $B$ բազմությունների հավասարությունը ստուգող ցանկացած ալգորիթմում միարժեքորեն ընտրում է որևէ ճյուղ: Ավելին, այդ ճյուղին համապատսխանում է “այո” (այսինքն՝ $A = B$) պատասխանը: Իրոք, քանի որ ընտրվածը ճյուղ է, ապա նրան համապատասխանում է մեկ պատասխան: Մյուս կողմից, ըստ գուշակի սահմանման այն միշտ թողնում է $n$ անկախ, թույլատրելի վանդակների հնարավորությունը, իսկ այդպիսի վանդակներին համապատասխանում է “այո” պատասխան, հետևաբար՝ ճյուղին համապատասխանում է “այո” պատասխան:

Դիտարկենք գուշակի կողմից լրացված աղյուսակը մեր խնդիրը լուծող որևէ ալգորիթմի աշխատանքի ավարտի պահին: Աղյուսակի $n$ անկախ վանդակներում գրված է “այո” (գուշակի վարվելակերպից հետևում է, որ “այո” պարունակող վանդակներն անկախ են, և քանի որ ալգորիթմը ավարտվել է իր աշխատանքը $A = B$ պատասխանով, ապա յուրաքանչյուր տողում պետք է լինի “այո” պարունակող վանդակ), որոշ վանդակներում՝ “ոչ”, ընդ որում թույլատրելի վանդակներից այլ եղանակով հնարավոր չէ ընտրել $n$ անկախ վանդակ:

**Լեմ**: Եթե գուշակի կողմից լրացված աղյուսակի ցանկացած տող և ցանկացած սյուն պարունակում է առնվազն երկու թույլատրելի վանդակ, ապա $n$ անկախ, թույլատրելի վանդակների ընտրության հնարավորությունը միարժեք չէ:

**Ապացույց**: Նախ նկատենք, որ գուշակի կողմից լրացված աղյուսակը միշտ պարունակում է $n$ անկախ, թույլատրելի վանդակ: Դիտարկենք $n$ անկախ, թույլատրելի վանդակների մի որևէ ընտրություն: Դիցուք առաջին տողից ընտրված է $(1, j_1)$ վանդակը: Դիտարկենք $j_1$-րդ սյունը: Այն պարունակում է մեկ այլ, դիցուք $(i_1, j_1)$ թույլատրելի վանդակը: Դիտարկենք $i_1$-րդ տողը: Այն պետք է պարունակի ընտրված անկախ վանդակներից մեկ, դիցուք՝ $(i_1, j_2)$ վանդակը: Շարունակելով այս դատողությունները, մենք կստանանք տողերի և սյուների հետևյալ հաջորդականությունը՝ $1, j_1, i_1, j_2, i_2, j_3, i_3, \ldots$, ընդ որում $(1, j_1), (i_1, j_2), (i_2, j_3), \ldots$ ընտրված անկախ վանդակներից են, իսկ $(i_1, j_1), (i_2, j_2), (i_3, j_3), \ldots$ թույլատրելի վանդակներ են: Պարզ է, որ նշված հաջորդականությունը չի կարող շարունակվել, հետևաբար՝ գոյություն կունենա տող կամ սյուն, որը կկրկնվի: Ենթադրենք առաջին կրկնվողը $i_k$ –րդ տողն է.

$$i_k, j_{k+1}, i_{k+1}, j_{k+2}, \ldots, i_{k+l}, j_{k+l}, i_k :$$

Հաջորդականության կառուցման եղանակից հետևում է, որ $(i_k, j_{k+1}), (i_{k+1}, j_{k+2}), \ldots, (i_{k+l}, j_{k+l})$ վանդակները պատկանում են ընտրված թույլատրելի անկախ վանդակների բազմությանը: Փոխարինենք այս վանդակները $(j_{k+1}, i_{k+1}), (j_{k+2}, i_{k+2}), \ldots, (j_{k+l}, i_k)$ վանդակներով (որոնք գտնվում են նույն տողերում և նույն սյուներում, ինչ նշվածները): Նկատենք, որ կստացվի թույլատրելի վանդակների մեկ այլ, նոր ընտրություն: Լեմն ապացուցված է:

Վերադառնանք թեորեմի ապացույցին: Լեմից հետևում է, որ ալգորիթմի աշխատանքի ավարտի պահին գուշակի կողմից լրացված աղյուսակում գոյություն ունի տող կամ սյուն, որի $n-1$ վանդակներում գրված է “ոչ”, իսկ մի վանդակում՝ “այո”: Ջնջենք “այո” վանդակով անցնող տողը և սյունը: Նկատենք, որ ջնջված տողը և սյունը պարունակում են առնվազն $n$ հարցի պատասխան: Համաձայն լեմի, ստացված աղյուսակում կրկին գոյություն ունի տող կամ սյուն, որի $n-2$ վանդակներում գրված է “ոչ”, իսկ մի վանդակում՝ “այո”: Ջնջենք “այո” վանդակով անցնող տողը և սյունը: Նկատենք, որ ջնջված տողը և սյունը պարունակում են առնվազն $n-1$ հարցի պատասխան: Շարունակելով պրոցեսը, կհանգենք $1\times1$ աղյուսակի, որի միակ վանդակում գրված է “այո”: Այսպիսով, գուշակի կողմից լրացված աղյուսակը պարունակում է առնվազն

$$\geq n + (n-1) + \ldots + 2 + 1 = \frac{n(n+1)}{2}$$

հարցի պատասխան, և հետևաբար՝ գուշակի կողմից ընտրված ճյուղի երկարությունը առնվազն $\frac{n(n+1)}{2}$ է: Թեորեմն ապացուցված է:

# Գրականություն

Գտնված սխալների, առաջարկությունների, ինչպես նաև դասախոսություն-ներն e-mail-ով ստանալու համար կարող էք դիմել vahanmkrtchyan2002@yahoo.com հասցեով:

Կոմբինատորային ալգորիթմներ և ալգորիթմների վերլուծություն
Վահան Վ. Մկրտչյան

Դասախոսություն 5: Մրցաշարային խնդիրներ. Մրցաշարի հաղթողի, հաղթողի և պարտվողի որոշումը:

**Խնդիրների մեկնաբանումը**: $\vec{G}=(V,E)$ օրգրաֆը կանվանենք մրցաշար, եթե նրա միմյանցից տարբեր երկու՝ $u$ և $v$ գագաթների համար տեղ ունի հետևյալ պնդումը.

$$(u,v)\in E \Leftrightarrow (v,u)\notin E:$$

Այս անվանումը կապված է սպորտային մրցումների արդյունքներն օրգրաֆների միջոցով ներկայացնելու հետ: Դիցուք ունենք մրցաշար, որին մասնակցում են $1,\ldots,n$ թիմերը: Սահմանենք $\vec{G}=(V,E)$ օրգրաֆը հետևյալ կերպ.

$V=\{1,\ldots,n\}$, և $(i,j)\in E$ եթե $i$ –րդ թիմը հաղթել է $j$ -րդ թիմին:

Նկատենք, որ այս ձևով սահմանված օրգրաֆը բավարարում է մրցաշար լինելու՝ վերը նշված պայմանին:

$\vec{G}=(V,E)$ մրցաշարը կանվանենք *տրանզիտիվ* մրցաշար, եթե նրա ցանկացած $u$, $v$ և $w$ գագաթների համար տեղ ունի հետևյալ պնդումը (նկար 1)

$$(u,v)\in E,(v,w)\in E \Rightarrow (u,w)\in E:$$

Մենք կդիտարկենք միայն տրանզիտիվ մրցաշարեր, այնպես որ ամեն անգամ տրանզիտիվ բառը չենք նշի: Նշենք նաև, որ այս խնդիրներն ունեն իրար համարժեք այլ մեկնաբանություններ.

ա) Դիտարկենք մրցաշար, որին մասնակցում են $a_1,\ldots,a_n$ խաղացողները: Ենթադրենք, որ նրանք տարբեր ուժի են, բայց մենք տեղեկություն չունենք նրանց ուժի մասին: Պահանջվում է հնարավորինս քիչ թվով խաղեր կազմակերպելով որոշել նրանցից ուժեղին և թույլին: Մենք կենթադրենք, որ ուժեղը միշտ հաղթում է թույլին:

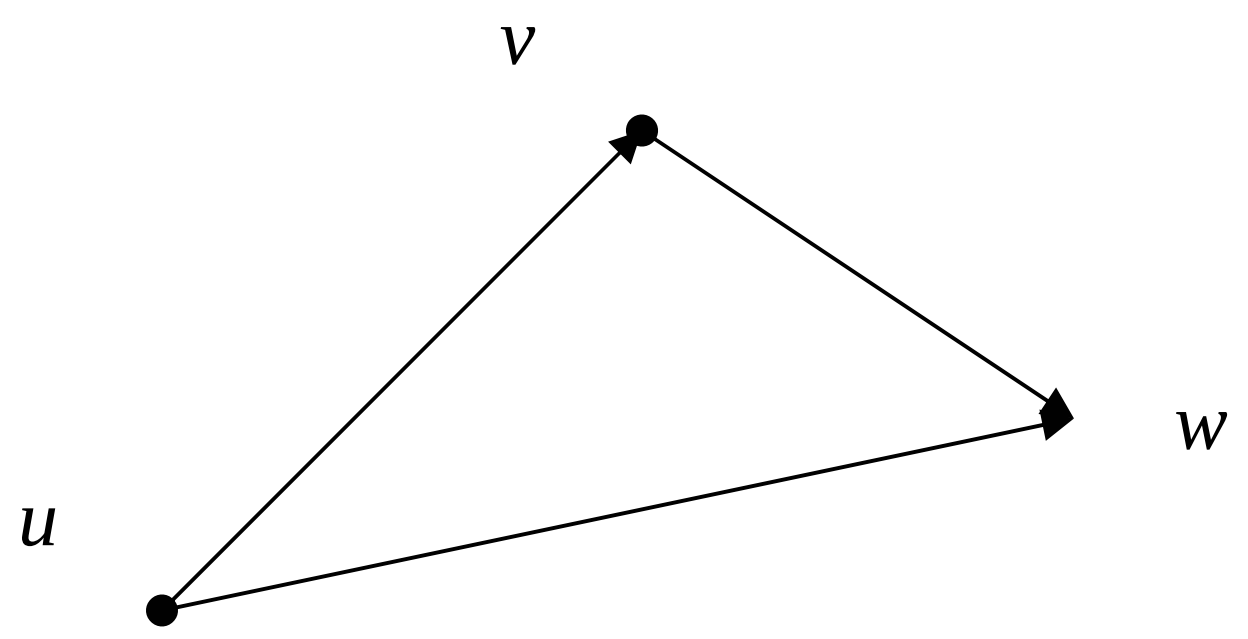

նկար 1

բ) դիտարկենք միանման $a_1,\ldots,a_n$ կշռաքարերը, որոնց քաշերը միմյանցից տարբեր են: Ունենք նժարավոր կշեռք, որի միջոցով կարող ենք համեմատել մեր կողմից ընտրված երկու կշռաքարերի քաշերը: Պահանջվում է հնարավորինս քիչ թվով կշռումների միջոցով գտնել ամենածանր և ամենաթեթև կշռաքարերը:

Դժվար չէ տեսնել, որ իրականում այս երկու մեկնաբանությունները նույն մրցաշարային խնդիրն են:

**Մրցաշարի հաղթողի որոշում**: Դիտարկենք մրցաշար, որին մասնակցում են $a_1,\ldots,a_n$ խաղացողները: Խնդիրը կայանում է մրցաշարի հաղթողին որոշելու մեջ կազմակերպելով նվազագույն թվով խաղեր:
Դիտարկենք խնդիրը լուծող հետևյալ երկու ալգորիթմները.

Ալգորիթմ 1 (սովորական մրցակարգ)
Քայլ 1: Որպես հավակնորդ ընտրենք $a_1$-ը, իսկ որպես հերթական մասնակից՝ $a_2$-ը:
Քայլ 2: Կազմակերպենք խաղ հավակնորդի և հերթական մասնակցի միջև, որի հաղթողին համարենք հավակնորդ:
Քայլ 3: Եթե հերթական մասնակիցը $a_n$-ն է, ապա ավարտենք ալգորիթմի աշխատանքը՝ "հավակնորդը հաղթողն է պատասխանով", հակառակ դեպքում՝ որպես հերթական մասնակից վերցնենք հաջորդ մասնակցին և անցնենք Քայլ 2-ին:

Պարզ է, որ այս ալգորիթմը $n-1$ խաղերից հետո կորոշի հաղթողին:

Ալգորիթմ 2 (զավաթային մրցակարգ)

Քայլ 1: Բոլոր մասնակիցներին համարենք հավակնորդներ:
Քայլ 2: Հավակնորդներին բաժանենք զույգերի (եթե նրանց թիվը կենտ է, ապա մեկին առանձնացնել) և որոշենք յուրաքանչյուր զույգի հաղթողին:
Զույգերում հաղթողներին համարենք հավակնորդներ:
Քայլ 3: Եթե հավակնորդների թիվը մեկ է, ապա "միակ հավակնորդին համարել հաղթող" և ավարտել ալգորիթմի աշխատանքը, հակառակ դեպքում՝ անցնել Քայլ 2-ին:

Նախ նկատենք, որ զավաթային մրցակարգում յուրաքանչյուր թիմ մասնակցում է ամենաշատը $\lceil \log_2 n \rceil$ խաղի: Իրոք, քանի որ $n \le 2^{\lceil \log_2 n \rceil}$ ապա զավաթային մրցակարգում փուլերի քանակը չի գերազանցում $\lceil \log_2 n \rceil$–ը, իսկ ամեն մի փուլում յուրաքանչյուր թիմ խաղում է ամենաշատը մեկ անգամ:
Ինդուկցիայով ապացուցենք, որ զավաթային մրցակարգում խաղերի քանակը կրկին $n-1$–է:
$n=1$ դեպքում պնդումն ակնհայտ է: Ենթադրենք, որ այն ճիշտ է $<n$ դեպքում, և փորձենք ապացուցել $n$–ի համար: Առաջին փուլում կանցկացվի $\left\lfloor \frac{n}{2} \right\rfloor$ խաղ, իսկ երկրորդ փուլում կլինեն $\left\lceil \frac{n}{2} \right\rceil$ հավակնորդներ, հետևաբար, ըստ ինդուկցիոն ենթադրության, նրանցից հաղթողին զավաթային մրցակարգը կորոշի $\left\lceil \frac{n}{2} \right\rceil - 1$ խաղերի արդյունքում: Ընդհանուր խաղերի քանակը կլինի՝

$$\left\lfloor \frac{n}{2} \right\rfloor + \left\lceil \frac{n}{2} \right\rceil - 1 = n-1:$$

$W_1(n)$-ով նշանակենք $n$ մասնակիցներից կազմված մրցաշարում հաղթողին որոշող լավագույն ալգորիթմում կազմակերպած խաղերի քանակը:

Թեորեմ 1: $W_1(n) = n-1$:
Ապացույց: Վերը նշված երկու ալգորիթմներից հետևում է, որ $W_1(n) \le n-1$: Մյուս կողմից, եթե մրցաշարի ավարտին հաղթողը որոշված է, ապա մնացած $n-1$ թիմերից յուրաքանչյուրը պետք է ունենա գոնե մեկ պարտություն, հետևաբար ցանկացած ալգորիթմ պետք է կազմակերպի առնվազն $n-1$ խաղ, այսինքն՝ $W_1(n) \ge n-1$: Թեորեմն ապացուցված է:

**Մրցաշարի հաղթողի և պարտվողի որոշումը**: Դիտարկենք $n$ թիմերից բաղկացած մրցաշար, և փորձենք նվազագույն թվով խաղեր կազմակերպելով որոշել հաղթողին և պարտվողին: $U(n)$-ով նշանակենք նվազագույն խաղերի քանակը, որն անհրաժեշտ է մրցաշարի հաղթողին և պարտվողին գտնելու համար:

**Թեորեմ 2**: $U(n)=\left\lceil \frac{3n}{2} \right\rceil - 2$:

**Ապացույց**: Նախ քննարկենք երկու դեպք.

$n=2k$ Այս դեպքում մրցաշարի հաղթողին և պարտվողին որոշենք հետևյալ կերպ.

1. մրցաշարի թիմերին տրոհենք $k$ զույգերի և նրանցից յուրաքանչյուրում անցկացնենք խաղ:
2. զույգերի $k$ հաղթողներից որոշենք հաղթողին ($k-1$ խաղերի միջոցով)
3. զույգերի $k$ պարտվողներից որոշենք պարտվողին ($k-1$ խաղերի միջոցով):

Նկատենք, որ այս դեպքում խաղերի քանակը կլինի $3k-2$, հետևաբար՝

$$U(n)=U(2k)\leq 3k-2=\left\lceil \frac{3n}{2} \right\rceil - 2:$$

$n=2k+1$ Այս դեպքում մրցաշարի հաղթողին և պարտվողին որոշենք հետևյալ կերպ.

1. առանձնացնենք մեկ մասնակցին, մրցաշարի մնացած թիմերին տրոհենք $k$ զույգերի և նրանցից յուրաքանչյուրում անցկացնենք խաղ:
2. զույգերի $k$ հաղթողների և առանձնացված մեկ մասնակցի միջից որոշենք հաղթողին ($k$ խաղերի միջոցով)
3. զույգերի $k$ պարտվողներից և առանձնացված մեկ մասնակցի միջից որոշենք պարտվողին ($k$ խաղերի միջոցով)

Նկատենք, որ այս դեպքում խաղերի քանակը կլինի $3k$, հետևաբար՝

$$U(n)=U(2k+1)\leq 3k=\left\lceil \frac{3n}{2} \right\rceil - 2:$$

Երկու դեպքերի քննարկման արդյունքում ունենք՝

$$U(n)\leq\left\lceil \frac{3n}{2} \right\rceil - 2:$$

$U(n)\geq\left\lceil \frac{3n}{2} \right\rceil - 2$ ստորին գնահատականի ապացույցի համար օգտվենք արդեն նշված զուշակի գաղափարից (կանոնների բազմություն, որոնք թույլ էին տալիս խնդիրը լուծող ցանկացած ալգորիթմից ընտրել ճյուղ): Խնդիրը լուծող ալգորիթմից ճյուղ ընտրելու ժամանակ, մենք թիմերին կվերագրենք հատկություններ՝

$A$-այն խաղացողներին, որոնք ոչ մի խաղի չեն մասնակցել;

$B$-այն խաղացողներին, որոնք ունեն միայն հաղթանակներ;
$C$-այն խաղացողներին, որոնք ունեն միայն պարտություններ;
$D$-այն խաղացողներին, որոնք ունեն ինչպես հաղթանակներ, այնպես էլ պարտություններ:

Գուշակը սահմանենք ստորև բերված աղյուսակի առաջին և երկրորդ սյան համապատասխան.

| Եթե հանդիպում են այս հատկությամբ օժտված թիմեր | Գուշակի ընտրած արդյունքը | Գուշակի ընտրության արդյունքում $A$, $B$, $C$, $D$ հատկություններով օժտված մասնակիցների քանակների փոփոխությունը | | | |
|---|---|---|---|---|---|
| | | $A$ | $B$ | $C$ | $D$ |
| $A, A$ | կամայական | -2 | +1 | +1 | 0 |
| $A, B$ | $A$ պարտվում է $B$ | -1 | 0 | +1 | 0 |
| $A, C$ | $A$ հաղթում է $C$ | -1 | +1 | 0 | 0 |
| $A, D$ | $A$ հաղթում է $D$ | -1 | +1 | 0 | 0 |
| $B, B$ | կամայական | 0 | -1 | 0 | 1 |
| $B, C$ | $B$ հաղթում է $C$ | 0 | 0 | 0 | 0 |
| $B, D$ | $B$ հաղթում է $D$ | 0 | 0 | 0 | 0 |
| $C, C$ | կամայական | 0 | 0 | -1 | 1 |
| $C, D$ | $C$ պարտվում է $D$ | 0 | 0 | 0 | 0 |
| $D, D$ | կամայական | 0 | 0 | 0 | 0 |

Ենթադրենք, որ մրցաշարի հաղթողին և պարտվողին որոշող որևէ Ա ալգորիթմում, խաղերի արդյունքները գուշակի վարվելակերպին համապատասխան ընտրելուց հետո կատարվել է $x_1$ - $A,A$, $x_2$ - $A,B$, $x_3$ - $A,C$, $x_4$ - $B,B$, $x_5$ - $C,C$, $x_6$ - $A,D$, և $y$ մնացած տիպի խաղ: Նկատենք, որ ալգորիթմով պահանջվող քայլերի քանակը կլինի $x_1+x_2+x_3+x_4+x_5+x_6+y$:

Ալգորիթմի աշխատանքի սկզբում

1. $A$ հատկությամբ օժտված մասնակիցների քանակը $n$-էր, իսկ վերջում՝ $0$, հետևաբար՝ $2x_1+x_2+x_3+x_6=n$;
2. $D$ հատկությամբ օժտված մասնակիցների քանակը $0$-էր, իսկ վերջում՝ $n-2$, հետևաբար՝ $x_4+x_5=n-2$;

Արդյունքում՝

$$2(x_1+x_2+x_3+x_4+x_5+x_6+y)\geq 2x_1+x_2+x_3+x_6+2(x_4+x_5)=$$
$$=n+2(n-2)=3n-4$$

հետևաբար՝

$$U(n) = x_1 + x_2 + x_3 + x_4 + x_5 + x_6 + y \geq \left\lceil \frac{3n}{2} \right\rceil - 2:$$

Թեորեմն ապացուցված է:

# Գրականություն

Գտնված սխալների, առաջարկությունների, ինչպես նաև դասախոսություն-ներն e-mail-ով ստանալու համար կարող էք դիմել vahanmkrtchyan2002@yahoo.com հասցեով:

Կոմբինատորային ալգորիթմներ և ալգորիթմների վերլուծություն
Վահան Վ. Մկրտչյան

Դասախոսություն 6: Մրցաշարային խնդիրներ. Առաջին և երկրորդ, առաջին, երկրորդ և երրորդ տեղերի որոշումը:

**Մրցաշարի 1-ին և 2-րդ տեղերը գրավողների որոշումը**: Դիտարկենք մրցաշար, որին մասնակցում են $n \geq 2$ թիմեր և քննարկենք նվազագույն թվով խաղերի միջոցով մրցաշարի առաջին և երկրորդ տեղերի գրավողների որոշման խնդիրը: $W_2(n)$-ով նշանակենք նվազագույն խաղերի քանակը, որն անհրաժեշտ է նշված խնդրի լուծման համար:

**Թեորեմ 1**: $W_2(n) = n - 2 + \lceil \log_2 n \rceil$:

**Ապացույց**: Մրցաշարում երկրորդ տեղը գրավողին պետք է փնտրել միայն առաջին տեղը գրավողին պարտվածների մեջ: Դիտարկենք հետևյալ ալգորիթմը.
ա) որոշել մրցաշարի հաղթողին՝ գավաթային մրցակարգով ($n-1$ խաղ)
բ) մրցաշարի հաղթողին պարտվողների միջից որոշենք ամենաուժեղին ($\lceil \log_2 n \rceil - 1$ խաղ): Պարզ է, որ այն կլինի երկրորդ տեղը գրավողը:
Այս ալգորիթմից հետևում է, որ

$$W_2(n) \leq n - 1 + \lceil \log_2 n \rceil - 1 = n - 2 + \lceil \log_2 n \rceil:$$

Ցույց տանք, որ

$$W_2(n) \geq n - 2 + \lceil \log_2 n \rceil:$$

Դիցուք Ա-ն մրցաշարի առաջին և երկրորդ տեղերը գրավողներին որոշող ալգորիթմ է: $a_i$-ով նշանակենք ալգորիթմի աշխատանքի ավարտի պահին առնվազն $i$ պարտություն ունեցող խաղացողների քանակը: Նկատենք, որ՝ անցկացված խաղերի քանակը=ընդհանուր պարտությունների քանակին=մեկ պարտություն ունեցող մասնակիցների քանակին+երկու պարտություն ունեցող մասնակիցների քանակին+ ... $= a_1 + a_2 + a_3 + ...$

Պարզ է նաև, որ մրցաշարի հաղթողից բացի մնացած մասնակիցները պետք է ունենան գոնե մեկ պարտություն, հետևաբար՝ $a_1 = n - 1$: $q$-ով նշանակենք մրցաշարի հաղթողի անցկացրած խաղերի քանակը: Նկատենք, որ նրան պարտվողներից մեկը պետք է զբաղեցնի երկրորդ տեղը, հետևաբար՝ $a_2 \geq q - 1$: Թեորեմն ապացուցելու համար բավական է կառուցել գուշակ, որը մրցաշարի 1-ին և 2-րդ տեղերը զրավողներին որոշող ցանկացած ալգորիթմում խաղերի արդյունքն ընտրում է այնպես, որ մրցաշարի հաղթողն անցկացնի առնվազն $\lceil \log_2 n \rceil$ խաղ:

Գուշակի վարվելակերպը սահմանենք հետևյալ կերպ՝
ա) պարտություն չունեցողը հաղթում է պարտություն ունեցողին
բ) պարտություն չունեցողներից հաղթում է նա, ով ավելի շատ հաղթանակներ ունի
գ) մնացած դեպքերում խաղերի արդյունքն ընտրել կամայապես:

Մրցաշարի հանդիպումների քանակը գնահատելու համար, սահմանենք $a \prec b$ ($b$-ն գերազանցում է $a$-ն) հարաբերությունը:

1. $a \prec a$
2. Եթե $a \succ b$ և $c$-ն առաջին անգամ պարտվել է $b$-ին, ապա $a \succ c$:

Լեմ: Եթե $a$-ն ունի միայն $p$ հաղթանակ, ապա այն գերազանցում է ոչ շատ, քան $2^p$ մասնակիցների, այսինքն՝

$$\left|\{x / a \succ x\}\right| \leq 2^p:$$

Ապացույց: Ապացույցը կատարենք ինդուկցիայով:
$p = 0$ դեպքում պնդումն ակնհայտ է: Ենթադրենք, որ այն ճիշտ է $p-1$ հաղթանակ ունեցող մասնակիցների համար, և դիցուք $a$ մաասնակիցն ունի $p$ հաղթանակ: Դիցուք նրա վերջին խաղը, ուր $a$–ն հաղթել է, կայացել է $b$ մասնակցի հետ (նկատենք, որ պարտությունները չեն ավելացնում նշված բազմության տարրերի քանակը): Եթե մինչ այդ խաղը $b$-ն արդեն ուներ պարտություն, ապա խաղից հետո այն մասնակիցների քանակը, որոնց $a$-ն գերազանցում էր, չի փոխվի: Իսկ եթե $b$-ն պարտություն չուներ, ապա նրա հաղթանակների քանակը$\leq p-1$, և հետևաբար, ըստ ինդուկցիոն ենթադրության, $b$-ն գերազանցում է$\leq 2^{p-1}$ մասնակիցների: Այդ դեպքում $a$-ն կգերազանցի $\leq 2^{p-1}+2^{p-1}=2^{p}$ մասնակիցների:

Եթե լեմում որպես $a$ մասնակից վերցնենք մրցաշարի հաղթողին, ապա այն պետք է գերազանցի բոլոր $n$ մասնակիցներին, հետևաբար, $q$-ն՝ մրցաշարի հաղթողի անցկացրած խաղերի քանակը, պետք է բավարարի հետևյալ անհավասարմանը

$$n=\left|\{x/a\succ x\}\right|\leq 2^{q} \text{ կամ } \lceil\log_{2}n\rceil\leq q:$$

**Մրցաշարի 1-ին, 2-րդ և 3-րդ տեղերի գրավողների որոշումը**: Դիտարկենք մրցաշար, որին մասնակցում են $n\geq 3$ թիմեր և քննարկենք նվազագույն թվով խաղերի միջոցով մրցաշարի առաջին, երկրորդ և երրորդ տեղերի գրավողների որոշման խնդիրը: $W_{3}(n)$-ով նշանակենք նվազագույն խաղերի քանակը, որն անհրաժեշտ է նշված խնդրի լուծման համար:

**Թեորեմ 2**: $W_{3}(n)\leq n+2\lceil\log_{2}n\rceil-3$:

**Ապացույց**: Առաջարկենք եղանակ մրցաշարի 1-ին, 2-րդ և 3-րդ տեղերի գրավողների որոշման համար: Մրցաշարի հաղթողին որոշենք զավաթային մրցակարգով ($n-1$ խաղ): Երկրորդ տեղի հավակնորդներն ամենաշատը $\lceil\log_{2}n\rceil$ են: Նրանց համարակալ ենք $1,...,\lceil\log_{2}n\rceil$ թվերով. հավակնորդին վերագրենք $k$ համարը, եթե նա հաղթողին պարտվել է $k$-րդ փուլում (նկատենք, որ եթե հաղթողը մրցաշարին սկսել է մասնակցել 2-րդ փուլից, ապա 1 համարն ունեցող հավակնորդ չի լինի):

Հավակնորդներից հաղթողին գտնենք հետևյալ փոքր մրցաշարի միջոցով. սկզբից մրցում են 1-ին և 2-րդ հավակնորդները, նրանց հաղթողի հետ մրցում է հերթական 3-րդ հավակնորդը և այսպես շարունակ:

Փոքր մրցաշարի խաղերի քանակը չի գերազանցում $\lceil\log_{2}n\rceil-1$-ը և այն որոշում է 2-րդ տեղը գրավողին: Դժվար չէ տեսնել, որ երկրորդ տեղը գրավողը հաղթել է ամենաշատը $\lceil\log_{2}n\rceil$ խաղում: Իրոք, դիցուք երկրորդ տեղը գրավողը եղել է $k$-րդ մասնակիցը: Դա նշանակում է, որ սկզբնական զավաթային փուլում նա հաղթել է $k-1$ խաղում, իսկ փոքր մրցաշարում՝ $\lceil\log_{2}n\rceil-k+1$, հետևաբար՝ երկրորդ տեղը գրավողի հաղթանակների քանակը կլինի՝ $\lceil\log_{2}n\rceil$: Հետևաբար՝ ոչ ավել քան $\lceil\log_{2}n\rceil-1$ խաղերի միջոցով՝ երկրորդ տեղը գրավողից պարտվողներից կորոշենք երրորդ տեղը գրավողին: Այսպիսով, առաջարկած եղանակով մրցաշարի 1-ին, 2-րդ, 3-րդ տեղերը զբաղեցնողներին կգտնենք $\leq n+2\lceil\log_{2}n\rceil-3$ խաղերի միջոցով: Թեորեմն ապացուցված է:

Թեորեմ 3: $W_3(n) \geq n + \lceil \log_2 n(n-1) \rceil - 3$:

Ապացույց: Սահմանենք ապացույցի համար անհրաժեշտ որոշ գաղափարներ: Թիմ անվանենք մասնակիցների կարգավոր զույգը՝ $(a,b)$: Մրցաշարի 1-ին, 2-րդ և 3-րդ տեղերի գրավողներին որոշելիս, որոշվում է նաև հաղթող թիմը՝ առաջին և երկրորդ տեղերը գրավողների զույգը:

$(a,b)$-ն հաղթող թիմի հավակնորդ է, եթե $a$-ն չունի պարտություն և $b$-ն չունի պարտություն, կամ եթե $a$-ն չունի պարտություն և $b$-ն ունի միակ պարտություն՝ $a$-ից: $a$ խաղացողի հաղթանակների քանակը նշանակենք $h(a)$ (այն մրցաշարի ընթացքում փոփոխվում է): $(a,b)$ թիմի բնութագրիչ կանվանենք $2^{h(a)+h(b)}$ թիվը:

Սահմանենք գուշակ, որը մրցաշարի 1-ին, 2-րդ և 3-րդ տեղերի գրավողներին որոշելիս ընտրում է առնվազն $n + \lceil \log_2 n(n-1) \rceil - 3$ երկարությամբ ճյուղ:

Գուշակի վարվելակերպը. երկու թիմերի խաղի արդյունքը որոշելիս ընտրվում է այն ճյուղը, որի համար հաղթողին հավակնորդ թիմերի բնութագրիչների գումարն ամենամեծն է:

Ցույց տանք, որ գուշակի ընտրած ճյուղի որևէ խաղից հետո հաղթողին հավակնորդ թիմերի բնութագրիչների գումարը չի նվազում:

Դիցուք կատարվում է $a:b$ խաղը և, մինչ այդ խաղը, հաղթողին հավակնորդ թիմերի բնութագրիչների գումարը $M$ է: Պարզ է, որ $M = M_1 + M_2 + M_3$, որտեղ $M_1$-ը հաղթողին հավակնորդ այն թիմերի բնութագրիչների գումարն է, որոնց մասնակից է $a$-ն, $M_2$-ը՝ այն թիմերինը, որոնց մասնակից է $b$-ն, իսկ $M_3$-ը՝ մնացած թիմերինը: Պայմանավորվենք $(a,b)$ թիմի բնութագրիչը հաշվել $M_1$-ում, $(b,a)$ թիմինը՝ $M_2$-ում (եթե նրանք հավակնորդ են):

Եթե $a$-ն հաղթում է $b$-ին, ապա հաղթողին հավակնորդ թիմերի բնութագրիչների գումարը կլինի $2M_1 + M_3$, իսկ $b$-ն՝ $a$-ին հաղթելու դեպքում՝ $2M_2 + M_3$: Այս երկու թվերի գումարը $2M$-է, հետևաբար նրանցից մեծագույնը կլինի առնվազն $M$ :

Ալգորիթմի սկզբում հաղթողին հավակնորդ թիմերի քանակը $n(n-1)$ է, իսկ նրանց բնութագրիչների գումարը՝ $n(n-1)$: Հետևաբար, ալգորիթմի ավարտի պահին հաղթողին հավակնորդ թիմերի բնութագրիչների գումարը կլինի առնվազն՝ $n(n-1)$: Բայց ալգորիթմի ավարտի պահին կա հաղթողին հավակնորդ մեկ թիմ՝ $(a,b)$-ն, որի բնութագրիչը՝ $2^{h(a)+h(b)}$-է: Հետևաբար,

$$h(a)+h(b)\geq\lceil\log_2 n(n-1)\rceil:$$

Հաղթող թիմի խաղերից զատ պետք է կազմակերպել առնվազն $n-3$ խաղ մնացած խաղացողներից 3-րդ տեղը գրավողին որոշելու համար: Հետևաբար,

$$W_3(n)\geq n+\lceil\log_2 n(n-1)\rceil-3:$$

Թեորեմն ապացուցված է:

Ամփոփելով երկու թեորեմները, կունենանք՝

$$n+\lceil\log_2 n(n-1)\rceil-3\leq W_3(n)\leq n+2\lceil\log_2 n\rceil-3:$$

Նկատենք, որ այս խնդրի համար լավագույն ալգորիթմ չնշեցինք: Այն հայտնի չէ, սակայն գնահատեցինք նրա բարդությունը, ընդ որում ՝ վերևի և ներքևի գնահատականները կարող են տարբերվել ամենաշատը 1-ով:

# Գրականություն

Գտնված սխալների, առաջարկությունների, ինչպես նաև դասախոսություն-ներն e-mail-ով ստանալու համար կարող էք դիմել vahanmkrtchyan2002@yahoo.com հասցեով:

Կոմբինատորային ալգորիթմներ և
ալգորիթմների վերլուծություն
Վահան Վ. Մկրտչյան

Դասախոսություն 7: Ով ով է:

**Ով ով է**: Դիտարկենք մարդկանց խմբակցություն, որի անդամները $1, 2, ..., n$, $n \geq 3$, անհատներն են: Ենթադրենք, որ խմբակցության անդամների կեսից ավելին օրինավոր մարդիկ են, իսկ մնացածները՝ անօրեններ են: Ենթադրենք նաև, որ խմբակցության անդամներից յուրաքանչյուրը գիտի, թե ով ով է, այսինքն՝ ովքեր են օրինավոր, իսկ ովքեր՝ անօրեն:

Մենք ցանկանում ենք իմանալ խմբակցության կազմը տալով հետևյալ տիպի հարցեր. $i$-րդ անդամին հարցնում ենք "Օրինավոր է $j$-րդ անդամը" $i, j = 1, 2, ..., n, i \neq j$: Կենթադրենք, որ եթե $i$-րդ անդամը օրինավոր է, ապա այն հարցերին տալիս է ճիշտ պատասխան, իսկ երբ $i$-րդ անդամը օրինավոր չէ, ապա այն հարցերին կարող է տալ ինչպես ճիշտ, այնպես էլ սխալ պատասխան: Մեր խնդիրն է որոշել խմբակցության կազմը տալով հնարավորին չափ քիչ հարցեր: Պայմանավորվենք $(i, j)$ հարց անվանել խմբակցության $i$-րդ անդամին տրված "Օրինավոր է $j$-րդ անդամը" հարցը:

$\alpha(n)$-ով նշանակենք խմբակցության կազմը որոշող լավագույն ալգորիթմի բարդությունը:

**Թեորեմ**: $\alpha(n) = \left\lceil \frac{3(n-1)}{2} \right\rceil$:

**Ապացույց**: Նախ ցույց տանք, որ $\alpha(n) \leq \left\lceil \frac{3(n-1)}{2} \right\rceil$: Ապացույցը կատարենք ինդուկցիայով:

$n = 3$: Որպես առաջին հարց վերցնենք $(1, 2)$–ը: Եթե այս հարցի պատասխանը եղել է "այո", ապա սա նշանակում է, որ 2-րդ անդամը օրինավոր է: Իրոք, եթե այն օրինավոր չէ, ապա օրինավոր չէ նաև 1-ը, ինչը հակասում է այն պայմանին, որ խմբակցության անդամների կեսից ավելին օրինավոր մարդիկ են: Մյուս կողմից, եթե այս հարցի պատասխանը եղել է

“ոչ”, ապա սա նշանակում է, որ 1 և 2 անդամներից գոնե մեկը անօրեն է, հետևաբար, 3-ը հաստատ օրինավոր է:

Արդյունքում, մեկ հարցից հետո մենք գտանք մեկ օրինավորի, հետևաբար, նրան տալով երկու հարց մենք կիմանանք խմբակցության կազմը, այնպես որ $\alpha(3) \le 3$:

Ենթադրենք, որ $\alpha(k) \le \left\lceil \frac{3(k-1)}{2} \right\rceil$, $k = 3,...,n-1$ համար, և ցույց տանք, որ $n$ անդամից բաղկացած խմբակցության կազմը կարելի է որոշել ոչ ավել, քան $\left\lceil \frac{3(n-1)}{2} \right\rceil$ հարցերի միջոցով:

Ընտրենք խմբակցության անդամներից որևէ մեկին և նրա մասին հերթով հարցնենք մյուսներից քանի դեռ

ա) “ոչ” պատասխանների քանակը չի գերազանցում “այո” պատասխանների քանակին, կամ

բ) “այո” պատասխանների քանակը փոքր է $\left\lfloor \frac{n-1}{2} \right\rfloor$–ից:

Քննարկենք երկու դեպք:

Դեպք 1: Դիցուք մենք կանգ ենք առել ա) կետով: Այդ դեպքում գոյություն ունի $j$, այնպես որ $j+1$ մարդ ասել է “ոչ” (գտնում են, որ ընտրված մարդը անօրեն է), իսկ $j$ մարդ՝ “այո” (գտնում են, որ ընտրված մարդը օրինավոր է): Այդ դեպքում հարցմանը մասնակցած $2j+2$ մարդկանցից առնվազն $j+1$-ը անօրեն է: Իրոք, եթե ընտրվածը օրինավոր է, ապա $j+1$ “ոչ” պատասխանողները անօրեն են, իսկ եթե ընտրվածը անօրեն է, ապա անօրեն են նաև $j$ “այո” պատասխանողները: Դիտարկենք մնացած $n-2j-2$ մարդկանց: Նախ նկատենք, որ նրանցում օրինավորների քանակը ավել է անօրենների քանակից: Մյուս կողմից,

- եթե $j = \left\lfloor \frac{n-1}{2} \right\rfloor - 1$, ապա $n-2j-2$ մարդկանցից յուրաքանչյուրը օրինավոր է, քանի որ եթե նրանց մեջ լիներ գոնե մեկ անօրեն, ապա մենք կունենայինք առնվազն $1 + j + 1 = \left\lfloor \frac{n-1}{2} \right\rfloor + 1 \ge \frac{n}{2}$ անօրեն, ինչը հնարավոր չէ:
- եթե $j \le \left\lfloor \frac{n-1}{2} \right\rfloor - 2$, ապա մնացած մարդկանց քանակը $\ge n - 2(\left\lfloor \frac{n-1}{2} \right\rfloor - 1) \ge 3$: Համաձայն ինդուկցիոն ենթադրության, $n-2j-2$ մարդկանցից կորոշենք թե ով ով է ոչ ավել քան $\left\lceil \frac{3(n-2j-2-1)}{2} \right\rceil$ հարցերի միջոցով:

Այնուհետև որոշենք թե ընտրվածն ով է, և վերջապես ամենաշատը $j+1$ հարցերով որոշենք մնացածներին (եթե ընտրվածը օրինավոր է, ապա $j+1$ “ոչ” պատասխանողները անօրեն են, իսկ եթե ընտրվածը անօրեն է, ապա անօրեն են նաև $j$ “այո” պատասխանողները): Այսպիսով, այս դեպքում քայլերի քանակը չի գերազանցի

$$2j+1+\left\lceil \frac{3(n-2j-2-1)}{2} \right\rceil +1+j+1=\left\lceil \frac{3(n-1)}{2} \right\rceil :$$

Դեպք 2: Դիցուք մենք կանգ ենք առել բ) կետով: Սա նշանակում է, որ մենք ստացել ենք $\left\lfloor \frac{n-1}{2} \right\rfloor$ “այո” պատասխան: $j$-ով նշանակենք “ոչ” պատասխանողների քանակը:

Նկատենք, որ այս դեպքում ընտրված օրինավոր է: Իրոք, եթե այն օրինավոր չլիներ, ապա $\left\lfloor \frac{n-1}{2} \right\rfloor$ “այո” պատասխանողները ևս կլինեին անօրեններ, և հետևաբար, անօրենների քանակը կլիներ առնվազն $\left\lfloor \frac{n-1}{2} \right\rfloor +1\geq \frac{n}{2}$, ինչը հնարավոր չէ: Այստեղից հետևում է, որ $j$ “ոչ” պատասխանողները անօրեն են:

Ընտրած անդամին տալով $n-1-j$ հարց մնացած անդամների մասին, մենք կորոշենք խմբակցության կազմը: Նկատենք, որ այս դեպքում հարցերի քանակը կլինի

$$\left\lfloor \frac{n-1}{2} \right\rfloor +j+n-1-j\leq \left\lceil \frac{3(n-1)}{2} \right\rceil :$$

Հետևաբար, $\alpha(n)\leq \left\lceil \frac{3(n-1)}{2} \right\rceil$: Հիմա ցույց տանք, որ $\alpha(n)\geq \left\lceil \frac{3(n-1)}{2} \right\rceil$: Նկատենք, որ պնդման ապացույցի համար բավական է ցույց տալ, որ խնդիրը լուծող ցանկացած ալգորիթմում կարելի է նշել առնվազն $\left\lceil \frac{3(n-1)}{2} \right\rceil$ երկարություն ունեցող ճյուղ:

Դիտարկենք խնդիրը լուծող ցանկացած ալգորիթմում ճյուղի ընտրման հետևյալ եղանակը՝ գուշակը:

Ալգորիթմի առաջին $(i_1,j_1),\ldots,(i_k,j_k)$, $k=\left\lceil \frac{n-1}{2} \right\rceil -1$ հարցերին պատասխանենք ‘ոչ”: Մնացած հարցերի պատասխաններն ընտրելուց առաջ դիտարկենք հետևյալ ձևով սահմանված $G=(V,E)$ գրաֆը.

$$V=\{i_1,j_1\}\cup\ldots\cup\{i_k,j_k\},$$

$$E=\{(i_1,j_1)\}\cup\ldots\cup\{(i_k,j_k)\}:$$

Դիցուք, $G=(V,E)$ գրաֆի կապակցվածության բաղադրիչները $G_i=(V_i,E_i)$, $i=1,...,r$, և դիցուք՝ $W=\{1,...,n\}\setminus V$:

Հաջորդ $h_{k+1},h_{k+2},...$ հարցերի պատասխաններն ընտրենք հետևյալ կերպ.

$(i,j)$ հարցի պատասխանը=

$$=\begin{cases}\text{"այո", եթե } j\in W;\\ \text{"ոչ", եթե } j\in V_l,\ 1\le l\le r \text{ և գոյություն ունի գագաթ } V_l\text{-ից, որի մասին}\\ h_{k+1},h_{k+2},... \text{ հարցերի ժամանակ հարցում չի եղել կամ հնչել է "այո"}\\ \text{պատասխան;}\\ \text{"այո", եթե } j\in V_l\ \ 1\le l\le r \text{ և } V_l\text{-ի ցանկացած գագաթի համար}\\ h_{k+1},h_{k+2},... \text{ հարցերի ժամանակ հարցում եղել է, և չի եղել}\\ \text{"այո" պատասխան:}\end{cases}$$

Գուշակի նկարագիրն ավարտված է:

$L$-ով նշանակենք խմբակցության այն անդամների բազմությունը, որոնց համար $h_{k+1},h_{k+2},...$ հարցերի ժամանակ “ոչ” պատասխան չի ստացվել: Նկատենք, որ $|L\cap V_l|\ge 1$ $1\le l\le r$:

Դիտարկենք հետևյալ կանոնների միջոցով սահմանված $L^*$ բազմությունը

1) $W$-ն ավելացնել $L^*$-ին;

2) եթե $j\in V_l$ $1\le l\le r$ և $j$–ի մասին $h_{k+1},h_{k+2},...$ հարցերից որևէ մեկում ասվել է “այո”, ապա $j\in L^*$;

3) եթե $V_l$–ի ոչ մի տարրի մասին $h_{k+1},h_{k+2},...$ հարցերից որևէ մեկում չի ասվել “այո”, ապա $L\cap V_l$ բազմության ամենափոքր $j$ տարրը ավելացնել $L^*$-ին, $j\in L^*$:

4) Այլ կանոններ չկան, $L^*$-ը որոշվում է վերը նկարագրված 1-3 կանոնների միջոցով:

Նկատենք, որ $L^*$-ի սահմանումից հետևում է, որ $|L^*\cap V_l|=1$, $1\le l\le r$: Մյուս կողմից նկատենք, որ քանի որ $G_i=(V_i,E_i)$ գրաֆները կապակցված են, ապա

$$|E_i|\ge|V_i|-1,\ 1\le i\le r, \text{ և հետևաբար՝}$$

$$k=|E|=|E_1|+...+|E_r|\ge|V_1|+...+|V_r|-r=|V|-r \text{ կամ, որ նույնն է}$$

$$|V|\le k+r:$$

Այստեղից հետևում է, որ

$$|V\setminus L^*|=|V|-|V\cap L^*|\le k+r-r=k \text{ հետևաբար՝}$$

$$|L^*|=|L^*\cap W|+|L^*\cap V|=n-|V|+|L^*\cap V|=n-|V|+|V|-|V\setminus L^*|\ge n-k:$$

Ցույց տանք, որ գուշակի ընտրած ճյուղի երկարությունը առնվազն $\left\lceil \frac{3(n-1)}{2} \right\rceil$ է: Ենթադրենք հակառակը: Քանի որ

$$\left\lceil \frac{3m}{2} \right\rceil - \left\lceil \frac{m}{2} \right\rceil = m \text{, ապա}$$

$$\left\lceil \frac{3(n-1)}{2} \right\rceil = k + \left\lceil \frac{3(n-1)}{2} \right\rceil - \left\lceil \frac{n-1}{2} \right\rceil + 1 = k + n,$$

և հետևաբար, $h_{k+1}, h_{k+2}, \ldots$ հարցերի քանակը չի գերազանցում $n-1$-ը: Այստեղից հետևում է, որ գոյություն ունի խմբակցության $a$ անդամ, որի մասին $h_{k+1}, h_{k+2}, \ldots$ հարցերի ժամանակ հարցում չի արվում: Պարզ է, որ $a \in L^*$: Ցույց տանք, որ առանց ընդհանրությունը խախտելու, կարելի է ենթադրել, որ $a \in L^*$: Իրոք, դիցուք, $a \notin L^*$: Այստեղից հետևում է, որ $a \notin W$ և հետևաբար՝ $a \in V_l$, $1 \le l \le r$, ընդ որում՝ $G_l = (V_l, E_l)$ բաղադրիչում ընդհանրապես չի եղել "այո" պատասխան: Սա նշանակում է, որ $L^*$ բազմության՝ $G_l = (V_l, E_l)$ բաղադրիչին պատկանող տարրն ընտրվել է $L^*$ բազմության սահմանման 3-կետով: Հետևաբար, եթե մենք կազմակերպեինք խմբակցության անդամների վերահամարակալում, ապա կարող էինք հասնել այն բանին, որ $L \cap V_l$ բազմության ամենափոքր տարրը դառնար $a$ անդամը, և հետևաբար՝ $a \in L^*$:

Դիտարկենք հետևյալ երկու իրավիճակները.

**Իրավիճակ Ա**: $L^*$-ին պատկանող խմբակցության անդամները օրինավոր են, իսկ մնացածները՝ անօրեն: Պարզ է, որ անօրենների քանակը այս դեպքում չի գերազանցում $\left\lceil \frac{n-1}{2} \right\rceil - 1$–ը:

**Իրավիճակ Բ**: $L^*$-ին պատկանող խմբակցության անդամները, բացի $a$ անդամից, օրինավոր են, իսկ մնացածները՝ անօրեն: Պարզ է, որ անօրենների քանակը այս դեպքում չի գերազանցում $\left\lceil \frac{n-1}{2} \right\rceil$–ը:

Ցույց տանք, որ այս երկու իրավիճակները բավարարում են գուշակի կողմից ընտրած ճյուղի բոլոր հարցերին:

Առաջին $k = \left\lceil \frac{n-1}{2} \right\rceil - 1$ հարցերի ժամանակ հանդես են գալիս միևնույն բաղադրիչի երկու զագաթներ: Քանի որ $\left| L^* \cap V_l \right| = 1$ և $\left| (L^* \setminus \{a\}) \cap V_l \right| \le 1$, $1 \le l \le r$, ապա քննարկվող իրավիճակները բավարարում են այս հարցերի պատասխաններին:

Դիտարկենք $h_{k+1}, h_{k+2}, \ldots$ հարցերի՝ գուշակի թելադրած պատասխանները: Եթե $j \in W$, ապա $(i, j)$ հարցի պատասխանը=”այո”: Բայց քննարկվող իրավիճակներում $W$ բազմության այն տարրերը, որոնց մասին հարցում եղել է, օրինավոր են, հետևաբար՝ Ա և Բ իրավիճակները բավարարում են այս հարցերի պատասխաններին:

Հիմա ենթադրենք $(i, j)$ հարցի պատասխանը=”ոչ”, սա նշանակում է, որ $j \notin L$ և հետևաբար, $j \notin L^*$ և $j \notin L^* \setminus \{a\}$: Բայց քննարկվող իրավիճակներում այսպիսի $j$-երը անօրեն են, հետևաբար՝ Ա և Բ իրավիճակները բավարարում են այս հարցերի պատասխաններին:

Վերջապես, ենթադրենք $(i, j)$ հարցի պատասխանը=”այո”, սա նշանակում է, որ $j \in L$ և քանի որ $j$-ի մասին հարցում եղել է, ապա, $j \in L^* \setminus \{a\}$: Քննարկվող իրավիճակներում այսպիսի $j$-երը օրինավոր են, հետևաբար՝ Ա և Բ իրավիճակները բավարարում են այս հարցերի պատասխաններին:

Այսպիսով, այս երկու, **իրարից տարբեր** իրավիճակները բավարարում են գուշակի կողմից ընտրած ճյուղի բոլոր հարցերին: Հետևաբար, խնդիրը լուծող ալգորիթմը չի տարբերում այս իրավիճակները: Ստացված հակասությունն ապացուցում է թեորեմը:

# Գրականություն

Գտնված սխալների, առաջարկությունների, ինչպես նաև դասախոսություն-ներն e-mail-ով ստանալու համար կարող էք դիմել vahanmkrtchyan2002@yahoo.com հասցեով:

Կոմբինատորային ալգորիթմներ և
ալգորիթմների վերլուծություն
Վահան Վ. Մկրտչյան

Դասախոսություն 8: Տեսակավորման խնդիրներ: Ներքևից գնահատական: Տեսակավորում տեղավորման եղանակով: Տեսակավորում շարքերի ձուլման եղանակով:

**Տեսակավորման խնդիրներ. ներքևից գնահատական**: Դիցուք ունենք կշռաքարեր, որոնց քաշերը մեզ անհայտ, միմյանցից տարբեր թվեր են՝ $a_1,\ldots,a_n$: Թույլատրվում է կատարել $(a_i,a_j)$՝ մեր կողմից ընտրված $a_i$ և $a_j$ քաշեր ունեցող երկու կշռաքարերի քաշերի համեմատում: Տեսակավորման խնդիրը կայանում է հնարավորին չափ քիչ համեմատումներ կատարելով կշռաքարերը քաշերի աճման կարգով դասավորելու մեջ:

Այս խնդիրը կարելի է մեկնաբանել նաև հետևյալ կերպ. հնարավորին չափ քիչ խաղեր կազմակերպելով տրանզիտիվ մրցաշարի մասնակիցներին դասավորել ուժերի աճման կարգով:

Դժվար չէ տեսնել, որ տեսակավորման խնդիրը իրականում որոնման խնդիր է. $n!$ տեղադրություններից գտնել որոնելի դասավորությանը համապատասխանող տեղադրությունը կատարելով հնարավորին չափ քիչ համեմատումներ:

Պարզ է, որ գոյություն ունի խնդիրը լուծող լավագույն ալգորիթմ: Նրա բարդությունը նշանակենք $S(n)$–ով:

**Թեորեմ 1**: $S(n)\geq\lceil\log_2 n!\rceil$:

**Ապացույց**: $a_1,\ldots,a_n$ տարրերը կարգավորող յուրաքանչյուր ալգորիթմին համապատասխանեցնենք 2-ծառ, որի գագաթներին վերագրվում են կշռումները, իսկ տերևներին՝ կշռաքարերի որոնելի դասավորությունները: Պարզ է, որ ծառը պետք է ունենա առնվազն $n!$ տերև, հետևաբար նրա $S(n)$ բարձրությունը պետք է բավարարի

$$2^{S(n)}\geq n!$$

անհավասարմանը, որտեղից հետևում է $S(n) \geq \lceil \log_2 n! \rceil$ առնչությունը:

Նախ փորձենք գտնել տեսակավորման լավագույն ալգորիթմ $n$–ի փոքր արժեքների դեպքում:
$n = 2$ դեպքում պարզ է, որ մեկ համեմատությամբ խնդիրը կլուծվի, հետևաբար՝ $S(2) = 1$:
$n = 3$ դեպքում ունենք $\lceil \log_2 3! \rceil = 3 \leq S(3)$: Ստորև բերված է 3 բարդությամբ ալգորիթմին համապատասխանող որոնման ծառը՝

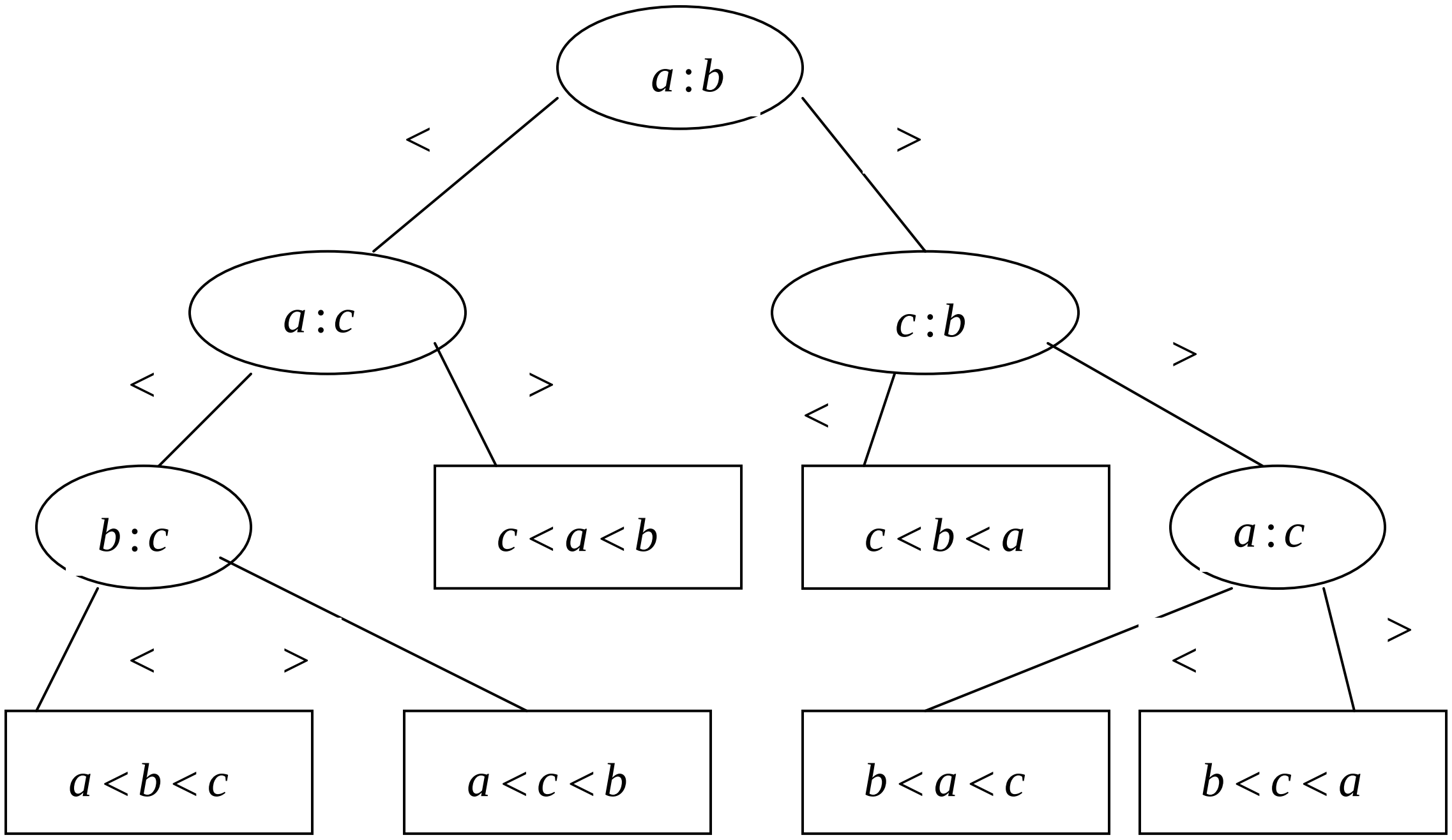

Նկար 1

Պարզ է, որ $n > 3$ դեպքում ալգորիթմին համապատասխանող որոնման ծառի նկարելը կախված է տեխնիկական դժվարությունների հետ: Բերենք այն պատկերելու մեկ այլ եղանակ:
Պայմանավորվենք ալգորիթմի յուրաքանչյուր քայլից հետո ստեղծված իրավիճակը պատկերել համապատասխան դիագրամի միջոցով: Օրինակ, նույն $n = 3$ դեպքում առաջին համեմատումից հետո կունենանք հետևյալ պատկերը.

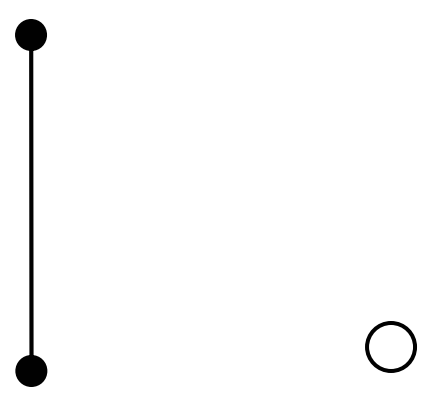

Նկար 2

իսկ երկրորդից հետո՝

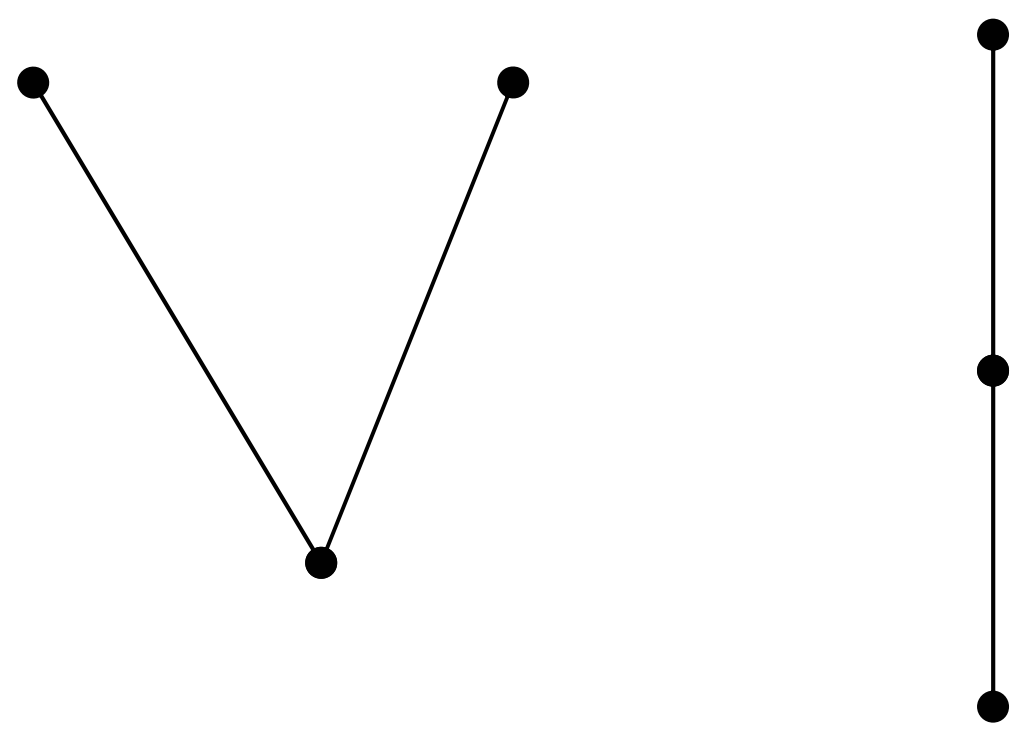

Նկար 3

Կարելի է ապացուցել, որ $n=4$ և $n=5$ դեպքերում կունենանք համապատասխանաբար՝ $S(4)=5$ և $S(5)=7$:

Դիցուք Ա-ն $a_1,...,a_n$ տարրերը տեսակավորող որևէ ալգորիթմ է, որի բարդությունը $\Phi(n)$ է: Կասենք, որ Ա-ն համարյա լավագույն ալգորիթմ է, եթե $\Phi(n) \sim S(n)$ այսինքն՝

$$\lim_{n\to+\infty}\frac{\Phi(n)}{S(n)}=1:$$

Համաձայն Ստիրլինգի բանաձևի՝ $n! \sim (\frac{n}{e})^n\sqrt{2\pi n}$, հետևաբար՝

$$\log_2 n! \sim n\log_2 n:$$

Մյուս կողմից

$$1\le\frac{\Phi(n)}{S(n)}\le\frac{\Phi(n)}{\lceil\log_2 n!\rceil}$$

հետևաբար՝ եթե $\Phi(n) \sim n\log_2 n$, ապա Ա-ն տեսակավորման լավագույն ալգորիթմ է:

**Տեսակավորում տեղավորման եղանակով**: Նախ քննարկենք մի օժանդակ խնդիր. ենթադրենք $a_1,...,a_{k-1}$ տարրերը դասավորված են շարքով՝ $a_1 < ... < a_{k-1}$ և անհրաժեշտ է առաջարկել մեզ անհայտ $a$ տարրը այդ շարքում դասավորելու ալգորիթմ:

$a$ տարրը կարող է զբաղեցնել $k$ տեղ ($a_1$-ից ձախ, $a_1,...,a_{k-1}$ արանքում, $a_{k-1}$-ից աջ), հետևաբար ալգորիթմի բարդությունը առնվազն $\lceil \log_2 k \rceil$-է: Մյուս կողմից, կարգավոր բազմության մեջ տարրի որոնում խնդրում մենք ցույց էինք տվել թե ինչպես կարելի է ամեն քայլում համարյա հավասար մասերի տրոհելու իդեայով գտնել $a$ տարրի տեղը $\lceil \log_2 k \rceil$ համեմատությունների միջոցով:

Ձևակերպենք տեսակավորման մի ալգորիթմ, որը հիմնված է տեղավորման այս խնդրի վրա:

Քայլ 1: $k := 2; X := a_1$

Քայլ 2: $a_k$ տարրը տեղավորել $X$ շարքում $\lceil \log_2 k \rceil$ համեմատությունների միջոցով:

Քայլ 3: Եթե $k < n$ ապա $k := k+1$ և անցնել Քայլ 2-ին, հակառակ դեպքում՝ ավարտ:

Նկատենք, որ այս ալգորիթմը կատարում է

$$A_n = \lceil \log_2 2 \rceil + \lceil \log_2 3 \rceil + ... + \lceil \log_2 n \rceil$$

համեմատություն: Հաշվի առնելով

$$2^{\lceil \log_2 n \rceil - 1} < n \le 2^{\lceil \log_2 n \rceil} \text{ և}$$

$$\lceil \log_2 i \rceil = k \text{ եթե } 2^{i-1} < k \le 2^i$$

կստանանք՝

$$A_n = \sum_{i=1}^{\lceil \log_2 n \rceil - 1} i2^{i-1} + (n - 2^{\lceil \log_2 n \rceil - 1})\lceil \log_2 n \rceil :$$

Ածանցելով երկրաչափական պրոգրեսիայի բանաձը, կստանանք՝

$$\sum_{i=1}^{k} ix^{i-1} = (\sum_{i=1}^{k} x^i)' = (\frac{x^{k+1}-1}{x-1})' = \frac{(k+1)(x-1)x^k - (x^{k+1}-1)}{(x-1)^2} :$$

Վերցնելով վերջին բանաձևում՝ $k = \lceil \log_2 n \rceil - 1$, $x = 2$, կստանանք՝

$$A_n = (\lceil \log_2 n \rceil - 1)2^{\lceil \log_2 n \rceil - 1} - 2^{\lceil \log_2 n \rceil - 1} + 1 + (n - 2^{\lceil \log_2 n \rceil - 1})\lceil \log_2 n \rceil =$$

$$= n\lceil \log_2 n \rceil - 2^{\lceil \log_2 n \rceil} + 1$$

Քանի որ $A_n \sim n\lceil \log_2 n \rceil$, ապա այս ալգորիթմը տեսակավորման համարյա լավագույն ալգորիթմ է:

**Տեսակավորում շարքերի ձուլման եղանակով**: Նախ դիտարկենք հետևյալ՝ շարքերի ձուլման խնդիրը: Տրված են $a_1 < \ldots < a_m$ և $b_1 < \ldots < b_n$ շարքերը և դիցուք այդ շարքերի բոլոր տարրերը միմյանցից տարբեր են: Անհրաժեշտ է հնարավորին չափ քիչ $(a_i : b_j)$ համեմատությունների միջոցով, այդ տարրերը դասավորել $m+n$ երկարությամբ մեկ շարքով: $M(m,n)$-ով նշանակենք շարքերի ձուլման լավագույն ալգորիթմի բարդությունը:

**Թեորեմ 2**: $\left\lceil \log_2 \binom{m+n}{n} \right\rceil \le M(m,n) \le m+n-1$:

**Ապացույց**: $m$ և $n$ երկարությամբ շարքերի ձուլման հնարավոր ելքերի քանակը $\binom{m+n}{n}$-է ($m+n$ տեղից ընտրենք $b_1,\ldots,b_n$-ի տեղերը), հետևաբար այդ խնդիրը լուծող ցանկացած ալգորիթմին համապատասխանող որոնման ծառի տերևների քանակն առնվազն պետք է լինի $\binom{m+n}{n}$, և հետևաբար այդ ծառի բարձրությունը, այսինքն՝ ալգորիթմի բարդությունը $\left\lceil \log_2 \binom{m+n}{n} \right\rceil$-ից պակաս լինել չի կարող:

Վերին գնահատականի ապացուցման համար դիտարկենք հետևյալ ալգորիթմը.

Շարքերի ձուլման ալգորիթմ
Քայլ 1: Որպես $X$ շարք ընդունենք $a_1 < \ldots < a_m$-ը, իսկ $Y$ շարք ընդունենք $b_1 < \ldots < b_n$-ը, $Z$- շարքը դատարկ է:
Քայլ 2: Համեմատենք $X$ և $Y$ շարքերի ամենափոքր տարրերը: Նրանցից ամենափոքրը հանենք շարքից և դնենք $Z$ շարքի հերթական տեղում:
Քայլ 3: Եթե $X$ և $Y$ շարքերից մեկն ու մեկը դատարկ է, ապա մյուսը կցենք $Z$- շարքին և ավարտենք ալգորիթմի աշխատանքը, հակառակ դեպքում՝ վերադարձ Քայլ 2-ին:

Նկատենք, որ այս ալգորիթմը վատագույն դեպքում կատարում է $m+n-1$ համեմատություն, հետևաբար՝ $M(m,n) \le m+n-1$:
Թեորեմն ապացուցված է:

Ապացուցված վերին և ստորին գնահատականները որակապես տարբերվում են միմյանցից, իսկ $M(m,n)$-ը կարող է մոտ լինել, ինչպես մեկին, այնպես էլ՝ մյուսին: Օրինակ, մեկ տարրի տեղավորման ալգորիթմից հետևում է, որ $M(1,n)=\lceil \log_2(n+1)\rceil$: Մյուս կողմից, տեղի ունի

Թեորեմ 3: $M(n,n)=2n-1$:

Ապացույց: Ապացույցի համար բավական է կառուցել գուշակ, որը $a_1<...<a_n$ և $b_1<...<b_n$ շարքերը ձուլող ցանկացած ալգորիթմից ընտրում է $2n-1$ երկարությամբ ճյուղ:

Գուշակի վարվելակերպը սահմանենք հետևյալ կերպ.

$(a_i:b_j)$ համեմատության դեպքում ընտրվում է $a_i<b_j$ ճյուղը, եթե $i<j$, և ընտրվում է $a_i>b_j$ ճյուղը, եթե $i\geq j$:

Նկատենք որ, փաստորեն, գուշակը ընտրում է

$$b_1<a_1<b_2<a_2<...<b_n<a_n$$

ճյուղը: Ցույց տանք, որ այս ճյուղում ցանկացած ալգորիթմ պետք է կատարի

$$(b_1:a_1),(a_1:b_2),(b_2:a_2),...,(b_n:a_n)$$

համեմատություններից յուրաքանչյուրը: Իրոք, եթե ալգորիթմը չկատարի, օրինակ, $(b_i,a_i)$ համեմատությունը, ապա այն չի տարբերի

$$b_1<a_1<b_2<a_2<...<b_n<a_n$$

և այս հաջորդականության մեջ $b_i$ և $a_i$ տարրերի փոխատեղումից ստացված հաջորդականությունը: Թեորեմն ապացուցված է:

Ստորև կնկարագրենք $a_1,...,a_n$ տարրերը տեսակավորող մի ալգորիթմ, որը հիմնված է շարքերի ձուլման պարզագույն ալգորիթմի վրա:

Նախ կնկարագրենք այն դեպքում, երբ $n=2^k$:

Քայլ 1: Որպես ձուլման ենթակա շարքեր ընտրել $a_1,...,a_n$ տարրերը:

Քայլ 2: Ձուլման ենթակա շարքերը տրոհել զույգերի և շարքերի ձուլման պարզագույն ալգորիթմի միջոցով յուրաքանչյուր զույգից ստանալ ձուլման ենթակա մեկ շարք:

Քայլ 3: Եթե ձուլման ենթակա շարքերի քանակը մեկ է, ապա ավարտել ալգորիթմի աշխատանքը, հակառակ դեպքում՝ վերադառնալ Քայլ 2-ին:

Նկատենք, որ $i$-րդ ձուլումից հետո ($i=1,...,k$) ստացվում են $2^{k-i}$ շարքեր, որոնցից յուրաքանչյուրը ունի $2^i$ երկարություն: Հետևաբար, $i$-րդ ձուլման ընթացքում կատարված համեմատությունների քանակը կլինի $(2^i-1)2^{k-i}$, իսկ ալգորիթմի բարդությունը $n=2^k$ դեպքում կլինի՝

$$B_n=\sum_{i=1}^{k}(2^i-1)2^{k-i}=k2^k-2^k+1:$$

Ալգորիթմի նկարագիրը ընդհանուր դեպքում:

Քայլ 1: $n$-ը ներկայացնել թվարկության երկուական համակարգում՝ 2-ի նվազող ցուցիչներով աստիճանների գումարի տեսքով.

$$n = 2^{k_1} + 2^{k_2} + \ldots + 2^{k_s},\ \text{որտեղ}\ k_1 > k_2 > \ldots > k_s \geq 0,$$

և օգտագործելով $n = 2^k$ դեպքում շարքերի ձուլման ալգորիթմը $a_1, \ldots, a_n$ տարրերից կառուցել $2^{k_1}, 2^{k_2}, \ldots, 2^{k_s}$ երկարությամբ շարքեր:

Քայլ 2: Ստացված շարքերից հերթականորեն ընտրել երկու ամենակարճ շարքերը և ձուլել շարքերի ձուլման պարզագույն ալգորիթմի միջոցով, մինչև մեկ շարքի ստանալը:

Առանց ապացույցի նշենք, որ տեղի ունի հետևյալ

**Թեորեմ 4**: Եթե $n = 2^{l_1} + 2^{l_2} + \ldots + 2^{l_k}$, որտեղ $l_1 > l_2 > \ldots > l_k$, ապա վերը նշված ալգորիթմի բարդությունը՝ $B_n$-ը որոշվում է

$$B_n = 1 - 2^{l_k} + \sum_{i=1}^{k} (l_i + i - 1) 2^{l_i}$$

բանաձևով:

# Գրականություն

Գտնված սխալների, առաջարկությունների, ինչպես նաև դասախոսություններն e-mail-ով ստանալու համար կարող էք դիմել vahanmkrtchyan2002@yahoo.com հասցեով:

Կոմբինատորային ալգորիթմներ և
ալգորիթմների վերլուծություն
Վահան Վ. Մկրտչյան

Դասախոսություն 9: Տեսակավորում ձուլման և տեղավորման եղանակով:

**Տեսակավորում ձուլման և տեղավորման եղանակով:** Կրկին դիտարկենք $a_1,...,a_n$ տարրերի տեսակավորման խնդիրը և նախքան ձուլման և տեղավորման եղանակով տեսակավորման ալգորիթմի ընդհանուր դեպքի նկարագրին անցնելը քննարկենք $n=5$ և $n=10$ դեպքերը: Նկատենք, որ

$$S(5)\geq\lceil\log_2 5!\rceil=7 \text{ և } S(10)\geq\lceil\log_2 10!\rceil=22:$$

Ցույց տանք, որ վերը նշված անհավասարություններում տեղի ունի հավասարություն:

Սկզբից դիտարկենք $n=5$ դեպքը: Առանձնացնենք այդ տարրերից որևէ մեկը, մնացածները տրոհենք զույգերի և յուրաքանչյուր զույգ կարգավորենք (նկար 1 ա) :

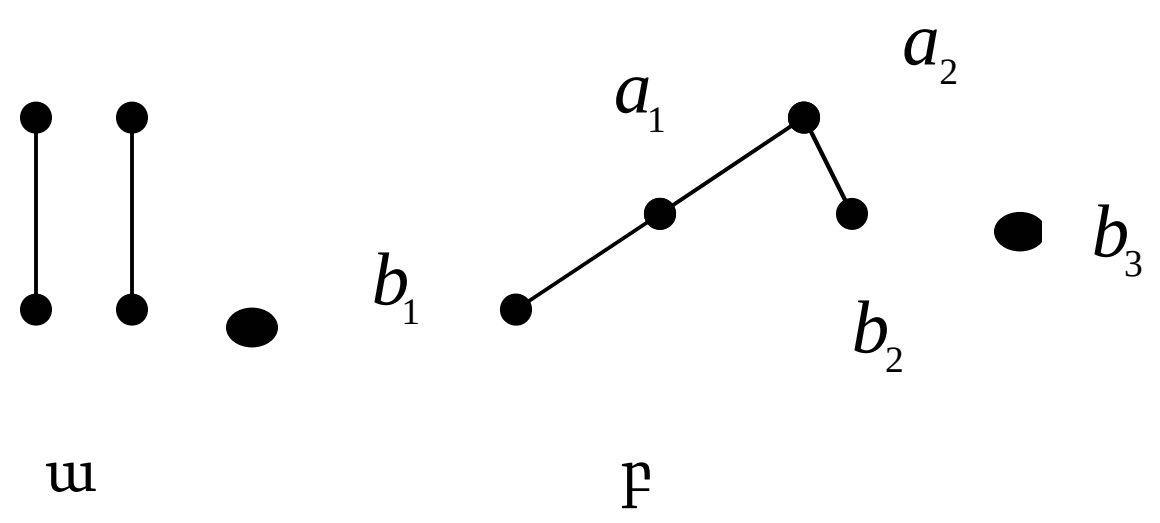

Նկար 1

Կարգավորենք նաև զույգերի ծանր կշռաքարերը: Այսպիսով 3 համեմատությունների հետո կունենանք նկար 1բ-ում նկարված իրավիճակը: Այժմ, եթե $b_2$-ը $\lceil\log_2 3\rceil=2$ համեմատությունների միջոցով տեղավորենք $b_1<a_1<a_2$ շարքում, իսկ այնուհետև՝ $b_3$-ը $\lceil\log_2 5\rceil=3$ համեմատությունների միջոցով տեղավորենք ստացված շարքում, ապա 5 տարրերը կկարգավորվեն 8 համեմատությունների միջոցով:

Իսկ եթե սկզբում տեղավորեինք $b_3$-ը $b_1 < a_1 < a_2$ շարքում $\lceil \log_2 4 \rceil = 2$ համեմատությունների միջոցով, իսկ այնուհետև՝ $b_2$-ը $\le \lceil \log_2 4 \rceil = 2$ համեմատությունների միջոցով տեղավորենք ստացված շարքում, ապա 5 տարրերը կկարգավորվեն արդեն 7 համեմատությունների միջոցով: Հետևաբար, $S(5) = 5$:

Քննարկենք $n = 10$ դեպքը: Կրկին, այդ տարրերը տրոհենք զույգերի և յուրաքանչյուր զույգ կարգավորենք: Այնուհետև, օգտագործելով $n = 5$ դեպքում արդեն նկարագրված ալգորիթմը, տեսակավորենք նաև զույգերի մեծ տարրերը: Արդյունքում կստանանք նկար 2-ում բերված պատկերը, որտեղ ենթադրված է, որ $a_1,...,a_5$-ը ծանր կշռաքարերն են, իսկ $b_1,...,b_5$-ը՝ թեթևները:

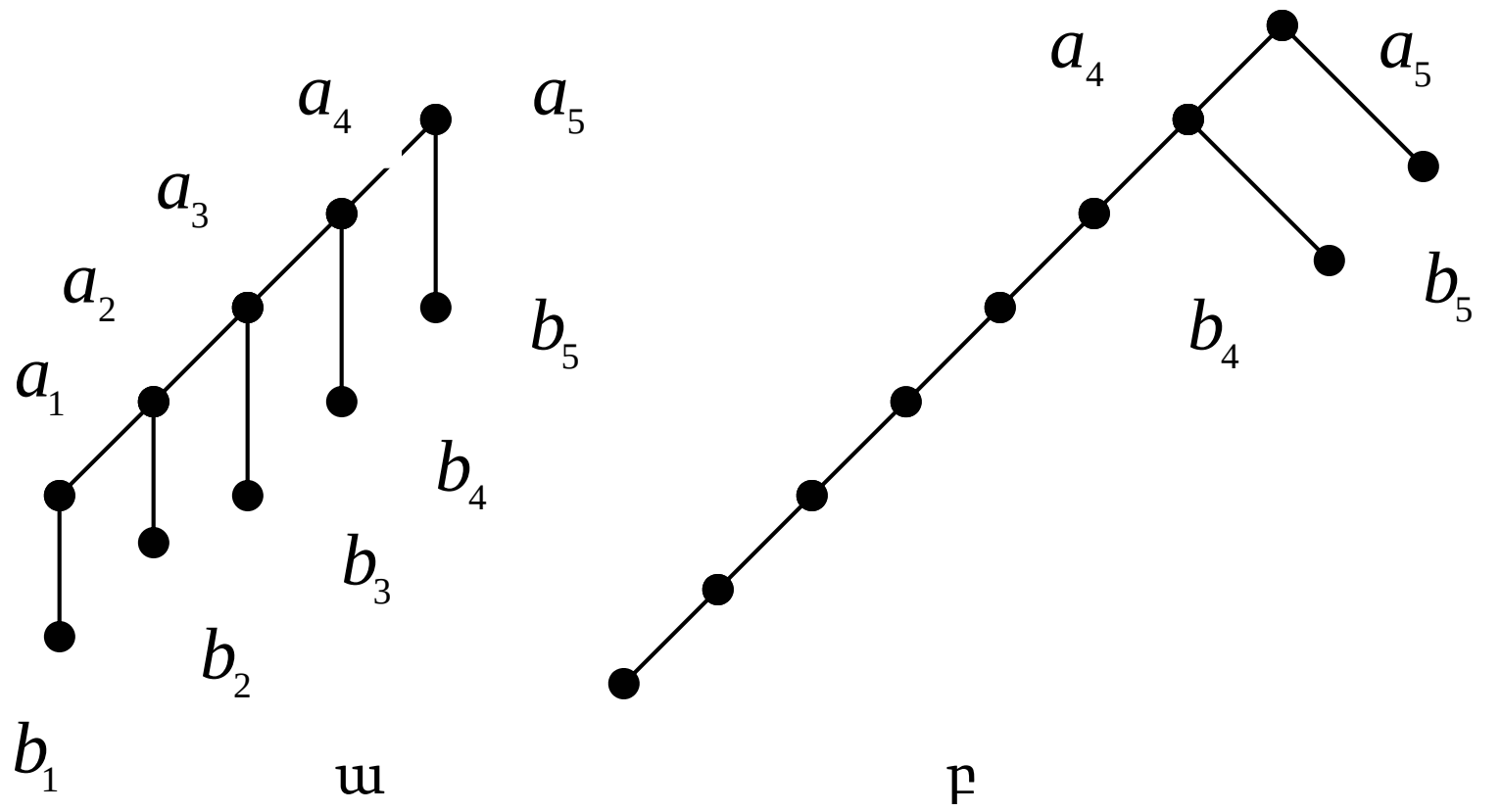

Նկար 2

Տեսակավորումն ավարտելու համար բավական է $b_2,...,b_5$ տարրերն տեղավորել $b_1 < a_1 < ... < a_5$ շարքում: Եթե սկզբից տեղավորենք $b_3$-ը, իսկ այնուհետև $b_2$-ը, ապա յուրաքանչյուր տեղավորման համար կօգտագործվի երկու համեմատություն և արդյուքում կստացվի նկար 2բ-ում պատկերված իրավիճակը:

Քանի որ ստացված շարքում $b_5$-ը կարող է զբաղեցնել 8 տեղերից որևէ մեկը, ապա եթե շարքում տեղավորենք $b_5$-ը, իսկ հետո՝ $b_4$-ը, ապա նրանցից յուրաքանչյուրի համար կօգտագործվի 3 համեմատություն: Այսպիսով, 10 տարրերը կարելի է տեսակավորել կատարելով, ոչ շատ, քան՝ $5 + 7 + 2 \cdot 2 + 2 \cdot 3 = 22$ համեմատություն: Հետևաբար, $S(10) = 22$:

Ստորև կնկարագրենք ձուլման և տեղավորման ալգորիթմը, որը ընդհանրացնում է 5 և 10 տարրերի տեսակավորման ժամանակ առաջարկված մոտեցումը:

*$n$ տարրերի տեսակավորման ձուլման և տեղավորման ալգորիթմի նկարագիրը*

Քայլ 1: $n$ տարրերը տրոհել զույգերի և յուրաքանչյուր զույգ կարգավորել: Եթե $n$–ը կենտ է, ապա մեկ տարր չի մասնակցում համեմատմանը:

Քայլ 2: Օգտվելով $\left\lfloor \frac{n}{2} \right\rfloor$ տարրերի ձուլման և տեղավորման ալգորիթմը, տեսակավորենք զույգերի մեծ տարրերը: Դիցուք, ինչպես վերևում, $a_1,\dots,a_{\left\lfloor \frac{n}{2} \right\rfloor}$-ը ծանր կշռաքարերն են, իսկ $b_1,\dots,b_{\left\lfloor \frac{n}{2} \right\rfloor}$-ը` թեթևները: Գլխավոր շղթա կանվանենք $b_1 < a_1 < \dots < a_{\left\lfloor \frac{n}{2} \right\rfloor}$ շղթան: Եթե $n$–ը կենտ է, ապա այն մեկ տարրը, որը սկզբում չէր մասնակցում համեմատմանը, կնշանակենք $b_{\left\lceil \frac{n}{2} \right\rceil}$-ով:

Քայլ 3: Ստորև նկարագրված եղանակով $b_2,\dots,b_{\left\lceil \frac{n}{2} \right\rceil}$ տարրերը տեղավորել գլխավոր շղթայի վրա:

Դիտարկենք թվերի $t_2,t_3,\dots,t_k,\dots$. հաջորդականությունը, որտեղ $t_2 = 3$ իսկ $t_k$–ն ցույց է տալիս այն ամենամեծ համարը, որ $b_{t_k}$-ն հնարավոր է $k$ համեմատությունների միջոցով տեղավորել գլխավոր շղթայի մեջ, երբ արդեն տեղավորված են` $b_2,b_3,\dots,b_{t_{k-1}}$-ը: Նկատենք, որ $t_3 = 5$ (տես $n = 5$ դեպքի վերլուծությունը վերևում):

ա) գլխավոր շղթայի վրա սկզբից տեղավորել $b_3$–ը, իսկ հետո` $b_2$-ը:

բ) եթե գլխավոր շղթայի վրա արդեն տեղավորված են $b_2,b_3,\dots,b_{t_{k-1}}$-ը, ապա նրանցից հետո հերթով դասավորել $b_{t_k},b_{t_k-1},\dots,b_{t_{k-1}+2},b_{t_{k-1}+1}$-ը: Եթե դասավորման ընթացքում պարզվեց, որ $\left\lceil \frac{n}{2} \right\rceil < t_k$, ապա դասավորել միայն $b_{\left\lceil \frac{n}{2} \right\rceil},\dots,b_{t_{k-1}+2},b_{t_{k-1}+1}$ տարրերը:

Փորձենք գտնել $t_2,t_3,\dots,t_k,\dots$. հաջորդականության ընդհանուր անդամի արժեքը: Նկատենք, որ եթե գլխավոր շղթայի վրա արդեն տեղավորված են $b_2,b_3,\dots,b_{t_{k-1}}$-ը, ապա այդ շղթայի վրա` $a_{t_k}$-ից փոքր տարրերն են` $a_1,\dots,a_{t_k-1}$ և $b_1,\dots,b_{t_{k-1}}$, հետևաբար $b_{t_k}$-ն կարող է զբաղեցնել այդ շղթայի $t_k + t_{k-1}$ տեղերից որևէ մեկը, և որպեսզի $t_k$-ն լինի ամենամեծ համարը, ապա որի դեպքում $b_{t_k}$-ն հնարավոր լինի տեղավորել $k$ համեմատությունների միջոցով, պետք է տեղի ունենա

$$t_k + t_{k-1} = 2^k$$

հավասարությունը: Նկատենք, որ եթե սահմանենք $t_0 = t_1 = 1$, ապա այս հավասարությունը չի խախտվի: Ունենք

$$t_1 = 1$$
$$t_2 + t_1 = 2^2$$
$$t_3 + t_2 = 2^3$$
$$\dots$$
$$t_k + t_{k-1} = 2^k:$$

Եթե $i$-րդ հավասարությունը բազմապատկենք $(-1)^{k-i}$-ով և բոլորը գումարենք, ապա կստանանք՝

$$t_k = \frac{2^{k+1} + (-1)^k}{3}:$$

Գնահատենք առաջարկված ալգորիթմի բարդությունը, այսինքն տրված $n$ տարրերը ձուլման և տեղավորման եղանակով տեսակավորող ալգորիթմում համեմատությունների առավելագույն $F(n)$ քանակը: $G(m)$-ով նշանակենք $b_2, b_3, \dots, b_m$ տարրերը ալգորիթմի համաձայն գլխավոր շղթայում տեղավորելիս կատարած համեմատությունների քանակը: Նկատենք, որ եթե $t_{k-1} < m \le t_k$, ապա

$$G(m) = \sum_{j=1}^{k-1} j(t_j - t_{j-1}) + k(m - t_{k-1}) = km - \left\lceil \frac{2^{k+1}}{3} \right\rceil:$$

Ալգորիթմի նկարագրից երևում է, որ $F(1) = G(1) = 0$ և

$$F(n) = \left\lfloor \frac{n}{2} \right\rfloor + F(\left\lfloor \frac{n}{2} \right\rfloor) + G(\left\lceil \frac{n}{2} \right\rceil) \text{ երբ } n \ge 2:$$

**Թեորեմ**: Եթե $\frac{2^{k+1}}{3} < n < \frac{2^{k+2}}{3}$ ապա $F(n) - F(n-1) = k$:

**Ապացույց** կատարենք ինդուկցիայով: $k = 1$ ունենք $n = 2$ և $F(n) - F(n-1) = F(2) - F(1) = 1 = k$: Ենթադրենք, որ պնդումը ճիշտ է այն դեպքում, երբ տարրերի քանակը փոքր է $n$–ից, և փորձենք ապացուցել այն $n$–ի դեպքում: Քննարկենք երկու դեպք.

ա) $n = 2p$: Այս դեպքում ունենք՝

$$F(2p) = p + F(p) + G(p)$$
$$F(2p-1) = p - 1 + F(p-1) + G(p),$$

և հետևաբար ըստ ինդուկցիոն ենթադրության՝

$$F(2p)-F(2p-1)=F(p)-F(p-1)+1=(k-1)+1=k,$$

քանի որ՝

$$\frac{2^k}{3}<p<\frac{2^{k+1}}{3}:$$

բ) $n=2p+1$: Այս դեպքում ունենք՝

$$F(2p+1)=p+F(p)+G(p+1)$$
$$F(2p)=p+F(p)+G(p),$$

հետևաբար

$$F(n)-F(n-1)=G(p+1)-G(p):$$

$\frac{2^{k+1}}{3}<n=2p+1<\frac{2^{k+2}}{3}$ պայմանից ունենք, որ $\frac{2^k}{3}+\frac{1}{2}<p+1<\frac{2^{k+1}}{3}+\frac{1}{2}$, և հետևաբար՝

$$\frac{2^k}{3}+\frac{(-1)^{k-1}}{3}<\frac{2^k}{3}+\frac{1}{2}<p+1<\frac{2^{k+1}}{3}+\frac{1}{2}:$$

Այսպիսով, $t_{k-1}<p+1<t_k+1$, և հետևաբար՝ $t_{k-1}<p+1\le t_k$: Այստեղից հետևում է, որ $b_{p+1}$ տարրը գլխավոր շղթայի վրա տեղավորելու համար անհրաժեշտ է $k$ համեմատություն, այսինքն՝ $G(p+1)-G(p)=k$: Թեորեմն ապացուցված է:

Նկատենք, որ $\frac{2^{k+1}}{3}<n<\frac{2^{k+2}}{3}$ պայմանը համարժեք է $\left\lceil \log_2\frac{3}{4}n\right\rceil=k$, և հետևաբար, թեորեմից հետևում է, որ

$$F(n)-F(n-1)=\left\lceil \log_2\frac{3}{4}n\right\rceil:$$

Հաշվի առնելով, որ $F(1)=0$, վերջնականապես կստանանք

$$F(n)=\sum_{k=2}^{n}\left\lceil \log_2\frac{3}{4}k\right\rceil:$$

Նկատենք, որ $F(n)\le n\log_2 n$, հետևաբար ձուլման և տեղավորման ալգորիթմը տեսակավորման համարյա լավագույն ալգորիթմ է:

## Գրականություն

Գտնված սխալների, առաջարկությունների, ինչպես նաև դասախոսություն-ներն e-mail-ով ստանալու համար կարող էք դիմել vahanmkrtchyan2002@yahoo.com հասցեով:

Կոմբինատորային ալգորիթմներ և ալգորիթմների վերլուծություն
Վահան Վ. Մկրտչյան

Դասախոսություն 10: Հաջորդականության միջնակետի որոնումը:

**Հաջորդականության միջնակետի որոնումը**: Դիտարկենք $a_1,...,a_n$ տարրերի տեսակավորման խնդրին առնչվող հետևյալ խնդիրը. տրված $a_1,...,a_n$ տարրերից գտնել $t$-րդ ամենամեծը: Նշված խնդիրը լուծող լավագույն ալգորիթմի ստուգումների քանակը նշանակենք $W_t(n)$–ով: Նկատենք, որ

$$W_t(n) = W_{n+1-t}(n):$$

Ինչպես արդեն նշել ենք, այս խնդիրը կարող ենք ձևակերպել նաև որպես մրցաշարում $t$-րդ տեղը գրավողի որոշման խնդիր: Հետևաբար՝

$W_1(n) = n-1$ (մրցաշարի հաղթողին որոշող լավագույն ալգորիթմ)

$W_2(n) = n + \lceil \log_2 n \rceil - 2$ (մրցաշարի առաջին և երկրորդ տեղերն որոշող լավագույն ալգորիթմ):

**Թեորեմ 1**: Եթե $n \geq t$ ապա

$$W_t(n) \leq n - t + (t-1)\lceil \log_2(n+2-t) \rceil:$$

Նկատենք, որ գրված վերին գնահատականը ճշգրիտ է $t=1$ և $t=2$ դեպքերում:

**Ապացույց**: Ցույց տանք, որ $n$ մասնակիցներից բաղկացած տրանզիտիվ մրցաշարում $t$-րդ տեղը գրավողին կարող ենք որոշել՝ կազմակերպելով ոչ ավել, քան $n-t+(t-1)\lceil \log_2(n+2-t) \rceil$ խաղ: Նկատենք, որ մրցաշարի $t$-րդ տեղը գրավողը նա է, ով ուժեղ է $n-t$ մասնակցից և թույլ՝ $t-1$-ից:

Խաղերը կազմակերպենք հետևյալ կերպ.

**Քայլ 1**: Կազմակերպել գավաթային մրցաշար առաջին $n-t+2$ մասնակիցների միջև և, կազմակերպելով $n-t+1$ խաղ, որոշել այս փոքր մրցաշարի հաղթողին:

**Քայլ 2**: Նկատենք, որ այս փոքր մրցաշարի հաղթողի չի կարող լինել $t$-րդ տեղը գրավողը, քանի որ այն ուժեղ է $n-t+1$ մասնակցից: Նրան հեռացնենք, և փոխարենը խաղացնենք դեռևս չխաղացած $t-2$ մասնակիցներից որևէ մեկին: Կրկին որոշենք այս նոր փոքր մրցաշարի հաղթողին: Նկատենք, որ մեզ բավական է կազմակերպել ամենաշատը $\lceil \log_2(n+2-t) \rceil$ նոր խաղ, քանի որ $n+2-t$ մասնակիցներից բաղկացած զավաթային մրցաշարում փուլերի քանակը չի գերազանցում $\lceil \log_2(n+2-t) \rceil$-ը: Կրկին նշենք, որ այս նոր փոքր մրցաշարի հաղթողը չի կարող լինել $t$-րդ տեղը գրավողը, քանի որ այն ուժեղ է $n-t+1$ մասնակցից: Վարվենք նույն ձևով, ինչպես վերևում այնքան ժամանակ, մինչև չխաղացած մասնակից չմնա:

**Քայլ 3**: Նկատենք, որ այս ձևով մենք կհեռացնենք $t-1$ մասնակիցների, որոնք ուժեղ են $n+1-t$ մասնակիցներից և հետևաբար նրանցից ոչ մեկը չի կարող լինել մրցաշարի $t$-րդ տեղը գրավողը: $t$-րդ տեղը գրավողին որոշելու համար կազմակերպենք մրացաշար մնացած $n+1-t$ մասնակիցների միջև: Պարզ է, որ այս մրցաշարի հաղթողը կլինի հենց $t$-րդ տեղը գրավողը: Վերջին մրցաշարում նոր խաղերի քանակը չի գերազանցի $\lceil \log_2(n+2-t) \rceil - 1$-ը:

Այսպիսով, գումարային խաղերի քանակը չի գերազանցի
$$(n-t+1)+(t-2)\lceil \log_2(n+2-t) \rceil + \lceil \log_2(n+2-t) \rceil - 1 = n-t+(t-1)\lceil \log_2(n+2-t) \rceil$$
Թեորեմն ապացուցված է:

**Սահմանում**: Եթե $n=2q+1$, ապա $q+1$-րդ ամենամեծ տարրին կանվանենք $a_1,\ldots,a_n$ հաջորդականության միջնակետ:

**Թեորեմ 2**: Եթե $n \geq t$ և $n > 32$, ապա
$$W_t(n) \leq 15n - 163:$$

**Ապացույց**: Ցույց տանք, որ տրված $a_1,\ldots,a_n$ տարրերից $t$-րդ ամենամեծը կարող ենք գտնել ոչ ավել, քան $15n-163$ համեմատությունների միջոցով:

Նախ նկատենք, որ թեորեմը ճիշտ է, երբ $32 < n \leq 2^{10}$: Իրոք,
$$W_t(n) \leq S(n) \leq F(n) = \sum_{k=2}^{n} \left\lceil \log_2 \frac{3}{4}k \right\rceil \leq 10n \leq 15n - 163, \text{ երբ } 32 < n \leq 2^{10}:$$
Այնպես, որ առանց ընդհանրությունը խախտելու կարող ենք ենթադրել, որ $n > 2^{10}$: Ավելացնելով ամենաշատը 13 հատ $-\infty$ տարրեր (այսինքն, տարրեր, որոնք միշտ փոքր են մյուսներից), մենք, առանց ընդհանրությունը խախտելու, կարող ենք ենթադրել, որ $n = 7(2q+1)$, որտեղ $q \geq 73$:

Դիտարկենք տրված $a_1,\ldots,a_n$ տարրերից $t$-րդ ամենամեծը որոշող հետևյալ ալգորիթմը.

**Քայլ 1**: $a_1,...,a_n$ տարրերը տրոհել $2q+1$ խմբի յուրաքանչյուրում 7 տարր: Կարգավորել յուրաքանչյուր խմբի տարրերը: Նկատենք, որ

$$13=\lceil \log_2 7! \rceil \le S(7) \le F(7)=13,$$

այնպես, որ մենք կկատարվի ոչ շատ, քան $13(2q+1)$ համեմատություն:

**Քայլ 2**: Գտնել յուրաքանչյուր խմբի միջնակետերից կազմված հաջորդականության $x$ միջնակետը: Նկատենք, որ ըստ ինդուկցիոն ենթադրության, $x$-ի գտնելը մեզանից կպահանջի ոչ շատ քան

$$W_{q+1}(2q+1) \le 15(2q+1)-163 = 30q-148$$

համեմատություն:

**Քայլ 3**: Նկատենք, որ $a_1,...,a_n$ տարրերից մնացած $n-1$-ը կարելի է տրոհել երեք խմբի (նկար 1)

$4q+3$-ը $x$-ից մեծ են (տիրույթ 2)

$4q+3$-ը $x$-ից փոքր են (տիրույթ 3)

$6q$ հատ, որոնց հարաբերությունը $x$-ի հետ մեզ անհայտ է (տիրույթներ 1 և 4)

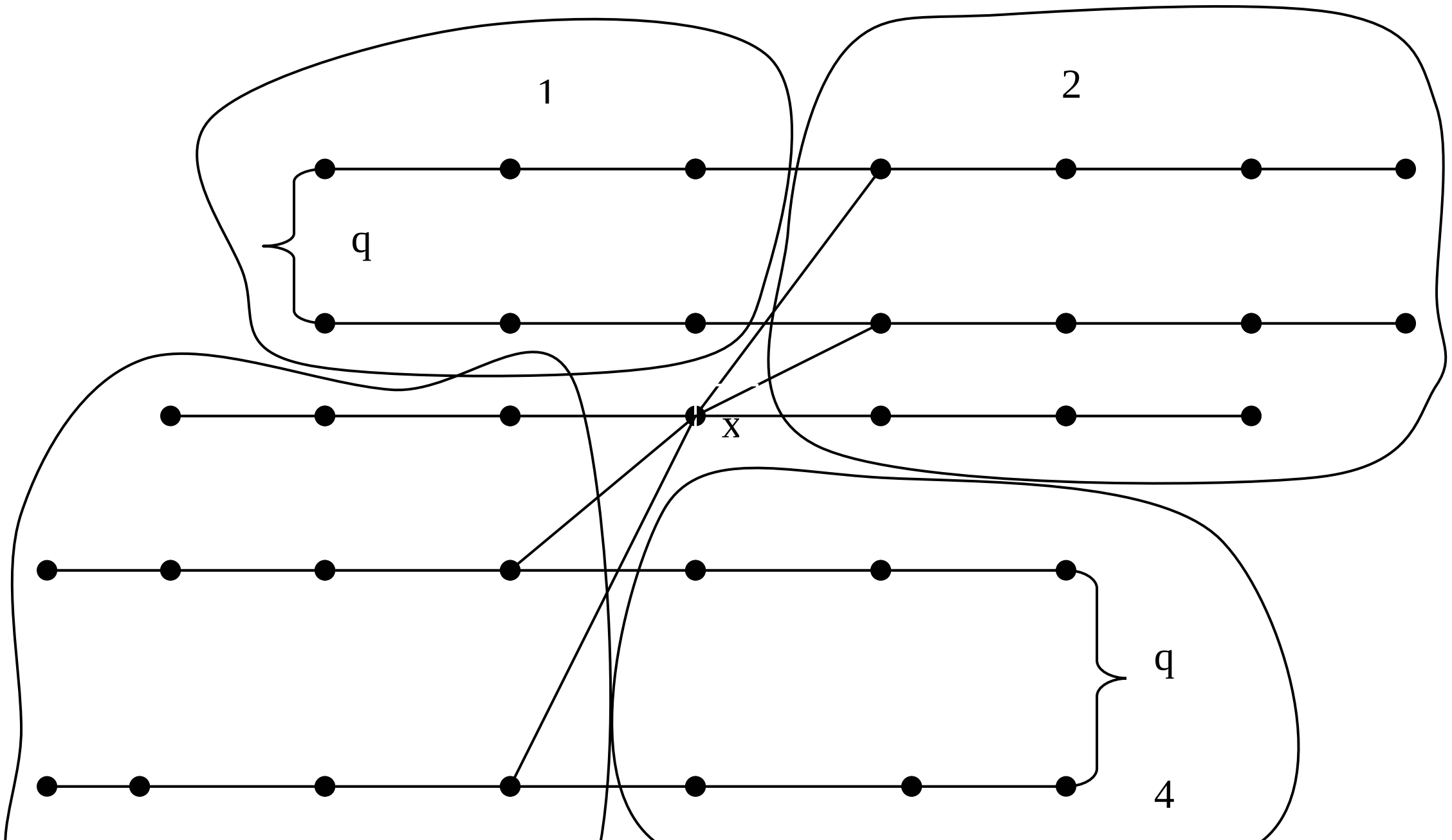

նկար 1

Կատարելով լրացուցիչ $4q$ համեմատություններ, պարզել, թե 1 և 4 տիրույթների որ տարրերն են մեծ կամ փոքր $x$-ից (սկզբում համեմատել յուրաքանչյուր եռյակի միջին տարրը $x$-ի հետ):

**Քայլ 4**: Արդյունքում մենք գտանք $r$ տարրեր, որոնք մեծ են $x$-ից, և $n-1-r$-ը՝ փոքր $x$-ից: Նկատենք, որ եթե $t = r + 1$, ապա $x$-ը պատասխանն է,
եթե $t < r + 1$, ապա մենք պետք է գտնենք $t$–րդ ամենամեծ տարրը $r$ տարրերի մեջ,
եթե $t > r + 1$, ապա մենք պետք է գտնենք $t - 1 - r$–րդ ամենամեծ տարրը $n - 1 - r$ տարրերի մեջ:

Քանի որ և $r$-ը և $n - 1 - r$-ը չեն գերազանցում $10q + 3$-ը (1 և 4 տիրույթների չափերին գումարել 2 կամ 3 տիրույթների չափերը), ապա համաձայն ինդուկցիոն ենթադրության, այս քայլը կարելի է իրականացնել ոչ ավել, քան $15(10q + 3) - 163$ համեմատությունների միջոցով:

Արդյունքում ոչ շատ, քան

$$13(2q + 1) + 30q - 148 + 4q + 15(10q + 3) - 163 = 15(14q - 6) - 163$$

համեմատությունների միջոցով, մենք կորոշենք $a_1, \ldots, a_n$ տարրերից $t$-րդ ամենամեծը: Հաշվի առնելով, որ ամենասկզբում մենք սկսել էինք $t$-րդ ամենամեծը տարրի որոնումն առնվազն $14q - 6$ տարրերից, թեորեմն ապացուցված է:

## Գրականություն

Գտնված սխալների, առաջարկությունների, ինչպես նաև դասախոսություն-ներն e-mail-ով ստանալու համար կարող էք դիմել vahanmkrtchyan2002@yahoo.com հասցեով:

Կոմբինատորային ալգորիթմներ և ալգորիթմների վերլուծություն
Վահան Վ. Մկրտչյան

Դասախոսություն 11: Գրաֆի լայնությամբ շրջանցում: Էյլերյան ցիկլ: Գրաֆի կապակցվածության բաղադրիչներ:

Գրաֆի լայնությամբ շրջանցում և կապակցվածության բաղադրիչներ: Նախ հիշենք գրաֆների տեսության որոշ հասկացություններ: Դիցուք $V=\{v_1,...,v_p\}$ վերջավոր բազմություն է, իսկ $E$–ն $V$–ի երկու տարր պարունակող ենթաբազմությունների բազմության ինչ-որ ենթաբազմություն է, այսինքն՝

$$E \subseteq \{\{u,v\}/u,v \in V\}:$$

Այդ դեպքում $G=(V,E)$ կարգավոր զույգին կանվանենք գրաֆ: $V$–ի տարրերին կանվանենք $G$ գրաֆի գագաթներ, իսկ $E$–ի տարրերին $G$ գրաֆի կողեր:

Եթե $e=\{u,v\}\in E$ ապա կասենք, որ $e$ կողը ինցիդենտ է $u,v$ գագաթներին, իսկ $u,v$ գագաթներին կանվանենք կից:

Բացի գրաֆների՝ հարթության վրա պատկերման քաջ հայտնի եղանակից, մենք շատ հաճախ կօգտվենք հետևյալ երեք ներկայացումներից.

Ա) $G=(V,E)$ գրաֆին համապատասխանեցնենք $A=(a_{ij})$ $p\times p$ չափի մատրիցը, որտեղ

$$a_{ij}=\begin{cases}1, \text{ եթե } \{v_i,v_j\}\in E\\ 0, \text{ եթե } \{v_i,v_j\}\notin E\end{cases}$$

$A=(a_{ij})$ մատրիցին կանվանենք $G=(V,E)$ գրաֆի կցության մատրից:

Բ) $G=(V,E)$ գրաֆին համապատասխանեցնենք $B=(b_{ij})$ $p\times q$ չափի մատրիցը, որտեղ $E=\{e_1,...,e_q\}$ և

$$b_{ij} = \begin{cases} 1, \text{ եթե } v_i \text{ գագաթը ինցիդենտ է } e_j \text{ կողին} \\ 0, \text{ եթե } v_i \text{ գագաթը ինցիդենտ չէ } e_j \text{ կողին} \end{cases}$$

$B = (b_{ij})$ մատրիցին կանվանենք $G = (V, E)$ գրաֆի ինցիդենտության մատրից:

Գ) $G = (V, E)$ գրաֆին համապատասխանեցնենք $Adj$ մասիվը, որի յուրաքանչյուր $Adj[v]$ էլեմենտ իրենից ներկայացնում է ցուցակ՝ կազմված $v \in V$ գագաթին կից գագաթներից գրված կամայական կարգով: Գրաֆների այս ներկայացմանը կանվանենք գրաֆի կից գագաթներով ներկայացում:

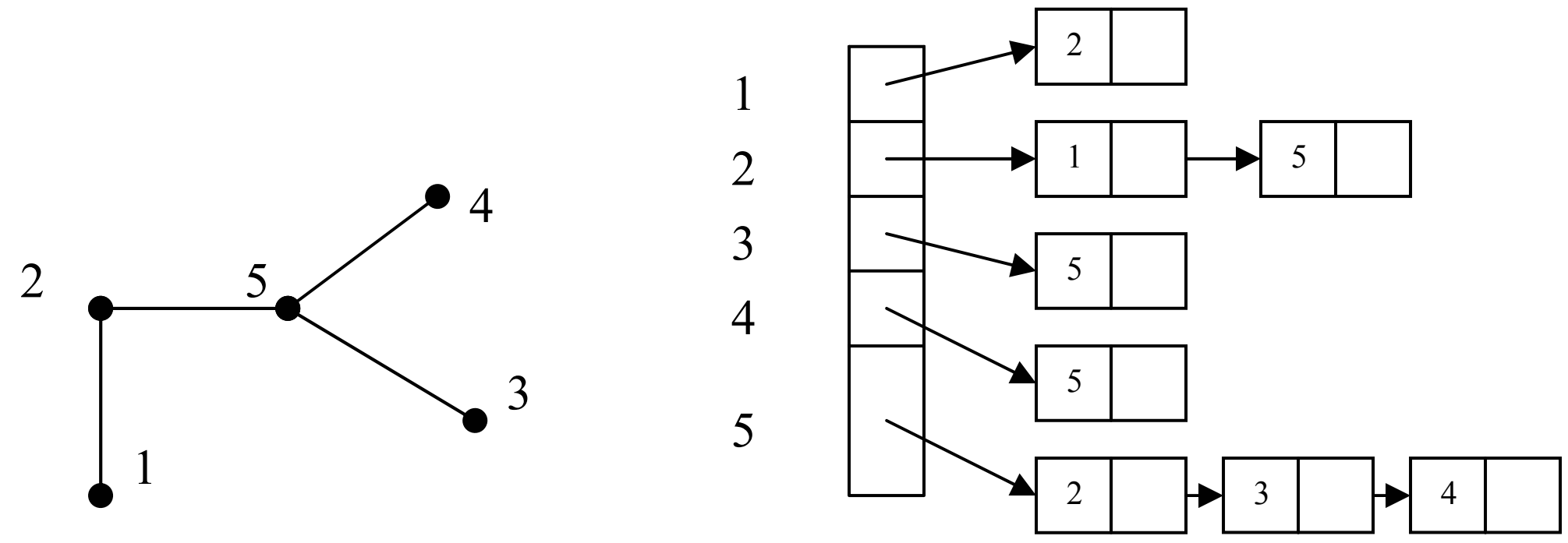

Նկար 1

Նկատենք, որ այս ներկայացումը պահանջում է $O(|V| + |E|)$ հիշողություն ($|V|$ հատ էլեմենտ պարունակում է մասիվը, իսկ ցուցակներում էլեմենտների քանակը հավասար է երկու անգամ կողերի քանակին): Այս ներկայացումից մենք կօգտվենք նաև կողմնորոշված գրաֆների ներկայացման համար:

Նկատենք նաև, որ $a_{ii} = 0 \quad i = 1, ..., p$: $v \in V$ գագաթի համար $d_G(v)$-ով նշանակենք $v$ գագաթի աստիճանը, այսինքն՝ նրան ինցիդենտ կողերի քանակը: Օգտվելով կցության կամ ինցիդենտության մատրիցներից, հեշտությամբ կարելի է ապացուցել հետևյալ

**Թեորեմ 1**: $2|E| = \sum_{v \in V} d_G(v)$:

**Հետևանք**: Ցանկացած գրաֆում կենտ աստիճան ունեցող գագաթների քանակը զույգ է:

Դիցուք տրված են $G = (V, E)$ և $G_1 = (V_1, E_1)$ գրաֆները: Կասենք, որ $G_1$ գրաֆը $G$ գրաֆի ենթագրաֆ է, եթե $V_1 \subseteq V, E_1 \subseteq E$: Դիցուք $V' \subseteq V$: $G$ գրաֆի այն ենթագրաֆը, որի գագաթների բազմությունը $V'$-ն է և որը պարունակում է առավելագույն թվով կողեր, կոչվում է $G$ գրաֆի $V'$ բազմությամբ ծնված ենթագրաֆ:

$G=(V,E)$ գրաֆի գագաթների $u_1,u_2,...,u_{k-1},u_k$ հաջորդականությունը կանվանենք $u_1$ –ից $u_k$ ճանապարհ, եթե $\{u_1,u_2\},\{u_2,u_3\},...,\{u_{k-2},u_{k-1}\},\{u_{k-1},u_k\}$ -ն $G$ գրաֆի միմյանցից տարբեր կողեր են: Ճանապարհի երկարություն կանվանենք նրա մեջ մտնող կողերի քանակը: Ճանապարհը կանվանենք պարզ, եթե նրանում գագաթները չեն կրկնվում: $u_1$ –ից $u_k$ ճանապարհը կանվանենք ցիկլ, եթե $u_1=u_k$: Ցիկլը կանվանենք պարզ, եթե համապատասխան ճանապարհը պարզ է:

Նկատենք, որ եթե գրաֆում գոյություն ունի $u$ –ից $v$ ճանապարհ, ապա գոյություն ունի նաև $v$ –ից $u$ ճանապարհ, ավելին գոյություն ունի նաև $u$ –ից $v$ պարզ ճանապարհ:

$G=(V,E)$ գրաֆը կանվանենք կապակցված, եթե նրա ցանկացած երկու գագաթների միջև գոյություն ունի այդ գագաթները միացնող ճանապարհ:

Կգրենք $u \to v$, եթե $G=(V,E)$ գրաֆում գոյություն ունի $u$ –ից $v$ ճանապարհ: Նկատենք, որ

ա) $u \to u$

բ) եթե $u \to v$, ապա $v \to u$ ;

գ) եթե $u \to v$, $v \to w$, ապա $u \to w$:

Փաստորեն, $u \to v$ բինար հարաբերությունը համարժեքության հարաբերություն է, և հետևաբար, այս հարաբերությունը $G=(V,E)$ գրաֆի գագաթների $V$ բազմությունը տրոհում է միմյանց հետ հատում չունեցող ենթաբազմությունների՝

$$V=V_1 \cup ... \cup V_k,\ V_i \cap V_j = \varnothing$$

այնպես, որ $V_i$ բազմության պատկանող գագաթները միմյանց կապակցված են, իսկ տարբեր $V_i$ -երին պատկանող գագաթները միմյանց կապակցված չեն:

$G=(V,E)$ գրաֆի $V_1,...,V_k$ բազմություններով ծնված ենթագրաֆներին կանվանենք կապակցվածության բաղադրիչներ:

Դիտարկենք գրաֆի գագաթների լայնությամբ շրջանցման (breadth-first search) ալգորիթմը, որը հիմք է հանդիսանում գրաֆների տեսության մի շարք խնդիրների լուծման համար:

Ալգորիթմը մուտքին ստանում է $G=(V,E)$ գրաֆը և վերադարձնում նրա $BFS(G)$ ենթագրաֆը: Աշխատանքի ընթացքում գագաթները ստանում են նշումներ, որը թույլ է տալիս յուրաքանչյուր գագաթ դիտարկել մեկ անգամ:

**Քայլ 1**: Վերցնել մի որևէ $s \in V$ գագաթ, նշել նրան և ավելացնել $Q$-նախապես դատարկ հերթին:

**Քայլ 2**: Քանի դեռ $Q \neq \varnothing$ կատարել

$Q$-ից հանել նրա առաջին $v$ տարրը:

Դիցուք $v_1,...,v_k$-ն $v$-ին կից այն գագաթներն են, որոնք դեռևս նշված չեն:
Նշել $v_1,...,v_k$ գագաթները և նրանց ավելացնել $Q$ հերթին:
$(v,v_1),...,(v,v_k)$ կողերն ավելացնել $BFS(G)$ գրաֆին:
**Քայլ 3**: $G=(V,E)$ գրաֆից հեռացնել $BFS(G)$ գրաֆի գագաթների բազմությունը: Եթե $V=\varnothing$ ապա վերադարձնել $BFS(G)$ գրաֆը, հակառակ դեպքում՝ անցնել Քայլ 1-ին:

Ալգորիթմի աշխատանքից երևում է, որ $BFS(G)$ գրաֆը իրականում անտառ է: Գնահատենք ալգորիթմի բարդությունը, ենթադրելով, որ այն ներկայացված է կից գագաթների ցուցակի միջոցով:
Նկատենք, որ յուրաքանչյուր $v$ գագաթ $Q$ հերթի մեջ լինում է մեկ անգամ: Հետևաբար քայլ 2-ում գրված ցիկլը կկատարվի ոչ շատ քան $|V|$ անգամ:
Հերթի հետ առնչվող գործողությունները պահանջում են $O(1)$ ժամանակ: Կից գագաթների ցուցակը դիտարկվում է այն ժամանակ, երբ գագաթ է հանվում այնտեղից: Այս ցուցակների երկարությունը $2|E|$-է, հետևաբար նրա մշակման համար անհրաժեշտ ժամանակը $O(|E|)$ է: Արդյունքում՝ ալգորիթմի բարդությունը կլինի $O(|V|+|E|)$:

Օգտվելով գրաֆի լայնությամբ շրջանցման ալգորիթմից կարելի է լուծել հետևյալ խնդիրը. տրված է $G=(V,E)$ գրաֆը, պահանջվում է գտնել նրա կապակցվածության բաղադրիչները:
Դիտարկենք այդ խնդիրը լուծող հետևյալ ալգորիթմը.

**Քայլ 1**: Կառուցել $BFS(G)$ գրաֆը:
**Քայլ 2**: Կառուցված $BFS(G)$ գրաֆի գագաթների բազմությունը կլինի $G$ գրաֆի կապակցվածության բաղադրիչների գագաթների բազմությունը:

**Էյլերյան ցիկլ**: $G=(V,E)$ գրաֆի ցիկլը կանվանենք էյլերյան, եթե այն անցնում է գրաֆի բոլոր կողերով, ընդ որում յուրաքանչյուր կողով ճիշտ մեկ անգամ: $G=(V,E)$ գրաֆը կանվանենք էյլերյան, եթե այն պարունակում է էյլերյան ցիկլ:
Ստորև բերված նկարներից առաջինում պատկերված է էյլերյան, իսկ երկրորդում՝ ոչ էյլերյան գրաֆ:

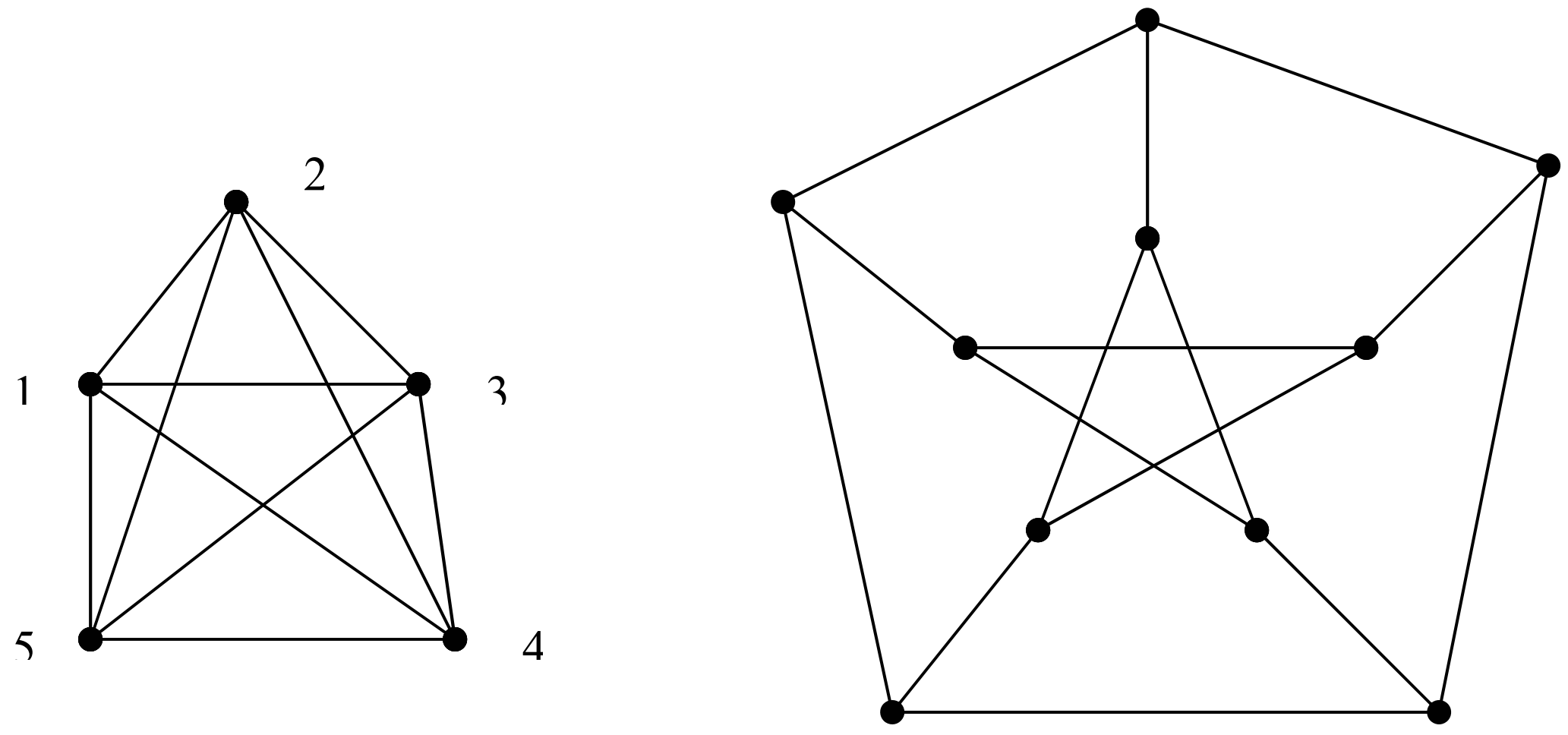

նկար 2

Առաջին գրաֆում էյլերյան ցիկլի օրինակ է 1,2,3,4,5,1,3,5,2,4,1 ցիկլը: Հիմա փորձենք նկարագրել էյլերյան գրաֆները: Նշենք, որ այս նկարագրից, մասնավորապես կհետևի նկար 2-ում բերված գրաֆի ոչ էյլերյան լինելը:

**Թեորեմ 2**: Որպեսզի $G=(V,E)$ գրաֆը լինի էյլերյան, անհրաժեշտ է և բավարար, որ այն լինի կապակցված և նրա բոլոր գագաթների աստիճանները լինեն զույգ թիվ:

**Ապացույց**: Նախ նկատենք, որ եթե գրաֆը էյլերյան է, այսինքն՝ պարունակում է էյլերյան ցիկլ, ապա այն կապակցված է: Դիտարկենք նրա մի որևէ էյլերյան ցիկլ: Նկատենք, որ այդ ցիկլի վրա յուրաքանչյուր գագաթ ինցիդենտ է զույգ թվով կողերի (ցիկլի վրա բոլոր կողերը տարբեր են): Քանի որ գրաֆի բոլոր կողերը գտնվում են ցիկլի վրա, ապա պարզ է, որ գրաֆի բոլոր գագաթների աստիճանները կլինեն զույգ:

Ցույց տանք, որ հակառակն էլ է ճիշտ. եթե ունենք կապակցված $G=(V,E)$ գրաֆը, որի բոլոր աստիճանները զույգ են, ապա այն պարունակում է էյլերյան ցիկլ:

Քանի որ $G=(V,E)$ գրաֆի բոլոր աստիճանները զույգ են, ապա նրա ցանկացած կող պատկանում է գոնե մեկ ցիկլի, որի բոլոր կողերը միմյանցից տարբեր են: Դիտարկենք $G=(V,E)$ գրաֆի այդպիսի ցիկլերից այն $C$ ցիկլը, որը ներառում է առավելագույն թվով կողեր: Ցույց տանք, որ գրաֆի բոլոր կողերը պատկանում են այդ ցիկլին: Ենթադրենք հակառակը, դիցուք գոյություն ունի $G=(V,E)$ գրաֆի կող, որը չի պատկանում $C$ ցիկլին: Քանի որ գրաֆը կապակցված է, ապա գոյություն կունենա $e=(u,v)$ կող, որը չի պատկանում $C$ ցիկլին, բայց $u,v$ գագաթներից գոնե մեկը պատկանում է ցիկլին: Դիտակենք $G\setminus E(C)$ գրաֆը, $G$-ից հանենք $C$–ի կողերը: Նկատենք, որ ստացված գրաֆում ցանկացած գագաթի աստիճանը զույգ է, հետևաբար, ըստ վերը ասվածի, $e=(u,v)$ կողը կպատկանի մի որևէ $C'$ ցիկլի, որի կողերը միմյանցից տարբեր են: Դիտարկենք $G$ գրաֆի $C''$ ցիկլը, որը ստացվում է հետևյալ կերպ. սկսելով $u$-ից շրջանցենք $C$–ն իսկ հետո՝ $C'$-ը:

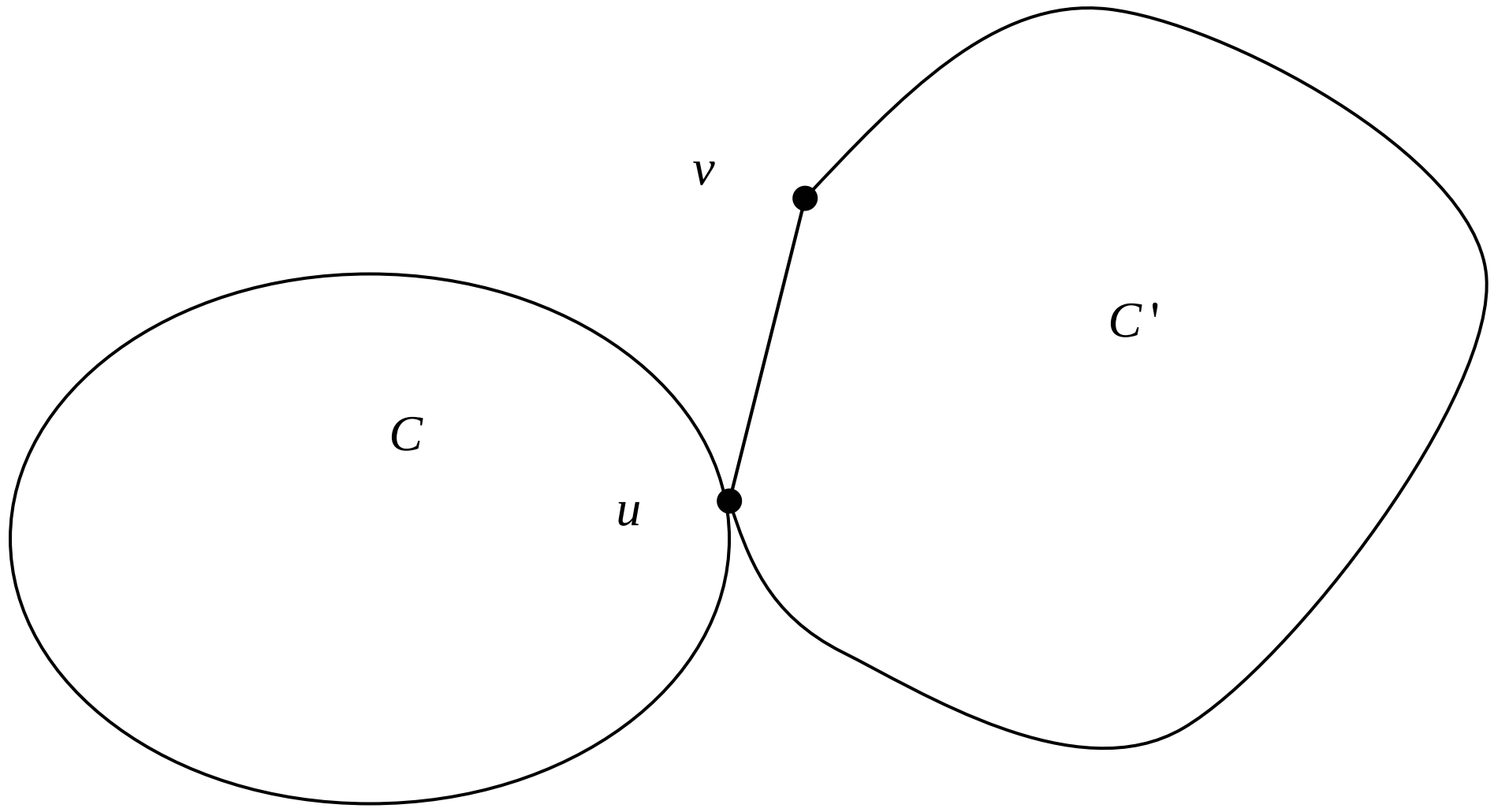

Նկար 3

Նկատենք, որ այն պարունակում է $C$–ից շատ կող, ինչը հակասում է $C$–ի ընտրությանը: Հետևաբար, գրաֆի բոլոր կողերը պատկանում են $C$ ցիկլին: Թեորեմն ապացուցված է:

Նկատենք, որ ապացուցված թեորեմը փաստորեն իր մեջ պարունակում է ալգորիթմ հետևյալ խնդիրը լուծելու համար. տրված է $G=(V,E)$ էյլերյան գրաֆը, պահանջվում է կառուցել նրա մի որևէ էյլերյան ցիկլ: Իրոք, դիտարկենք հետևյալ ալգորիթմը.

**Քայլ 1**: Վերցնել $G=(V,E)$ գրաֆի մի որևէ $C$ ցիկլ, որի կողերը միմյանցից տարբեր են:
**Քայլ 2**: Եթե գրաֆի բոլոր կողերը պատկանում են $C$ ցիկլին, ապա ավարտել ալգորիթմի աշխատանքը, հակառակ դեպքում՝ գտնել գրաֆի $e=(u,v)$ կող, որը չի պատկանում այդ ցիկլին, և որին ինցիդենտ գագաթներից գոնե մեկը գտնվում է ցիկլի վրա:
**Քայլ 3**: Կառուցել $G\setminus E(C)$ գրաֆի $e=(u,v)$ կողը պարունակող $C'$ ցիկլ, որի կողերը միմյանցից տարբեր են:
**Քայլ 4**: Այս երկու ցիկլերի միջոցով կառուցել $C''$ ցիկլը, և $C:=C''$: Անցնել **Քայլ 1**-ի կատարմանը:

Նկատենք, որ ալգորիթմի բարդությունը չի գերազանցում $|E| \cdot O(|V| + |E|)$:

Վերջում նշենք, որ Ֆլյորին առաջարկել է չափազանց պարզ մի ալգորիթմ հետևյալ խնդիրը լուծելու համար. տրված է $G = (V, E)$ գրաֆը, պահանջվում է պարզել էյլերյան այն թե ոչ, և եթե այո, ապա կառուցել նրա մի որևէ էյլերյան ցիկլ: Ալգորիթմը հետևյալն է.

Վերցնել մի որևէ գագաթ, և յուրաքանչյուր անգամ անցած կողը հեռացնել: Չանցնել կողով, եթե նրա հեռացումը առաջացնում է գրաֆի երկու կապակցվածության բաղադրիչ չհաշված մեկուսացված գագաթները:

## Գրականություն

Գտնված սխալների, առաջարկությունների, ինչպես նաև դասախոսություն-ներն e-mail-ով ստանալու համար կարող էք դիմել vahanmkrtchyan2002@yahoo.com հասցեով:

Կոմբինատորային ալգորիթմներ և ալգորիթմների վերլուծություն
Վահան Վ. Մկրտչյան

Դասախոսություն 12: Գրաֆի խորությամբ շրջանցում: Կողմնորոշված գրաֆի ուժեղ կապակցվածության բաղադրիչների որոնում:

**Գրաֆի խորությամբ շրջանցում: Կողմնորոշված գրաֆի ուժեղ կապակցվածության բաղադրիչների որոնում:** Նախ հիշենք կողմնորոշված գրաֆների տեսությանն առնչվող որոշ հասկացություններ: Դիցուք $V=\{v_1,...,v_p\}$ վերջավոր բազմություն է, իսկ $E$–ն $V$–ի կարգավոր զույգերի բազմության ինչ-որ ենթաբազմություն է, այսինքն՝

$$E \subseteq \{(u,v)/u,v\in V, u\neq v\}:$$

Այդ դեպքում $G=(V,E)$ կարգավոր զույգին կանվանենք գրաֆ կողմնորոշված գրաֆ: $V$–ի տարրերին կանվանենք $G$ գրաֆի գագաթներ, իսկ $E$–ի տարրերին $G$ գրաֆի կողեր կամ աղեղներ:

Եթե $e=(u,v)\in E$ ապա կասենք, որ $e$ կողը (աղեղը) միացնում է $u$ գագաթը $v$ գագաթին, իսկ $u,v$ գագաթներին կանվանենք կից:

Բացի կողմնորոշված գրաֆների՝ հարթության վրա պատկերման քաջ հայտնի եղանակից, մենք շատ հաճախ կօգտվենք հետևյալ ներկայացումներից.

Ա) $G=(V,E)$ գրաֆին համապատասխանեցնենք $A=(a_{ij})$ $p\times p$ չափի մատրիցը, որտեղ

$$a_{ij}=\begin{cases}1, \text{ եթե } (v_i,v_j)\in X\\ 0, \text{ եթե } (v_i,v_j)\notin X\end{cases}$$

$A=(a_{ij})$ մատրիցին կանվանենք $G=(V,E)$ գրաֆի կցության մատրից:

Բ) $G=(V,E)$ գրաֆին համապատասխանեցնենք $Adj$ մասիվը, որի յուրաքանչյուր $Adj[v]$ էլեմենտ իրենից ներկայացնում է ցուցակ՝կազմված $v\in V$ գագաթին կից գագաթներից գրված կամայական կարգով: Գրաֆների այս ներկայացմանը կանվանենք կողմնորոշված գրաֆի կից գագաթներով ներկայացում:

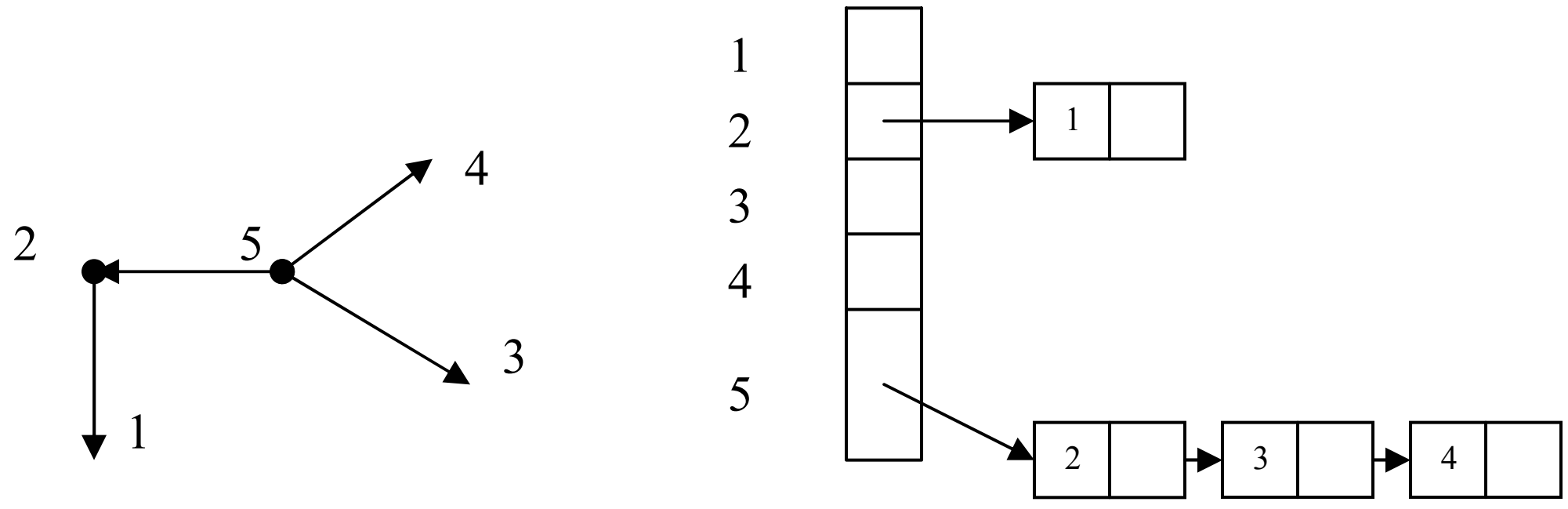

Նկար 1

Նկատենք, որ այս ներկայացումը պահանջում է $O(|V|+|E|)$ հիշողություն ($|V|$ հատ էլեմենտ պարունակում է մասիվը, իսկ ցուցակներում էլեմենտների քանակը հավասար է երկու անգամ կողերի քանակին): Այս ներկայացումից մենք կօգտվենք նաև կողմնորոշված գրաֆների ներկայացման համար:

Նկատենք նաև, որ $a_{ii}=0 \quad i=1,...,p$:

$G=(V,E)$ կողմնորոշված գրաֆի տրանսպոնացված գրաֆ ասելով կհասկանանք $G^T=(V,E^T)$ գրաֆը, որտեղ

$$E^T=\{(u,v)/(v,u)\in E\},$$

այլ կերպ ասած կողերի կողմնորոշման շուռ տալուց ստացված գրաֆը:

Նկատենք, որ $G=(V,E)$ գրաֆից $G^T=(V,E^T)$ գրաֆը կարելի է ստանալ $O(|V|+|E|)$ ժամանակում (այստեղ ենթադրվում է, որ $G=(V,E)$ գրաֆը ներկայացված է կից գագաթների ցուցակի միջոցով):

$G=(V,E)$ գրաֆի գագաթների $u_1,u_2,...,u_{k-1},u_k$ հաջորդականությունը կանվանենք $u_1$–ից $u_k$ ուղի, եթե $(u_1,u_2),(u_2,u_3),...,(u_{k-2},u_{k-1}),(u_{k-1},u_k)$-ն $G$ գրաֆի միմյանցից տարբեր կողեր (աղեղներ) են: Ուղու երկարություն կանվանենք նրա մեջ մտնող կողերի քանակը: Ուղին կանվանենք պարզ, եթե նրանում գագաթները չեն կրկնվում: $u_1$–ից $u_k$ ուղին կանվանենք ցիկլ կամ կոնտուր, եթե $u_1=u_k$: Ցիկլը (կոնտուրը) կանվանենք պարզ, եթե համապատասխան ուղին պարզ է:

Նկատենք, որ եթե գրաֆում գոյություն ունի $u$–ից $v$ ուղի, ապա կարող է գոյություն չունենալ $v$–ից $u$ ուղի:

Կգրենք $u\to v$ ($v$–ն հասանելի է $u$-ից), եթե $G=(V,E)$ գրաֆում գոյություն ունի $u$–ից $v$ ուղի: Նկատենք, որ

ա) $u\to u$

բ) եթե $u\to v$, $v\to w$, ապա $u\to w$:

$G=(V,E)$ գրաֆի ուժեղ կապակցվածության բաղադրիչ ասելով կհասկանանք գագաթների մաքսիմալ $U \subset V$ ենթաբազմություն, որում ցանկացած երկու $u \in U, v \in U$ գագաթների համար $u \to v$ և $v \to u$:

**Գրաֆի խորությամբ շրջանցում:** Գրաֆի խորությամբ շրջանցման ստրատեգիան կայանում է հետևյալում. "իջնել" այնքան, ինչքան հնարավոր է, և վերադառնալ ուման գալ նոր ճանապարհ, եթե այդպիսի կողեր չկան:

Նախքան ալգորիթմի նկարագիրը նշենք, որ աշխատանքի ընթացքում մենք գագաթներին կվերագրենք գույներ և ժամանակային նիշեր: Ալգորիթմի աշխատանքի սկզբում բոլոր գագաթներին կվերագրենք 0 գույնը: Հենց, որ մի որևէ $u$ գագաթ հայտնաբերվի, ապա նրան կվերագրենք 1 գույնը, և $d[u]$ թիվը (հայտնաբերման պահը): Հենց գագաթը լրիվությամբ դիտարկվի, ապա նրան կվերագրենք 2 գույնը և $f[u]$ թիվը՝ ավարտման պահ: Այս նշումները շատ հաճախ են օգտագործվում գրաֆների վերաբերյալ տարբեր խնդիրներ լուծման ժամանակ: Ստորև կդիտարկենք այդպիսի մի խնդիր. գտնել տրված կողմնորոշված գրաֆի ուժեղ կապակցվածության բաղադրիչները:

Վերոհիշյալ թվերը և գույները կբավարարեն հետևյալ պայմաններին. ցանկացած $u$ գագաթ կլինի

- 0 գույնով ներկված (0-անոց) մինչև $d[u]$ պահը,
- 1 գույնով ներկված (1-անոց) $d[u]$ պահից մինչև $f[u]$ պահը,
- 2 գույնով ներկված (2-անոց) $f[u]$ պահից հետո:

Նշենք նաև, որ այս ալգորթմը կիրառելի է ինչպես կողմնորոշված, այնպես էլ ոչ կողմնորոշված գրաֆների համար:

**Խորությամբ շրջանցման ալգորիթմի նկարագիրը**

Ալգորիթմը մուտքին ստանում է $G=(V,E)$ գրաֆը, և վերադարձնում է նրա $DFS(G)$ ենթագրաֆը, որին մենք կանվանենք նաև խորությամբ շրջանցման անտառ (ալգորիթմի նկարագրից հետևում է, որ $DFS(G)$ գրաֆն իրականում անտառ է):

**Քայլ** 1: $G=(V,E)$ գրաֆի բոլոր գագաթներին վերագրել 0 գույնը: $time := 0$ (սկզբնական պահի ֆիքսում)

**Քայլ 2**: FOR $u \in V(G)$ ($G$ –ի բոլոր գագաթների համար) DO
IF ($u$ -ն 0-անոց է) THEN ԱՅՑԵԼԵԼ($u$)

ԱՅՑԵԼԵԼ($u$)

**Քայլ 1**: $u$-ն դարձնել 1-անոց; $time := time+1$; $d[u] := time$;
**Քայլ 2**: FOR $v \in Adj[u]$ ($u$-ի բոլոր կից գագաթների համար) DO
IF ($v$-ն 0-անոց է) THEN
$(u,v)$ կողը ավելացնել $DFS(G)$ գրաֆին;
ԱՅՑԵԼԵԼ($v$);
**Քայլ 3**: $u$-ն դարձնել 2-անոց; $time := time+1$; $f[u] := time$;

Գնահատենք ալգորիթմի բարդությունը: Առաջին երկու քայլը պահանջում են $O(|V|)$ ժամանակ: ԱՅՑԵԼԵԼ($u$)-ի քայլ 1,3-ը կկատարվեն հաստատուն ժամանակում, իսկ քայլ 2-ում գրված ցիկլը կատարվում է $|Adj[u]|$ անգամ, և քանի որ

$$\sum_{v \in V} |Adj[v]| = O(|E|)$$

ապա կստանանք, որ ալգորիթմի աշխատանքի ընդհանուր ժամանակը կլինի $O(|V|+|E|)$:

Նշենք խորությամբ շրջանցման մի քանի հատկություններ՝

1) Ցանկացած $u \in V, v \in V$ գագաթների համար $[d[u], f[u]]$ և $[d[v], f[v]]$ հատվածները կամ չեն հատվում կամ մեկն ընկած է մյուսի մեջ: Իրոք, դիցուք $d[u] < d[v]$: Սա նշանակում է, որ $u$-ն ավելի շուտ է հայտնաբերվել, քան $v$-ն, հետևաբար համաձայն ԱՅՑԵԼԵԼ($u$)-ի նկարագրի, նախ կավարտվի $v$-ի դիտարկումը (որի արդյունքում $v$-ն կդառնա 2-նոց և կստանա $f[v]$ նշումը), որից հետո կավարտվի $u$-ի քննարկումը, այնպես որ՝ $[d[v], f[v]] \subset [d[u], f[u]]$:
2) $v$ գագաթը հանդիսանում է $u$ գագաթի հետնորդ $DFS(G)$ անտառում, այն և միայն այն դեպքում, երբ $d[u] < d[v] < f[v] < f[u]$:
3) $v$ գագաթը հանդիսանում է $u$ գագաթի հետնորդ $DFS(G)$ անտառում, այն և միայն այն դեպքում, երբ $d[u]$ պահին գոյություն ունի $u$-ից $v$ ուղի, որի բոլոր գագաթները 0-անոց են:

Դիտարկենք կողմնորոշված գրաֆի ուժեղ կապակցվածության բաղադրիչները գտնելու խնդիրը:

Ուժեղ կապակցվածության բաղադրիչներ($G$)
**Քայլ 1**: Կառուցել $DFS(G)$ անտառը ցանկացած $v$ գագաթում գրելով $f[v]$ թիվը
**Քայլ 2**; Կառուցել $G^T = (V, E^T)$ գրաֆը
**Քայլ 3**: Կառուցել $DFS(G^T)$ անտառը, ընդ որում խորությամբ շրջանցման ալգորիթմում գագաթները դիտարկել $f[v]$-ի արժեքների նվազմանը համապատասխան (սկզբում դիտարկել այն գագաթը, որի համար $f[v]$-ի արժեքն ավելի մեծ է)
**Քայլ 4**: $G=(V,E)$ գրաֆի ուժեղ կապակցվածության բաղադրիչները կլինեն քայլ 3-ում կառուցված $DFS(G^T)$ անտառի կոմպոնենտները:

Նկատենք, որ առաջարկված ալգորիթմի բարդությունը $O(|V|+|E|)$–է: Նախքան ալգորիթմի կոռեկտության ապացույցին անցնելը, ապացուցենք մի քանի օժանդակ հատկություններ:

**Լեմմա 1**: Եթե երկու գագաթ պատկանում են գրաֆի միևնույն ուժեղ կապակցվածության բաղադրիչին, ապա գոյություն չունի այդ գագաթները միացնող ուղի, որը դուրս է գալիս այդ բաղադրիչից:
**Ապացույց**: Դիցուք $S$–ը $G=(V,E)$ գրաֆի ուժեղ կապակցվածության բաղադրիչ է, և $u\in S, v\in S$, իսկ $w$-ն $u$-ն $v$-ին միացնող ուղու մի որևէ գագաթ է: Նկատենք, որ $u\to w$, $w\to u$, հետևաբար $w\in S$:
**Լեմմա 2**: Գրաֆի խորությամբ շրջանցման դեպքում միևնույն ուժեղ կապակցվածության բաղադրիչին պատկանող գագաթները մտնում են $DFS(G)$ անտառի նույն բաղադրիչի մեջ:
**Ապացույց**: Վերցնենք գրաֆի մի որևէ ուժեղ կապակցվածության բաղադրիչ: $r$-ով նշանակենք այդ բաղադրիչի այն գագաթը, որը առաջինն է հայտնաբերվում խորությամբ շրջանցման դեպքում ($d[r]\to \min$): Այդ պահին բաղադրիչի բոլոր գագաթները 0-անոց են: Համաձայն վերը նշված 3)-հատկության այս բաղադրիչի բոլոր գագաթները կհանդիսանան $r$-ի հետնորդներ, և հետևաբար կմտնեն $DFS(G)$ անտառի նույն բաղադրիչի մեջ:

$G=(V,E)$ գրաֆի ցանկացած $u$ գագաթի համար $\varphi(u)$–ով նշանակենք նրա նախահայրը, այսինքն՝

$$\varphi(u)=w,\ \ u\to w \text{ և } f[w]\to \max:$$

Նկատենք, որ քանի որ $u\to u$, ապա $f[u]\le f[\varphi(u)]$: Ցույց տանք, որ $\varphi(\varphi(u))=\varphi(u)$: Իրոք, $f[\varphi(u)]\le f[\varphi(\varphi(u))]$: Մյուս կողմից, քանի որ $u\to\varphi(\varphi(u))$, ապա $f[\varphi(\varphi(u))]\le f[\varphi(u)]$, հետևաբար՝ $f[\varphi(\varphi(u))]=f[\varphi(u)]$ և $\varphi(\varphi(u))=\varphi(u)$ քանի որ եթե երկու գագաթի համար $f$-ի արժեքները համընկնում են, ապա այդ գագաթները նույնն են:

Դիցուք $S$–ը $G=(V,E)$ գրաֆի ուժեղ կապակցվածության բաղադրիչ է: Դիտարկենք $v\in S$ գագաթը, որը բավարարում է $d[v]\to\min$ պայմանին: Նշենք որոշ հատկություններ.

- $S$–ի բոլոր գագաթները 0-անոց են (հակառակ դեպքում $d[v]\to\min$ չի)

- $S$ –ի բոլոր գագաթները հասանելի են ուղիով, որի բոլոր գագաթները 0-անոց են (ըստ լեմմա 1-ի $v$ -ից $S$ -ի մեկ այլ գագաթ տանող ուղի անցնում է $S$ -ով որի բոլոր գագաթները 0-անոց են)
- գոյություն չունի 1-անոց գագաթ, որը հասանել է $v$ -ից (իրոք, 1-անոց գագաթները կազմում են շղթա ծառի արմատից դեպի $v$, և եթե մի որևէ գագաթ այդ շղթայից լիներ հասանելի $v$ -ից, մենք կունենայինք, որ $S$ -ում կա գագաթ, որը 0-անոց չէ)
- $v$ -ից հասանելի ցանկացած $w$ 0-անոց գագաթ հասանելի է ուղիով, որի բոլոր գագաթները 0-անոց են ( իրոք, ուղու վրա չկան 1-անոց գագաթներ, մյուս կողմից խորությամբ շրջանցման դեպքում չեն ծագում կողեր, որոնք միացնում են 2-նոցը 0-անոցին)
- $S$ –ի բոլոր գագաթները կդառնան 2-անոց գագաթներ ԱՅՑԵԼԵԼ($v$)-ի կանչի ժամանակ, և հետևաբար կլինեն $v$ –ի հետնորդներ,
- $\varphi(v) = v$, իրոք, $v$ –ից հասանելի 2-անոց գագաթների համար ակնհայտորեն $f$ -ի արժեքը փոքր է $f[v]$ -ից, իսկ $v$ –ից հասանելի 0-անոց գագաթները կմշակվեն մինչև $f[v]$ պահը
- Ցանկացած $u \in S$ համար $\varphi(u) = v$: Իրոք, $S$ -ում $u$ -ից և $v$ –ից հասանելի գագաթների բազմությունը նույնն է:

Այս հատկություններից հետևում է, որ ցանկացած ուժեղ կապակցվածության բաղադրիչում գոյություն ունի գագաթ, որը հայտնաբերվում է առաջինը, մշակվում է վերջինը, և որը հանդիսանում է կոմպոնենտի բոլոր գագաթների նախահայրը:

**Թեորեմ 1**: $G = (V, E)$ կողմնորոշված գրաֆում ցանկացած $u$ գագաթի $\varphi(u)$ նախահայրը հանդիսանում է $u$ -ի նախնի $DFS(G)$ անտառում:

**Ապացույց**: Իրոք, եթե $u \in S$ ուժեղ կապակցվածության բաղադրիչին, ապա $S$ -ի այն գագաթը, որը բավարարում է $d[v] \to \min$ պայմանին հանդիսանում է նախահայր $S$ -ի բոլոր գագաթների համար:

**Հետևանք 1**: $G = (V, E)$ կողմնորոշված գրաֆի ցանկացած խորությամբ շրջանցման դեպքում $u$ և $\varphi(u)$ գագաթները պատկանում են միևնույն կոմպոնենտին:

**Ապացույց**: Ըստ լեմմա 2-ի միևնույն ուժեղ կապակցվածության բաղադրիչին պատկանող գագաթները մտնում են $DFS(G)$ անտառի նույն բաղադրիչի մեջ, իսկ $u$ և $\varphi(u)$ գագաթները այդպիսին են:

**Թեորեմ 2**: $G = (V, E)$ կողմնորոշված գրաֆում երկու գագաթ պատկանում են միևնույն ուժեղ կապակցվածության բաղադրիչին այն և միայն այն դեպքում, երբ նրանք ունեն ընդհանուր նախահայր խորությամբ շրջանցման դեպքում:

**Ապացույց**: Իրոք, եթե $u \in S, v \in S$ և $S$-ը ուժեղ կապակցվածության բաղադրիչ է, ապա եթե $w$-ով նշանակենք այն գագաթը $S$-ից, որը բավարարում է $d[w] \to \min$ պայմանին, ապա $\varphi(u) = \varphi(v) = w$: Մյուս կողմից, պարզ է, որ եթե $u, v$-ն ունեն ընդհանուր նախահայր, ապա նրանք պատկանում են միևնույն ուժեղ կապակցվածության բաղադրիչին:

Մենք պատրաստ ենք ապացուցելու, որ ձևակերպված ալգորիթմը լուծում է խնդիրը, այսինքն ցանկացած $G = (V, E)$ կողմնորոշված գրաֆի համար այն կառուցում է նրա ուժեղ կապակցվածության բաղադրիչները:
Նկատենք, որ վերը նշված հատկություններից հետևում է, որ ուժեղ կապակցվածության բաղադրիչների գտնելը հանգում է բոլոր գագաթների նախահայրերին գտնելուն: Հենց սրանում է կայանում քայլ 3-ի իմաստը:

**Թեորեմ 3**: Ձևակերպված ալգորիթմը կոռեկտ է:
**Ապացույց**: Ալգորիթմն ընտրում է $r$ գագաթը, որը բավարարում է $f[r] \to \max$ պայմանին: Այն կլինի բոլոր այն գագաթների նախահայրը, որոնցից այն հասանելի է (ավելի մեծ $f$ արժեք ունեցող գագաթ չկա): Համաձայն թեորեմ 1-ի բոլոր այդ գագաթներն էլ կլինեն $r$-ի հետնորդներ, հետևաբար առաջին ուժեղ կապակցվածության բաղադրիչը գտնված է:
Հեռացնելով գտնված ուժեղ կապակցվածության բաղադրիչը, ալգորիթմը վերցնում է $r'$ գագաթը, որը բավարարում է $f[r'] \to \max$ պայմանին: Մնացած գագաթներից բոլոր այն $u$ գագաթները, որոնցից հասանելի է $r'$-ը, կլինեն $r'$-ի հետնորդներ (նկատենք, որ հեռացված գագաթներից ոչ մեկը չի կարող հասանելի լինել $u$-ից, քանի որ այդ դեպքում $r$-ը կլիներ հասանելի $u$-ից, իսկ այդպիսինները հեռացված են): Հետևաբար, ըստ թեորեմ 1-ի բոլոր այդ գագաթներն էլ կլինեն $r'$-ի հետնորդներ, և երկրորդ ուժեղ կապակցվածության բաղադրիչը գտնված է, և այլն:

# Գրականություն

Գտնված սխալների, առաջարկությունների, ինչպես նաև դասախոսություն-ներն e-mail-ով ստանալու համար կարող էք դիմել vahanmkrtchyan2002@yahoo.com հասցեով:

Կոմբինատորային ալգորիթմներ և
ալգորիթմների վերլուծություն
Վահան Վ. Մկրտչյան

Դասախոսություն 13: Կողմնորոշված գրաֆում կարճագույն ուղու և գրաֆում կարճագույն ճանապարհի գտնելու խնդիրներ: Դեյկստրայի և Ֆլոյդի ալգորիթմների նկարագիրը: Կողմնորոշված գրաֆի տրանզիտիվ փակում:

Օրգրաֆում կարճագույն ուղու խնդիրը: Դիտարկենք $G=(V,E)$ կողմնորոշված գրաֆը, և ենթադրենք, որ նրա յուրաքանչյուր $(u,v)$ աղեղին համապատասխանեցված է $d(u,v)\geq 0$ թիվ, որին կանվանենք աղեղի երկարություն կամ կշիռ: $G=(V,E)$ գրաֆի $v_0,v_1,...,v_k$ ուղու երկարություն ասելով կհասկանանք $d(v_0,v_1)+d(v_1,v_2)+...+d(v_{k-1},v_k)$ թիվը:

Օրգրաֆում կարճագույն ուղու խնդիրը ձևակերպվում է հետևյալ կերպ. տրված $u$ և $v$ գագաթների համար գտնել $u$ գագաթը $v$ գագաթին միացնող այնպիսի ուղի որի երկարությունն ամենափոքրն է:

Ստորև կառաջարկենք ալգորիթմ գրաֆի ցանկացած $u$ և $v$ գագաթները միացնող կարճագույն ուղու և նրա երկարության գտնելու համար: Ալգորիթմի աշխատանքի ընթացքում յուրաքանչյուր $v\in V$ գագաթի կհամապատասխանեցնեք $l(v)$ թիվը, որը ցույց կտա $u$ և $v$ գագաթները միացնող արդեն գտած ուղու երկարությունը: Այս թիվը անընդհատ կփոխվի ավելի կարճ ուղի գտնելիս: Ինչ-որ պահից սկսած ալգորիթմը կեզրակացնի, որ արդեն գտել է $u$ և $v$ գագաթները միացնող կարճագույն ուղին, և այդ ժամանակ $v$ գագաթին համապատասխանող $l(v)$ նշումը կհամարվի հիմնական և ալգորիթմի հետագա աշխատանքի ընթացքում այն չի փոփոխվի:

*$G=(V,E)$ կողմնորոշված գրաֆում $u$ և $v$ գագաթները միացնող կարճագույն ուղու երկարության որոնման Դեյկստրայի ալգորիթմը*

**Քայլ 1** (Սկզբնական արժեքների վերագրում): $u$ գագաթին վերագրենք $l(u)=0$ նշումը: Այս նշումը կհամարենք **հիմնական** և ալգորթմի հետագա աշխատանքի ընթացքում այն չի փոփոխվի: Գրաֆի մնացած գագաթներին վերագրենք $l(w)$ **ժամանակավոր** նշումը, որտեղ

$$l(w)=\begin{cases} d(u,w), \text{ եթե } (u,w)\in E \\ +\infty, \text{ եթե } (u,w)\notin E : \end{cases}$$

**Քայլ 2**: Դիտարկել գրաֆի բոլոր **ժամանակավոր** նշում ունեցող գագաթները և նրանցից ընտրել այն $x^*$ գագաթը, որի $l(x^*)$ նշումը ամենափոքրն է: $x^*$ գագաթի $l(x^*)$ նշումը համարել **հիմնական** և որպես հաջորդ դիտարկվող գագաթ ընդունել $x^*$-ը, այսինքն՝ $p=x^*$:

**Քայլ 3**: Դիտարկել $p$ գագաթից դուրս եկող աղեղների բազմությունը, և $\{x/(p,x)\in E\}$ բազմությանը պատկանող և **ժամանակավոր** նշում ունեցող յուրաքանչյուր $x$ գագաթի նշումը փոխել **ժամանակավոր** $\min\{l(x),l(p)+d(p,x)\}$ նշումին:

**Քայլ 4** (Ավարտի պահի որոշում): Եթե $p=v$, ապա ավարտել ալգորիթմի աշխատանքը՝ $l(v)$-ն $u$ և $v$ գագաթները միացնող կարճագույն ուղու երկարությունն է պատասխանով, հակառակ դեպքում՝ վերադառնալ քայլ 2-ին:

**Թեորեմ 1**: Դեյկստրայի ալգորիթմը գտնում է $u$ և $v$ գագաթները միացնող կարճագույն ուղու երկարությունը:

**Ապացույց**: Ապացույցի համար վարվենք հետևյալ կերպ. մենք կապացուցենք ավելի ուժեղ պնդում:

$V_1$-ով նշանակենք **հիմնական** նշում ստացած գագաթների բազմությունը (սկզբում՝ $V_1=\{u\}$): Ինդուկցիայով ըստ $V_1$ բազմության հզորության, ապացուցենք, որ ա) $V_1$ բազմության ցանկացած $x$ գագաթի համար տեղի ունի $l(x)=\text{կարճ}(u,x)$ հավասարությունը, բ) ցանկացած $x\in V\setminus V_1$ համար $l(x)$ **ժամանակավոր** նշումը ցույց է տալիս միայն $V_1$ բազմության գագաթներից կազմված այն ուղու երկարությունը, որը միացնում է $u$ գագաթը $x$–ին և որի երկարությունը ամենափոքրն է:

Նախ նկատենք, որ պնդումը (վերը նշված ա և բ կետերը) ճիշտ է $|V_1|=1$ դեպքում (քայլ 1): Ենթադրենք, որ այն ճիշտ է $|V_1|=k$ դեպքում, և ապացուցենք, որ հերթական $x^*$ գագաթի ավելացումից հետո նշված ա) և բ) պնդումները մնում են ճիշտ:

Նախ ցույց տանք, որ ա) պնդումը մնում է ճիշտ $x^*$ գագաթի ավելացումից հետո, այսինքն՝ $l(x^*)=\text{կարճ}(u,x^*)$, որտեղ $x^*$ գագաթը բավարարում է $l(x^*)\to\min$:

Վերցնենք մի որևէ $P$՝ $u-x^*$ կարճագույն ուղի: Նկատենք, որ նախքան $x^*$ գագաթի $V_1$ բազմությանն ավելացնելը, տեղի ուներ $x^* \notin V_1$ առնչությունը, և հետևաբար՝ գոյություն կունենա $P$ շղթայի առաջին $x_1 \notin V_1$ գագաթ (նկար 1):

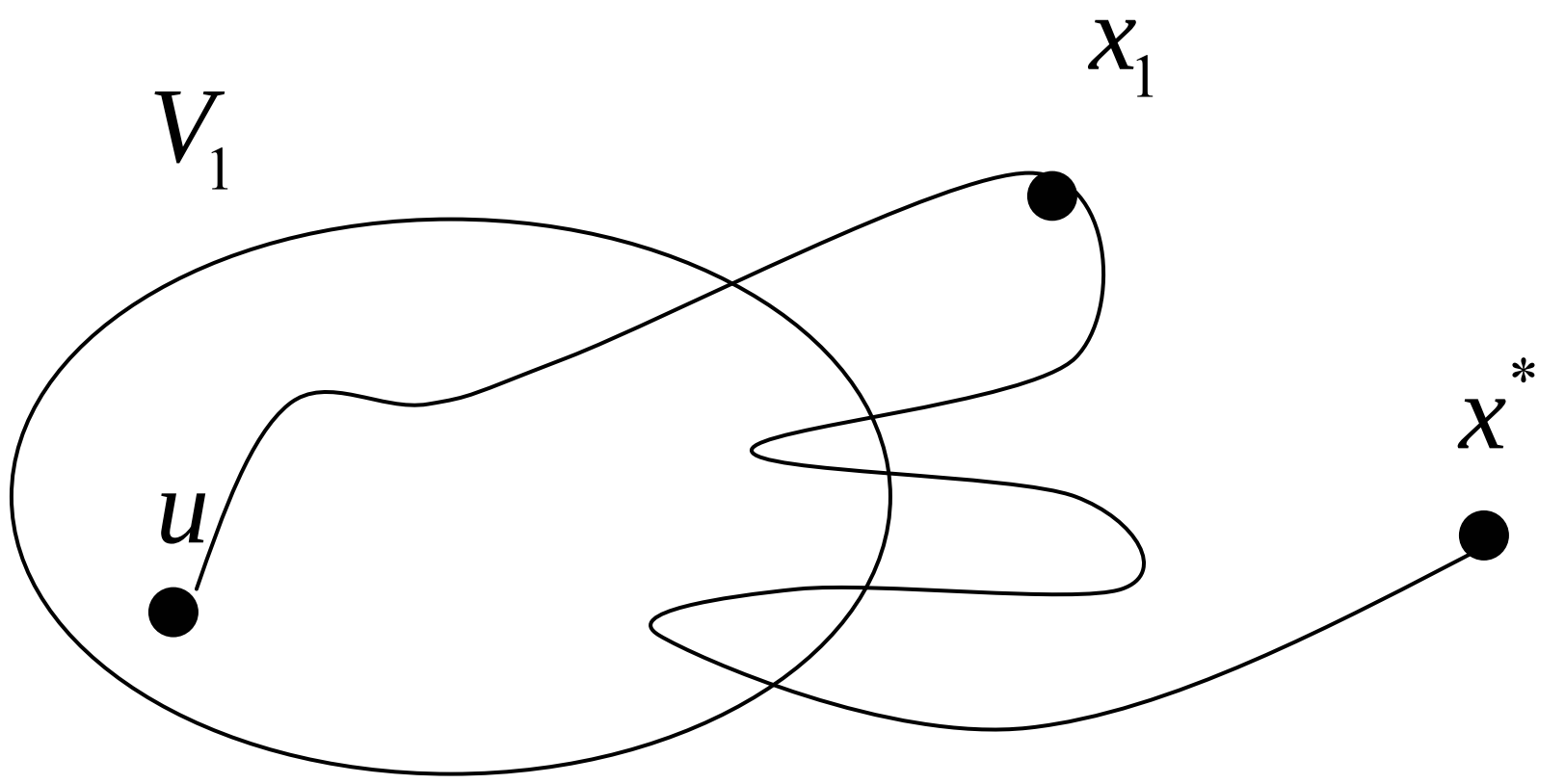

Նկար 1

Նկատենք, որ համաձայն ինդուկցիոն ենթադրության $l(x^*) \geq \text{կարճ}(u,x^*)$, $\text{երկ}(u,x_1) = l(x_1)$: Քանի որ՝

$$\text{կարճ}(u,x^*) = \text{երկ}(u,x_1) + \text{երկ}(x_1,x^*) \geq \text{երկ}(u,x_1) = l(x_1) \geq l(x^*),$$

ապա

$$l(x^*) = \text{կարճ}(u,x^*):$$

Հիմա ցույց տանք, որ $x^*$ գագաթի $V_1$ բազմությանն ավելացումից հետո բ) պնդումը ևս մնում է ճիշտ: Իրոք, $x^*$ գագաթի ավելացումից հետո **ժամանակավոր** նշում ունեցող գագաթներից նրանք, որոնք հարևան էին $x^*$ գագաթին, կարող էին $V_1$ բազմության գագաթներով ավելի կարճ ճանապարհով հասանելի լինել $x^*$ գագաթի միջոցով, բայց քայլ 3-ում այդ $x$ գագաթների նիշը փոխվում է $\min\{l(x), l(p)+d(p,x)\}$ **ժամանակավոր** նշումին՝ ապահելով $l(x)$-ի՝ բ) պնդմանը բավարարելը: Թեորեմն ապացուցված է:

Նկատենք, որ թեորեմի ապացույցից հետևում է, որ ալգորիթմի աշխատանքի ընթացքում **հիմնական** նշում ստացած գագաթներին համապատասխանող նշումները ցույց են տալիս $u$-ից այդ գագաթները միացնող կարճագույն ուղու երկարությունները, հետևաբար եթե մեզ պետք լինի գտնել $u$-ից մնացած բոլոր գագաթներ տանող կարճագույն ուղու երկարությունները, ապա ալգորիթմի քայլ 4-ում բավական է վարվել հետևյալ կերպ. կանգ առնել միայն այն դեպքում, երբ բոլոր գագաթների նշումները հիմնական են:

Նշենք նաև, որ եթե մեզանից պահանջվեր գտնել ոչ միայն $u$ և $v$ գագաթները միացնող կարճագույն ուղու երկարությունը, այլն գտնել այդպիսի մի ուղի, ապա Դեյկստարի ալգորիթմում կատարելով ստորև բերված ձևափոխությունը, մենք կկարողանանք լուծել նաև այս խնդիրը.

**Քայլ 5**: Որպես հերթական $w$ գագաթ ընտրել $v$ գագաթը, $w = v$:
**Քայլ 6**: Դիտարկել $w$ գագաթը, որի արդյունքում $w$-ին նախորդող գագաթների $\{x/(x,w)\in E\}$ բազմությունից ընտրել այնպիսի $x^*$ գագաթ, որի համար $l(x^*)+d(x^*,w)=l(w)$ և որպես հերթական $w$ գագաթ ընտրել $x^*$ գագաթը, $w = x^*$:
**Քայլ 7**: Եթե $w = u$, ապա ավարտել (ճանապարհը գտնված է), հակառակ դեպքում՝ անցնել Քայլ 6-ին:

Գնահատենք նաև Դեյկստրայի ալգորիթմի բարդությունը, այսինքն՝ ալգորիթմի աշխատանքի համար պահանջվող գործողությունների քանակը: Նկատենք, որ Քայլ 2 և 3-ում գրաֆի գագաթներից մեկը ստանում է **հիմնական** նշում, հետևաբար պարզ է, որ մուտքային $G=(V,E)$ գրաֆի համար ալգորիթմը կկատարի ոչ ավել քան $|V|$ քայլ (իտերացիա): Յուրաքանչյուր քայլում (իտերացիայում) ալգորիթմի գործողությունների քանակը $O(|V|)$–է, հետևաբար Դեյկստրայի ալգորիթմի բարդությունը կլինի $O(|V|^2)$:

**Օրգրաֆում կարճագույն ուղու խնդրի մեկ ընդհանրացման մասին:** Կրկին դիտարկենք $G=(V,E)$ կողմնորոշված գրաֆը, և ենթադրենք, որ պահանջվում է գտնել գրաֆի բոլոր գագաթների զույգերի միջև կարճագույն ուղիների երկարությունները և հենց այդ ուղիները: Պարզ է, որ ձևակերպած խնդիրը կարել է լուծել օգտվելով Դեյկստրայի ալգորիթմից՝ պարզապես գագաթների բոլոր զույգերի համար աշխատացնելով այդ ալգորիթմը: Ստորև կդիտարկենք նշված խնդրի լուծման ավելի էֆեկտիվ ալգորիթմ, այն է Ֆլոյդի ալգորիթմը:
Նշենք, որ եթե $G=(V,E)$ կողմնորոշված գրաֆի աղեղների բազմությանը համապատասխանեցված թվերի մեջ լինեն բացասականները, ապա Դեյկստրայի ալգորիթմը կիրառելի չէ: Նշենք նաև, որ եթե գրաֆի $u$ և $v$ գագաթների միջև կա ուղի, որը պարունակում է այպիսի կողմնորոշված ցիկլ (կոնտուր), որին պատկանող աղեղների երկարությունների գումարը բացասական թիվ է, ապա $u$ և $v$ գագաթների միջև կարճագույն ուղու որոնման խնդիրն ընդհանրապես լուծում չունի:
Ֆլոյդի ալգորիթմը կիրառելի է այն գրաֆների դեպքում, երբ գրաֆի աղեղների երկարությունները կարող են լինեն բացասական թվեր, սակայն որոնք չեն պարունակում կոնտուր, որի երկարությունը բացասական թիվ է:

Դիցուք տրված է $G=(V,E)$ կողմնորոշված գրաֆը, որտեղ $V=\{1,...,n\}$ և ենթադրենք, որ նրա յուրաքանչյուր $(i,j)$ աղեղին համապատասխանեցված է $d(i,j)$ թիվ, որին կանվանենք աղեղի երկարություն կամ կշիռ: Նշենք, որ $d(i,j)$ թիվը կարող է լինել բացասական: Ենթադրենք նաև, որ գրաֆը չի պարունակում կոնտուր, որի երկարությունը բացասական թիվ է: Այս ենթադրության դեպքում, պարզ է, որ եթե $i$ և $j$ գագաթների միջև կա ուղի, ապա կա նաև կարճագույն ուղի:

Դիտարկենք $n \times n$ չափի $D = (d(i, j))$ մատրիցը, որտեղ $d(i,i) = 0$ և $d(i, j) = \infty$ երբ $(i, j) \notin E$, $i, j = 1,...,n$: Ալգորիթմի աշխատանքի ընթացքում կկառուցվեն $D^{(0)}, D^{(1)},..., D^{(n)}$ մատրիցների հաջորդականություն այնպես, որ $D^{(n)}$ մատրիցի $d^{(n)}(i, j)$ տարրը $G = (V, E)$ կողմնորոշված գրաֆում $i$ և $j$ գագաթների միջև կարճագույն ուղու երկարությունն է:

Վերցնենք՝ $D^{(0)} = D$ և $k = 1,...,n$ համար $D^{(k)} = (d^{(k)}(i, j))$ մատրիցի տարրերը սահմանենք հետևյալ անրադարձ (ռեկուռենտ) առնչության միջոցով.

$$d^{(k)}(i, j) = \min\{d^{(k-1)}(i, j), d^{(k-1)}(i,k) + d^{(k-1)}(k, j)\},\ i, j = 1,...,n:$$

**Թեորեմ 2**: $D^{(k)}$ մատրիցի $d^{(k)}(i, j)$ տարրը ցույց է տալիս $i$ և $j$ գագաթների միջև այն ուղու երկարությունը, որը բոլոր ներքին գագաթները պատկանում են $\{1,...,k\}$ բազմությանը և որի երկարությունը ամենափոքրն է:

**Ապացույց**: Նկատենք, որ պնդումը ճիշտ է $k = 0$ դեպքում: Ենթադրենք, որ պնդումը ճիշտ է $k < l$ դեպքում և ցույց տանք, որ այն մնում է ճիշտ $k = l$ դեպքում: Ցանկացած $i, j, l = 1,...,n$ համար դիցուք $P_{ij}^{(l)}$-ը $i$ և $j$ գագաթները միացնող ուղի է, որի բոլոր ներքին գագաթները պատկանում են $\{1,...,l\}$ բազմությանը և որի երկարությունը ամենափոքրն է: Նկատենք, որ

$$\text{երկ}(P_{ij}^{(l)}) = \min\{\text{երկ}(P_{ij}^{(l-1)}), \text{երկ}(P_{il}^{(l-1)}) + \text{երկ}(P_{lj}^{(l-1)})\} =$$

$$= \min\{d^{(l-1)}(i, j), d^{(l-1)}(i,l) + d^{(l-1)}(l, j)\} = d^{(l)}(i, j):$$

**Հետևանք**: $D^{(n)}$ մատրիցի $d^{(n)}(i, j)$ տարրը $G = (V, E)$ կողմնորոշված գրաֆում $i$ և $j$ գագաթների միջև կարճագույն ուղու երկարությունն է:

$G = (V, E)$ կողմնորոշված գրաֆում $i$ և $j$ գագաթների միջև կարճագույն ուղին գտնելու համար սահմանենք նաև մատրիցների $Z^{(0)}, Z^{(1)},..., Z^{(n)}$ հաջորդականությունը, որտեղ $Z^{(k)}$ մատրիցի $z_{ij}^{(k)}$ տարրը ցույց է տալիս այն գագաթը, որը անմիջապես հաջորդում է $i$ գագաթին $P_{ij}^{(k)}$ ուղիում: Նկատենք, որ $z_{ij}^{(0)} = j$, երբ $d(i, j) \neq \infty$: Պայմանավորվենք $d(i, j) = \infty$ դեպքում ընդունել, որ $z_{ij}^{(0)} = 0$: Նկատենք, որ $Z^{(k)}$ մատրիցի $z_{ij}^{(k)}$ տարրը $Z^{(k-1)}$ մատրիցից որոշվում է հետևյալ կերպ.

$$z_{ij}^{(k)} = \begin{cases} z_{ij}^{(k-1)} & \text{եթե } d^{(k)}(i, j) = d^{(k-1)}(i, j) \\ z_{ik}^{(k-1)} & \text{եթե } d^{(k)}(i, j) < d^{(k-1)}(i, j): \end{cases}$$

Ունենալով $Z^{(n)}$ մատրիցը, $G = (V, E)$ կողմնորոշված գրաֆում $i$ և $j$ գագաթների միջև կարճագույն $i, i_1, i_2,..., i_p, j$ ուղին որոշվում է հետևյալ կերպ.

$$i_1 = z_{ij}^{(n)},\ i_2 = z_{i_1 j}^{(n)}, \ldots, i_p = z_{i_{p-1} j}^{(n)},\ j = z_{i_p j}^{(n)}:$$

*$G=(V,E)$ կողմնորոշված գրաֆում ցանկացած երկու գագաթների զույգերի միջև կարճագույն ուղին գտնող Ֆլոյդի ալգորիթմի նկարագիրը*

**Քայլ 1**: Կառուցել մատրիցների $D^{(0)},D^{(1)},...,D^{(n)}$ և $Z^{(0)},Z^{(1)},...,Z^{(n)}$ հաջորդականությունը:
**Քայլ 2**: Եթե գոյություն ունի $D^{(n)}$ մատրիցի $d^{(n)}(i,i)<0$, ապա $G=(V,E)$ գրաֆի $i$ տարրը պատկանում է բացասական երկարությամբ ցիկլի, հակառակ դեպքում՝ գրաֆում $i$ և $j$ գագաթների միջև կարճագույն $i,i_1,i_2,...,i_p,j$ ուղին որոշվում է վերը նշված բանաձևով:

Նկատենք, որ քանի որ ալգորիթմը պարունակում է 3 ներդրված ցիկլեր, ապա նրա բարդությունը $O(n^3)$ –է:

**Գրաֆում կարճագույն ճանապարհի խնդիրներ**: Դիտարկենք $G=(V,E)$ սովորական գրաֆը, և ենթադրենք, որ նրա յուրաքանչյուր $\{u,v\}$ կողին վերագրված է ոչ բացասական $d(\{u,v\})\geq 0$ թիվ: Գրաֆի $v_0,v_1,...,v_k$ ճանապարհի երկարություն ասելով կհասկանանք $d(v_0,v_1)+d(v_1,v_2)+...+d(v_{k-1},v_k)$ թիվը:
Գրաֆում կարճագույն ճանապարհի խնդիրը ձևակերպվում է հետևյալ կերպ. գտնել $u$ և $v$ գագաթները միացնող այնպիսի ճանապարհ, որի երկարությունն ամենափոքրն է:
Ցույց տանք, որ նշված խնդիրը կարելի է հանգեցնել կողմնորոշված գրաֆում կարճագույն ուղու խնդրին: $G=(V,E)$ սովորական գրաֆի համար դիտարկենք $\vec{G}=(V,\vec{E})$ կողմնորոշված գրաֆը, որտեղ

$$(u,v)\in\vec{E}\Leftrightarrow\{u,v\}\in E \text{ (նկար 2):}$$

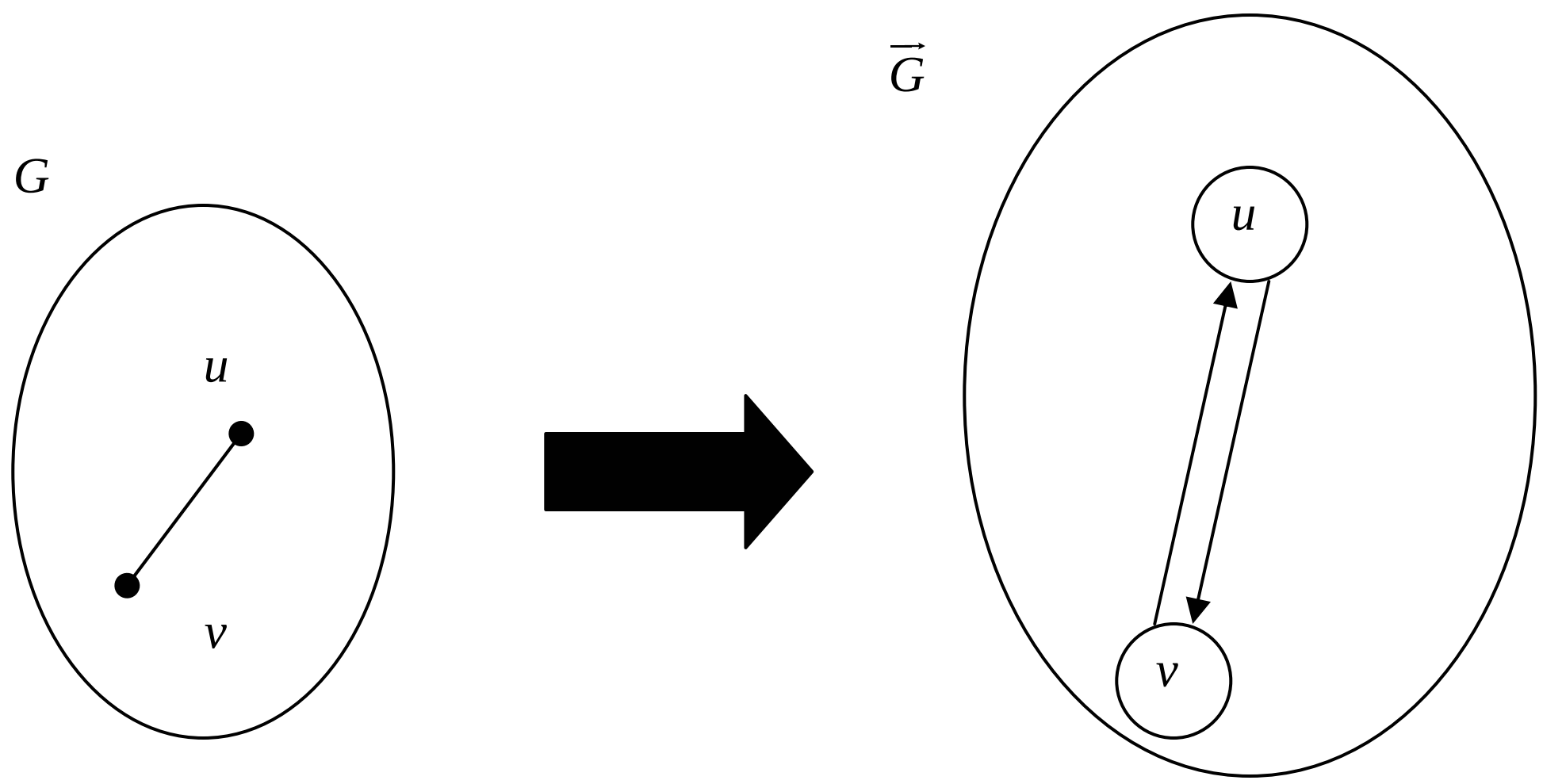

Նկար 2

Պարզ է, որ $G=(V,E)$ գրաֆում $u$ և $v$ գագաթները միացնող յուրաքանչյուր ճանապարհի կհամապատասխանի $\vec{G}=(V,\vec{E})$ կողմնորոշված գրաֆի նույն երկարություն ունեցող և $u$ գագաթը $v$ գագաթին միացնող ուղի: Հետևաբար՝ $G=(V,E)$ գրաֆում $u$ և $v$ գագաթները միացնող կարճագույն ճանապարհը գտնելու համար բավական է $\vec{G}=(V,\vec{E})$ կողմնորոշված գրաֆում գտնել $u$ գագաթը $v$ գագաթին միացնող ուղի, ինչը կարող ենք անել, օրինակ, Դեյկստրայի ալգորիթմի միջոցով:

Վերջում նշենք նաև, որ կողմնորոշված գրաֆների համար վերը նկարագրված ալգորիթմները օգտագործելով կարելի է

- գտնել սովորական գրաֆում տրված գագաթից մինչև մնացած գագաթները տանող կարճագույն ճանապարհի երկարությունը և որևէ կարճագույն ճանապարհ,
- գտնել գրաֆում ցանկացած երկու գագաթի միջև կարճագույն ճանապարհի երկարությունը և որևէ կարճագույն ճանապարհ, նույնիսկ այն դեպքում, երբ կողերի երկարությունները բացասական թվեր են:

**Կողմնորոշված գրաֆի տրանզիտիվ փակում**: $G=(V,E)$ կողմնորոշված գրաֆը կանվանենք **տրանզիտիվ**, եթե $(u,v)\in E$, $(v,w)\in E$, ապա $(u,w)\in E$: Նկատենք, որ ցանկացած $G=(V,E)$ կողմնորոշված գրաֆի համար գոյություն ունի $G^*=(V,E^*)$ տրանզիտիվ գրաֆ, որը բավարարում է $E\subseteq E^*$ պայմանին: Իրոք, բավարար է որպես $E^*$ վերցնել բոլոր աղեղների բազմությունը: Դժվար չէ տեսնել, որ եթե $E^*,E^{**}$ այնպիսին են, որ $G^*=(V,E^*)$ և $G^{**}=(V,E^{**})$ տրանզիտիվ գրաֆներ են, որոնք բավարարում են $E\subseteq E^*$, $E\subseteq E^{**}$ պայմաններին, ապա $G^{***}=(V,E^*\cap E^{**})$ գրաֆը կլինի տրանզիտիվ և $E\subseteq E^*\cap E^{**}$: Այստեղից հետևում է, որ ցանկացած $G=(V,E)$ կողմնորոշված գրաֆի համար գոյություն ունի **միակ մինիմալ** $G^+=(V,E^+)$ տրանզիտիվ գրաֆ, որը բավարարում է հետևյալ երկու պայմաններին.

- $E\subseteq E^+$
- ցանկացած $G^*=(V,E^*)$ տրանզիտիվ գրաֆի համար, որը բավարարում է $E\subseteq E^*$ պայմանին, տեղի ունի $E^+\subseteq E^*$:

Ընդունված է $G^+=(V,E^+)$ տրանզիտիվ գրաֆին անվանել $G=(V,E)$ կողմնորոշված գրաֆի տրանզիտիվ փակում: Դժվար չէ տեսնել, որ իրականում ցանկացած $G=(V,E)$ կողմնորոշված գրաֆի համար

$$E^+=\{(u,v)\in E\,/\,G \text{ գրաֆում գոյություն ունի } u \text{ գագաթը } v\text{-ին միացնող ուղի}\}:$$

Կողմնորոշված գրաֆի տրանզիտիվ փակման գտնելու խնդիրը ձևակերպվում է հետևյալ կերպ. տրված է $G=(V,E)$ կողմնորոշված գրաֆը, պահանջվում է գտնել նրա $G^+=(V,E^+)$ տրանզիտիվ փակումը:

Նկատենք, որ օգտագործելով Ֆլոյդի ալգորիթմը, կարելի է գտնել կողմնորոշված գրաֆի տրանզիտիվ փակումը: Իրոք, $G=(V,E)$ կողմնորոշված գրաֆի կողերին վերագրենք 1 երկարություն, և օգտագործելով Ֆլոյդի ալգորիթմը, գտնենք նրա ցանկացած երկու գագաթների միջև կարճագույն ուղու երկարությունը: Նկատենք, որ $G=(V,E)$ գրաֆի $u$ և $v$ գագաթների համար

$(u,v)\in E^{+}$ այն և միայն այն դեպքում, երբ $u$ գագաթը $v$-ին միացնող կարճագույն ուղու երկարությունը փոքր է $|V|$-ից:

Հաշվի առնելով այն հանգամանքը, որ գործնականում տրամաբանական գործողությունները մեքենաներում կատարվում են ավելի արագ, քան թվաբանականները, դիտարկենք գրաֆի տրանզիտիվ փակումը գտնելու մեկ այլ ալգորիթմ: $i,j,k=1,...,n$ համար, որտեղ $|V|=n$, $t_{ij}^{(k)}$ վերցնենք հավասար 1-ի, եթե $G=(V,E)$ գրաֆում գոյություն ունի $i$ գագաթը $j$ գագաթին միացնող ուղի է, որի բոլոր ներքին գագաթները պատկանում են $\{1,...,k\}$ բազմությանը, և վերցնենք 0՝ հակառակ դեպքում: Նկատենք, որ $(i,j)$ կողը կպատկանի $G=(V,E)$ գրաֆի տրանզիտիվ փակմանը, այն և միայն այն դեպքում, երբ $t_{ij}^{(n)}=1$: Եթե վերցնենք

$$t_{ij}^{(0)}=\begin{cases}0 & \text{եթե } i\neq j \text{ և } (i,j)\notin E\\ 1 & \text{եթե } i=j \text{ կամ } (i,j)\in E,\end{cases}$$

ապա $k\geq 1$ համար կունենանք՝

$$t_{ij}^{(k)}=t_{ij}^{(k-1)}\vee(t_{ik}^{(k-1)}\,\&\,t_{kj}^{(k-1)}):$$

Այսպիսով, գտնելով $\left(t_{ij}^{(n)}\right)$ մատրիցը մենք $O(n^3)=O(|V|^3)$ ժամանակում կգտնենք $G=(V,E)$ կողմնորոշված գրաֆի տրանզիտիվ փակումը:

## Գրականություն

Գտնված սխալների, առաջարկությունների, ինչպես նաև դասախոսություն-ներն e-mail-ով ստանալու համար կարող էք դիմել vahanmkrtchyan2002@yahoo.com հասցեով:

Կոմբինատորային ալգորիթմներ և ալգորիթմների վերլուծություն
Վահան Վ. Մկրտչյան

Դասախոսություն 14: Կապակցված գրաֆի մինիմալ կմախքային ծառը գտնելու Կրասկալի և Պրիմի ալգորիթմների նկարագիրը և վերլուծությունը:

**Գրաֆի մինիմալ կմախքային ծառը գտնելու խնդիրը**: Կապակցված, ցիկլ չպարունակող գրաֆին կանվանենք ծառ: Դիտարկենք $G=(V,E)$ կապակցված գրաֆը: $G$ գրաֆի $H=(V,E')$ ենթագրաֆին կանվանենք կմախքային ենթածառ, եթե $H$–ը ծառ է : Նկատենք, որ ցանկացած կապակցած գրաֆ պարունակում է կմախքային ենթածառ: Իրոք, եթե գրաֆը արդեն ծառ է, ապա ապացույցն ավարտված է, հակառակ դեպքում գրաֆը պարունակում է ցիկլ: Հեռացնենք, ցիկլի մի որևէ կող: Ստացված գրաֆը կրկին կապակցված է: Ըստ ինդուկցիոն ենթադրության այն պարունակում է կմախքային ծառ, որն իհարկե կհանդիսանա սկզբնական գրաֆի կմախքային ենթածառ:

Ենթադրենք $G=(V,E)$ կապակցված գրաֆի յուրաքանչյուր $e=\{u,v\}\in E$ կողին համապատասխանեցված է $d(e)$, հնարավոր է նաև բացասական, թիվը: $G=(V,E)$ գրաֆի կմախքային ծառի երկարություն ասելով կհասկանանք նրան պատկանող կողերի երկարությունների գումարը: Դիտարկենք հետևյալ խնդիրը. գտնել տրված գրաֆի այնպիսի կմախքային ծառ, որի երկարությունը հնարավորինս փոքր է:

Ստորև կդիտարկենք այս խնդիրը լուծող երկու ալգորիթմ:

**Պրիմի ալգորիթմը**: Դիցուք $G=(V,E)$ կապակցված գրաֆի գագաթների բազմությունը $V=\{v_1,...,v_n\}$-ն է, $d(v_i,v_j)$-ն $(v_i,v_j)$ կողի երկարությունն է (եթե $(v_i,v_j)\notin E$ ապա $d(v_i,v_j)=\infty$): Ալգորիթմի աշխատանքի ընթացքում արդեն կառուցված $(U,Y)$ ծառին հերթականորեն ավելացվում է մեկ գագաթ և մեկ կող, որը նոր ավելացված գագաթը միացնում է արդեն կառուցված ծառի որևէ գագաթի հետ:

*Պրիմի ալգորիթմի նկարագիրը*

**Քայլ 1**: Վերցնել $U=\{v_1\}, Y=\varnothing$:

**Քայլ 2**: Յուրաքանչյուր $u\in V\setminus U$ գագաթի համար գտնել $u^*\in U$ գագաթ այնպիսին, որ $d(u,u^*)=\min_{v\in U} d(u,v)$ և $u$ գագաթը նշել $(u^*,\beta(u))$ զույգով, որտեղ $d(u,u^*)=\beta(u)$: Եթե այդպիսի գագաթ գտնել հնարավոր չէ, ապա $u\in V\setminus U$ գագաթը նշում ենք $(-,\infty)$ պայմանանշանով:

**Քայլ 3**: Ընտրել այնպիսի $v(v^*,\beta(v))\in V\setminus U$ գագաթ, որի նիշը ամենափոքրն է, այսինքն՝ $\beta(v)=\min_{u\in V\setminus U}\beta(u)$ և $v$ գագաթն ավելացնել $U$ բազմությանը, իսկ $\{v,v^*\}$ կողը՝ $Y$ բազմությանը:

**Քայլ 4**: Եթե $|U|=n$ ապա ընտրված գագաթների և կողերի $(U,Y)$ բազմությունը կկազմի կմախքային ծառ, և ալգորիթմն ավարտում է աշխատանքը, հակառակ դեպքում՝ վերադարձ քայլ 2-ին:

**Թեորեմ 1**: Պրիմի ալգորիթմը կառուցում է մինիմալ կմախքային ծառ:

**Ապացույց**: Նախ նկատենք, որ ալգորիթմը կառուցում է ծառ: Իրոք, այն պարունակում է $n$ գագաթ, $n-1$ կող և կապակցված է: Ցույց տանք, որ կառուցված ծառը մինիմալ է: Ենթադրենք հակառակը. ալգորիթմի աշխատանքի արդյունքում ստացված կմախքային ծառը չունի մինիմալ երկարություն: Դիցուք $y_1,y_2,...,y_{n-1}$-ը կառուցված ծառի կողերն են՝ գրված այն հերթականությամբ, որով նրանք ավելացվել են $Y$ բազմությանը:

Դիտարկենք մինիմալ կմախքային ծառերի բազմությունը: Ամեն մի մինիմալ կմախքային ծառի համապատասխանեցնեք այն ամենափոքր $i$ համարը, որի դեպքում $y_i$ կողը չի պատկանում այդ ծառին: Նկատենք, որ այդպիսի կող միշտ գոյություն ունի:

$T_0$-ով նշանակենք այն մինիմալ ծառը, որի համապատասխան համարը ամենամեծն է: Դիցուք այդ համարը $k$–ն է, $1\le k\le n-1$: Այդ դեպքում $T_0$-ն պարունակում է $y_1,y_2,...,y_{k-1}$ կողերը և չի պարունակում $y_k$ կողը: $U_1$-ով նշանակենք այն ծառի գագաթների բազմությունը, որի կողերի բազմությունը $y_1,y_2,...,y_{k-1}$-ն է: Դիցուք $y_k=\{u,v\}$, $u\in U_1$, $v\notin U_1$: Քանի որ $T_0$-ն ծառ է, ապա նրանում գոյություն ունի $u\in U_1$ և $v\notin U_1$ գագաթները միացնող շղթա: Նկատենք, որ այդ շղթան պարունակում է այնպիսի $\{u',v'\}$ կող, որ $u'\in U_1$ և $v'\notin U_1$: Դիտարկենք հետևյալ ձևով սահմանված $T_1$ ծառը.

$$T_1=(T_0\setminus\{u',v'\})\cup\{u,v\}:$$

Նկատենք, որ $y_1,y_2,...,y_k$ կողերն արդեն պատկանում են $T_1$ ծառին: Ավելին, $(U,Y)$ կմախքային ծառի կառույցի ժամանակ դիտարկվել են $\{u,v\}$ և $\{u',v'\}$ կողերը, և $Y$-ի մեջ մտցվել է $y_k=\{u,v\}$ կողը, հետևաբար՝ $d(\{u,v\})\le d(\{u',v'\})$, որտեղից կունենանք, որ $T_1$ ծառի երկարությունը չի գերազանցում $T_0$-ի երկարությանը, հետևաբար՝ $T_1$ ծառը ևս մինիմալ կմախքային ծառ է: Նկատենք, որ այս պայմանը հակասում է $T_0$-ի ընտրությանը՝ որպես այն մինիմալ ծառ, որի համապատասխան համարը ամենամեծն է: Թեորեմն ապացուցված է:

Գնահատենք Պրիմի ալգորիթմի բարդությունը $n$ գագաթ պարունակող գրաֆի համար: Ենթադրենք, որ ալգորիթմի հերթական քայլում ստացված $(U,Y)$ ծառը պարունակում է $k$ գագաթ: Քայլ 2-ում ալգորիթմը յուրաքանչյուր $u \in V \setminus U$ գագաթի համար գտնում է $u^* \in U$ գագաթ և $u$ գագաթը նշում $(u^*, \beta(u))$ զույգով, որն անելու համար անհրաժեշտ է ոչ շատ քան $O(k)$ գործողություն, հետևաբար՝ $O(k(n-k))$ գործողությունների միջոցով հաշվվում է բոլոր $u \in V \setminus U$ գագաթները գնահատող զույգերը: Հետևաբար, քայլ 2-ի ը 3-ի համար անհրաժեշտ է $O(k(n-k))$ կարգի գործողություն: Արդյունքում կարող ենք պնդել, որ ալգորիթմի բարդությունը $O(n^3)$ –է:

Պարզվում է, որ ալգորիթմի քայլերի քանակը կարելի է էապես պակասեցնել նրա աշխատանքի հարմար կազմակերպման դեպքում: Եթե քայլ 2-ը կատարելուց հետո հիշենք յուրաքանչյուր $u \in V \setminus U$ գագաթի նշումը, ինչպես նաև այն վերջին $v$ գագաթը, որը քայլ 3-ում մտցվել է գագաթների $U$ բազմության մեջ, ապա քայլ 2-ի կատարման ժամանակ յուրաքանչյուր $u \in V \setminus U$ գագաթի նշումը կարող ենք հաշվել հետևյալ կերպ.

եթե $\beta(u) > d(u,v)$ ապա ընդունել $\beta(u) = d(u,v), u^* = v$:

Արդյունքում $O(n-k)$ գործողությունների միջոցով կարող ենք հաշվել բոլոր $u \in V \setminus U$ գագաթների նշումները: Պարզ է, որ այդ դեպքում ալգորիթմի բարդությունը կլինի $O(n^2)$:

**Կրասկալի ալգորիթմը**: Կրկին դիցուք $G = (V,E)$ կապակցված գրաֆի գագաթների բազմությունը $V = \{v_1,...,v_n\}$-ն է, իսկ կողերի բազմությունը՝ $E = \{e_1,...,e_q\}$-ն է: Դիցուք $d\{v_i,v_j\} = d(x)$-ն $x = \{v_i,v_j\}$ կողի երկարությունն է: Ալգորիթմի աշխատանքի ընթացքում արդեն կառուցված գրաֆին հերթականորեն ավելացվում են կողեր այնպես, որ արդյունքում ստացվում է մինիմալ կմախքային ծառ:

*Կրասկալի ալգորիթմի նկարագիրը*

**Քայլ 1**: Որպես կառուցվելիք ծառի գագաթների բազմություն ընդունել $V = \{v_1,...,v_n\}$-ը, իսկ կողերի բազմությունը՝ $Y = \varnothing$:

**Քայլ 2**: $G = (V,E)$ գրաֆի կողերը դասավորել երկարությունների չնվազման կարգով. $d(y_1) \le ... \le d(y_q)$:

**Քայլ 3**: Հերթականորեն դիտարկել $y_1,...,y_q$ կողերը և դիտարկված կողը մտցնել $Y$ բազմության մեջ, եթե այդ կողն ավելացնելուց հետո $(V,Y)$ գրաֆը ցիկլ չի պարունակում:

Նախ նկատենք, որ ալգորիթմը կառուցում է կմախքային ծառ: Իրոք, ալգորիթմի աշխատանքի ժամանակ դեն են նետվում կողեր, որոնք չեն ազդում գրաֆի կապակցվածության վրա:
Նշենք նաև, որ կառուցված կմախքային ծառի մինիմալության ապացույցը կատարվում է նույն ձևով, ինչպես Պրիմի ալգորիթմի դեպքում:
Գնահատենք Կրասկալի ալգորիթմի գործողությունների քանակը $n$ գագաթ և $q$ կող ունեցող գրաֆների համար: Քայլ 2-ն իրականացնելու համար կարող ենք օգտագործել արդեն ուսումնասիրված տեսակավորման ալգորիթմներից մեկը: Գործողությունների կարգն այստեղ $O(q\log_2 q)$-է:
Քայլ 3-ը կազմակերպենք հետևյալ կերպ. յուրաքանչյուր $u$ գագաթի հետ հիշենք $l(u)$-ն՝ այն կապակցվածության բաղադրիչի համարը, որին պատկանում է այդ գագաթը $(V,Y)$ գրաֆում: Սկզբնական պահին կապակցվածության բաղադրիչների քանակը հավասար է գրաֆի գագաթների քանակին:
$y=\{u,v\}$ կողը դիտարկելիս կատարվում է հետևյալը. Եթե $l(u)=l(v)$, ապա այդ կողը չի մտցվում $Y$ բազմության մեջ, իսկ եթե այդ գագաթները պատկանում են տարբեր կապակցվածության բաղադրիչների, այսինքն՝ եթե $l(u)<l(v)$, ապա $y=\{u,v\}$ մտցվում է $Y$ բազմության մեջ, որից հետո բոլոր այն գագաթներում, որոնց կապակցվածության բաղադրիչի համարը $l(v)$-է, փոխարինում ենք $l(u)$–ով: Նկատենք, որ քայլ 3-ի այսպիսի իրականացման դեպքում գործողությունների քանակը կլինի $O(n^2)$:
Այսպիսով, Կրասկալի ալգորիթմի բարդությունը $O(q\log_2 q+n^2)$–է: Ապացուցված է, որ քայլ 3-ը կարելի է իրականացնել $O(q\log_2 q)$ ժամանակում, այնպես որ Կրասկալի ալգորիթմի իրական բարդությունը $O(q\log_2 q)$–է:
Վերջում նշենք նաև, որ քանի որ կողերի երկարությունները ցանկացած թվեր էին, ապա այս ալգորիթմները կարելի է օգտագործել մաքսիմալ կմախքային ծառ խնդրի լուծման համար, այն է տրված $G=(V,E)$ կապակցված գրաֆի համար կառուցել ամենաերկար կմախքային ծառը:
Դրա համար բավական է փոխել երկարությունների նշանները և լուծել մինիմալ կմախքային ծառ գտնելու խնդիրը: Պարզ է, որ արդյունքում կստանանք սկզբնական գրաֆի մաքսիմալ կմախքային ծառ:

**Հանձնարարություն**: Ապացուցել, որ եթե կապակցված գրաֆի կողերի երկարությունները զույգ առ զույգ տարբեր թվեր են, ապա այն պարունակում է ճիշտ **մեկ** մինիմալ կմախքային ծառ:

# Գրականություն

Գտնված սխալների, առաջարկությունների, ինչպես նաև դասախոսություն-ներն e-mail-ով ստանալու համար կարող էք դիմել vahanmkrtchyan2002@yahoo.com հասցեով:

Կոմբինատորային ալգորիթմներ և ալգորիթմների վերլուծություն
Վահան Վ. Մկրտչյան

Դասախոսություն 15: Դինամիկ ծրագրման մեթոդ: Ռեսուրսների բաշխման խնդիր, ուսապարկի խնդիր, երկու հաջորդականությունների ամենաերկար ընդհանուր ենթահաջոր-դականության գտնելու խնդիր:

**Դինամիկ ծրագրավորում:** Դիսկրետ օպտիմիզիցիայի խնդիրների լուծման եղանակներից մեկն այդ խնդրի լուծման հանգեցումն է նույնատիպ, ավելի պարզ խնդիրների հաջորդականության լուծմանը: Առանձին-առանձին այդ խնդիրներից յուրաքանչյուրի լուծելը համարյա նույն դժվարությունն ունի, ինչ քննարկվողինը: Սակայն որոշակի եղանակով ընտրած նույնատիպ խնդիրների հաջորդականության լուծելը հաճախ ավելի հեշտ է իրականացվում, քան առանձին խնդրի լուծումը: Առաջին հայացքից տարօրինակ և անհավանական թվացող այս փաստն է ընկած օպտիմիզացման խնդիրների լուծման շատ տարածված մի եղանակի՝ դինամիկ ծրագրավորման հիմքում: Որպես օրինակ, դիտարկենք հետևյալ խնդիրը. տրված են $P_i : Z^+ \to R^+$ (ոչ բացասական ամբողջ և իրական թվերի բազմությունը) մոնոտոն աճող ֆունկցիաները և $L \in Z^+$ թիվը, պահանջվում է գտնել $P_1(x_1)+...+P_N(x_N)$ արտահայտության առավելագույն արժեքը, երբ $x_1+...+x_N = L$ և $x_1,...,x_N \in Z^+$: Պարզության համար կենթադրենք, որ $P_i(0)=0$, $i=1,...,N$: Այս խնդրին կարճ կանվանենք $(N,L)$ խնդիր ($N$-ը և $L$-ը խնդրի մեջ առկա պարամետրերն են):

$n, 0 \le n \le N$ և $l, 0 \le l \le L$ թվերի համար $F(n,l)$-ով նշանակենք $(n,l)$ խնդրի առավելագույն արժեքը: Նկատենք, որ

$$F(1,l) = P_1(l) \quad 0 \le l \le L:$$

Մյուս կողմից, $x_n$ փափոխականի ֆիքսած արժեքի համար նպատակային ֆունկցիան իր առավելագույն արժեքին կհասնի, եթե $x_1+...+x_{n-1} = l - x_n$ և նրանց համար $P_1(x_1)+...+P_{n-1}(x_{n-1})$ արտահայտությունն ստանա իր առավելագույն արժեքը: Հետևաբար,

$$F(n,l)=\max_{0\le x_n\le l}(P_n(x_n)+F(n-1,l-x_n)):$$

Արդյունքում, մենք ստացանք ալգորիթմ որոնելի $F(N,L)$ մեծությունը հաշվելու համար (հերթով հաշվել $F(1,L),...,F(N,L)$ )

*Դինամիկ ծրագրավորման ալգորիթմի նկարագիրը* $(N,L)$ *խնդրի համար*

**Մուտք**: $P_i(x)$ ֆունկցիաների արժեքները $x=0,...,L$ կետերում:

**Քայլ 1**: Ընդունել $F(1,l)=P_1(l)$ , $l=0,...,L$ և $i=1$:

**Քայլ 2**: Ունենալով $F(i,l)$ և $P_{i+1}(l)$ ֆունկցիաների արժեքները $l=0,...,L$ կետերում, հաշվել

$$F(i+1,l)=\max_{0\le j\le l}(P_{i+1}(j)+F(i,l-j))$$

արժեքը $l=0,...,L$ դեպքում:

**Քայլ 3**: Եթե $i+1=N$, ապա ավարտել ալգորիթմի աշխատանքը, հակառակ դեպքում՝ $i$-ի արժեքն ավելացնել մեկով և անցնել Քայլ 2-ին:

Ալգորիթմի գործողությունների քանակի գնահատման համար նկատենք, որ $F(i+1,l)$ արժեքը հաշվվում է $l+1$ գումարման և համեմատման գործողությունների միջոցով, և հետևաբար՝ տրված $i$-ի դեպքում քայլ 2-ի գործողությունների քանակը կլինի $O(L^2)$, իսկ ալգորիթմի գործողությունների քանակը՝ $O(NL^2)$: Նկատենք, որ ալգորիթմի բարդությունը էապես քիչ է $\binom{N+L-1}{L}$–ից՝ $x_1+...+x_N=L$ հավասարման ոչ բացասական ամբողջ լուծումների քանակից:

**Ռեսուրսների բաշխման խնդիր**: Դիցուք ունենք որոշակի $k$ քանակությամբ ռեսուրս, որն անհրաժեշտ է բաշխել՝ տարբեր տեսակի արտադրատեսակներ թողարկելու համար:

Ենթադրենք պետք է արտադրել $N$ տեսակի արտադրանք և $i=1,...,N$ համար

1. $i$-րդ արտադրանքի $x_i$ քանակը չպետք է գերազանցի $b$ թիվը,
2. $i$-րդ արտադրանքն արտադրելու համար անհրաժեշտ է $c_i(x_i)$ քանակությամբ ռեսուրս,
3. $i$-րդ արտադրանքի սպառումից ստացված շահույթը $p_i(x_i)$–է:

Խնդիրը կայանում է հետևյալում. պահանջվում է կազմել թողարկման այնպիսի $(x_1,...,x_N)$ պլան, որ օգտագործագործված ռեսուրսի ընդհանուր ծախքը չգերազանցի $k$-ն, իսկ սպասվելիք ընդհանուր շահույթը լինի առավելագույնը: Նկարագրված խնդրի մաթեմատիկական մոդելը կլինի հետևյալը. անհրաժեշտ է գտնել $x_1,...,x_N$ թվեր, որոնք բավարարում են

$$c_1(x_1)+...+c_N(x_N)\le k$$
$$0\le x_i\le b,\ i=1,...,N$$

պայմաններին, և որոնց համար $p_1(x_1)+...+p_N(x_N)$ արտահայտությունն ընդունում է իր առավելագույն արժեքը:

Այստեղ $k$-ն և $b$-ն հայտնի թվեր են, իսկ տրված $c_i(x_i)$ և $p_i(x_i)$ ֆունկցիաները մոնոտոն աճող և ոչ բացասական են, ընդ որում $i=1,...,N$ համար $c_i(0)=p_i(0)=0$: Պարզության համար մենք կենթադրենք նաև, որ $c_i(x_i)$ և $p_i(x_i)$ ֆունկցիաներն ընդունում են ամբողջ արժեքներ:

Ինչպես վերևում, հիմա էլ դիտարկենք համանման խնդիրների բազմությունը: $i=1,...,N$ և $j=0,...,k$ համար սահմանենք $(i,j)$ խնդիր հետևյալ կերպ.

$$c_1(x_1)+...+c_i(x_i)\le j$$
$$0\le x_1,...,x_i\le b,\ x_1,...,x_i\in Z$$
$$p_1(x_1)+...+p_i(x_i)\to \max$$

$i=1,...,N$ և $j=0,...,k$ համար $F(i,j)$-ով նշանակենք այս խնդրի արժեքը, այսինքն՝ $p_1(x_1)+...+p_i(x_i)$ արտահայտության առավելագույն արժեքը: Նկատենք, որ մեր նպատակը $F(N,k)$-ն գտնելն է: Դժվար չէ տեսնել, որ

$$j=0,...,k \text{ համար } F(1,j)=\max\{p_1(x)/x\in\{z/0\le z\le b, c_1(z)\le j\}\}:$$

Օգտվելով վերևում բերված դատողություններից, $i>1$ համար կստանանք՝

$$F(i,j)=\max\{p_i(x_i)+F(i-1,j-c_i(x_i))\},$$

որտեղ $\max$–ը վերցվում է $0\le x_i\le b$, $c_i(x_i)\le j$ պայմաններին բավարարող ամբողջաթիվ $x_i$-երից:

Ստացված անրադարձ առնչությունը թույլ է տալիս հերթականորեն գտնել $F(1,b),...,F(N,k)$ արժեքները:

**Ուսապարկի խնդիր**: Ունենք որոշ թվով առարկաներ: Հայտնի է նրանցից յուրաքանչյուրի գինն ու ծավալը: Անհրաժեշտ է որոշակի տարողություն ունեցող ուսապարկով տեղափոխել այս առարկաներից այնպիսիները, որոնց ծավալների գումարը չգերազանցի ուսապարկի ծավալը և որոնց գումարային գինը լինի հնարավորին չափ մեծ:

Խնդիրը կարելի է մեկնաբանել նաև հետևյալ կերպ. գողը փորձում է վերցնել այնպիսի իրեր, որոնք կտեղավորվեն իր ուսապարկում և գողացած իրերի վերավաճառքից նա կստանա առավելագույն շահույթ:

Տանք խնդրի մաթեմատիկական ձևակերպումը. Համարակալենք իրերը $1,...,n$ թվերով և դիցուք $i=1,...,n$ համար $c_i$-ն $i$-րդ առարկայի գինն է, իսկ $v_i$-ն՝ ծավալը: $V$-ով նշանակենք ուսապարկի տարողությունը: Դիտարկենք $x_1,...,x_n$ փոփոխականները, որտեղ

$$x_i = \begin{cases} 1, & \text{եթե որոշել ենք վերցնել } i\text{-րդ առարկան,} \\ 0, & \text{հակառակ դեպքում:} \end{cases}$$

Պարզ է, որ $(x_1v_1 + \ldots + x_nv_n)$–ը կլինի վերցրած առարկաների ծավալների գումարը, իսկ $(x_1c_1 + \ldots + x_nc_n)$–ը՝ վերցրած առարկաների գների գումարը: Արդյունքում ուսապարկի խնդրին համապատասխանող մաթեմատիկական մոդելը կլինի հետևյալը.

տրված են $c_1,\ldots,c_n$, $v_1,\ldots,v_n$, և $V$ ոչ բացասական թվերը: Անհրաժեշտ է $x_1,\ldots,x_n$ փոփախականների համար ընտրել 0 կամ 1 արժեքներ, որ բավարարվի $x_1v_1 + \ldots + x_nv_n \le V$ պայմանը և $(x_1c_1 + \ldots + x_nc_n)$ արտահայտությունը ստանա իր առավելագույն հնարավոր արժեքը:

Նկատենք, որ $x_1,\ldots,x_n$ փոփախականներին արժեքներ տալու բոլոր հնարավոր եղանակները քննարկելը կպահանջի $2^n$ դիտարկում:

Նշենք նաև, որ առաջին հայացքից լավ թվացող ալգորիթմները ընդհանուր դեպքում չեն գտնում լավագույն լուծումը: Որպես այդպիսի օրինակ դիտարկենք հետևյալ ալգորիթմը.

**Քայլ 1**: առարկաները վերադասավորենք միավոր ծավալի գների նվազման կարգով, այսինքն՝

$$\frac{c_{i_1}}{v_{i_1}} \ge \ldots \ge \frac{c_{i_n}}{v_{i_n}}$$

**Քայլ 2**: հերթականորեն դիտարկել $i_1,\ldots,i_n$ առարկաները և դիտարկված առարկան տեղավորել ուսապարկում, եթե նրա ավելացումից հետո ուսապարկի ուսապարկի մեջ առկա առարկաների ծավալների գումարը չի գերազանցում $V$–ն:

Նշված ալգորիթմը չի գտնում լավագույն լուծումը: Իրոք դիտարկենք հետևյալ օրինակը: Դիցուք ունենք 85 տարողությամբ ուսապարկ և 4 առարկաներ, որոնց գինը և ծավալը հետևյալ թվերն են.

| | | | | |
|---|---|---|---|---|
| գին | 160 | 250 | 180 | 30 |
| ծավալ | 40 | 50 | 40 | 20 |

Նկատենք, որ ալգորիթմը առարկաները կդասավորի

$$\frac{250}{50} \ge \frac{180}{40} \ge \frac{160}{40} \ge \frac{30}{20}$$

հերթականությամբ: Քայլ 2-ում այն կընտրի երկրորդ և չորրորդ առարկաները, որոնց գումարային ծավալը 70-է, իսկ գինը՝ 280: Նկատենք, որ առաջին և երրորդ առարկաների գումարային ծավալը 80-է, իսկ գինը՝340:

Նախ դժվար չէ համոզվել, որ ուսապարկի խնդիրն իրենից ներկայացնում է ռեսուրսների բաշխման խնդրի մասնավոր դեպքը: Իրոք, վերցնենք $c_i(x)=v_i x$, $p_i(x)=c_i x$, և $b=1$: Այստեղից հետևում է, որ ուսապարկի խնդիրը կարելի է լուծել կիրառելով ռեսուրսների բաշխման լուծման ալգորիթմը:

Ստորը կդիտարկենք դինամիկ ծրագրավորման մեթոդով ուսապարկի խնդրի լուծման մեկ այլ ալգորիթմ:

Առարկաների ընտրությունը կկատարենք $n$ փուլերի միջոցով: $k$-րդ փուլում ընտրությունը կկատարենք $1,\ldots,k$ համարն ունեցող առարկաներից: Յուրաքանչյուր ընտրությանը համապատասխանեցնենք $(S,c,w)$ եռյակը, որտեղ $S$–ն ընտրված առարկաների բազմությունն է, $c$-ն՝ նրանց արժեքը, իսկ $w$-ն՝ նրանց ծավալների գումարը:

Նկատենք, որ $k$-րդ փուլում առարկաների ընտրությունների քանակը $2^k$–է: Մենք կփորձենք դիտարկել հնարավորին չափ քիչ ընտրություններ: Նախ նկատենք, որ իմաստ չունի դիտարկել այն $(S,c,w)$ ընտրությունը, որում $w>V$:

Կասենք, որ $(S,c,w)$ ընտրությունը հեռանկարային է, եթե նրանից հնարավոր է ստանալ լավագույն ընտրություն՝ ավելացնելով $k+1,\ldots,n$ առարկաներից որոշները:

Նկատենք, որ եթե $M_k$-ն $1,\ldots,k$ համարն ունեցող առարկաների կամայական բազմություն է, որը պարունակում է գոնե մեկ հեռանկարային ընտրություն և $(S,c,w)\in M_k$, $(S',c',w')\in M_k$ $c\geq c'$, $w\leq w'$, ապա $M_k\setminus\{(S',c',w')\}$ բազմությունը ևս կպարունակի հեռանկարային ընտրություն:

Դիտարկենք հետևյալ ալգորիթմը

**Քայլ** 1:Վերցնել $M_0=\{(\varnothing,0,0)\}$

**Քայլ** 2: $k=1,\ldots,n$ համար կատարել

Ա) վերցնել $M_k=\varnothing$

Բ) $M_k := M_{k-1}\cup\{(S\cup\{k\},c+c_k,w+v_k)/(S,c,w)\in M_{k-1}, w+v_k\leq V\}$

Գ) եթե գոյություն ունեն $(S,c,w)\in M_k$, $(S',c',w')\in M_k$ այնպես, որ $c\geq c'$, $w\leq w'$, ապա $M_k := M_k\setminus\{(S',c',w')\}$

**Քայլ** 3: $M_n$-ից ընտրել այն $(S,c,w)$ ընտրությունը, որի համար $c$–ն ամենամեծն է:

Գնահատենք ալգորիթմի բարդությունը: Ենթադրենք $c$–ն նպատակային ֆունկցիայի արժեքն է լավագույն լուծման դեպքում: Նկատենք, որ $k=1,\ldots,n$ համար $M_k$ բազմության տարրերի երկրորդ բաղադրիչները $c$–ին չգերազանցող ամբողջ թվեր են, իսկ ալգորիթմի Բ) գործողության ընթացքում երկու հավասար երկրորդ բաղադրիչ ունեցող տարրերից մեկը դեն է նետվում: Հետևաբար, $M_k$ բազմությունը պարունակում է ամենաշատը $c$ տարր: Նկատենք, որ $c \le nc_0$, որտեղ $c_0 = \max\limits_{1\le i\le n} c_i$: Այստեղից հետևում է, որ տրված $k$-ի համար 2-րդ քայլի գործողությունների քանակը չի գերազանցում $O(c^2)$, հետևաբար՝ ալգորիթմի բարդությունը կլինի՝ $O(nc^2) = O(n^3{c_0}^2)$:

**Երկու հաջորդականությունների ամենաերկար ընդհանուր ենթահաջորդականության գտնելու խնդիր**: Ենթադրենք, որ $X = (x_1,...,x_n)$-ը վերջավոր հաջորդականություն է, որի էլեմենտների: Այսուհետ մենք կդիտարկենք միայն վերջավոր հաջորդականություններ այնպես, որ ամեն անգամ չենք նշի “վերջավոր” բառը: $Z = (z_1,...,z_k)$ հաջորդականությունը կանվանենք $X = (x_1,...,x_n)$ հաջորդականության ենթահաջորդականություն, եթե գոյություն ունի ինդեքսների մոնոտոն աճող $(i_1,...,i_k)$ հաջորդականություն այնպես, որ $z_1 = x_{i_1},...,z_k = x_{i_k}$: Օրինակ, $Z = (B,C,B,B,D)$ հաջորդականությունը հանդիսանում է $X = (A,A,C,B,D,C,D,B,A,B,D)$ հաջորդականության ենթահաջորդականությունը: Կասենք, որ $Z$ հաջորդականությունը հանդիսանում է $X$ և $Y$ հաջորդականությունների ընդհանուր ենթահաջորդականությունը, եթե այն ինչպես $X$, այնպես էլ $Y$ հաջորդականության ենթահաջորդականությունն է: Երկու հաջորդականությունների ամենաերկար ընդհանուր ենթահաջորդականությունը գտնելու խնդիրը կայանում է հետևյալում.

տրված են է $X$ և $Y$ հաջորդականությունները: Պահանջվում է գտնել $X$ և $Y$ հաջորդականությունների այնպիսի ընդհանուր ենթահաջորդականություն, որի երկարությունը հնարավորին չափ մեծ է:

Եթե $Z$ հաջորդականությունը հանդիսանում է $X$ և $Y$ հաջորդականությունների ամենաերկար ընդհանուր ենթահաջորդականությունը, ապա այդ փաստը կգրենք $Z = LCS(X,Y)$ (Largest Common Subsequence):

Եթե փորձենք խնդիրը լուծել հատարկման եղանակով, այսինքն եթե հերթով դիտարկենք $X$ հաջորդականության ենթահաջորդականությունները և յուրաքանչյուրի համար ստուգենք, թե արդյոք այն հանդիսանում է $Y$ հաջորդականության ենթահաջորդականություն, ապա այս ալգորիթմը կաշխատի էքսպոնենցիալ ժամանակում, քանի որ $m$ երկարություն ունեցող հաջորդականությունն ունի $2^m$ ենթահաջորդականություններ:

Ստորև ապացուցված թեորեմը թույլ է տալիս կիրառել դինամիկ ծրագրման մեթոդը, և շատ ավելի արագ լուծել խնդիրը:

Ենթադրենք $X=(x_1,...,x_n)$-ը հաջորդականություն է: $i=0,...,n$ համար $X_i=(x_1,...,x_i)$ հաջորդականությանը կանվանենք $X$ հաջորդականության $i$ երկարությամբ նախածանց: $X_0$-ն դատարկ հաջորդականություն է:

**Թեորեմ**: Ենթադրենք $Z=(z_1,...,z_k)$-ն հանդիսանում է $X=(x_1,...,x_n)$ և $Y=(y_1,...,y_m)$-ը հաջորդականությունների ինչ-որ մի ամենաերկար ենթահաջորդականություն: Այդ դեպքում

1. եթե $x_n=y_m$, ապա $z_k=x_n=y_m$ և $Z_{k-1}=LCS(X_{n-1},Y_{m-1})$;
2. եթե $x_n\neq y_m$ և $z_k\neq y_m$, ապա $Z=LCS(X,Y_{m-1})$;
3. եթե $x_n\neq y_m$ և $z_k\neq x_n$, ապա $Z=LCS(X_{n-1},Y)$:

**Ապացույց**: Եթե $x_n=y_m$, ապա պարզ է, որ եթե $z_k\neq x_n=y_m$, ապա ավելացնելով $x_n=y_m$ տարրը $Z=(z_1,...,z_k)$ հաջորդականությանը, մենք կստանանք ավելի երկար հաջորդականություն:

Իսկ եթե $z_k\neq y_m$ և $x_n\neq y_m$, ապա պարզ է, որ $Z=(z_1,...,z_k)$ հաջորդականությունը հանդիսանում է $X$ և $Y_{m-1}$ հաջորդականությունների ամենաերկար ենթահաջորդականություն, այսինքն՝ $Z=LCS(X,Y_{m-1})$: 3 կետն ապացուցվում է համանման ձևով:

Ապացուցված թեորեմը թույլ է տալիս $X=(x_1,...,x_n)$ և $Y=(y_1,...,y_m)$ հաջորդականությունների ամենաերկար ենթահաջորդականության գտնելու խնդիրն հանգեցնել մեկ կամ երկու համանման խնդիրների: Իրոք, եթե $x_n=y_m$, ապա բավական է գտնել $LCS(X_{n-1},Y_{m-1})$-ը, իսկ եթե $x_n\neq y_m$, ապա բավական է գտնել $LCS(X,Y_{m-1})$ և $LCS(X_{n-1},Y)$ հաջորդականություններից ամենաերկարը: Հետևաբար, եթե $i=0,...,n$ և $j=0,...,m$ համար $c[i,j]$-ով նշանակենք $X_i$ և $Y_j$ հաջորդականությունների ամենաերկար ենթահաջորդականության երկարությունը, ապա

$$c[i,j]=\begin{cases}0, & \text{եթե} \quad i=0 \quad \text{կամ} \quad j=0,\\ 1+c[i-1,j-1], & \text{եթե} \quad i,j>0 \quad \text{և} \quad x_i=y_j,\\ \max\{c[i,j-1],c[i-1,j]\}, & \text{եթե} \quad i,j>0 \quad \text{և} \quad x_i\neq y_j:\end{cases}$$

Որպեսզի գտնենք $X=(x_1,...,x_n)$ և $Y=(y_1,...,y_m)$ հաջորդականությունների ինչ-որ մի ամենաերկար ընդհանուր ենթահաջորդականության, ապա $i=1,...,n$ և $j=1,...,m$ համար $b[i,j]$-ում հիշենք, թե $c[i,j]$-ի արժեքը ինչպես է հաշվվել, այսինքն, վերցնենք

$$b[i,j]=\begin{cases}\nwarrow & \text{եթե} \quad c[i,j]=1+c[i-1,j-1]\\ \uparrow & \text{եթե} \quad c[i,j]=c[i-1,j]\\ \leftarrow & \text{եթե} \quad c[i,j]=c[i,j-1]\end{cases}$$

Ստորև նկարագրված ալգորիթմը $O(nm)$ ժամանակում տրված երկու $X=(x_1,...,x_n)$ և $Y=(y_1,...,y_m)$ հաջորդականությունների համար գտնում է $c[i,j]$ և $b[i,j]$ մատրիցները:

$LCS(X,Y)$
for $i:=1$ to $n$ do $c[i,0]:=0$;
for $j:=0$ to $m$ do $c[0,j]:=0$;
for $i:=1$ to $n$ do
  for $j:=1$ to $m$ do
    if $(x_i=y_j)$ then $c[i,j]:=c[i-1,j-1]+1$, $b[i,j]:=\nwarrow$;
      else
        if $c[i-1,j]\geq c[i,j-1]$ then
          $c[i,j]:=c[i-1,j]$, $b[i,j]:=\uparrow$;
          else
            $c[i,j]:=c[i,j-1]$, $b[i,j]:=\leftarrow$;
return $c,b$

Ունենալով $c[i,j]$ և $b[i,j]$ մատրիցները, մենք կարող ենք գտնել ինչպես $Z=LCS(X,Y)$-ը, այնպես էլ նրա երկարությունը (այն $c[n,m]$-ն է): Դիտարկենք մի օրինակ: Դիցուք $X=(A,A,B)$ և $Y=(B,A,A)$: Այդ դեպքում $c,b$ մատրիցները կունենան հետևյալ տեսքը.

| | | $B$ | $A$ | $A$ |
|---|---|---|---|---|
| | 0 | 0 | 0 | 0 |
| $A$ | 0 | $0\uparrow$ | $1\nwarrow$ | $1\nwarrow$ |
| $A$ | 0 | $0\uparrow$ | $1\nwarrow$ | $2\nwarrow$ |
| $B$ | 0 | $1\nwarrow$ | $1\uparrow$ | $2\uparrow$ |

Կարմիր գույնով ցույց է տրված, որ $X=(A,A,B)$ և $Y=(B,A,A)$ հաջորդականությունների ամենաերկար հաջորդականության օրինակ է հանդիսանում $Z=(A,A)$ հաջորդականությունը, որի երկարությունը 2-է:

## Գրականություն

Գտնված սխալների, առաջարկությունների, ինչպես նաև դասախոսություն-ներն e-mail-ով ստանալու համար կարող էք դիմել vahanmkrtchyan2002@yahoo.com հասցեով:

Կոմբինատորային ալգորիթմներ և ալգորիթմների վերլուծություն
Վահան Վ. Մկրտչյան

Դասախոսություն 16: Դինամիկ ծրագրման մեթոդ: Մի քանի մատրիցների բազմապատկման խնդիր: Բազմանկյան օպտիմալ տրիանգուլյացիայի խնդիր:

**Մի քանի մատրիցների բազմապատկման խնդիր**: Ենթադրենք տրված են $A(p\times q)=\| a_{ij} \|$ և $B(q\times r)=\| b_{jk} \|$ մատրիցները: Ինչպես գիտենք, $C(p\times r)=\| c_{ik} \|$ մատրիցը, որտեղ

$$c_{ik}=\sum_{j=1}^{q} a_{ij}b_{jk} ,$$

կոչվում է $A$ և $B$ մատրիցների արտադրյալ: Նկատենք, որ $A$ և $B$ մատրիցների արտադրյալը հաշվելու համար անհրաժեշտ է կատարել $pqr$ բազմապատկում և գումարում: Այս թիվը կանվանենք երկու մատրիցների բազմապատկման բարդություն:

Ինչպես գիտենք, ցանկացած $A$, $B$, $C$ մատրիցների համար ճիշտ է հետևյալ հավասարությունը

$$A(BC)=(AB)C$$

այլ կերպ ասած, արդյունքը կախված չէ փակագծերի տեղադրման հերթականությունից: Հետևյալ օրինակը ցույց է տալիս, որ փակագծերի տեղադրման հերթականությունը կարող է էապես ազդել բազմապատկման արդյունքի հաշման համար անհրաժեշտ գործողությունների քանակի վրա:

Դիտարկենք $A(10\times100)$, $B(100\times5)$, $C(5\times50)$ մատրիցները և ենթադրենք, որ պահանջվում է հաշվել այս մատրցների արտադրյալը: Նկատենք, որ եթե արտադրյալը հաշվելու համար մենք $A$-ն բազմապատկենք $B$-ով, իսկ հետո՝ արդյունքը բազմապատկենք $C$-ով, ապա մեզ անհրաժեշտ կլինի $10\times100\times5+10\times5\times50=7500$ բազմապատկում և գումարում, իսկ եթե մենք նախ $B$-ն բազմապատկենք $C$-ով, իսկ հետո՝ $A$-ն բազմապատկենք նրանց արտադրյալով, ապա մեզ անհրաժեշտ գործողությունների քանակը կլինի՝ $100\times5\times50+10\times100\times50=75000$: Եվ հետևաբար, մի քանի մատրցներ բազմապատկելուց առաջանում է բնական խնդիր. ինչպիսի հերթականությամբ բազմապատկել մատրիցները, որպեսզի գումարային գործողությունների քանակը լինի հնարավորին չափ փոքր, ավելի ճիշտ դիտարկենք հետևյալ խնդիրը

**Մատրիցների բազմապատկման խնդիր**: Տրված են $A_1(p_0 \times p_1)$, $A_2(p_1 \times p_2)$, $\ldots, A_n(p_{n-1} \times p_n)$ մատրիցները: Պահանջվում է հաշվել նրանց արտադրյալը՝ կատարելով հնարավորին չափ քիչ գործողություններ:

Նախքան դինամիկ ծրագրավորման մեթոդով խնդրի լուծելը, ցույց տանք, որ հատարկման (բոլոր դեպքերի քննարկման) մեթոդը իրոք պիտանի չէ: $P(n)$-ով նշանակենք $n$ մատրիցների բազմապատկման ժամանակ փակագծերի դասավորման եղանակների քանակը: Նկատենք, որ եթե $n=1$, ապա $P(n)=1$, իսկ եթե $n>1$, ապա $P(n)=\sum_{k=1}^{n} P(k)P(n-k)$ (փակագծեր կարող ենք դնել $k$–րդ և $k+1$-րդ մատրիցների միջև): Կարելի է ապացուցել, որ այս անրադարձ առնչության $P(n)$ լուծումը բավարարում է $P(n)=O(\frac{4^n}{n\sqrt{n}})$ առնչությանը, և հետևաբար դեպքերի քանակը իրոք էքսպոնենցիալ է:

Կարելի է նաև այլ կերպ համոզվել, որ դեպքերի քանակը էքսպոնենցիալ է: $n$ մատրիցները տրոհենք $\left[\frac{n}{3}\right]$ եռյակների,

$$(A_1A_2A_3)(A_4A_5A_6)(A_7A_8A_9)\ldots$$

Նկատենք, որ յուրաքանչյուր եռյակում փակագծեր կարելի է դնել երկու եղանակով, հետևաբար՝

$$P(n) \geq 2^{\left[\frac{n}{3}\right]}:$$

$A_1(p_0 \times p_1)$, $A_2(p_1 \times p_2)$, $\ldots, A_n(p_{n-1} \times p_n)$ մատրիցների արտադրյալը դինամիկ ծրագրավորման մեթոդով հաշվելու համար վարվենք հետևյալ կերպ, $1 \leq i \leq j \leq n$ համար $A_{i\ldots j}$-ով և $m[i,j]$-ով նշանակենք

$$A_{i\ldots j} \equiv A_iA_{i+1}\ldots A_j,$$

$m[i,j]$- $A_{i\ldots j}$ արտադրյալը հաշվելու համար մինիմալ բազմապատկումների քանակը:

Փաստորեն մեր նպատակը $A_{1\ldots n}$-ը և $m[1,n]$-ը գտնելն է: Նկատենք, որ $m[i,j]$ թվերը բավարարում են հետևյալ անրադարձ առնչությանը.

$$m[i,j]=\begin{cases}0, & \text{եթե} \quad i=j; \\ \min\limits_{i\leq k<j}\{m[i,k]+m[k+1,j]+p_{i-1}p_kp_j\} & \text{եթե} \quad i<j:\end{cases}$$

$i<j$ համար $s[i,j]$–ով նշանակենք այն $k$–ն, որի համար $m[i,j]$-ն ընդունում է իր փոքրագույն արժեքը: Փաստորեն $s[1,n]$-ը ցույց է տալիս, թե որտեղ պետք է դրվի վերջին փակագիծը մեր որոնելի $A_{1...n}$ արտադրյալը հաշվելիս, այսինքն՝

$$A_{1...n}=(A_1...A_k)(A_{k+1}...A_n)\text{, որտեղ՝ } s[i,j]=k:$$

Հաշվի առնելով վերը նշված բանաձը կարելի է առաջարկել $m[i,j]$ և $s[i,j]$ թվերը որոշելու հետևյալ ալգորիթմը.

Քայլ 1: for $i:=1$ to $n$ do $m[i,i]:=0$

Քայլ 2:

```
        for l := 2 to n do (հաշվել l երկարությամբ արտադրյալները)
            for i := 1 to n − l + 1 do
        begin
            j := i + l − 1;
            m[i, j] := +∞;
                for k := i to j − 1 do
            begin
                q := m[i,k] + m[k+1, j] + p_{i−1} p_k p_j;
                if q < m[i, j] then
                begin
                    m[i, j] := q; s[i, j] := k
                end
            end
        end
```

Քայլ 3: return $m,s$

Նկատենք, որ քանի որ $m,s$ մատրիցները որոշելու համար կատարվում է 3 ներդրված for ցիկլեր, ապա բերված ալգորիթմի բարդությունը կլինի՝ $O(n^3)$: Ունենալով $m,s$ մատրիցները, մենք կարող ենք գտնել փակագծերի այն դասավորությունը, որի դեպքում $A_{1...n}$ արտադրյալը հաշվելու համար կպահանջվի նվազագույն, այսինքն՝ $m[1,n]$ գործողություններ: $s[1,n]=k$-ն ցույց է տալիս, որ մեր որոնելի $A_{1...n}$ արտադրյալը հաշվելիս վերջին փակագիծը պետք է դրվի $A_k$ և $A_{k+1}$ մատրիցների միջև, $s[1,k]=l$-ը ցույց է տալիս, որ $A_{1...l}$ արտադրյալը հաշվելիս վերջին փակագիծը պետք է դրվի $A_l$ և $A_{l+1}$ մատրիցների միջև, $s[k+1,n]=r$-ը ցույց է տալիս, որ $A_{k+1...n}$ արտադրյալը հաշվելիս վերջին փակագիծը պետք է դրվի $A_r$ և $A_{r+1}$ մատրիցների միջև, և այլն:

**Բազմանկյան օպտիմալ տրիանգուլյացիայի խնդիր**: Բազմանկյունը փակ կոր է, որը կազմված է հատվածներից: Այդ հատվածներին ընդունված է անվանել բազմանկյան կողեր: Այն կետը, որտեղ հատվում են երկու կող, կոչվում է բազմանկյան գագաթ: Ինքնահատումներ չունեցող բազմանկյունը կանվանենք պարզ: Քանի որ մենք միշտ դիտարկելու ենք պարզ բազմանկյուններ, ապա ամեն անգամ “պարզ” բառը չենք նշի, և բազմանկյուն ասելով կհասկանանք պարզ բազմանկյուն:

Բազմանկյանը կանվանենք ուռուցիկ, եթե նրա ցանկացած երկու կետերի համար, որոնք պատկանում են բազմանկյան եզրին կամ ներքին տիրույթին, այդ կետերը միացնող հատվածի բոլոր կետերը պատկանում են բազմանկյան եզրին կամ ներքին տիրույթին:

Բազմանկյունը մենք կարող ենք ներկայացնել $P=(v_0,v_1,...,v_n)$ տեսքով, որտեղ՝ $v_0,v_1,...,v_n$-ը $P$ բազմանկյան գագաթներն են $(v_0=v_n)$, $\overline{v_0v_1},\ldots,\overline{v_{n-1}v_n}$-ը՝ $P$ բազմանկյան կողերն են:

Եթե $P$ բազմանկյան $v_i$ և $v_j$ գագաթները հարևան չեն, ապա $\overline{v_iv_j}$ հատվածին կանվանենք անկյունագիծ: Նկատենք, որ $\overline{v_iv_j}$ անկյունագիծը բազմանկյունը բաժանում է երկու՝ $P_1=(v_i,v_{i+1},...,v_j)$ և $P_2=(v_j,v_{j+1},...,v_i)$ բազմանկյունների: Նկատենք, որ $n$ գագաթ պարունակող բազմանկյունը պարունակում է $n(n-3)/2$ անկյունագիծ:

Բազմանկյան տրիանգուլյացիա ասելով կհասկանանք բազմանկյան անկյունագծերի այնպիսի ենթաբազմություն, որոնք բազմանկյունը տրոհում են եռանկյունների: Նկատենք, որ $n$ գագաթ պարունակող բազմանկյան ցանկացած տրիանգուլյացիա պարունակում է միևնույն թվով անկյունագծեր:

Իրոք, $k$-ով նշանակենք բազմանկյան մի որևէ տրիանգուլյացիայում առկա անկյունագծերի քանակը: Քանի որ յուրաքանչյուր անկյունագիծ բազմանկյունը տրոհում է երկու մասի, ապա տրիանգուլյացիայի բոլոր անկյունագծերը տանելուց հետո, կստանանք $k+1$ եռանկյուն: Նկատենք, որ այս $k+1$ եռանկյունների ներքին անկյունների գումարը հավասար է բազմանկյան ներքին անկյունների գումարին (նկար 1), հետևաբար՝

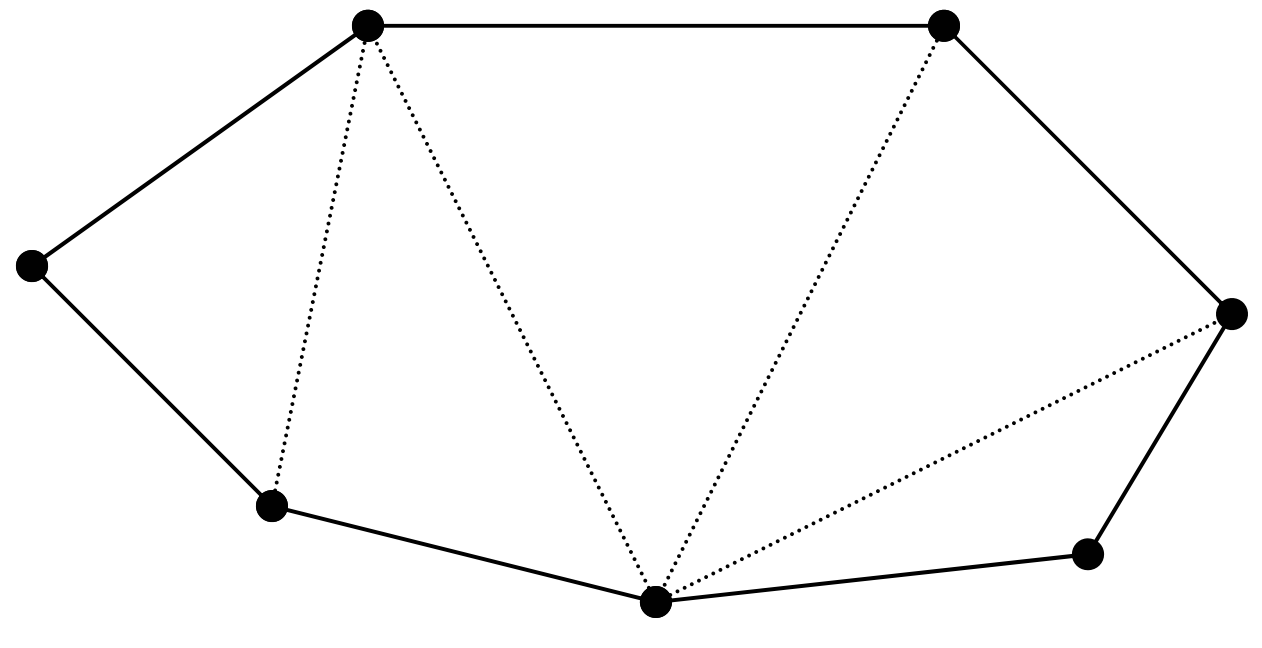

նկար 1

$$(k+1)\cdot 180^\circ=(n-2)\cdot 180^\circ$$

կամ

$$k = n - 3:$$

Բազմանկյան օպտիմալ տրիանգուլյացիայի խնդիրը ձևակերպվում է հետևյալ կերպ. տրված է $P = (v_0, v_1, ..., v_n)$ բազմանկյունը, և $P$ բազմանկյան գագաթները որպես գագաթներ պարունակող եռանկյունների վրա որոշված $w$ կշռային ֆունկցիան: Պահանջվում է գտնել $P$ բազմանկյան այնպիսի տրիանգուլյացիայի, որին պատկանող անկյունագծերը տանելուց առաջացած եռանկյունների կշիռների գումարը հնարավորին չափ փոքր է:

Նկատենք, որ կշռային ֆունկցիայի օրինակ կարող է ծառայել օրինակ եռանկյան մակերեսը, պարագիծը: Նշենք նաև, որ եթե $w$ կշռային ֆունկցիան սահմանված լինի որպես եռանկյան մակերեսը, ապա $P$ բազմանկյան ցանկացած տրիանգուլյացիա կլինի օպտիմալը:

Դինամիկ ծրագրավորման մեթոդով խնդիրը լուծելու համար վարվենք հետևյալ կերպ, $1 \le i \le j \le n$ համար $m[i, j]$-ով նշանակենք $(v_{i-1}, v_i, ..., v_j)$ բազմանկյան օպտիմալ տրիանգուլյացիայի արժեքը: Նկատենք, որ մեր նպատակը $m[1, n]$-ը գտնելն է: Մենք կենթադրենք, որ $m[i, i] = 0$ ($(v_{i-1}, v_i)$ “երկանկյան” օպտիմալ տրիանգուլյացիայի արժեքը հավասար է զրոյի):

Նկատենք, որ $m[i, j]$ համար ճիշտ է հետևյալ անրադարձ առնչությունը (նկար 2),

$$m[i, j] = \begin{cases} 0, & \text{եթե } i = j; \\ \min\limits_{i \le k < j} \{m[i, k] + m[k+1, j] + w(\Delta v_{i-1} v_k v_j)\} & \text{եթե } i < j: \end{cases}$$

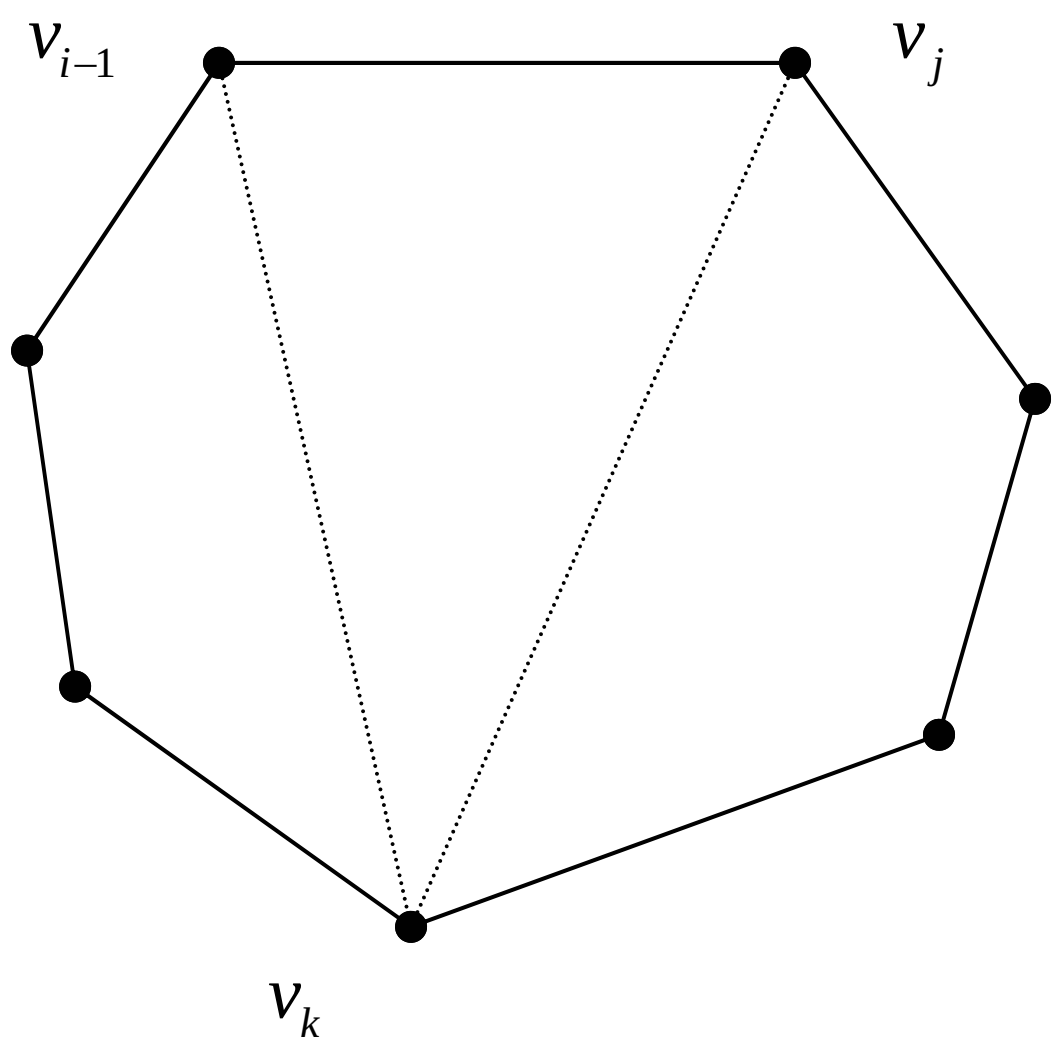

նկար 2

Քանի որ $m[i, j]$ թվերի համար մենք ստացանք նույն տեսքի անրադարձ առնչություն, ինչ-որ մատրիցների բազմապատկման խնդիրի համար, ապա կարող ենք պնդել, որ $O(n^3)$ ժամանակում մենք կարող ենք գտնել օպտիմալ տրիանգուլյացիայի արժեքը, և եթե պետք լինի, ապա հենց ինքը տրիանգուլյացիան՝ դիտարկելով համանման $s[i, j]$ մատրից:

Վերջում նշենք, որ բազմանկյան տրիանգուլյացիայի խնդիրը հանդիսանում է մատրիցների բազմապատկման խնդրի ընդհանրացումը:

Իրոք, դիցուք պահանջվում է գտնել փակագծերի օպտիմալ դասավորությունը $A_1(p_0 \times p_1)$, $A_2(p_1 \times p_2), \ldots, A_n(p_{n-1} \times p_n)$ մատրիցների արտադրյալը հաշվելու համար: Դիտարկենք $P = (v_0, v_1, ..., v_n)$ $n+1$ անկյունը, որի կողերին համապատասխանեցված են $A_1, A_2, \ldots, A_n$ մատրիցները (նկար 3), և որտեղ $w$ կշռային ֆունկցիան սահմանված է հետևյալ կերպ.

$$w(\Delta v_i v_k v_j) = p_i p_k p_j$$

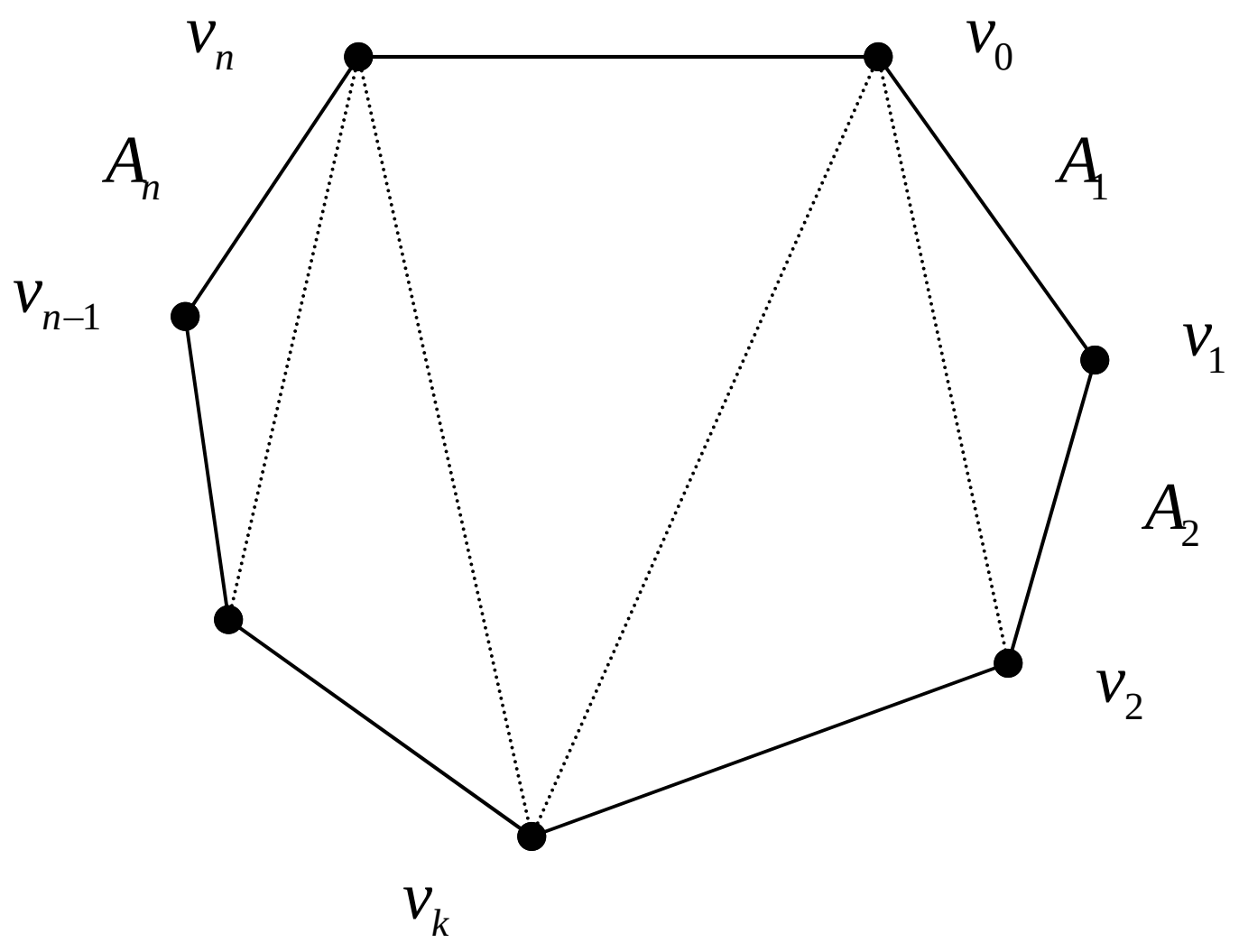

նկար 3

Նկատենք, որ այս $n+1$ անկյան օպտիմալ տրիանգուլյացիայի արժեքը կլինի հավասար հենց $A_1, A_2, \ldots, A_n$ մատրիցների արտադրյալը հաշվելու համար անհրաժեշտ նվազագույն գործողությունների քանակին: Ավելին, ունենալով $n+1$ անկյան օպտիմալ տրիանգուլյացիան, մենք կարող ենք գտնել փակագծերի օպտիմալ դասավորությունը: Դրա համար առաջնորդվենք հետևյալ կանոնով. եթե եռանկյան երկու կողմերին համապատասխանում են երկու արտահայտություններ, ապա երրորդ կողմին համապատասխանեցնենք այդ երկու արտահայտությունների արտադրյալը: Արդյունքում, $n+1$ անկյան $\overline{v_0 v_n}$ կողին կհամապատասխանի հենց $A_1, A_2, \ldots, A_n$ մատրիցների արտադրյալը:

Դիտարկենք օրինակ: Ենթադրենք ունենք $A_1(p_0 \times p_1)$, $A_2(p_1 \times p_2), \ldots, A_6(p_5 \times p_6)$ մատրիցները: Ենթադրենք, այս խնդրին համապատասխանող 7-անկյան օպտիմալ տրիանգուլյացիան նկար 4-ում ցույց տրվածն է:

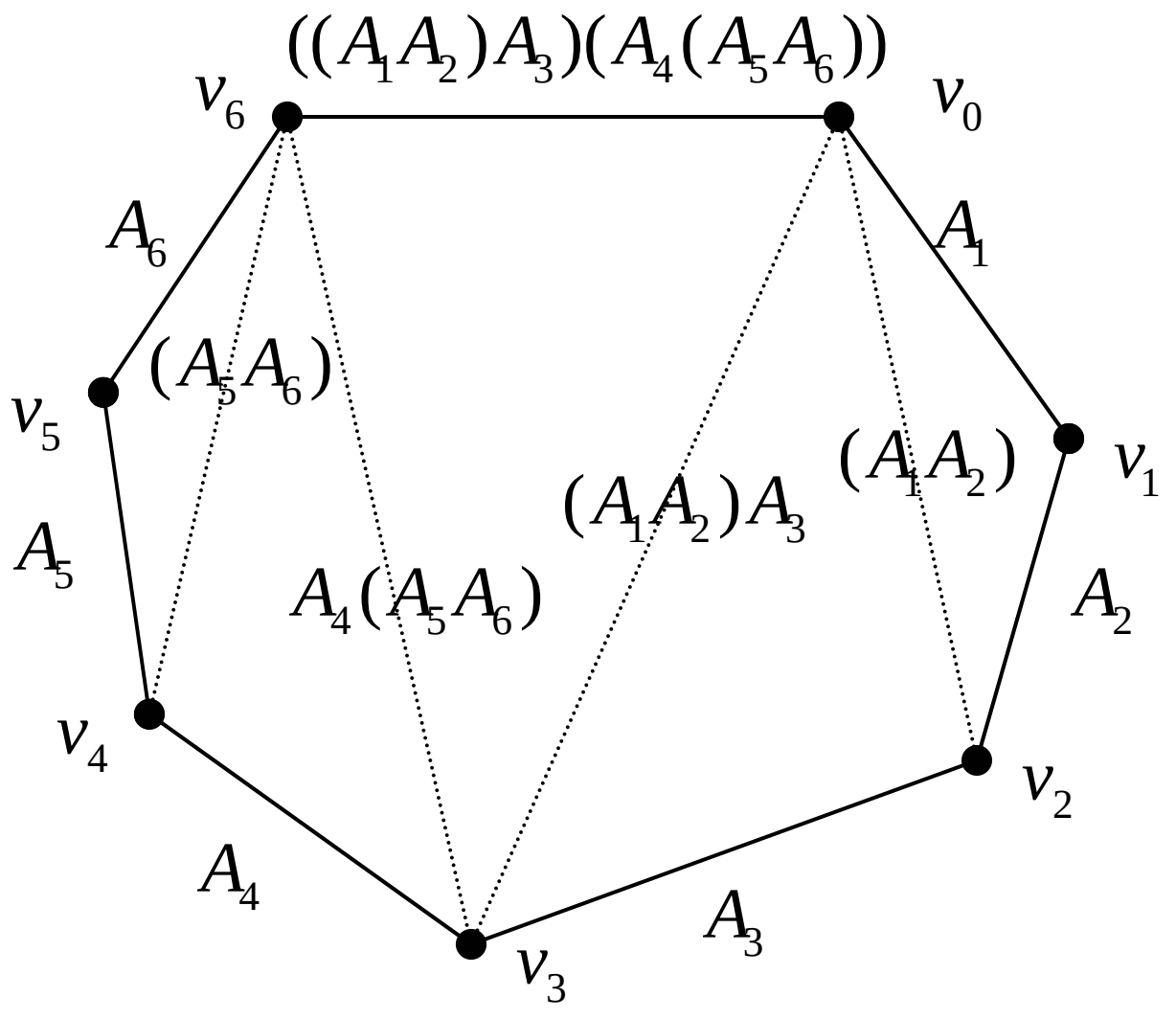

նկար 4

Այդ դեպքում $A_1(p_0 \times p_1)$, $A_2(p_1 \times p_2),\ldots, A_6(p_5 \times p_6)$ մատրիցների արտադրյալը հաշվելու համար փակագծերի օպտիմալ դասավորությունը կլինի՝ $((A_1A_2)A_3)(A_4(A_5A_6))$:

# Գրականություն

Գտնված սխալների, առաջարկությունների, ինչպես նաև դասախոսություն-ներն e-mail-ով ստանալու համար կարող էք դիմել vahanmkrtchyan2002@yahoo.com հասցեով:

Կոմբինատորային ալգորիթմներ և ալգորիթմների վերլուծություն
Վահան Վ. Մկրտչյան

Դասախոսություն 17: $P, NP$ դասերը: $NP$-լրիվություն: Բերումներ:

**Ալգորիթմական խնդիրների $P, NP$ դասերը:** Ինչպես տեսանք գրաֆում կարճագույն ճանապարհ գտնելու խնդիրը կարելի է հանգեցնել օրգրաֆում կարճագույն ուղի գտնելու խնդրին: Բնական է տալ այսպիսի հարց, ինչ ի նկատի ունենք, երբ ասում ենք, որ "ալգորիթմական մի խնդիրը կարելի է հանգեցնել մեկ այլ խնդրի" կամ "ալգորիթմական մի խնդիրը ավելի դժվար է, քան մեկ այլ խնդիրը": Սրանք են այն հիմնական հարցադրումները, որոնց մենք փորձելու ենք տալ մաթեմատիկական ձևակերպում և այդ ձևակերպումների շրջանակներում տալ հարցադրումների ինչ-որ իմաստով սպառիչ պատասխան:

Դիցուք $\Sigma=\{a_1,...,a_k\}$–ն մի որևէ վերջավոր բազմություն է: Այդ դեպքում $\Sigma$ բազմությանը կանվանենք այբուբեն: $\Sigma$ այբուբենի վերջավոր հաջորդականություններին կանվանենք բառեր, որոնց բազմությունը կնշանակենք $\Sigma^*$–ով: $n=1,2,3,4,...$ համար նշանակենք

$$\Sigma_n=\{x: x\in\Sigma^*, |x|=n\},$$

որտեղ $|x|$–ով նշանակված է $x$ բառի երկարությունը, այսինքն նրանում առկա տառերի քանակը: Նկատենք, որ $|\Sigma_n|=k^n$:

$\Sigma^*$ բազմության ցանկացած $L\subseteq\Sigma^*$ ենթաբազմությանը կանվանենք լեզու $\Sigma$ այբուբենում:

Ինչպես գիտենք Թյուրինգի $M$ մեքենան բաղկացած է վիճակների վերջավոր բազմությունից, կարդացող/գրող գլխիկից, որը կարող է շարժվել երկու կողմից անվերջ ձգվող ժապավենի վրա և անցման ֆունկցիայից (ծրագիր): Ժապավենը բաժանված է քառակուսիների, որոնցից յուրաքանչյուրը կարող է պարունակել նախապես տրված և $\Lambda$ դատարկ սիմվոլը պարունակող $\Gamma$ այբուբենից մի որևէ տառ: Յուրաքանչյուր Թյուրինգի $M$ մեքենա ունի որոշակի մուտքային $\Sigma$ այբուբեն, որը $\Gamma$ այբուբենի $\Lambda$ դատարկ սիմվոլը չպարունակող ենթաբազմություն է: Ժամանակի ցանկացած պահին $M$ մեքենան գտնվում է նախապես տրված $Q$ վիճակների բազմությանը պատկանող մի ինչ-որ $q$ վիճակում: Մեքենան աշխատում է հետևյալ կերպ. ժապավենի հաջորդական վանդակներում գրված է $x \in \Sigma^*$ մուտքային բառը, մեքենան գտնվում է սկզբնական $q_0 \in Q$ վիճակում, և մեքենայի կարդացող/գրող գլխիկը ցույց է տալիս $x \in \Sigma^*$ մուտքային բառի ամենաձախ սիմվոլի տառի վրա: Յուրաքանչյուր քայլում եթե մեքենան գտնվում է մի ինչ-որ $q$ վիճակում, և մեքենայի կարդացող/գրող գլխիկը ցույց է տալիս $s \in \Sigma$ տառի վրա, ապա մեքենան իր անցման ֆունկցիայի միջոցով որոշում է թե ինչ տառով պետք է փոխարինել ժապավենի այն վանդակում գրած $s \in \Sigma$ տառը, որի վրա ցույց էր տալիս կարդացող/գրող գլխիկը, ինչ վիճակի անցնել և կարդացող/գլխիկը տեղափոխել մեկ վանդակով ձախ թե աջ:

Եթե ավելի ֆորմալ խոսելու լինենք, ապա կարող ենք ասել, որ Թյուրինգի $M$ մեքենան իրենից ներկայացնում է $(\Sigma, \Gamma, Q, \delta)$ քառյակ, որտեղ $\Sigma, \Gamma, Q$–ն ոչ դատարկ, վերջավոր բազմություններ են, որոնք բավարարում են $\Sigma \subseteq \Gamma$ և $\Lambda \in \Gamma \setminus \Sigma$ պայմաններին: $Q$–ն պարունակում է երեք $q_0, q_{\text{ընդունում}}, q_{\text{մերժում}}$ վիճակներ: Անցման $\delta : (Q \setminus \{q_{\text{ընդունում}}, q_{\text{մերժում}}\}) \times \Gamma \to Q \times \Gamma \times \{-1, 1\}$ ֆունկցիան ունի հետևյալ իմաստը. եթե $\delta(q, s) = (q', s', h)$, ապա սա նշանակում է որ եթե մեքենան գտնվում է $q$ վիճակում, և մեքենայի կարդացող/գրող գլխիկը ցույց է տալիս $s \in \Sigma$ տառի վրա, ապա մեքենան իր վիճակը պետք է փոխի $q'$-ի, այն վանդակում, ուր գրված է $s$-ը, պետք է գրել $s'$ և կարդացող/գրող գլխիկը պետք է տեղաշարժել մեկ վանդակով ձախ կամ աջ կախված այն բանից $h = -1$ թե $h = 1$:

Կոնֆիգուրացիա ասելով կհասկանանք ցանկացած $xqy$ տող, որտեղ $x, y \in \Gamma^*$, $y \neq \Lambda$ և $q \in Q$: $xqy$ կոնֆիգուրացիայի իմաստը կայանում է նրանում, որ $M$ մեքենան գտնվում է $q \in Q$ վիճակում, որի ժապավենին գրված է $xy$ տողը և մեքենայի կարդացող/գրող գլխիկը ցույց է տալիս $y$ տողի ամենաձախ տառի վրա:

Դիցուք $C$–ն և $C'$-ը երկու կոնֆիգուրացիաներ են: Կգրենք $C \overset{M}{\rightarrow} C'$, եթե $C$–կոնֆիգուրացիայից $M$ մեքենայի միջոցով կարելի է անցնել $C'$-կոնֆիգուրացիային: $xqy$ կոնֆիգուրացիան կանվանենք վերջնական, եթե $q \in \{q_{\text{ընդունում}}, q_{\text{մերժում}}\}$: Նկատենք, որ եթե $M$ մեքենային ժամանակի ինչ-որ պահին համապատասխանում է որոշակի, ոչ վերջնական $C$ կոնֆիգուրացիա, ապա գոյություն ունի ***ճիշտ մեկ*** $C'$ կոնֆիգուրացիա, այնպես, որ $C \overset{M}{\rightarrow} C'$:

$M$ մեքենայի աշխատանքը $w \in \Sigma^*$ մուտքային բառի վրա իրենից ներկայացնում է միակ $C_0, C_1, \ldots$ հաջորդականություն, որտեղ $C_0 = q_0 w$ և $C_i \xrightarrow{M} C_{i+1}$: Եթե այս հաջորդականությունը վերջավոր է, ապա $M$ մեքենայի կատարած քայլերի քանակը $w \in \Sigma^*$ մուտքային բառի վրա հավասար է $C_0, C_1, \ldots$ հաջորդականության անդամների քանակ հանած մեկ, և անվերջ է՝ հակառակ դեպքում:

Կասենք, որ $M$ մեքենան ճանաչում է $w \in \Sigma^*$ մուտքային բառը, եթե $C_0, C_1, \ldots$ հաջորդականությունը վերջավոր է և վերջնական կոնֆիգուրացիան պարունակում է $q_{\text{ընդունում}}$ վիճակը: Նկատենք, որ $M$ մեքենան չի ճանաչում $w \in \Sigma^*$ մուտքային բառը, եթե $C_0, C_1, \ldots$ հաջորդականությունը անվերջ է կամ այն վերջավոր է և նրա վերջնական կոնֆիգուրացիան պարունակում է $q_{\text{մերժում}}$ վիճակը:

$M$ մեքենայի համար $L(M)$ -ով նշանակենք

$$L(M) = \{w \in \Sigma^* : M \text{ ճանաչում է } w\}:$$

$t_M(w)$-ով նշանակենք $M$ մեքենայի կատարած քայլերի քանակը $w \in \Sigma^*$ մուտքային բառի վրա: Եթե $M$ մեքենան ընդհանրապես կանգ չի առնում $w \in \Sigma^*$ մուտքային բառի վրա, ապա $t_M(w) = +\infty$: $n \in N$ բնական թվի համար $t_M(n)$-ով նշանակենք $M$ մեքենայի ***վատագույն*** դեպքում կատարած քայլերի քանակը, այսինքն՝

$$t_M(n) = \max\{t_M(w) : w \in \Sigma^n\}:$$

Կասենք, որ $M$ մեքենան աշխատում է բազմանդամային ժամանակում, եթե գոյություն ունի $p(n)$ բազմանդամ այնպես, որ $t_M(n) \le p(n)$: Դիտարկենք լեզուների $P$ դասը

$$P = \{L : \exists M \text{ բազմանդային Թյուրինգի մեքենա, որ } L = L(M)\}:$$

$NP$ նշանակումն իրենից ներկայացնում է Nondetermenistic Polynomial time հապավումը, քանի որ $NP$ դասի սահմանումը առաջին անգամ տրվել է այսպես կոչված “ոչ դետերմինացված Թյուրինգի” մեքենաների միջոցով: Այդպիսի մեքենաները տարբերվում են “սովորական” Թյուրինգի մեքենաներից միայն նրանով, որ եթե մեքենան գտնվում է $q$ վիճակում, և մեքենայի կարդացող/գրող գլխիկը ցույց է տալիս $s \in \Sigma$ տառի վրա, ապա $\delta(q, s)$-ն իրենից ներկայացնում է ***բազմություն***, այսինքն՝ **տրված կոնֆիգուրացիայից ոչ դետերմինացված Թյուրինգի մեքենան կարող է անցնել մի քանի կոնֆիգուրացիաների ի տարբերություն “սովորական” Թյուրինգի մեքենաների:**

Մենք կդիտարկենք $NP$ դասի մեկ այլ սահմանում: Դիցուք $R \subseteq \Sigma^* \times \Sigma^*$ բինար հարաբերություն է: Դիտարկենք $L_R \subseteq \Sigma^*$ լեզուն, որտեղ՝

$$L_R = \{(w, y) : R(w, y) = \text{ճիշտ}\}:$$

Կասենք, որ $R \subseteq \Sigma^* \times \Sigma^*$ բինար հարաբերությունը բազմանդամային է, եթե $L_R \in P$: $NP$ դասը բաղկացած է բոլոր այն $L \subseteq \Sigma^*$ լեզուներից, որոնց համար գոյություն ունեն $R \subseteq \Sigma^* \times \Sigma^*$ բազմանդամային բինար հարաբերություն և $p(n)$ բազմանդամ այնպես, որ ցանկացած $w \in \Sigma^*$ մուտքային բառի համար

$w \in L$ այն և միայն այն դեպքում, երբ *գոյություն* ունի $y \in \Sigma^*$ այնպես, որ $|y| \leq p(|w|)$ և $R(w, y) =$ ճիշտ:

Մեր ժամանակների հավանաբար ամենակարևոր պրոբլեմը ձևակերպվում է հետևյալ կերպ.

**$P, NP$ պրոբլեմը**: Արդյոք $P = NP$:

Նշենք, որ $P \subseteq NP$ այնպես, որ հիմնական խնդիրը $NP \subseteq P$ հարցի պատասխանը պարզելու մեջ է: Տես [2]-ը պրոբլեմի կարևության, ինչպես նաև խնդրի ուղղությամբ առկա արդյունքներին ծանոթանալու համար:

Փաստորեն, $P, NP$ դասերն իրենցից ներկայացնում են դասեր, որոնք բաղկացած են լեզուներից: Առաջանում է բնական հարց. իսկ ինչ կապ ունի այս ամենը խնդիրների, ավելի ճիշտ ալգորիթմական խնդիրների հետ:

Դիտարկենք մի օրինակ: Դիցուք ունենք բնական թվերի մի ինչ-որ $(a_1,...,a_n)$ հավաքածու ($a$-երից ոմանք կարող են լինել հավասար): Պահանջվում է պարզել, թե հնարավոր է տրոհել այս հավաքածուն երկու մասի այնպես, որ մի մասի գումարը հավասար լինի մյուս մասի գումարին: Այս խնդիրը կարճ կանվանենք ՏՐՈՀՈՒՄ:

Դիցուք $\Sigma$ վերջավոր այբուբեն է, որը բավարարում է $\{0,1,*\} \subseteq \Sigma$ պայմաններին: Ցանկացած $(a_1,...,a_n)$ հավաքածուի համապատասխանեցնենք $l(a_1)*l(a_2)*...*l(a_n)$ բառը $\Sigma$ այբուբենում, որտեղ $l(a_i)$-ով նշանակված է $a_1$–ի ներկայացումը թվարկության 2-ական համակարգում: Նշանակենք՝

$$L_{ՏՐՈՀՈՒՄ} = \{l(a_1)*l(a_2)*...*l(a_n) : (a_1,...,a_n) \text{ հավաքածուն կարելի է տրոհել երկու մասի այնպես, որ մի մասի գումարը հավասար լինի մյուս մասի գումարին}\}$$

Նկատենք, որ $L_{ՏՐՈՀՈՒՄ} \subseteq \Sigma^*$, այսինքն՝ $L_{ՏՐՈՀՈՒՄ}$-ը լեզու է $\Sigma$ այբուբենում: Ենթադրենք, որ $M$-ը Թյուրինգի մեքենա է, որը

- $\Sigma^*$-ին պատկանող ցանկացած մուտքային բառի վրա կանգ է առնում, և

- ճանաչում է $L_{\text{ՏՐՈՀՈՒՄ}}$-ը:

Նկատենք, որ $M$-ի միջոցով մենք կարող ենք ցանկացած $(a_1,...,a_n)$ հավաքածուի համար պարզել, թե երբ է այն հնարավոր տրոհել երկու մասի այնպես, որ մի մասի գումարը հավասար լինի մյուս մասի գումարին: Իրոք, բավական է $(a_1,...,a_n)$ հավաքածուից կառուցել $l(a_1)*l(a_2)*...*l(a_n)$ բառը, և աշխատեցնել $M$-մեքենան՝ որպես մուտքային բառ ընդունելով հենց այս բառը: Արդյունքում, եթե $M$ մեքենան կանգ առնի $q_{\text{ընդունում}}$ վիճակում, ապա $(a_1,...,a_n)$ հավաքածուն հնարավոր կլինի տրոհել երկու մասի այնպես, որ մի մասի գումարը հավասար լինի մյուս մասի գումարին, և հնարավոր չի լինի՝ եթե $M$ մեքենան կանգ առնի $q_{\text{մերժում}}$ վիճակում:

Հետևաբար, բնական է ասել, որ այդպիսի $M$ Թյուրինգի մեքենան լուծում է ՏՐՈՀՈՒՄ խնդիրը: Հենց այս մոտեցումն է, որ թույլ է տալիս $P, NP$ դասերը մեկնաբանել որպես խնդիրների բազմություն: Փաստորեն, խնդիրներին համապատասխանում են այդ խնդիրների դրական պատասխան ունեցող ենթախնդիրների կոդերից կազմված լեզուները, իսկ խնդիրները լուծող ալգորիթմներին՝ այդ խնդիրներին համապատասխանող լեզուները ճանաչող Թյուրինգի մեքենաները: Այս մեկնաբանության դեպքում կասենք, որ խնդիրը պատկանում է $P(NP)$ դասին, եթե այդ խնդրին համապատասխանող լեզուն $P(NP)$ դասից է:

Փորձենք պարզել, թե ինչ մեկնաբանություն ունի $NP$ դասը: Հիշենք $NP$ դասի սահմանումը: $L \in NP$ այն և միայն այն դեպքում, երբ գոյություն ունեն $R \subseteq \Sigma^* \times \Sigma^*$ բազմանդամային բինար հարաբերություն և $p(n)$ բազմանդամ այնպես, որ ցանկացած $w \in \Sigma^*$ մուտքային բառի համար

$w \in L$ այն և միայն այն դեպքում, երբ ***գոյություն*** ունի $y \in \Sigma^*$ այնպես, որ

$$|y| \le p(|w|) \text{ և } R(w, y) = \text{ճիշտ}:$$

Եթե խոսելու լինենք խնդիրների լեզվով, ապա $\Pi \in NP$, եթե նրա դրական պատասխան ունեցող $I$ ենթախնդիրների, և միայն նրանց համար, գոյություն ունի որոշակի կառուցվածք ($y \in \Sigma^*$ այդ կառուցվածքին համապատասխանող կոդն է), որի չափը չի գերազանցում $p(\text{չափ}(I))$, ($|y| \le p(|w|)$), և որը եթե մեզ տրված լիներ, ապա մենք բազմանդամային ժամանակում կարող էինք համոզվել, որ $I$ ենթախնդրի պատասխանը դրական է ($R(w, y) = \text{ճիշտ}$):

Նկատենք, որ $P$ դասի խնդիրների համար բավական է վերցնել $y = \Lambda$, այնպես, որ $P \subseteq NP$:

Փորձենք օրինակների վրա պարզաբանել ասվածը: Ցույց տանք, որ $\text{ՏՐՈՀՈՒՄ} \in NP$: Դիցուք $I = (a_1,...,a_n) \in \text{ՏՐՈՀՈՒՄ}$: Այդ դեպքում՝

$$\text{չափ}(I) = |w| = \lceil \log_2 a_1 \rceil + ... + \lceil \log_2 a_n \rceil + n - 1:$$

Ենթադրենք $I$ ենթախնդրի պատասխանը դրական է և գոյություն ունի $S, \overline{S}$, տրոհում այնպես, որ $\sum_{a \in S} a = \sum_{a \notin S} a$: Եթե մեզ տրված լիներ $S, \overline{S}$ տրոհումը, ապա չափ$(S, \overline{S}) \leq p($չափ$(I))$ $(|y| \leq p(|w|))$, և ունենալով $S, \overline{S}$ տրոհումը մենք կարող էինք բազմանդամային ժամանակում համոզվել, որ $S$-ին պատկանող տարրերի գումարը հավասար է $\overline{S}$-ին պատկանող տարրերի գումարին ($R(w, y) =$ ճիշտ ):

Դիտարկենք մեկ այլ օրինակ: Դիցուք $X = \{x_1, ..., x_n\}$–ը բուլյան փափախականների բազմություն է, և $f(x_1, ..., x_n) = D_1 \& ... \& D_r$-ը կոնյունկտիվ նորմալ ձև է: Պահանջվում է պարզել, թե գոյություն ունի $(\alpha_1, ..., \alpha_n)$ հավաքածու այնպես, որ $f(\alpha_1, ..., \alpha_n) = 1$: Այս խնդիրն ընդունված է անվանել ԻՐԱԳՈՐԾԵԼԻՈՒԹՅՈՒՆ: Ցույց տանք, որ ԻՐԱԳՈՐԾԵԼԻՈՒԹՅՈՒՆ $\in NP$: Դիցուք՝

$$I = (X = \{x_1, ..., x_n\}, f(x_1, ..., x_n) = D_1 \& ... \& D_r) \in \text{ԻՐԱԳՈՐԾԵԼԻՈՒԹՅՈՒՆ}$$

Այդ դեպքում՝ չափ$(I) = nr$: Նկատենք, որ եթե մեզ տրված լիներ $(\alpha_1, ..., \alpha_n)$ հավաքածուն, որի վրա $f(x_1, ..., x_n) = D_1 \& ... \& D_r$-ը կոնյունկտիվ նորմալ ձևն ընդունում է 1 արժեք, ապա մենք կարող էինք բազմանդամային ժամանակում համոզվել, որ $f(\alpha_1, ..., \alpha_n) = 1$:

Խնդիրների $NP$ դասին պատկանելը ցույց տալուց, երբեմն ստեղծվում է այն տպավորությունը, որ բավական է մեզ տրված լիներ այն, ինչ մենք փնտրում էինք, և մենք կարող էինք բազմանդամային ժամանակում համոզվել, որ խնդրի պատասխանը դրական է: Դիտարկենք հետևյալ օրինակը, որը ցույց է տալիս, որ միշտ չէ, որ այդ տպավորությունը ճիշտ է:

Ինչպես գիտենք, $G$ գրաֆի կողերի $E_0 \subseteq E(G)$ բազմությունը կոչվում է զուգակցում, եթե $E_0$–ում չկան կից կողեր: Նկատենք, որ եթե $E_0$–ն զուգակցում է, ապա $|E_0| \leq \frac{|V(G)|}{2}$: Զուգակցումը կանվանենք կատարյալ, եթե $|E_0| = \frac{|V(G)|}{2}$: Դիտարկենք հետևյալ խնդիրը. տրված է $G$ գրաֆը: Պահանջվում է պատասխանել հետևյալ հարցին. ճիշտ է արդյոք, որ $G$ գրաֆը չունի կատարյալ զուգակցում: Պարզվում է, որ այս խնդիրը պատկանում է $NP$ դասին (իրականում այս խնդիրը $P$ դասից է), չնայած առաջին հայացքից պարզ չէ, թե ինչ պետք է մեզ տրված լինի, որպեսզի կարողանանք բազմանդամային ժամանակում համոզվել, որ խնդրի պատասխանը դրական է: Տատի հետևյալ թեորեմը տալիս է հարցի պատասխանը

**Թեորեմ** (Տատ): Որպեսզի $G$ գրաֆը պարունակի կատարյալ զուգակցում, անհրաժեշտ է և բավարար, որ ցանկացած $S \subseteq V(G)$ համար տեղի ունենա $c_0(G - S) \leq |S|$ անհավասարությունը, որտեղ $c_0(H)$-ով նշանակված է $H$ գրաֆի այն կապակցվածության բաղադրիչների քանակը, որոնք պարունակում են կենտ թվով գագաթներ:

Օրինակ նկար 1-ում ցույց տրված գրաֆը չունի կատարյալ զուգակցում, քանի որ եթե նրանից հեռացնենք $v$ գագաթը, ապա ստացված գրաֆը կպարունակի երեք կապակցվածության բաղադրիչ, որոնք ունեն կենտ թվով գագաթներ:

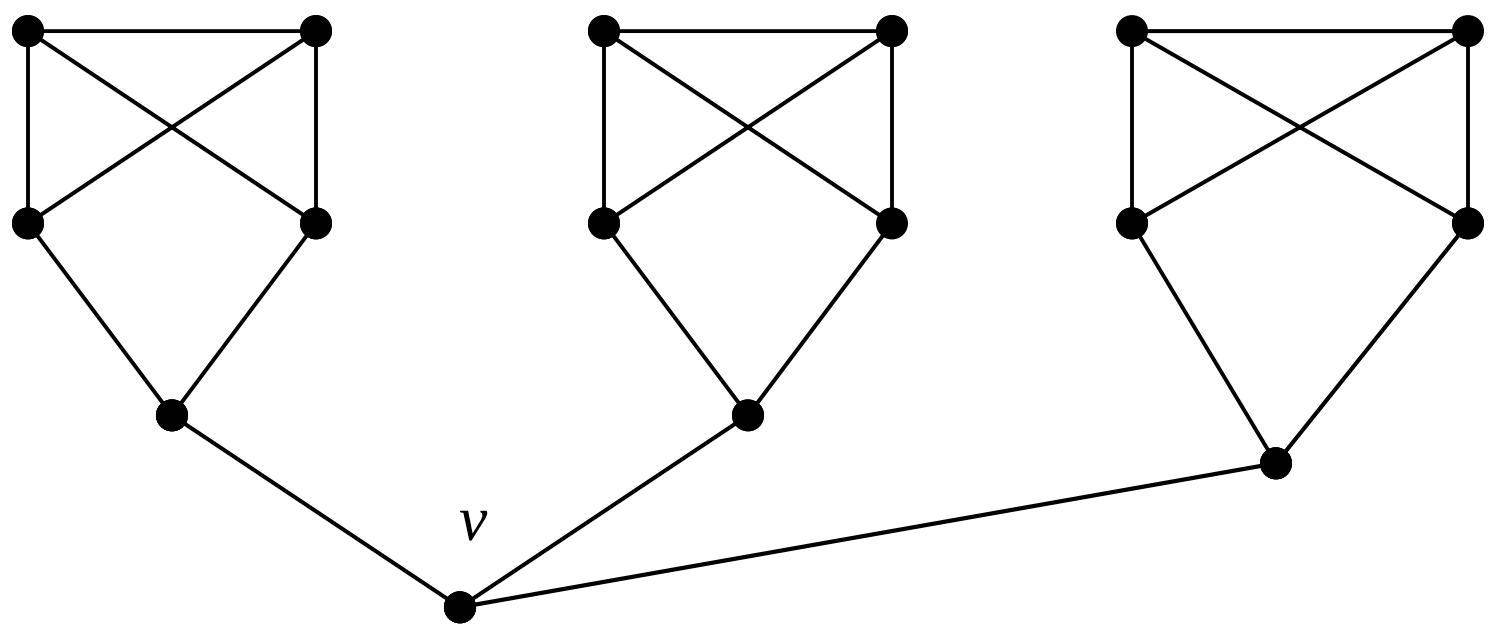

նկար 1

Փաստորեն, եթե $G$ գրաֆը չունի կատարյալ զուգակցում, ապա բավական է մեզ տրված լինի գագաթների այն $S\subseteq V(G)$ ենթաբազմությունը, որի համար $c_0(G-S)>|S|$, և մենք կարող ենք բազմանդամային ժամանակում համոզվել, որ $G$ գրաֆը չունի կատարյալ զուգակցում:

Դիցուք, $f:\Sigma^*\to\Sigma^*$ արտապատկերում է: Կասենք, որ $f:\Sigma^*\to\Sigma^*$ արտապատկերումը բազմանդամորեն հաշվարկելի է, եթե գոյություն ունի $M$ բազմանդամային բարդություն ունեցող Թյուրինգի մեքենա այնպես, որ ցանկացած $x\in\Sigma^*$ համար $M$ մեքենան կանգ է առնում $x$ բառի վրա և այդ ժամանակ $M$ մեքենայի ժապավենի վրա գրված է լինում $f(x)$ բառը:

Դիցուք $L,L'\subseteq\Sigma^*$: Կասենք, որ $L$ լեզուն բերվում է $L'$-ին և կգրենք $L\prec L'$, եթե գոյություն ունի բազմանդամորեն հաշվարկելի $f:\Sigma^*\to\Sigma^*$ արտապատկերում այնպես, որ ցանկացած $x\in\Sigma^*$ համար

$$x\in L \text{ այն և միայն այն դեպքում, երբ } f(x)\in L':$$

Կգրենք Խնդիր $1\prec$ Խնդիր $2$, եթե $L_{\text{Խնդիր 1}}\prec L_{\text{Խնդիր 2}}$: Դիտարկենք մի օրինակ: Ենթադրենք Խնդիր $1$–ը գրաֆում կարճագույն ճանապարհը գտնելու խնդիրն է, իսկ Խնդիր $2$–ը՝ օրգրաֆում կարճագույն ուղի գտնելու խնդիրն է: Մենք ցույց ենք տվել, թե ինչպես $G$ գրաֆին կարող ենք համապատասխանեցնել $f(G)$ օրգրաֆը այնպես, որ $G$ գրաֆի ցանկացած կարճագույն ճանապարհի համապատասխանի $f(G)$ օրգրաֆի կարճագույն ուղի, և հակառակը:

Դիցուք $L\in NP$: Կասենք, որ $L$ լեզուն $NP$-լրիվ է, եթե ցանկացած $L'\in NP$ համար $L'\prec L$: Նկատենք, որ տեղի ունեն հետևյալ հատկությունները.

1. $L_1\prec L_2$, $L_2\in P$, ապա $L_1\in P$,

2. եթե $L_1$-ը $NP$-լրիվ է, $L_2 \in NP$ և $L_1 \prec L_2$, ապա $L_2$-ը $NP$-լրիվ է,
3. եթե $L_1$-ը $NP$-լրիվ է, և $L_1 \in P$, ապա $P = NP$:

Փաստորեն (3)-ից հետևում է, որ $NP$-լրիվ խնդիրները $NP$ դասի "ամենադժվար" խնդիրներն են: Նկատենք նաև, որ (2)-ից հետևում է, որ որպեսզի ցույց տանք տրված $\Pi$ խնդրի $NP$-լրիվությունը, բավական է ցույց տալ, որ $\Pi \in NP$ և վերցնել մեկ այլ $\Pi'$ $NP$-լրիվ խնդիր, որը բավարարում է $\Pi' \prec \Pi$ պայմանին: Իհարկե, խնդրի $NP$-լրիվության ապացույցի այս եղանակը կիրառելու համար անհրաժեշտ է ունենալ գոնե մեկ $NP$-լրիվ խնդիր: Նշենք նաև առաջին հայացքից ընդհանրապես պարզ չէ, թե գոյություն ունի արդյոք գոնե մեկ $NP$-լրիվ խնդիր: Կուկի 1971 թվականին ապացուցված թեորեմը ցույց է տալիս առաջին $NP$-լրիվ խնդիրը:

**Թեորեմ** (Կուկ, 1971): ԻՐԱԳՈՐԾԵԼԻՈՒԹՅՈՒՆ խնդիրը $NP$-լրիվ է:

Նշենք,որ ներկայումս հայտնի են ավելի քան 10.000 $NP$-լրիվ խնդիրներ:

Վերջում նշենք, որ յուրաքանչյուր $L \in NP$ խնդրին զուգահեռ, բնական է դիտարկել համապատասխան կառուցման խնդիրը, այսինքն՝ պարզել, թե գոյություն ունի $y \in \Sigma^*$ այնպես, որ $|y| \le p(|w|)$ և $R(w, y) =$ ճիշտ, և եթե այո, ապա կառուցել այդպիսի $y$–երից գոնե մեկը: Նկատենք, որ կառուցման խնդիրը ավելի դժվար է, քան $L$-լեզվի ճանաչման խնդիրը: Պարզվում է, որ եթե $L$-ը $NP$-լրիվ է, ապա այս խնդիրները համարժեք են: Ասվածը պարզաբանենք ԻՐԱԳՈՐԾԵԼԻՈՒԹՅՈՒՆ խնդրի օրինակի վրա: Դիցուք ինչ-որ մեկին հաջողվել է գտնել գտնել ԻՐԱԳՈՐԾԵԼԻՈՒԹՅՈՒՆ խնդիրը լուծող բազմանդամային $A$ ալգորիթմ: Ցույց տանք, որ օգտագործելով $A$ ալգորիթմը, մենք կարող ենք բազմանդամային ժամանակում ուղակի կառուցել $f(x_1,...,x_n) = D_1 \& ... \& D_r$-ը կոնյունկտիվ նորմալ ձևն իրագործող հավաքածուն, եթե իհարկե այն գոյություն ունի:

Դիցուք տրված է $f(x_1,...,x_n) = D_1 \& ... \& D_r$-ը կոնյունկտիվ նորմալ ձև է: Նախ $A$-ի միջոցով պարզենք թե $f(x_1,...,x_n) = D_1 \& ... \& D_r$-ը իրագործելի է: Եթե ոչ, ապա ավարտել ալգորիթմի աշխատանքը, հակառակ դեպքում՝ դիտարկել $f(1, x_2,...,x_n)$ կոնյունկտիվ նորմալ ձևը: $A$-ի միջոցով պարզենք թե իրագործելի է այն: Եթե այն իրագործելի չէ, ապա $f(x_1,...,x_n) = D_1 \& ... \& D_r$-ը իրագործող հավաքածույում պետք է վերցնել $x_1 = 0$: Իսկ եթե այն իրագործելի է, ապա $f(x_1,...,x_n) = D_1 \& ... \& D_r$-ը իրագործող հավաքածույում վերցնել $x_1 = 1$: Այսպես շարունակել մինչև պարզվեն $x_1,...,x_n$ փոփոխականների արժեքները:

# Գրականություն

Գտնված սխալների, առաջարկությունների, ինչպես նաև դասախոսություն-ներն e-mail-ով ստանալու համար կարող էք դիմել vahanmkrtchyan2002@yahoo.com հասցեով:

Կոմբինատորային ալգորիթմներ և
ալգորիթմների վերլուծություն
Վահան Վ. Մկրտչյան

Դասախոսություն 18: 3-Իրագործելիություն, Խմբավորում, Գրաֆի ներկում, Անկախ բազմություն, Գագաթներով ծածկույթ խնդիրների $NP$-լրիվությունը:

Օգտվելով ԻՐԱԳՈՐԾԵԼԻՈՒԹՅՈՒՆ խնդրի $NP$–լրիվությունից, ստորև կապացուցվեն գրաֆների տեսության որոշ խնդիրների $NP$–լրիվությունը:

**3-ԻՐԱԳՈՐԾԵԼԻՈՒԹՅՈՒՆ**: Նախ ցույց տանք ԻՐԱԳՈՐԾԵԼԻՈՒԹՅՈՒՆ խնդրի մի մասնավոր դեպքի $NP$–լրիվությունը, որը շատ հաճախ է օգտագործվում այլ խնդիրների բարդության ուսումնասիրման ժամանակ:

Դիցուք $x$–ը բուլյան փոփոխական է: $\sigma \in \{0,1\}$ համար նշանակենք

$$x^{\sigma} = \begin{cases} x & \text{եթե} \quad \sigma = 1, \\ \bar{x} & \text{եթե} \quad \sigma = 0: \end{cases}$$

$x^{\sigma}$ արտահայտությանը կանվանենք $x$ բուլյան փոփոխականի լիտերալ: Դիցուք $X = \{x_1, ..., x_n\}$-ը բուլյան փոփոխականների բազմություն է, և $f(x_1, ..., x_n) = D_1 \& ... \& D_r$-ը կոնյունկտիվ նորմալ ձև է, որում յուրաքանչյուր տարրական $D_i$ դիզյունկցիա պարունակում է $x_1, ..., x_n$ փոփոխականների ճիշտ երեք լիտերալ: Պահանջվում է պարզել, թե գոյություն ունի $(\alpha_1, ..., \alpha_n)$ հավաքածու այնպես, որ $f(\alpha_1, ..., \alpha_n) = 1$: Այս խնդիրն ընդունված է անվանել 3-ԻՐԱԳՈՐԾԵԼԻՈՒԹՅՈՒՆ:

**Թեորեմ**: 3-ԻՐԱԳՈՐԾԵԼԻՈՒԹՅՈՒՆ խնդիրը $NP$–լրիվ է:

**Ապացույց**: Նախ նկատենք, որ 3-ԻՐԱԳՈՐԾԵԼԻՈՒԹՅՈՒՆ խնդիրը պատկանում է $NP$ դասին: Իրոք, դա հետևում է այն բանից, որ ԻՐԱԳՈՐԾԵԼԻՈՒԹՅՈՒՆ խնդիրն ինքը պատկանում է $NP$ դասին, իսկ 3-ԻՐԱԳՈՐԾԵԼԻՈՒԹՅՈՒՆ խնդիրը հանդիսանում է ԻՐԱԳՈՐԾԵԼԻՈՒԹՅՈՒՆ խնդրի մասնավոր դեպքը:

Ցույց տանք, որ ԻՐԱԳՈՐԾԵԼԻՈՒԹՅՈՒՆ $\prec$ 3-ԻՐԱԳՈՐԾԵԼԻՈՒԹՅՈՒՆ: Դիցուք $X = \{x_1, ..., x_n\}$-ը բուլյան փոփոխականների բազմություն է, և $f(x_1, ..., x_n) = D_1 \& ... \& D_r$-ը կոնյունկտիվ նորմալ ձև է: Դիտարկենք $f(x_1,...,x_n,z_1^{(1)},z_1^{(2)},...,z_n^{(1)},z_n^{(2)})$ կոնյունկտիվ նորմալ ձևը, որը ստացվում է $f(x_1,...,x_n)=D_1\,\&\ldots\&\,D_r$ կոնյունկտիվ նորմալ ձևից հետևյալ կերպ.

Եթե $D_i=x^{\sigma}$ ապա $D_i$-ն փոխարինենք իրեն համարժեք

$(x^{\sigma}\vee z_i\vee w_i)\,\&\,(x^{\sigma}\vee z_i\vee \overline{w_i})\,\&\,(x^{\sigma}\vee \overline{z_i}\vee w_i)\,\&\,(x^{\sigma}\vee \overline{z_i}\vee \overline{w_i})$ արտահայտությամբ

որտեղ $z_i$–ն և $w_i$-ն նոր բուլյան փոփոխականներ են,

Եթե $D_i=x^{\sigma}\vee y^{\tau}$ ապա $D_i$-ն փոխարինենք իրեն համարժեք

$(x^{\sigma}\vee y^{\tau}\vee w_i)\,\&\,(x^{\sigma}\vee y^{\tau}\vee \overline{w_i})$ արտահայտությամբ

որտեղ $w_i$-ն նոր բուլյան փոփոխական է,

իսկ եթե $D_i=x_{i_1}^{\sigma_{i_1}}\vee x_{i_2}^{\sigma_{i_2}}\vee\ldots\vee x_{i_k}^{\sigma_{i_k}}$, $k\geq 4$ ապա $D_i$-ն փոխարինենք իրեն համարժեք $(x_{i_1}^{\sigma_{i_1}}\vee x_{i_2}^{\sigma_{i_2}}\vee w_i)\,\&\,(\overline{w_i}\vee x_{i_3}^{\sigma_{i_3}}\ldots\vee x_{i_k}^{\sigma_{i_k}})$ արտահայտությամբ: Նկատենք, որ գրված կոնյունկցիայում երկրորդ անդամն արդեն պարունակում է $k-1$ լիտերալ: Այնուհետև, $(x_{i_1}^{\sigma_{i_1}}\vee x_{i_2}^{\sigma_{i_2}}\vee w_i)\,\&\,(\overline{w_i}\vee x_{i_3}^{\sigma_{i_3}}\ldots\vee x_{i_k}^{\sigma_{i_k}})$ արտահայտությունը փոխարինենք $(x_{i_1}^{\sigma_{i_1}}\vee x_{i_2}^{\sigma_{i_2}}\vee w_i)\,\&\,(\overline{w_i}\vee x_{i_3}^{\sigma_{i_3}}\vee z_i)\,\&\,(\overline{z_i}\vee x_{i_4}^{\sigma_{i_4}}\ldots\vee x_{i_k}^{\sigma_{i_k}})$ արտահայտությամբ: Նկատենք, որ գրված կոնյունկցիայում երրորդ անդամն արդեն պարունակում է $k-2$ լիտերալ: Այստեղ $z_i$–ն և $w_i$-ն նոր բուլյան փոփոխականներ են: Այսպիսով, $k$ փոխարինումների արդյունքում մենք $D_i$-ն կփոխարինենք մի արտահայտությամբ, որում ցանկացած դիզյունկցիա կպարունակի ճիշտ երեք լիտերալ:

Դիտարկենք $f(x_1,...,x_n,...)$ նոր կոնյունկտիվ նորմալ ձևը: Նկատենք, որ այն կազմված է ոչ ավել, քան $nr+n=n(r+1)$ փոփոկանների լիտերալներից: Քանի որ յուրաքանչյուր $D_i$- դիզյունկցիա փոխարինվում է ոչ ավել քան $n$ դիզյունկցիաներով, ապա $f(x_1,...,x_n,...)$ նոր կոնյունկտիվ նորմալ ձևը կազմված կլինի ոչ ավել, քան $nr$ դիզյունկցիաներից: Նկատենք, որ եթե որպես կոնյունկտիվ նորմալ ձևի չափ վերցնենք փոփոխականների և դիզյունկցիաների քանակների արտադրյալը, ապա սկզբնական կոնյունկտիվ նորմալ ձևի չափը կլինի $nr$-ը, իսկ վերջնական կոնյունկտիվ նորմալ ձևի չափը՝ $n(r+1)\cdot nr$, ինչը փոփոխվում է $nr$-ի նկատմամբ բազմանդամով: Հետևաբար՝ բերման բարդությունը բազմանդամային է:

Վերջում նկատենք, որ քանի որ յուրաքանչյուր դիզյունկցիա փոխարինվում է իրեն համարժեք արտահայտությամբ, ապա $f(x_1,...,x_n)=D_1\,\&\ldots\&\,D_r$-ը կոնյունկտիվ նորմալ ձևը կլինի իրագործելի այն և միայն այն դեպքում, երբ $f(x_1,...,x_n,...)$ նոր կոնյունկտիվ նորմալ ձևը կլինի իրագործելի: Թեորեմն ապացուցված է:

Նշենք նաև, որ ապացուցված է, որ 2-ԻՐԱԳՈՐԾԵԼԻՈՒԹՅՈՒՆ $\in P$ :

**Խմբավորում**: CLIQUE: $G$ գրաֆի համար $\omega(G)$-ով նշանակենք նրա ամենամեծ լրիվ ենթագրաֆի գագաթների քանակը: ԽՄԲԱՎՈՐՈՒՄ խնդիրը ձևակերպվում է հետևյալ կերպ.

ԽՄԲԱՎՈՐՈՒՄ
Տրված է $G$ գրաֆը և $k$ բնական թիվը:
Պահանջվում է պարզել $\omega(G) \geq k$ թե ոչ:

**Թեորեմ**: ԽՄԲԱՎՈՐՈՒՄ խնդիրը $NP$ –լրիվ է:
**Ապացույց**: Նախ նկատենք, որ ԽՄԲԱՎՈՐՈՒՄ խնդիրը պատկանում է $NP$ դասին: Իրոք, եթե մեզ տրված լիներ որոնելի գագաթների բազմությունը, ապա մենք կարող էինք համոզվել, որ այն պարունակում է առնվազն $k$ գագաթ և այն կազմում է լրիվ ենթագրաֆ: Նկատենք, որ այս ստուգումը կարելի է իրականացնել բազմանդամային ալգորիթմի միջոցով, այնպես որ ԽՄԲԱՎՈՐՈՒՄ խնդիրը պատկանում է $NP$ դասին:

ԽՄԲԱՎՈՐՈՒՄ խնդրի լրիվությունը ցույց տալու համար, ապացուցենք, որ ԻՐԱԳՈՐԾԵԼԻՈՒԹՅՈՒՆ $\prec$ ԽՄԲԱՎՈՐՈՒՄ:

Դիտարկենք $f(x_1,...,x_n) = D_1 \& ... \& D_r$՝ ԻՐԱԳՈՐԾԵԼԻՈՒԹՅՈՒՆ խնդրի անհատ խնդիրը և հետևյալ ձևով սահմանված $G$ գրաֆը, $k$ բնական թիվը:

$$V(G) \equiv \{(x^\sigma, D_j) : x^\sigma \in F_j, 1 \leq j \leq r\},$$

$$E(G) \equiv \{((x^\sigma, D_i),(y^\tau, D_j)) : i \neq j, x^\sigma \neq \overline{y^\tau}\},$$

$$k \equiv r:$$

Ցույց տանք, որ ԽՄԲԱՎՈՐՈՒՄ խնդրին պատկանող $G,k$ անհատ խնդրում $\omega(G) \geq k$ այն և միայն այն դեպքում, երբ $f(x_1,...,x_n) = D_1 \& ... \& D_r$ կոնյունկտիվ նորմալ ձևն իրագործելի է:

Ենթադրենք, որ $\omega(G) \geq k = r$: Համաձայն $G$ գրաֆի սահմանման նրանում չկան $((x^\sigma, D_i),(y^\tau, D_j))$ կողեր, որոնց երկրորդ բաղադրիչները նույնն են, հետևաբար՝ $r$ խմբավորման գագաթների երկրորդ կոմպոնենտները կլինեն $D_1,...,D_r$ դիզյունկցիաները: Նկատենք, որ ըստ $G$ գրաֆի սահմանման այս $r$ խմբավորման գագաթների առաջին կոմպոնենտներից ոչ մեկը չի լինի մյուսի ժխտումը, հետևաբար $r$ խմբավորման գագաթները կլինեն հետևյալ տեսքի՝ $(x_{i_1}^{\sigma_{i_1}}, D_1),...,(x_{i_r}^{\sigma_{i_r}}, D_r)$: Եթե $x_1,...,x_n$ փոփոխականների արժեքներն ընտրենք այնպես, որ $x_{i_1}^{\sigma_{i_1}},...,x_{i_r}^{\sigma_{i_r}}$ արտահայտություններն ընդունեն մեկ արժեք, ապա պարզ է, որ այդպիսի հավաքածուի համար $f(x_1,...,x_n) = D_1 \& ... \& D_r$ կոնյունկտիվ նորմալ ձևը կընդունի մեկ արժեք:

Հակառակը, ենթադրենք, որ $f(x_1,...,x_n)=D_1\&...\&D_r$ կոնյունկտիվ նորմալ ձևն իրագործելի է: Այդ դեպքում գոյություն ունի $x_1,...,x_n$ փոփոխականների այնպիսի արժեքներ, որոնց համար $D_1,...,D_r$ դիզյունկցիաներից յուրաքանչյուրն ընդունում է մեկ արժեք: Նկատենք, որ այստեղից հետևում է, որ $D_1,...,D_r$ դիզյունկցիաներին պատկանող լիտերալներից կարելի է ընտրել $x_{i_1}^{\sigma_{i_1}},...,x_{i_r}^{\sigma_{i_r}}$-ը այնպես, որ ընտրվածներից ոչ մեկը մյուսի ժխտումը չէ: Նկատենք, որ այդ դեպքում $G$ գրաֆի $(x_{i_1}^{\sigma_{i_1}},D_1),...,(x_{i_r}^{\sigma_{i_r}},D_r)$ գագաթները կկազմեն $k=r$ խմբավորում:

Վերջում նկատենք, որ քանի որ $|V(G)|\le nr$, ապա նկարագրված բերումը բազմանդամային է, հետևաբար՝ ԽՄԲԱՎՈՐՈՒՄ խնդիրն իրոք $NP$–լրիվ է: Թեորեմն ապացուցված է:

Օգտվելով ԽՄԲԱՎՈՐՈՒՄ խնդրի $NP$–լրիվությունից ցույց տանք ԱՆԿԱԽ ԲԱԶՄՈՒԹՅՈՒՆ և ԳԱԳԱԹՆԵՐՈՎ ԾԱԾԿՈՒՅԹ խնդիրների $NP$–լրիվությունը, որոնց ձևակերպումը կայանում է հետևյալում

ԱՆԿԱԽ ԲԱԶՄՈՒԹՅՈՒՆ
Տրված է $G$ գրաֆը և $k$ բնական թիվը:
Պահանջվում է պարզել $\alpha(G)\ge k$ թե ոչ:

ԳԱԳԱԹՆԵՐՈՎ ԾԱԾԿՈՒՅԹ
Տրված է $G$ գրաֆը և $k$ բնական թիվը:
Պահանջվում է պարզել $\beta(G)\le k$ թե ոչ:

որտեղ $\alpha(G)$-ով նշանակված է $G$ գրաֆի առավելագույն թվով անկախ գագաթների քանակը (երկու գագաթ կոչվում են անկախ, եթե նրանք կից չեն), իսկ $\beta(G)$-ով նշանակված է $G$ գրաֆի նվազագույն գագաթային ծածկույթի հզորությունը ($G$ գրաֆի գագաթների բազմության $V'\subseteq V(G)$ ենթաբազմությունը կոչվում է գագաթային ծածկույթ, եթե $G\setminus V'$ գրաֆը կողեր չի պարունակում, այլ կերպ ասած $G$ գրաֆի ցանկացած կող ինցիդենտ է $V'\subseteq V(G)$ ենթաբազմության ինչ-որ գագաթի):

**Թեորեմ**: ԱՆԿԱԽ ԲԱԶՄՈՒԹՅՈՒՆ և ԳԱԳԱԹՆԵՐՈՎ ԾԱԾԿՈՒՅԹ խնդիրները $NP$–լրիվ են:
**Ապացույց**: Նախ նկատենք, որ երկու խնդիրներն էլ $NP$ դասից են: Այնուհետև նկատենք, որ քանի որ ցանկացած $G$ գրաֆի համար $\omega(G)=\alpha(\overline{G})$ , որտեղ $\overline{G}$-ով նշանակված է $G$ գրաֆի լրացում գրաֆը, ապա կարող ենք ասել, որ

ԽՄԲԱՎՈՐՈՒՄ $\prec$ ԱՆԿԱԽ ԲԱԶՄՈՒԹՅՈՒՆ, և հետևաբար՝ ԱՆԿԱԽ ԲԱԶՄՈՒԹՅՈՒՆ խնդիրը $NP$–լրիվ է: Մյուս կողմից, քանի որ $G$ գրաֆում $V' \subseteq V(G)$ ենթաբազմությունը հանդիսանում է անկախ բազմություն այն և միայն այն դեպքում, երբ $V(G) \setminus V'$ ենթաբազմությունը հանդիսանում է գագաթային ծածկույթ, ապա կարող ենք ասել, որ $G$ գրաֆում $\alpha(G) \geq k$ այն և միայն այն դեպքում, երբ $\beta(G) \leq n-k$, հետևաբար՝ ԱՆԿԱԽ ԲԱԶՄՈՒԹՅՈՒՆ $\prec$ ԳԱԳԱԹՆԵՐՈՎ ԾԱԾԿՈՒՅԹ և ԳԱԳԱԹՆԵՐՈՎ ԾԱԾԿՈՒՅԹ խնդիրը ևս $NP$–լրիվ է: Թեորեմն ապացուցված է:

Ինչպես գիտենք, $G$ գրաֆի համար $f: V(G) \to \{1,...,k\}$ արտապատկերումը կոչվում է ճշգրիտ $k$-ներկում, եթե գրաֆի ցանկացած $e=(u,v) \in E(G)$ կողի համար $f(u) \neq f(v)$: Նկատենք, որ ցանկացած $G$ գրաֆ ունի $V(G)$-ներկում: Ընդունված է այն նվազագույն $k$-ն, որի համար $G$ գրաֆն ունի $k$-ներկում, անվանել $G$ գրաֆի քրոմատիկ ինդեքս կամ ներկման թիվ և նշանակել $\chi(G)$-ով: Դիտարկենք գրաֆների ներկումներին առնչվող հետևյալ խնդիրը

ԳՐԱՖԻ ՆԵՐԿՈՒՄ
Տրված է $G$ գրաֆը և $k$ բնական թիվը:
Պահանջվում է պարզել $\chi(G) \leq k$ թե ոչ:

Օգտվելով 3-ԻՐԱԳՈՐԾԵԼԻՈՒԹՅՈՒՆ խնդրի $NP$–լրիվությունից, ապացուցենք հետևյալ թեորեմը

**Թեորեմ**: ԳՐԱՖԻ ՆԵՐԿՈՒՄ խնդիրը $NP$–լրիվ է:
**Ապացույց**: Անմիջապես նկատենք, որ ԳՐԱՖԻ ՆԵՐԿՈՒՄ խնդիրը պատկանում է $NP$ դասին: Իրոք, եթե մեզ տրված լիներ գրաֆի որոնելի ներկումը, ապա մենք կարող էինք համոզվել, որ այն հանդիսանում է գրաֆի ճշգրիտ $k$-ներկում: Նկատենք նաև, որ այս ստուգումը կարելի է իրականացնել բազմանդամային ալգորիթմի միջոցով, այնպես որ ԳՐԱՖԻ ՆԵՐԿՈՒՄ խնդիրը պատկանում է $NP$ դասին:

ԳՐԱՖԻ ՆԵՐԿՈՒՄ խնդրի լրիվությունը ցույց տալու համար, ապացուցենք, որ 3-ԻՐԱԳՈՐԾԵԼԻՈՒԹՅՈՒՆ $\prec$ ԳՐԱՖԻ ՆԵՐԿՈՒՄ:

Դիտարկենք $f(x_1,...,x_n) = D_1 \& ... \& D_r$՝ 3-ԻՐԱԳՈՐԾԵԼԻՈՒԹՅՈՒՆ խնդրի անհատ խնդիրը: Նկատենք, որ ավելացնելով ոչ էական փոփոխականներ, մենք միշտ կարող ենք ենթադրել, որ $n \geq 4$: Դիտարկենք հետևյալ ձևով սահմանված $G$ գրաֆը, $k$ բնական թիվը:

$$k = n+1$$
$$V(G) \equiv \{x_1,...,x_n,\overline{x}_1,...,\overline{x}_n,D_1,...,D_r,v_1,...,v_n\},$$

իսկ գրաֆի կողերի $E(G)$ բազմությունը որոշվում է հետևյալ կանոններով

$$\{v_i, v_j\} \in E(G), 1 \le i < j \le n,$$
$$\{v_i, x_j\} \in E(G), 1 \le i \ne j \le n,$$
$$\{v_i, \overline{x_j}\} \in E(G), 1 \le i \ne j \le n,$$
$$\{x_i, \overline{x_i}\} \in E(G), 1 \le i \le n,$$

եթե $x_i^{\sigma_i} \notin D_j$ ապա $\{x_i, D_j\} \in E(G), 1 \le i \le n, 1 \le j \le r$ (նկար 1):

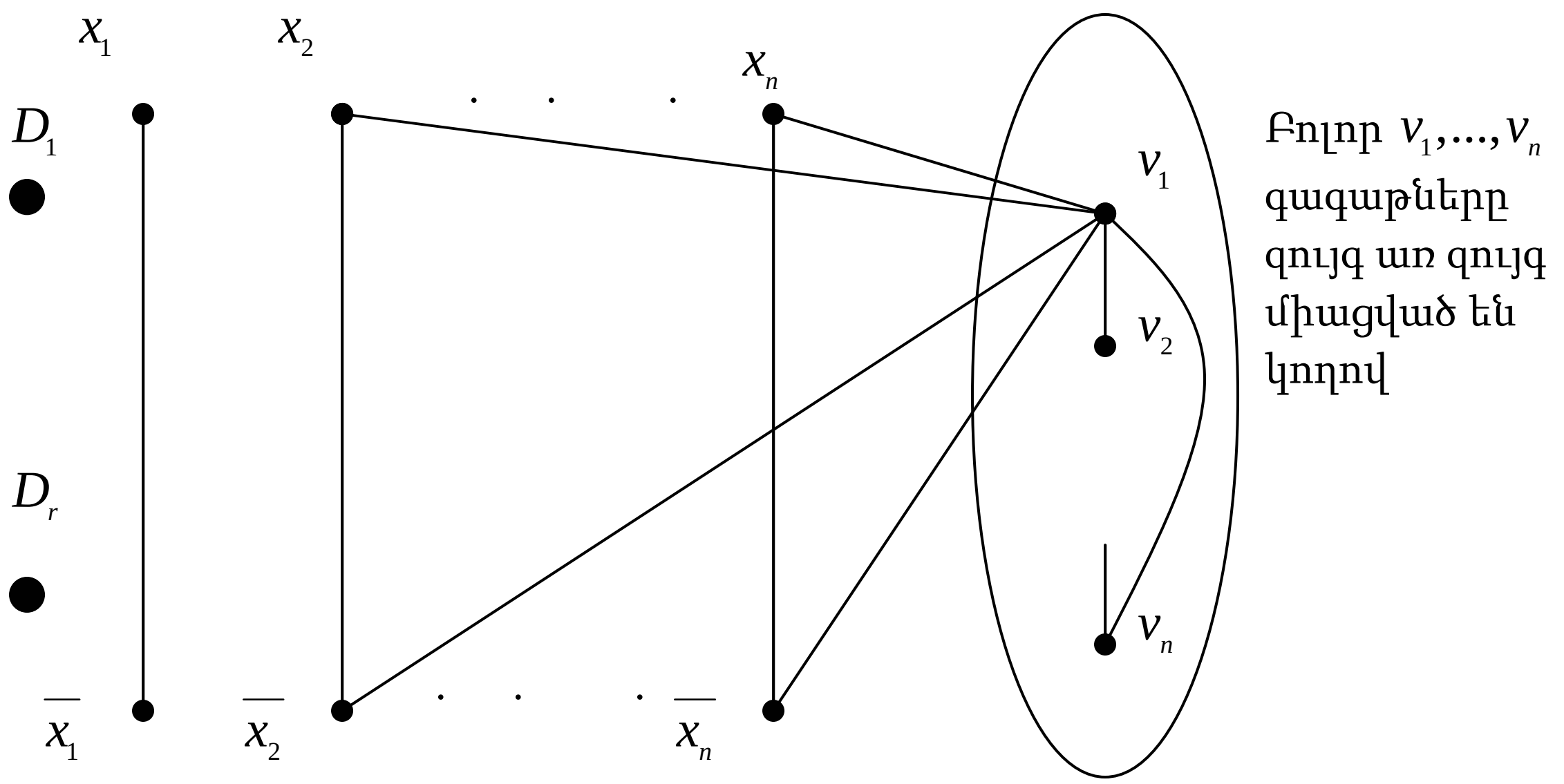

նկար 1

Նախ նկատենք, որ $|V(G)| = 3n + r$, հետևաբար նկարագրված բերումը բազմանդամային է: Ավելին, նկատենք նաև, որ $v_2,...,v_n, \overline{x_1}, x_1$ գագաթները կազմում են $G$ գրաֆի $n+1$ խմբավորում, հետևաբար $\chi(G) \ge n+1$:

Ցույց տանք, որ ԳՐԱՖԻ ՆԵՐԿՈՒՄ խնդրին պատկանող $G, k$ անհատ խնդրում $\chi(G) = n+1$ այն և միայն այն դեպքում, երբ $f(x_1,...,x_n) = D_1 \& ... \& D_r$ կոնյունկտիվ նորմալ ձևն իրագործելի է:

Ենթադրենք, որ գոյություն ունի $G$ գրաֆի ճշգրիտ ներկում $1,...,n+1$ գույների միջոցով: Ենթադրենք, որ այդ ներկման մեջ $v_1,...,v_n$ գագաթները ներկված են համապատասխանաբար $1,...,n$ գույների միջոցով: Քանի որ $G$ գրաֆում $v_i$ գագաթը կից է բոլոր փոփոխականների լիտերալներին բացառությամբ $x_i$ փոփոխականի լիտերալից, ապա կարող ենք պնդել, որ այդ ներկման մեջ $\overline{x_i}, x_i$ գագաթները ներկված են $\{i, n+1\}$ գույների միջոցով: Մյուս կողմից, քանի որ $n \ge 4$, ապա ցանկացած $D_j$ դիզյունկցիայի համար գոյություն ունի $x_i$ փոփոխական, որի լիտերալները չեն մասնակցում $D_j$-ում: Հետևաբար, համաձայն $G$ գրաֆի սահմանման $D_j$ գագաթը կից է $\overline{x_i}, x_i$ գագաթներին և հետևաբար վերոհիշյալ ճշգրիտ ներկման մեջ $D_j$ գագաթը չի կարող ներկված լինել $n+1$ գույնի միջոցով: Նկատենք, որ իրականում $D_j$ գագաթը կարող է ներկված լինել միայն իր մեջ առկա փոփոխականների համարի գույնով:

Սահմանենք $x_1,...,x_n$ փոփոխականների արժեքները հետևյալ կանոնով.

եթե $x_i$ գագաթը ներկվել է $i$ գույնի միջոցով, ապա վերցնել $x_i = 1$, իսկ եթե $x_i$ գագաթը ներկվել է $n+1$ գույնի միջոցով, ապա վերցնել $x_i = 0$: Ցույց տանք, որ $D_1,...,D_r$ դիզյունկցիաներից յուրաքանչյուրը այս հավաքածուի վրա ընդունում է մեկ արժեք: Դիտարկենք $D_j$ դիզյունկցիան և ենթադրենք, որ այն ներկվել է $i$ գույնով: Ինչպես նշեցինք վերևում, այստեղից հետևում է, որ $D_j$ դիզյունկցիան պարունակում է $x_i$ փոփոխականի լիտերալ: Քննարկենք երկու դեպք

**Դեպք 1**: $x_i = 1$: Սա նշանակում է, որ $x_i$ գագաթը ներկվել է $i$ գույնի միջոցով: Քանի որ $D_j$ դիզյունկցիան ինքն էլ ներկվել է $i$ գույնով, ապա կարող ենք ասել, որ $D_j$ և $x_i$ գագաթները միացված չեն կողով $G$ գրաֆում, հետևաբար՝ $x_i$ փոփոխականն ինքն է մասնակցում $D_j$ դիզյունկցիայում: Հետևաբար, եթե $x_i$ փոփոխականին տրվել է մեկ արժեք, ապա $D_j$-ն կընդունի մեկ արժեք:

**Դեպք 2**: $x_i = 0$: Սա նշանակում է, որ $x_i$ գագաթը ներկվել է $n+1$ գույնի միջոցով, և հետևաբար $\overline{x_i}$ գագաթը ներկվել է $i$ գույնով: Քանի որ $D_j$ դիզյունկցիան ինքն էլ ներկվել է $i$ գույնով, ապա կարող ենք ասել, որ $D_j$ և $\overline{x_i}$ գագաթները միացված չեն կողով $G$ գրաֆում, հետևաբար՝ $\overline{x_i}$ լիտերալն ինքն է մասնակցում $D_j$ դիզյունկցիայում: Հետևաբար, եթե $x_i$ փոփոխականին տրվել է զրո արժեք, ապա $D_j$-ն կընդունի մեկ արժեք:

Երկու դեպքերի քննարկման արդյունքում ունենք, որ $f(x_1,...,x_n) = D_1 \& ... \& D_r$ կոնյունկտիվ նորմալ ձևն իրագործվում է $x_1,...,x_n$ փոփոխականների արժեքների վերոհիշյալ սահմանման դեպքում:

Հիմա ենթադրենք, որ $f(x_1,...,x_n) = D_1 \& ... \& D_r$ կոնյունկտիվ նորմալ ձևն իրագործվում է $x_1,...,x_n$ փոփոխականների $\alpha_1,...,\alpha_n$ արժեքների դեպքում: Ցույց տանք, որ այդ դեպքում $G$ գրաֆը կարելի է ներկել $n+1$ գույների միջոցով: Դիտարկենք հետևյալ ձևով սահմանված $c: V(G) \to \{1,...,n+1\}$ արտապատկերումը:

$$c(v_i) = i, 1 \le i \le n,$$

$$c(x_i)=\begin{cases} i, & \text{եթե} \quad \alpha_i=1, \\ n+1, & \text{եթե} \quad \alpha_i=0, \end{cases}$$

$$c(\overline{x_i})=\begin{cases} i, & \text{եթե} \quad \alpha_i=0, \\ n+1, & \text{եթե} \quad \alpha_i=1: \end{cases}$$

Եվ վերջապես, քանի որ $f(x_1,...,x_n)=D_1\&...\&D_r$ կոնյունկտիվ նորմալ ձևն իրագործվում է $x_1,...,x_n$ փոփոխականների $\alpha_1,...,\alpha_n$ արժեքների դեպքում, ապա ցանկացած $D_j$ դիզյունկցիայի համար գոյություն ունի $x_{i_j}$ փոփոխական այնպես, որ $D_j$ դիզյունկցիայի մեջ մասնակցող $x_{i_j}$ փոփոխականի լիտերալն ընդունում է մեկ արժեք: $c:V(G)\to\{1,...,n+1\}$ արտապատկերման արժեքը $G$ գրաֆի $D_1,...,D_r$ գագաթների վրա սահմանենք հետևյալ կերպ $c(D_j)=i_j$, $1\le i_j\le n$:

Ցույց տանք, որ այս ձևով սահմանված $c:V(G)\to\{1,...,n+1\}$ արտապատկերումը հանդիսանում է $G$ գրաֆի ճշգրիտ ներկում:

Նկատենք, որ $c:V(G)\to\{1,...,n+1\}$ արտապատկերման սահմանումից հետևում է, որ $G$ գրաֆի $\{v_i,v_j\}$, $\{v_i,x_j\}$, $\{v_i,\overline{x_j}\}$, $\{x_i,\overline{x_i}\}$ կողերի ծայրակետերում $c$-ն ընդունում է տարբեր արժեքներ: Ցույց տանք, որ $G$ գրաֆի $D_j$ գագաթից դուրս եկող կողերի ծայրակետերում $c$-ն ևս ընդունում է տարբեր արժեքներ:

Ենթադրենք, որ $x_{i_j}$ փոփոխականից զատ $D_j$ դիզյունկցիայի մեջ մասնակցում են նաև $x_k,x_l$ փոփոխականների լիտերալներ ($k\ne l, k\ne i_j, l\ne i_j$): Քանի որ $x_k,x_l$ փոփոխականների լիտերալներին համապատասխանող գագաթները ներկվում են համապատասխանաբար $\{k,n+1\}$ և $\{l,n+1\}$ գույներով, ապա պարզ է, որ $G$ գրաֆի $D_j$ գագաթը այդ փոփոխականների լիտերալներին միացնող կողի ծայրակետերում $c$-ն ընդունում է տարբեր արժեքներ ($D_j$ գագաթը ներկված է $c(D_j)=i_j$ գույնով): Ցույց տանք, որ $D_j$ գագաթը $x_{i_j}$ փոփոխականին միացնող կողի ծայրակետերում $c$-ն ընդունում է տարբեր արժեքներ: Քննարկենք երկու դեպք:

**Դեպք 1**: $D_j$ դիզյունկցիայի մեջ մասնակցում է հենց $x_{i_j}$ փոփոխականը: Այդ դեպքում պարզ է, որ $x_{i_j}$ փոփոխականն ընդունել է մեկ արժեք, հետևաբար $G$ գրաֆի $x_{i_j}$ գագաթը ներկվել է $i_j$ գույնով, իսկ $\overline{x_{i_j}}$ գագաթը՝ $n+1$ գույնով: Համաձայն $G$ գրաֆի սահմանման $D_j$ գագաթը կից է $\overline{x_{i_j}}$ գագաթին, հետևաբար

$D_j$ զազաթը $x_{i_j}$ փոփոխականին միացնող կողի ծայրակետերում $c$-ն ընդունում է տարբեր արժեքներ:

**Դեպք 2**: $D_j$ դիզյունկցիայի մեջ մասնակցում է հենց $\overline{x_{i_j}}$-ը: Այդ դեպքում պարզ է, որ $x_{i_j}$ փոփոխականն ընդունել է զրո արժեք, հետևաբար $G$ գրաֆի $x_{i_j}$ գագաթը ներկվել է $n+1$ գույնով, իսկ $\overline{x_{i_j}}$ գագաթը՝ $i_j$ գույնով: Համաձայն $G$ գրաֆի սահմանման $D_j$ գագաթը կից է $x_{i_j}$ գագաթին, հետևաբար $D_j$ գագաթը փոփոխականին միացնող կողի ծայրակետերում $c$-ն ընդունում է տարբեր արժեքներ:

Այսպիսով, վերոհիշյալ ձևով սահմանված $c:V(G)\to\{1,\ldots,n+1\}$ արտապատկերումը հանդիսանում է $G$ գրաֆի ճշգրիտ ներկում: Թեորեմն ապացուցված է:

# Գրականություն

Գտնված սխալների, առաջարկությունների, ինչպես նաև դասախոսություն-ներն e-mail-ով ստանալու համար կարող եք դիմել vahanmkrtchyan2002@yahoo.com հասցեով:

Կոմբինատորային ալգորիթմներ և ալգորիթմների վերլուծություն
Վահան Վ. Մկրտչյան

Դասախոսություն 19: Բազմության և կատարյալ ծածկույթ, Ուսապարկ, 0-1 Ուսապարկ, Ներկայացուցիչների համակարգ, Տրոհում, Ստուգող թեստ խնդիրների $NP$-լրիվությունը:

Ստորև կդիտարկվեն բազմություններին առնչվող որոշ խնդիրներ և կապացուցվեն նրանց $NP$–լրիվությունը: Դիտարկենք ԿԱՏԱՐՅԱԼ ԾԱԾԿՈՒՅԹ խնդիրը

ԿԱՏԱՐՅԱԼ ԾԱԾԿՈՒՅԹ
Տրված է $A=\{a_1,...,a_n\}$ բազմությունը և նրա ենթաբազմությունների $\boldsymbol{A}=\{A_1,...,A_m\}$ ընտանիքը, որը ծածկում է $A=\{a_1,...,a_n\}$ բազմությունը, այսինքն՝

$$\bigcup_{i=1}^{m} A_i = A:$$

Պահանջվում է պարզել, թե գոյություն ունի արդյոք $\boldsymbol{A}=\{A_1,...,A_m\}$ բազմության ճշգրիտ ենթածածկույթ, այսինքն՝ զույգ առ զույգ չհատվող $A_{i_1},...,A_{i_k}$ տարրեր, որոնք ծածկում են $A=\{a_1,...,a_n\}$ բազմությունը

$$\bigcup_{j=1}^{k} A_{i_j} = A:$$

**Թեորեմ**: ԿԱՏԱՐՅԱԼ ԾԱԾԿՈՒՅԹ խնդիրը $NP$–լրիվ է:
**Ապացույց**: Նախ նկատենք, որ ԿԱՏԱՐՅԱԼ ԾԱԾԿՈՒՅԹ խնդիրը պատկանում է $NP$ դասին: Իրոք, եթե մեզ տրված լինեին որոնելի բազմությունները, ապա մենք կարող էինք համոզվել, որ նրանք զույգ առ զույգ չհատվող են և ծածկում են $A=\{a_1,...,a_n\}$ բազմությունը: Նկատենք, որ այս ստուգումը կարելի է իրականացնել բազմանդամային ալգորիթմի միջոցով, այնպես որ ԿԱՏԱՐՅԱԼ ԾԱԾԿՈՒՅԹ խնդիրը պատկանում է $NP$ դասին:
ԿԱՏԱՐՅԱԼ ԾԱԾԿՈՒՅԹ խնդրի լրիվությունը ցույց տալու համար, ապացուցենք, որ ԳՐԱՖԻ ՆԵՐԿՈՒՄ $\prec$ ԿԱՏԱՐՅԱԼ ԾԱԾԿՈՒՅԹ:

Դիտարկենք $G$ գրաֆը, $k$ բնական թիվը՝ ԳՐԱՖԻ ՆԵՐԿՈՒՄ խնդրի անհատ խնդիրը: Նշանակենք

$$A=V(G)\cup\{(i,x):1\leq i\leq k, x\in E(G)\}:$$

Դիտարկենք $A$ բազմության ենթաբազմությունների՝ հետևյալ ձևով սահմանված ընտանիքը

$$S_{vi}=\{v\}\cup\{(i,x): x \text{ կողը ինցիդենտ է } v \text{ գագաթին}\},$$

$$T_{xi}=\{(i,x)\}, \text{ որտեղ } x\in E(G):$$

Նկատենք, որ այս ձևով սահմանված ընտանիքը հանդիսանում է $A$ բազմության ծածկույթ: Ցույց տանք, որ այն պարունակում է ճշգրիտ ենթածածկույթ այն և միայն այն դեպքում, երբ $G$ գրաֆն ունի ճշգրիտ $k$-ներկում:

Ենթադրենք $G$ գրաֆն ունի $\varphi:V(G)\to\{1,...,k\}$ ճշգրիտ $k$-ներկում: Նկատենք, որ ցանկացած $u,v\in V(G)$ $(u\neq v)$ համար $S_{u\varphi(u)}\cap S_{v\varphi(v)}=\varnothing$: Իրոք, եթե $S_{u\varphi(u)}\cap S_{v\varphi(v)}\neq\varnothing$, ապա քանի որ $u\neq v$, կարող ենք եզրակացնել, որ գոյություն ունի $(i,x)$ զույգ այնպես, որ $(i,x)\in S_{u\varphi(u)}\cap S_{v\varphi(v)}$: Նկատենք, որ այդ դեպքում, $x=\{u,v\}$, $i=\varphi(u)=\varphi(v)$, ինչը հնարավոր չէ, քանի որ $\varphi$-ն ճշգրիտ ներկում է: Դիտակենք $A$ բազմության ենթաբազմությունների հետևյալ ընտանիքը, որն ստացվում է $\{S_{v\varphi(v)}:v\in V(G)\}$ ընտանիքին ավելացնելով մնացած $T_{xi}$ բազմությունները, այսինքն՝ այն $T_{xi}$-երը, որոնց համապատասխան $(i,x)$ զույգը չի մասնակցում $S_{v\varphi(v)}$-երի մեջ: Նկատենք, որ այս ընտանիքը հանդիսանում է $A$ բազմության ճշգրիտ ծածկույթ:

Հիմա ենթադրենք, որ գոյություն ունի $A$ բազմության՝ վերևում նկարագրված ծածկույթի ճշգրիտ ենթածածկույթ: Նկատենք, որ այդ դեպքում ցանկացած $v\in V(G)$ համար գոյություն ունի այդ ծածկույթի միակ $S_{vi}$ բազմություն այնպես, որ $v\in S_{vi}$ $1\leq i\leq k$: Դիտարկենք $f:V(G)\to\{1,...,k\}$ արտապատկերումը, որտեղ $f(v)=i$: Ցույց տանք, որ $f$-ը հանդիսանում է $G$ գրաֆի ճշգրիտ $k$-ներկում: Ենթադրենք, որ $x=\{u,v\}\in E(G)$: Ցույց տանք, որ $f(u)\neq f(v)$: Իրոք, եթե $f(u)=f(v)=i$, ապա $x=\{u,v\}\in S_{ui}$, $x=\{u,v\}\in S_{vi}$, և հետևաբար՝ $S_{ui}\cap S_{vi}\neq\varnothing$, ինչը հակասում է ծածկույթի ճշգրիտ լինելուն: Հետևաբար, $f(u)\neq f(v)$ և $f$-ը հանդիսանում է $G$ գրաֆի ճշգրիտ $k$-ներկում:

Վերջում նկատենք, որ $S_{vi}$ և $T_{xi}$ բազմությունները կառուցվում են ըստ $G$ գրաֆի և $k$ բնական թվի բազմանդամային ալգորիթմի միջոցով, հետևաբար՝ թեորեմն ապացուցված է:

Հիմա դիտարկենք բազմություններին առնչվող հետևյալ խնդիրը

ՆԵՐԿԱՅԱՑՈՒՑԻՉՆԵՐԻ ՀԱՄԱԿԱՐԳ

Տրված է $A=\{a_1,...,a_n\}$ և նրա ենթաբազմությունների $A_1,...,A_m$ ընտանիքը: Պահանջվում է պարզել, թե գոյություն ունի արդյոք $A_1,...,A_m$ համակարգի ներկայացուցիչների համակարգ, այսինքն այնպիսի $W\subseteq\{a_1,...,a_n\}$ ենթաբազմություն, որի համար $\left|W\cap A_i\right|=1,\ 1\le i\le m$:

**Թեորեմ**: ՆԵՐԿԱՅԱՑՈՒՑԻՉՆԵՐԻ ՀԱՄԱԿԱՐԳ խնդիրը $NP$–լրիվ է:

**Ապացույց**: Նախ նկատենք, որ ՆԵՐԿԱՅԱՑՈՒՑԻՉՆԵՐԻ ՀԱՄԱԿԱՐԳ խնդիրը պատկանում է $NP$ դասին: Իրոք, եթե մեզ տրված լիներ որոնելի $W\subseteq\{a_1,...,a_n\}$ ենթաբազմությունը, ապա մենք կարող ենք համոզվել, որ $\left|W\cap A_i\right|=1,\ 1\le i\le m$: Նկատենք, որ այս ստուգումը կարելի է իրականացնել բազմանդամային ալգորիթմի միջոցով, այնպես որ ՆԵՐԿԱՅԱՑՈՒՑԻՉՆԵՐԻ ՀԱՄԱԿԱՐԳ խնդիրը պատկանում է $NP$ դասին:

ՆԵՐԿԱՅԱՑՈՒՑԻՉՆԵՐԻ ՀԱՄԱԿԱՐԳ խնդրի լրիվությունը ցույց տալու համար ապացուցենք, որ ԿԱՏԱՐՅԱԼ ԾԱԾԿՈՒՅԹ≺ ՆԵՐԿԱՅԱՑՈՒՑԻՉՆԵՐԻ ՀԱՄԱԿԱՐԳ:

Դիտարկենք ԿԱՏԱՐՅԱԼ ԾԱԾԿՈՒՅԹ խնդրի մի որևէ անհատ խնդիր. $A=\{a_1,...,a_n\}$ բազմությունը և նրա ենթաբազմությունների $\boldsymbol{A}=\{A_1,...,A_m\}$ ընտանիքը, որը ծածկում է $A=\{a_1,...,a_n\}$ բազմությունը, այսինքն՝

$$\bigcup_{i=1}^{m}A_i=A:$$

$i=1,...,n$ համար նշանակենք $B(a_i)=\{A_j\in\boldsymbol{A}:a_i\in A_j\}$: Նկատենք, որ քանի որ $\boldsymbol{A}=\{A_1,...,A_m\}$ ընտանիքը հանդիսանում է $A=\{a_1,...,a_n\}$ բազմության ծածկույթ, ապա $B(a_i)\neq\varnothing$: Դիտարկենք $\boldsymbol{A}=\{A_1,...,A_m\}$ բազմությունը և նրա ենթաբազմությունների $B(a_1),...,B(a_n)$ համակարգը: Ցույց տանք, որ այն ունի ներկայացուցիչների համակարգ այն և միայն այն դեպքում, երբ $\boldsymbol{A}=\{A_1,...,A_m\}$ ընտանիքը պարունակում է կատարյալ ենթածածկույթ:

Ենթադրենք $\boldsymbol{A}=\{A_1,...,A_m\}$ ընտանիքը պարունակում է $A_{i_1},...,A_{i_k}$ կատարյալ ենթածածկույթ: Նկատենք, որ այդ դեպքում $A=\{a_1,...,a_n\}$ բազմության ցանկացած $a_i$ տարր պատկանում է նրանցից ճիշտ մեկին, հետևաբար՝ $\left|B(a_i)\cap\{A_{i_1},...,A_{i_k}\}\right|=1$, և հետևաբար $\boldsymbol{A}=\{A_1,...,A_m\}$ բազմության $W=\{A_{i_1},...,A_{i_k}\}$ ենթաբազմությունը հանդիսանում է $B(a_1),...,B(a_n)$ համակարգի ներկայացուցիչների համակարգ:

Հակառակը, ենթադրենք $\boldsymbol{A}=\{A_1,...,A_m\}$ բազմության $W=\{A_{i_1},...,A_{i_k}\}$ ենթաբազմությունը հանդիսանում է $B(a_1),...,B(a_n)$ համակարգի ներկայացուցիչների համակարգ: Այստեղից հետևում է, որ $A=\{a_1,...,a_n\}$ բազմության ցանկացած $a_i$ տարրի համար $\left|B(a_i)\cap\{A_{i_1},...,A_{i_k}\}\right|=1$, հետևաբար, $a_i$ տարրը պատկանում է նրանցից ճիշտ մեկին և հետևաբար $A_{i_1},...,A_{i_k}$ բազմությունները զույգ առ զույգ չեն հատվում: Այստեղից հետևում է, որ $A=\{a_1,...,a_n\}$ բազմության ենթաբազմությունների $A_{i_1},...,A_{i_k}$ համակարգը հանդիսանում է $\boldsymbol{A}=\{A_1,...,A_m\}$ ընտանիքի կատարյալ ենթածածկույթ:

Վերջում նկատենք, որ $B(a_1),...,B(a_n)$ համակարգը կառուցվում է ըստ $A=\{a_1,...,a_n\}$ բազմության և նրա ենթաբազմությունների $\boldsymbol{A}=\{A_1,...,A_m\}$ ընտանիքի բազմանդամային ալգորիթմի միջոցով, հետևաբար՝ թեորեմն ապացուցված է:

Հիշենք ուսապարկի խնդիրը: Ունենք որոշ թվով առարկաներ: Հայտնի է նրանցից յուրաքանչյուրի գինն ու ծավալը: Անհրաժեշտ է որոշակի տարողություն ունեցող ուսապարկով տեղափոխել այս առարկաներից այնպիսիները, որոնց ծավալների գումարը չգերազանցի ուսապարկի ծավալը և որոնց գումարային գինը լինի հնարավորին չափ մեծ: Այս խնդրի մաթեմատիկական մոդելը հետևյալն էր տրված են $c_1,...,c_n$, $v_1,...,v_n$, և $V$ ոչ բացասական թվերը: Անհրաժեշտ է $x_1,...,x_n$ փոփախականների համար ընտրել 0 կամ 1 արժեքներ, որ բավարարվի $x_1v_1+...+x_nv_n\le V$ պայմանը և $(x_1c_1+...+x_nc_n)$ արտահայտությունը ստանա իր առավելագույն հնարավոր արժեքը: Մենք կդիտարկենք այս խնդրի ճանաչման տարբերակը, որը կանվանենք ՈՒՍԱՊԱՐԿ

ՈՒՍԱՊԱՐԿ

տրված են $c_1,...,c_n$, $v_1,...,v_n$, և $V$ ,$C$ ոչ բացասական թվերը:

Հնարավոր է $x_1,...,x_n$ փոփախականների համար ընտրել 0 կամ 1 արժեքներ այնպես, որ բավարարվի հետևյալ համակարգը

$$\begin{cases} x_1v_1+...+x_nv_n\le V \\ x_1c_1+...+x_nc_n\ge C \end{cases}$$

Նկատենք, որ ՈՒՍԱՊԱՐԿ խնդիրը պատկանում է *NP* դասին: Իրոք, եթե մեզ տրված լիներ $x_1,...,x_n$ փոփախականների որոնելի արժեքները, ապա մենք կարող ենք համոզվել, որ նրանք բավարարում են վերոհիշյալ համակարգին: Նկատենք նաև, որ այս ստուգումը կարելի է իրականացնել բազմանդամային ալգորիթմի միջոցով, այնպես որ ՈՒՍԱՊԱՐԿ խնդիրը պատկանում է *NP* դասին:

Մենք կդիտարկենք նաև ՈՒՍԱՊԱՐԿ խնդրի հետևյալ մասնավոր դեպքը

0-1 ՈՒՍԱՊԱՐԿ
տրված են $a_1,...,a_n$, և $b$ ոչ բացասական թվերը:
Հնարավոր է $x_1,...,x_n$ փոփոխականների համար ընտրել 0 կամ 1 արժեքներ այնպես, որ բավարարվի $a_1x_1+...+a_nx_n=b$ հավասարումը:

Նկատենք, որ 0-1 ՈՒՍԱՊԱՐԿ խնդիրը հանդիսանում է ՈՒՍԱՊԱՐԿ խնդրի մասնավոր դեպքը: Իրոք, 0-1 ՈՒՍԱՊԱՐԿ խնդիրը կարելի է ստանալ ՈՒՍԱՊԱՐԿ խնդրից, եթե վերցնենք $c_1=v_1=a_1,...,c_n=v_n=a_n$, $V=C=b$: Այստեղից անմիջապես հետևում է, որ 0-1 ՈՒՍԱՊԱՐԿ խնդիրը ևս պատկանում է $NP$ դասին:
Ստորև կապացուցենք, որ 0-1 ՈՒՍԱՊԱՐԿ խնդիրը հանդիսանում է $NP$–լրիվ խնդիր: Նկատենք, որ քանի որ 0-1 ՈՒՍԱՊԱՐԿ խնդիրը հանդիսանում է ՈՒՍԱՊԱՐԿ խնդրի մասնավոր դեպքը, ապա այստեղից անմիջապես կհետևի նաև ՈՒՍԱՊԱՐԿ խնդրի $NP$–լրիվ լինելը:

**Թեորեմ**: 0-1 ՈՒՍԱՊԱՐԿ խնդիրը $NP$–լրիվ է:
**Ապացույց**: Արդեն նշել ենք, որ 0-1 ՈՒՍԱՊԱՐԿ խնդիրը պատկանում է $NP$ դասին:
0-1 ՈՒՍԱՊԱՐԿ խնդրի լրիվությունը ցույց տալու համար, ապացուցենք, որ ԿԱՏԱՐՅԱԼ ԾԱԾԿՈՒՅԹ $\prec$ 0-1 ՈՒՍԱՊԱՐԿ:
Դիտարկենք ԿԱՏԱՐՅԱԼ ԾԱԾԿՈՒՅԹ խնդրի մի որևէ անհատ խնդիր. $A=\{a_1,...,a_n\}$ բազմությունը և նրա ենթաբազմությունների $\mathbf{A}=\{A_1,...,A_m\}$ ընտանիքը, որը ծածկում է $A=\{a_1,...,a_n\}$ բազմությունը, այսինքն՝

$$\bigcup_{i=1}^{m} A_i = A:$$

Դիտարկենք $H=(h_{ij})$ $m\times n$ մատրիցը (նկար 1), որտեղ

$$h_{ij}=\begin{cases}1 & \text{եթե} \quad a_j\in A_i,\\ 0 & \text{եթե} \quad a_j\notin A:\end{cases}$$

| | | $a_1$ | $a_2$ | . | $a_j$ | . | $a_n$ |
|---|---|---|---|---|---|---|---|
| $x_1$ | $A_1$ | $h_{11}$ | $h_{12}$ | . | $h_{1j}$ | . | $h_{1n}$ |
| $x_2$ | $A_2$ | $h_{21}$ | $h_{22}$ | . | $h_{2j}$ | . | $h_{2n}$ |
| . | . | . | | . | | . | |
| $x_i$ | $A_i$ | $h_{i1}$ | $h_{i2}$ | . | $h_{ij}$ | . | $h_{in}$ |
| . | . | . | | . | | . | |
| $x_m$ | $A_m$ | $h_{m1}$ | $h_{m2}$ | . | $h_{mj}$ | . | $h_{mn}$ |

նկար 1

Նկատենք, որ քանի որ $\boldsymbol{A}=\{A_1,...,A_m\}$ ընտանիքը ծածկում է $A=\{a_1,...,a_n\}$ բազմությունը, ապա $H=(h_{ij})$ մատրիցի յուրաքանչյուր սյան մեջ գոյություն ունի առնվազն մեկ հատ մեկ:

$\boldsymbol{A}=\{A_1,...,A_m\}$ ընտանիքի ցանկացած $A_{i_1},...,A_{i_k}$ ենթաընտանիքի համապատասխանեցնենք $x_1,...,x_m$ թվերը, որտեղ

$$x_i=\begin{cases}1 & \text{եթե}\quad A_i\in\{A_{i_1},...,A_{i_k}\},\\ 0 & \text{եթե}\quad A_i\notin\{A_{i_1},...,A_{i_k}\}:\end{cases}$$

Այս ձևով, մենք ցանկացած $A_{i_1},...,A_{i_k}$ ենթաընտանիքի համապատասխանեցրինք 0,1 թվերի ինչ-որ $m$-յակ: Նկատենք, որ տեղի ունի նաև հակառակը, 0,1 թվերի ցանկացած $m$-յակի համապատասխանում է $\boldsymbol{A}=\{A_1,...,A_m\}$ ընտանիքի ինչ-որ մի ենթաընտանիք: Ավելին, $A_{i_1},...,A_{i_k}$ ենթաընտանիքը կկազմի $\boldsymbol{A}=\{A_1,...,A_m\}$ ընտանիքի կատարյալ ենթածածկույթ այն և միայն այն դեպքում, երբ $H=(h_{ij})$ մատրիցի $i_1,...,i_k$ տողերով ծնված մատրիցի ցանկացած սյան մեջ կա ճիշտ մեկ հատ մեկ, այլ կերպ ասած $A_{i_1},...,A_{i_k}$ ենթաընտանիքին համապատասխանող $x_1,...,x_m$ թվերը կբավարարեն

$$\begin{aligned}&h_{11}x_1+...+h_{m1}x_m=1\\&h_{12}x_1+...+h_{m2}x_m=1\\&.\\&.\qquad\qquad\qquad\qquad(1)\\&.\\&h_{1n}x_1+...+h_{mn}x_m=1\end{aligned}$$

համակարգին:

$i=1,...,m$ համար $h^{(i)}$-ով նշանակենք $H=(h_{ij})$ մատրիցի $i$-րդ տողը: Նկատենք, որ (1) համակարգը վեկտորական տեսքով կարելի է արտագրել հետևյալ կերպ

$$h^{(1)}x_1+...+h^{(m)}x_m=\bar{1}=(1,1,...,1):$$

$i=1,...,m$ համար $h^{(i)}$ վեկտորին համապատասխանեցնենք $z(h^{(i)})$ թիվը, որտեղ

$$z(h^{(i)})=h_{in}+h_{in-1}(m+1)+...+h_{i2}(m+1)^{n-2}+h_{i1}(m+1)^{n-1}:$$

Դիտարկենք 0-1 ՌԻՍՍԱՊԱՐԿ խնդրի $a_1=z(h^{(1)}),a_2=z(h^{(2)}),...,a_m=z(h^{(m)})$, $b=1+(m+1)+...+(m+1)^{n-2}+(m+1)^{n-1}$ անհատ խնդիրը:

Ցույց տանք, որ $x_1,...,x_m$ փոփախականների համար հնարավոր է ընտրել 0 կամ 1 արժեքներ այնպես, որ բավարարվի $a_1x_1+...+a_mx_m=b$ հավասարումը այն և միայն այն դեպքում, երբ $A=\{a_1,...,a_n\}$ բազմության ենթաբազմությունների $\boldsymbol{A}=\{A_1,...,A_m\}$ ընտանիքը պարունակում է կատարյալ ենթածածկույթ:

Ենթադրենք, որ $A=\{a_1,...,a_n\}$ բազմության ենթաբազմությունների $\boldsymbol{A}=\{A_1,...,A_m\}$ ընտանիքը պարունակում է $A_{i_1},...,A_{i_k}$ կատարյալ ենթածածկույթ: Դիտարկենք $A_{i_1},...,A_{i_k}$ ենթաընտանիքին համապատասխանող $x_1,...,x_m$ թվերը: Ինչպես արդեն նշել ենք այդ թվերը բավարարվում են (1) համակարգին: (1) համակարգի առաջին հավասարումը բազմապատկենք $(m+1)^{n-1}$-ով, երկրորդը՝ $(m+1)^{n-2}$-ով,..., $n-1$-րդը՝ $(m+1)$-ով, $n$-րդը՝ 1-ով, և գումարենք

$$h_{11}x_1+...+h_{m1}x_m=1:(m+1)^{n-1}$$
$$h_{12}x_1+...+h_{m2}x_m=1:(m+1)^{n-2}$$
.
.
.
$$h_{1n}x_1+...+h_{mn}x_m=1:(m+1)^0=1$$

Ընդհանուր հանելով $x_1,...,x_m$-ը կստանանք՝

$$z(h^{(1)})x_1+...+z(h^{(m)})x_m=b=1+(m+1)+...+(m+1)^{n-2}+(m+1)^{n-1}:$$

Հակառակը, ենթադրենք $x_1,...,x_m$ փոփախականների համար հնարավոր է ընտրել 0 կամ 1 արժեքներ այնպես, որ բավարարվի

$$z(h^{(1)})x_1+...+z(h^{(m)})x_m=b$$

հավասարումը: Նկատենք, որ այդ դեպքում $x_1,...,x_m$ թվերը բավարարում են (1) համակարգին: Իրոք, քանի որ

$$z(h^{(1)})=h_{1n}+(m+1)\gamma_1,$$
$$z(h^{(2)})=h_{2n}+(m+1)\gamma_2,$$
$$\dots$$
$$z(h^{(m)})=h_{mn}+(m+1)\gamma_m,$$
$$b=1+(m+1)\alpha$$

ապա պարզ է, որ $h_{1n}x_1+h_{2n}x_2+...+h_{mn}x_m-1$ թիվը պետք է պատիկ լինի $(m+1)$-ին: Մյուս կողմից քանի որ $-1\le h_{1n}x_1+h_{2n}x_2+...+h_{mn}x_m-1\le m-1$, ապա այստեղից հետևում է, որ $h_{1n}x_1+h_{2n}x_2+...+h_{mn}x_m=1$: Համանման ձևով կարելի է ցույց տալ, որ $x_1,...,x_m$ թվերը բավարարում են (1) համակարգի նախավերջին, ..., երկրորդ, առաջին հավասարումներին:

Նկատենք, որ այդ դեպքում $x_1,...,x_m$ թվերին համապատասխանող $A_{i_1},...,A_{i_k}$ ենթաընտանիքը կկազմի $A=\{a_1,...,a_n\}$ բազմության կատարյալ ենթածածկույթ: Թեորեմն ապացուցված է:

ՏՐՈՀՈՒՄ

Տրված են $a_1,...,a_n$ թվերը:

Գոյություն ունի արդյոք $I\subseteq\{1,...,n\}$ այնպես, որ $\sum_{i\in I}a_i=\sum_{i\notin I}a_i$ :

**Թեորեմ**: ՏՐՈՀՈՒՄ խնդիրը $NP$ –լրիվ է:

**Ապացույց**: Նախ նկատենք, որ ՏՐՈՀՈՒՄ խնդիրը պատկանում է $NP$ դասին: Իրոք, եթե մեզ տրված լիներ մուտքային թվերի բազմության որոնելի ենթաբազմությունը, ապա մենք կարող էինք համոզվել, որ նրան պատկանող թվերի գումարը հավասար է այդ ենթաբազմությունը չպատկանող թվերի գումարին: Նկատենք, որ այս ստուգումը կարելի է իրականացնել բազմանդամային ալգորիթմի միջոցով, այնպես որ ՏՐՈՀՈՒՄ խնդիրը պատկանում է $NP$ դասին:

ՏՐՈՀՈՒՄ խնդրի լրիվությունը ցույց տալու համար, ապացուցենք, որ 0-1 ՈՒՍԱՊԱՐԿ$\prec$ ՏՐՈՀՈՒՄ:

Դիտարկենք $a_1,...,a_n$ և $b$ ոչ բացասական թվերը՝ 0-1 ՈՒՍԱՊԱՐԿ խնդրի անհատ խնդիրը, և նրան համապատասխանեցնենք ՏՐՈՀՈՒՄ խնդրի $a_1,...,a_n,2b,\sum_{i=1}^{n}a_i$ անհատ խնդիրը: Նկատենք, որ այս խնդիրը կարելի է կառուցել ըստ $a_1,...,a_n$ և $b$ թվերի բազմանդամային ժամանակում:

Ցույց տանք, որ $x_1,...,x_n$ փոփոխականների համար հնարավոր է ընտրել 0 կամ 1 արժեքներ այնպես, որ բավարարվի $a_1x_1+...+a_nx_n=b$ հավասարումը այն և միայն այն դեպքում, երբ $a_1,...,a_n,2b,\sum_{i=1}^{n}a_i$ թվերը հնարավոր է տրոհել երկու մասի այնպես, որ մի մասի գումարը լինի հավասար մյուս մասի գումարին:

Ենթադրենք, որ գոյություն ունեն $x_1,...,x_n$, $x_i\in\{0,1\}$ այնպես, որ $a_1x_1+...+a_nx_n=b$: Նշանակենք $I=\{i:x_i=1\}$: Նկատենք, որ $\sum_{i\in I}a_i=b$ և հետևաբար՝

$$\sum_{i=1}^{n}a_i+\sum_{i\in I}a_i=2b+\sum_{i\notin I}a_i\ ,$$

որտեղից հետևում է, որ ՏՐՈՀՈՒՄ խնդրի $a_1,...,a_n,2b,\sum_{i=1}^{n}a_i$ անհատ խնդրում պատասխանը դրական է:

Հիմա ենթադրենք, որ $a_1,...,a_n,2b,\sum_{i=1}^{n}a_i$ թվերը հնարավոր է տրոհել երկու մասի այնպես, որ մի մասի գումարը լինի հավասար մյուս մասի գումարին: Նկատենք, որ այդ դեպքում $2b$ և $\sum_{i=1}^{n}a_i$ թվերը գտնվում են տրոհման տարբեր կողմերում: Այստեղից հետևում է, որ գոյություն ունի $I\subseteq\{1,...,n\}$ այնպես, որ

$$\sum_{i=1}^{n}a_i+\sum_{i\in I}a_i=2b+\sum_{i\notin I}a_i$$

հետևաբար՝

$$2\sum_{i\in I}a_i=2b$$

կամ՝

$$\sum_{i\in I}a_i=b:$$

Նշանակենք՝

$$x_i=\begin{cases}1 & \text{եթե} \quad i\in I,\\ 0 & \text{եթե} \quad i\notin I:\end{cases}$$

Նկատենք, որ

$$a_1x_1+...+a_nx_n=\sum_{i\in I}a_i=b,$$

և հետևաբար՝ 0-1 ՈՒՍԱՊԱՐԿ խնդրի $a_1,...,a_n$ և $b$ անհատ խնդրում պատասխանը դրական է: Թեորեմն ապացուցված է:

Դիտարկենք բազմությունների ծածկույթներին առնչվող հետևյալ խնդիրը

ԲԱԶՄՈՒԹՅԱՆ ԾԱԾԿՈՒՅԹ

Տրված է $k$ բնական թիվը, $A=\{a_1,...,a_n\}$ բազմությունը և նրա ենթաբազմությունների $\boldsymbol{A}=\{A_1,...,A_m\}$ ընտանիքը, որը ծածկում է $A=\{a_1,...,a_n\}$ բազմությունը, այսինքն՝

$$\bigcup_{i=1}^{m}A_i=A:$$

Պահանջվում է պարզել, թե գոյություն ունի արդյոք $\boldsymbol{A}=\{A_1,...,A_m\}$ բազմության $k$ ենթածածկույթ, այսինքն՝ $A_{i_1},...,A_{i_k}$ տարրեր, որոնք ծածկում են $A=\{a_1,...,a_n\}$ բազմությունը

$$\bigcup_{j=1}^{k} A_{i_j} = A:$$

**Թեորեմ**: ԲԱԶՄՈՒԹՅԱՆ ԾԱԾԿՈՒՅԹ խնդիրը $NP$ –լրիվ է:

**Ապացույց**: Նախ նկատենք, որ ԲԱԶՄՈՒԹՅԱՆ ԾԱԾԿՈՒՅԹ խնդիրը պատկանում է $NP$ դասին: Իրոք, եթե մեզ տրված լինեին որոնելի բազմությունները, ապա մենք կարող էինք համոզվել, որ նրանք $k$ հատ են և ծածկում են $A = \{a_1, ..., a_n\}$ բազմությունը: Նկատենք, որ այս ստուգումը կարելի է իրականացնել բազմանդամային ալգորիթմի միջոցով, այնպես որ ԲԱԶՄՈՒԹՅԱՆ ԾԱԾԿՈՒՅԹ խնդիրը պատկանում է $NP$ դասին:

ԲԱԶՄՈՒԹՅԱՆ ԾԱԾԿՈՒՅԹ խնդրի լրիվությունը ցույց տալու համար ապացուցենք, որ ԳԱԳԱԹՆԵՐՈՎ ԾԱԾԿՈՒՅԹ $\prec$ ԲԱԶՄՈՒԹՅԱՆ ԾԱԾԿՈՒՅԹ:

Դիտարկենք $G$ գրաֆը, $k$ բնական թիվը՝ ԳԱԳԱԹՆԵՐՈՎ ԾԱԾԿՈՒՅԹ խնդրի անհատ խնդիրը: Ենթադրենք, որ

$$V(G) = \{v_1, ..., v_p\},$$

$$X_i = \{x \in E(G) : x\text{-ը և } v\text{-ն ինցիդենտ են}\},\ 1 \le i \le p:$$

Նկատենք, որ $E(G) = X_1 \cup ... \cup X_p$: Ցույց տանք, որ $G$ գրաֆում $\beta(G) \le k$ այն և միայն այն դեպքում, երբ $E(G)$ բազմության $\{X_1, ..., X_p\}$ ծածկույթը պարունակում է $k$ ենթածածկույթ:

Ենթադրենք, որ $\beta(G) \le k$ և $V' = \{v_{i_1}, ..., v_{i_k}\}$ գագաթների բազմությունը հանդիսանում է $G$ գրաֆի գագաթային ծածկույթ: Սա նշանակում է, որ $G$ գրաֆի ցանկացած կող ինցիդենտ է $V'$ բազմության գագաթներից մեկին, և հետևաբար՝

$$E(G) = X_{i_1} \cup ... \cup X_{i_k}:$$

Հակառակը, դիցուք $E(G)$ բազմության $\{X_1, ..., X_p\}$ ծածկույթը պարունակում է $\{X_{i_1}, ..., X_{i_k}\}$ $k$ ենթածածկույթ: Այդ դեպքում $G$ գրաֆի ցանկացած կող պատկանում է $X_{i_1}, ..., X_{i_k}$ բազմություններից գոնե մեկին, և հետևաբար՝ $G$ գրաֆի գագաթների $V' = \{v_{i_1}, ..., v_{i_k}\}$ բազմությունը հանդիսանում է $G$ գրաֆի գագաթային ծածկույթ և $\beta(G) \le k$:

Վերջում նկատենք, որ $E(G)$ բազմության $\{X_1, ..., X_p\}$ ծածկույթը կառուցվում է ըստ $G$ գրաֆի և $k$ բնական թվի բազմանդամային ալգորիթմի միջոցով, հետևաբար՝ թեորեմն ապացուցված է:

Վերջում դիտարկենք մատրիցներին առնչվող հետևյալ խնդիրը

ՏՏՈՒԻԳՈՂ ԹԵՍՏ

Տրված է $k$ բնական թիվը, $m \times n$ կարգի $T=(t_{ij})$ մատրիցը, որտեղ $t_{ij} \in \{0,1\}$ և որի առաջին սյունը տարբերվում է մնացած սյուներից:

Պահանջվում է պարզել, թե գոյություն ունեն արդյոք $T=(t_{ij})$ մատրիցի $k$ տողեր, որոնցով ծնված ենթամատրիցի առաջին սյունը լինի տարբեր մնացած սյուներից:

**Թեորեմ**: ՍՏՈՒԳՈՂ ԹԵՍՏ խնդիրը $NP$–լրիվ է:

**Ապացույց**: Նախ նկատենք, որ ՍՏՈՒԳՈՂ ԹԵՍՏ խնդիրը պատկանում է $NP$ դասին: Իրոք, եթե մեզ տրված լինեին որոնելի սյուները, ապա մենք կարող էինք համոզվել, որ նրանք $k$ հատ են և որոնցով ծնված $T=(t_{ij})$ մատրիցի ենթամատրիցում առաջին սյունը տարբեր է մնացած սյուներից: Նկատենք, որ այս ստուգումը կարելի է իրականացնել բազմանդամային ալգորիթմի միջոցով, այնպես որ ՍՏՈՒԳՈՂ ԹԵՍՏ խնդիրը պատկանում է $NP$ դասին:

ՍՏՈՒԳՈՂ ԹԵՍՏ խնդրի լրիվությունը ցույց տալու համար ապացուցենք, որ ԲԱԶՄՈՒԹՅԱՆ ԾԱԾԿՈՒՅԹ $\prec$ ՍՏՈՒԳՈՂ ԹԵՍՏ:

Դիտարկենք ԲԱԶՄՈՒԹՅԱՆ ԾԱԾԿՈՒՅԹ խնդրի մի որևէ անհատ խնդիր. $k$ բնական թիվը, $A=\{a_1,...,a_n\}$ բազմությունը և նրա ենթաբազմությունների $\boldsymbol{A}=\{A_1,...,A_m\}$ ընտանիքը, որը ծածկում է $A=\{a_1,...,a_n\}$ բազմությունը, այսինքն՝

$$\bigcup_{i=1}^{m} A_i = A:$$

Դիտարկենք $H=(h_{ij})$ $m \times n$ մատրիցը (նկար 1), որտեղ

$$h_{ij}=\begin{cases}1 & \text{եթե} \quad a_j \in A_i, \\ 0 & \text{եթե} \quad a_j \notin A:\end{cases}$$

Նկատենք, որ քանի որ $\boldsymbol{A}=\{A_1,...,A_m\}$ ընտանիքը ծածկում է $A=\{a_1,...,a_n\}$ բազմությունը, ապա նրա ցանկացած սյուն պարունակում է գոնե մեկ հատ մեկ: Դիտարկենք $\overline{H}$ մատրիցը, որը ստացվում է $H=(h_{ij})$ մատրիցի սկզբին ավելացնելով 0-ական սյուն: Նկատենք, որ $\overline{H}$ մատրիցում առաջին սյունը տարբեր է մյուս սյուներից: Ավելին, նկատենք որ $\boldsymbol{A}=\{A_1,...,A_m\}$ ընտանիքի $A_{i_1},...,A_{i_k}$ ենթաընտանիքը կկազմի ծածկույթ այն և միայն այն դեպքում, երբ երբ $\overline{H}$ մատրիցի $i_1,...,i_k$ տողերով ծնված ենթամատրիցում բոլոր սյուները տարբեր կլինեն առաջին (0-ական) սյունից:

Վերջում նկատենք, որ $\overline{H}$ մատրիցը կառուցվում է ըստ $k$ բնական թվի, $A=\{a_1,...,a_n\}$ բազմության և նրա ենթաբազմությունների $\boldsymbol{A}=\{A_1,...,A_m\}$ ընտանիքի բազմանդամային ալգորիթմի միջոցով, հետևաբար՝ թեորեմն ապացուցված է:

## Գրականություն

Գտնված սխալների, առաջարկությունների, ինչպես նաև դասախոսություններն e-mail-ով ստանալու համար կարող էք դիմել vahanmkrtchyan2002@yahoo.com հասցեով:

Կոմբինատորային ալգորիթմներ և
ալգորիթմների վերլուծություն
Վահան Վ. Մկրտչյան

Դասախոսություն 20: ՀԱՄԻԼՏՈՆՅԱՆ ԿՈՆՏՈՒՐ ԵՎ ՑԻԿԼ, ՇՐՋԻԿ ԳՈՐԾԱԿԱԼ, ԱՄԲՈՂՋԱԹԻՎ ԳԾԱՅԻՆ ԾՐԱԳՐԱՎՈՐՈՒՄ խնդիրների $NP$-լրիվությունը:

Դիտարկենք գրաֆում համիլտոնյան ցիկլերի գոյությանն առնչվող հետևյալ խնդիրը

ՀԱՄԻԼՏՈՆՅԱՆ ԿՈՆՏՈՒՐ

Տրված է $\vec{G}=(V,E)$ օրգրաֆը:

Պահանջվում է պարզել, թե պարունակում է այն համիլտոնյան կոնտուր, այսինքն՝ կոնտուր, որն անցնում է բոլոր գագաթներով, ընդ որում յուրաքանչյուրով մեկ անգամ:

**Թեորեմ**: ՀԱՄԻԼՏՈՆՅԱՆ ԿՈՆՏՈՒՐ խնդիրը $NP$–լրիվ է:

**Ապացույց**: Նախ նկատենք, որ ՀԱՄԻԼՏՈՆՅԱՆ ԿՈՆՏՈՒՐ խնդիրը պատկանում է $NP$ դասին: Իրոք, եթե մեզ տրված լիներ որոնելի կոնտուրը, ապա մենք կարող էինք համոզվել, որ այն անցնում է օրգրաֆի բոլոր գագաթներով, ընդ որում յուրաքանչյուրով մեկ անգամ: Նկատենք, որ այս ստուգումը կարելի է իրականացնել բազմանդամային ալգորիթմի միջոցով, այնպես որ ՀԱՄԻԼՏՈՆՅԱՆ ԿՈՆՏՈՒՐ խնդիրը պատկանում է $NP$ դասին:

ՀԱՄԻԼՏՈՆՅԱՆ ԿՈՆՏՈՒՐ խնդրի լրիվությունը ցույց տալու համար, ապացուցենք, որ ԳԱԳԱԹՆԵՐՈՎ ԾԱԾԿՈՒՅԹ $\prec$ ՀԱՄԻԼՏՈՆՅԱՆ ԿՈՆՏՈՒՐ:

Դիտարկենք $G=(V,E)$ գրաֆը, $k$ բնական թիվը՝ ԳԱԳԱԹՆԵՐՈՎ ԾԱԾԿՈՒՅԹ խնդրի անհատ խնդիրը: Ենթադրենք, որ $V=\{v_1,...,v_p\}$ և $E=\{x_1,...,x_q\}$, ընդ որում կենթադրենք, որ $G=(V,E)$ գրաֆի կողերի բազմությունը կարգավորված է նշված հերթականությամբ: Այս խնդրին համապատասխանեցնենք $\vec{G}=(U,\vec{E})$ օրգրաֆը, որի գագաթների բազմությունը սահմանվում է հետևյալ կերպ

$U=\{(u,x,\delta):\delta\in\{0,1\}$ և $x$ կողը ինցիդենտ է $u$ գագաթին$\}\cup\{a_1,...,a_k\}$:
Այստեղ $a_1,...,a_k$-ն նոր գագաթներ են՝ տարբեր $G=(V,E)$ գրաֆի գագաթներից: Նկատենք, որ $G=(V,E)$ գրաֆի ցանկացած $x=\{u,v\}$ կողի համապատասխանեցված է $\vec{G}=(U,\vec{E})$ օրգրաֆի չորս գագաթ՝ $(u,x,0)$, $(u,x,1)$, $(v,x,0)$, $(v,x,1)$, հետևաբար՝ $\vec{G}=(U,\vec{E})$ օրգրաֆը կպարունակի $4q+k$ գագաթ: $\vec{G}=(U,\vec{E})$ օրգրաֆի աղեղների բազմությունը սահմանենք հետևյալ կերպ

1. եթե $x_u^0$–ն $G=(V,E)$ գրաֆի $u$ գագաթին ինցիդենտ ամենափոքր համար ունեցող կողն է, ապա $(a_i,(u,x_u^0,0))\in\vec{E}$, $u\in V$, $i=1,...,k$;
2. եթե $x_u^*$–ն $G=(V,E)$ գրաֆի $u$ գագաթին ինցիդենտ ամենամեծ համար ունեցող կողն է, ապա $((u,x_u^*,1),a_i)\in\vec{E}$, $u\in V$, $i=1,...,k$;
3. եթե $x$–ը և $y$-ը $u$ գագաթին ինցիդենտ կողեր են, ընդ որում $y$-ն անմիջապես հաջորդում է $x$–ին ըստ վերը նշված կարգի, ապա $((u,x,1),(u,y,0))\in\vec{E}$, $u\in V$, $x,y\in E$;
4. եթե $x$–ը $u$ գագաթին ինցիդենտ կող է, ապա $((u,x,0),(u,x,1))\in\vec{E}$, $u\in V$, $x\in E$;
5. եթե $x=\{u,v\}\in E$, ապա $((u,x,\delta),(v,x,\delta))$, $\delta\in\{0,1\}$:

Նկար 1-ում պատկերված են $\vec{G}=(U,\vec{E})$ օրգրաֆի այն գագաթներն ու աղեղները, որոնք համապատասխանում են $G=(V,E)$ գրաֆի $u,v$ գագաթներին, ինչպես նաև $x=\{u,v\}\in E$ կողին:

Ցույց տանք, որ $G=(V,E)$ գրաֆը պարունակում է գագաթային $k$ ծածկույթ այն և միայն այն դեպքում, երբ $\vec{G}=(U,\vec{E})$ օրգրաֆը պարունակում է համիլտոնյան կոնտուր:

Ենթադրենք $G=(V,E)$ գրաֆի գագաթների $\{u_1,...,u_k\}$ բազմությունը հանդիսանում է ծածկույթ, և դիցուք՝ $u_i$ գագաթին ինցիդենտ կողերը $y_i(1),...,y_i(l_i)$-ն են՝ գրված $E=\{x_1,...,x_q\}$ բազմության վրա սահմանված կարգով:

Նկատենք, որ այդ դեպքում գագաթների ստորև բերված հաջորդականությունը կկազմի $\vec{G}=(U,\vec{E})$ օրգրաֆի կոնտուր

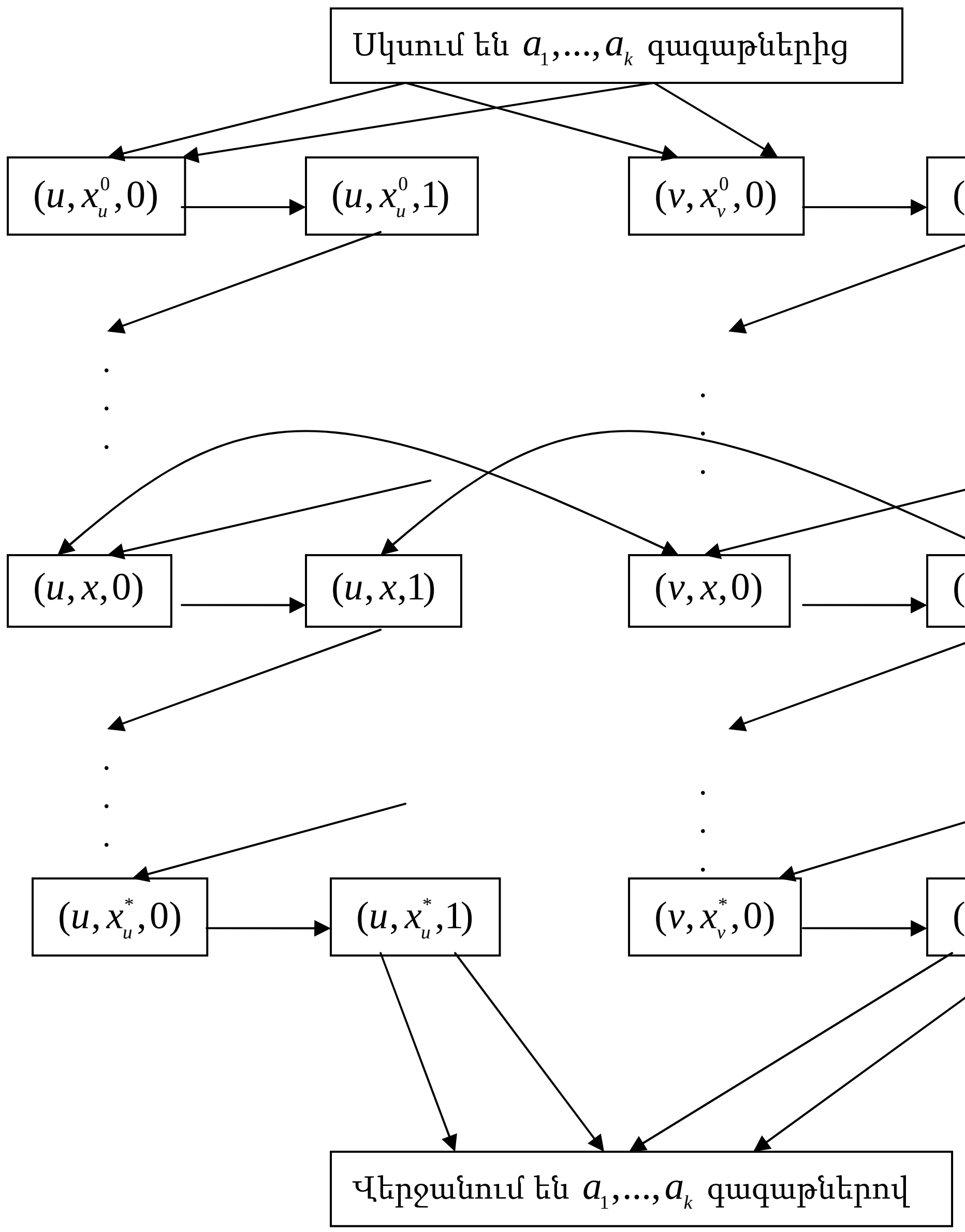

նկար 1

$a_1$, $(u_1,y_1(1),0)$, $(u_1,y_1(1),1)$, $(u_1,y_1(2),0)$,...,$(u_1,y_1(l_1),0)$, $(u_1,y_1(l_1),1)$,
$a_2$, $(u_2,y_2(1),0)$, $(u_2,y_2(1),1)$, $(u_2,y_2(2),0)$,...,$(u_2,y_2(l_2),0)$, $(u_2,y_2(l_2),1)$,
……………………………………………………………………………………
$a_k$, $(u_k,y_k(1),0)$, $(u_k,y_k(1),1)$, $(u_k,y_k(2),0)$,...,$(u_k,y_k(l_k),0)$, $(u_k,y_k(l_k),1)$, $a_1$,

որը պարունակում է բոլոր այն գագաթները, որոնք անցնում են այն $(u,x,\delta)$ գագաթներով, որոնց համար $u\in\{u_1,...,u_k\}$: Նկատենք նաև, որ այս կոնտուրում $x$ կողին համապատասխանող $(u,x,0),(u,x,1)$ գագաթները նշված կոնտուրում հաջորդում են միմյանց: Եթե $u\in\{u_1,...,u_k\}$, $v\notin\{u_1,...,u_k\}$ և $x=\{u,v\}\in E$, ապա այդ դեպքում կոնտուրը չի անցնում $(v,x,0),(v,x,1)$ գագաթներով: Կոնտուրի հարևան գագաթների միջև տեղավորենք այդ գագաթները, այսինքն՝

$$(u,x,0),(v,x,0),(v,x,1),(u,x,1):$$

Պարզ է, որ ստացված կոնտուրը արդեն կպարունակի $x=\{u,v\}\in E$ կողին համապատասխանող գագաթները: Պարզ է նաև, որ այս գործողության հաջորդական կիրառման արդյունքում մենք կստանանք $\vec{G}=(U,\vec{E})$ օրգրաֆի բոլոր գագաթներով մեկ անգամ անցնող կոնտուր, այսինքն՝ համիլտոնյան կոնտուր:

Այժմ ենթադրենք, որ $\vec{G}=(U,\vec{E})$ օրգրաֆում գոյություն ունի համիլտոնյան կոնտուր: $a_1,...,a_k$ գագաթները հեռացնելուց հետո այդ կոնտուրը կտրոհվի $k$ ուղիների: Դիտարկենք այդ ուղիներից որևէ մեկը: Ենթադրենք, որ նրա առաջին գագաթը $(u,x,0)$-է, իսկ վերջինը՝ $(v,y,1)$: Օրգրաֆի սահմանումից հետևում է, որ $x$-ը $u$-ին ինցիդենտ ամենափոքր համար ունեցող կողն է, իսկ՝ $y$-ը՝ $v$-ին ինցիդենտ ամենամեծ համար ունեցող կողը:

Նկատենք, որ իրականում $u=v$, քանի որ հակառակ դեպքում կարելի կլիներ նշել $(u,z,1)$ կամ $(w,y,0)$ գագաթ, որով չի անցնում կոնտուրը: Հետևաբար, եթե $a_i$ գագաթին հաջորդող գագաթը նշանակենք $v_i$-ով, ապա $v_1,...,v_k$ գագաթները կկազմեն $G=(V,E)$ գրաֆի գագաթային ծածկույթ: Թեորեմն ապացուցված է:

Օգտվելով ՀԱՄԻԼՏՈՆՅԱՆ ԿՈՆՏՈՒՐ խնդրի $NP$–լրիվությունից, ապացուցենք հետևյալ խնդրի $NP$–լրիվությունը

ՀԱՄԻԼՏՈՆՅԱՆ ՑԻԿԼ

Տրված է $G=(V,E)$ գրաֆը:

Պահանջվում է պարզել, թե պարունակում է այն համիլտոնյան ցիկլ, այսինքն՝ ցիկլ, որն անցնում է բոլոր գագաթներով, ընդ որում յուրաքանչյուրով մեկ անգամ:

**Թեորեմ**: ՀԱՄԻԼՏՈՆՅԱՆ ՑԻԿԼ խնդիրը $NP$–լրիվ է:

**Ապացույց**: Նախ նկատենք, որ ՀԱՄԻԼՏՈՆՅԱՆ ՑԻԿԼ խնդիրը պատկանում է $NP$ դասին: Իրոք, եթե մեզ տրված լիներ որոնելի ցիկլը, ապա մենք կարող էինք համոզվել, որ այն անցնում է գրաֆի բոլոր գագաթներով, ընդ որում յուրաքանչյուրով մեկ անգամ: Նկատենք, որ այս ստուգումը կարելի է իրականացնել բազմանդամային ալգորիթմի միջոցով, այնպես որ ՀԱՄԻԼՏՈՆՅԱՆ ՑԻԿԼ խնդիրը պատկանում է $NP$ դասին:

ՀԱՄԻԼՏՈՆՅԱՆ ՑԻԿԼ խնդրի լրիվությունը ցույց տալու համար, ապացուցենք, որ ՀԱՄԻԼՏՈՆՅԱՆ ԿՈՆՏՈՒՐ $\prec$ ՀԱՄԻԼՏՈՆՅԱՆ ՑԻԿԼ:

Դիտարկենք $\vec{G}=(U,\vec{E})$ օրգրաֆը՝ ՀԱՄԻԼՏՈՆՅԱՆ ԿՈՆՏՈՒՐ խնդրի անհատ խնդիրը, որում $U=\{u_1,...,u_p\}$: Նրան համապատասխանեցնենք $G=(V,E)$ գրաֆը, որում

$$V=\{u_1(1),u_1(2),u_1(3),u_2(1),...,u_p(1),u_p(2),u_p(3)\},$$

$$E=\{(u_i(1),u_i(2)),(u_i(2),u_i(3)):1\le i\le p\}\cup\{(u_i(3),u_j(1)):(u_i,u_j)\in E\}:$$

Ցույց տանք, որ $\vec{G}=(U,\vec{E})$ օրգրաֆը պարունակում է համիլտոնյան կոնտուր այն և միայն այն դեպքում, երբ $G=(V,E)$ գրաֆը պարունակում է համիլտոնյան ցիկլ:

Ենթադրենք $\vec{C}=u_{i_1},...,u_{i_p}$-ն հանդիսանում է $\vec{G}=(U,\vec{E})$ օրգրաֆի համիլտոնյան կոնտուր: Դիտարկենք $G=(V,E)$ գրաֆի հետևյալ ձևով սահմանված ցիկլը.

$$C=u_{i_1}(1),u_{i_1}(2),u_{i_1}(3),u_{i_2}(1),...,u_{i_p}(1),u_{i_p}(2),u_{i_p}(3):$$

Նկատենք, որ այն անցնում է $G=(V,E)$ գրաֆի բոլոր գագաթներով ընդ որում յուրաքանչյուրով ճիշտ մեկ անգամ, հետևաբար՝ $G=(V,E)$ գրաֆը համիլտոնյան է:

Հակառակը, ենթադրենք, որ $C'$-ը հանդիսանում է $G=(V,E)$ գրաֆի համիլտոնյան ցիկլ: Նկատենք, որ քանի որ ցանկացած $i, 1\le i\le p$ համար $u_i(2)$ գագաթի աստիճանը հավասար է երկուսի, ապա այն միանգամից է անցնում $u_i(1),u_i(2),u_i(3)$ գագաթներով: Հետևաբար, կարող ենք պնդել, որ

$$C'=u_{j_1}(1),u_{j_1}(2),u_{j_1}(3),u_{j_2}(1),...,u_{j_p}(1),u_{j_p}(2),u_{j_p}(3):$$

Նկատենք, որ այդ դեպքում $\vec{C}=u_{j_1},...,u_{j_p}$-ն հանդիսանում է $\vec{G}=(U,\vec{E})$ օրգրաֆի համիլտոնյան կոնտուր: Թեորեմն ապացուցված է:

Դիտարկենք հետևյալ խնդիրը: Դիցուք ունենք $n\ge 3$ բնակավայրեր, որոնցից մեկում գտնվում է գործակալը: Նա պետք է շրջագայի այդ բնակավայրերը և վերադառնա մեկնավայրը՝ յուրաքանչյուր բնակավայրում գտնվելով ճիշտ մեկ անգամ: Հայտնի է նաև բնակավայրերի միջև հեռավորությունը: Խնդիրը կայանում է հետևյալում. ինչ հերթականությամբ պետք է գործակալը շրջանցի բնակավայրերը, որպեսզի նրա անցած ճանապարհի երկարությունը լինի նվազագույնը:

Բնակավայրերը համարակալենք $1,2,...,n$ թվերով և ենթադրենք, որ գործակալը գտնվում է 1 բնակավայրում: $c_{ij}$-ով նշանակենք $i$–րդ բնակավայրից $j$–րդ բնակավայր տանող ճանապարհի երկարությունը:

Հատուկ նշենք, որ պարտադիր չէ, որ $c_{ij}$ թվերը բավարար են $c_{ij} = c_{ji}$ հավասարությանը կամ եռանկյան անհավասարությանը՝ $c_{ij} \le c_{ik} + c_{kj}$:

Եթե գործակալն ընտրել է բնակավայրերի շրջանցման $1, i_1, \dots, i_{n-1}, 1$ հերթականությունը, ապա նրա անցած ճանապարհի երկարությունը կլինի՝

$$c_{1,i_1} + c_{i_1,i_2} + \dots + c_{i_{n-1},1}:$$

Մենք կդիտարկենք այս խնդրի ճանաչման տարբերակը, որը կարճ կանվանենք

ՇՐՋԻԿ ԳՈՐԾԱԿԱԼ

Տրված է $C = (c_{ij})$, $c_{ij} \in Z^+$, $c_{ii} = 0$ մատրիցը և $L \in Z^+$ բնական թիվը:

Գոյություն ունի $2, \dots, n$ թվերի այնպիսի $i_1, \dots, i_{n-1}$ տեղափոխություն, որ

$$c_{1,i_1} + c_{i_1,i_2} + \dots + c_{i_{n-1},1} \le L:$$

**Թեորեմ**: ՇՐՋԻԿ ԳՈՐԾԱԿԱԼ խնդիրը $NP$ –լրիվ է:

**Ապացույց**: Նախ նկատենք, որ ՇՐՋԻԿ ԳՈՐԾԱԿԱԼ խնդիրը պատկանում է $NP$ դասին: Իրոք, եթե մեզ տրված լիներ որոնելի երթուղին, ապա մենք կարող էինք համոզվել, որ այն անցնում է գրաֆի բոլոր գագաթներով, ընդ որում նրա երկարությունը չի գերազանցում $L$-ը: Նկատենք, որ այս ստուգումը կարելի է իրականացնել բազմանդամային ալգորիթմի միջոցով, այնպես որ ՇՐՋԻԿ ԳՈՐԾԱԿԱԼ խնդիրը պատկանում է $NP$ դասին:

ՇՐՋԻԿ ԳՈՐԾԱԿԱԼ խնդրի լրիվությունը ցույց տալու համար, ապացուցենք, որ ՀԱՄԻԼՏՈՆՅԱՆ ՑԻԿԼ $\prec$ ՇՐՋԻԿ ԳՈՐԾԱԿԱԼ:

ՀԱՄԻԼՏՈՆՅԱՆ ՑԻԿԼ խնդրի $G = (V, E)$ անհատ խնդրին, որում $V = \{v_1, \dots, v_p\}$ համապատասխանեցնենք ՇՐՋԻԿ ԳՈՐԾԱԿԱԼ խնդրի $L = p$ և $C = (c_{ij})$ խնդիրը, որտեղ՝

$$c_{ij} = \begin{cases} 1 & \text{եթե} \quad (v_i, v_j) \in E, \\ 2 & \text{եթե} \quad (v_i, v_j) \notin E: \end{cases}$$

Նկատենք, որ $G = (V, E)$ գրաֆում գոյություն ունի համիլտոնյան ցիկլ այն և միայն այն դեպքում, երբ նրան համապատասխանող՝ ՇՐՋԻԿ ԳՈՐԾԱԿԱԼ խնդրի անհատ խնդրում գոյություն ունի $L = p$ երկարությամբ երթուղի: Թեորեմն ապացուցված է:

**Դիտողություն**: Նկատենք, որ վերջին բերման մեջ սահմանված $C = (c_{ij})$ մատրիցը բավարարում է եռանկյան անհավասարմանը՝

$c_{ij} \le c_{ik} + c_{kj}$ ցանկացած $i, j, k$ թվերի համար:

Հետևաբար կարող ենք ասել, որ ՇՐՋԻԿ ԳՈՐԾԱԿԱԼ խնդրի այն ենթախնդիրը, որում $C=(c_{ij})$ մատրիցը բավարարում է եռանկյան անհավասարմանը, ևս $NP$ –լրիվ է:

$a=(a_1,...,a_n)$ և $b=(b_1,...,b_n)$ վեկտորների համար նշանակենք նրանց սկալյար արտադրյալը, այսինքն՝

$$ab=a_1b_1+...+a_nb_n:$$

Վերջում դիտարկենք հետևյալ խնդիրը

ԱՄԲՈՂՋԱԹԻՎ ԳԾԱՅԻՆ ԾՐԱԳՐԱՎՈՐՈՒՄ (ԳԾ)

Տրված է $m\times n$ կարգի ամբողջաթիվ $A=(a_{ij})$ մատրիցը և $b=(b_1,...,b_m)$ վեկտորը:

Պահանջվում է պարզել, հետևյալ հարցի պատասխանը. գոյություն ունի ոչ բացասական, ամբողջաթիվ բաղադրիչներով $x=(x_1,...,x_n)$ վեկտոր, որը բավարարում է հետևյալ համակարգին

$$\overline{a_i}\,\overline{x_i}\le b_i,\ i=1,...,m_1,$$
$$\overline{a_i}\,\overline{x_i}=b_i,\ i=m_1+1,...,m_1+m_2,$$
$$\overline{a_i}\,\overline{x_i}\ge b_i,\ i=m_1+m_2+1,...,m,$$

որտեղ $\overline{a_i}$ -ով նշանակված է $A=(a_{ij})$ մատրիցի $i$ –րդ տողը:

**Թեորեմ**: ԱՄԲՈՂՋԱԹԻՎ ԳԾ խնդիրը $NP$ –լրիվ է:

**Ապացույց**: Ի տարբերություն մինչև այժմ դիտարկած խնդիրների, որոնց $NP$ դասին պատկանելը ցույց էր տրվում բավականին հեշտ ձևով, ԱՄԲՈՂՋԱԹԻՎ ԳԾ խնդրի $NP$ դասին պատկանելն այնքան էլ պարզ չէ: Նկատենք, որ բնական է, որ բնական է ԱՄԲՈՂՋԱԹԻՎ ԳԾ խնդրի դրական պատասխան ունեցող անհատ խնդիրների հավաստիացում ընտրել հենց ինքը, համակարգի լուծումը: Ընդհանուր դեպքում ասած, պարտադիր չէ, որ այդ լուծման երկարությունը լինի բազմանդամորեն սահմանափակ ամբողջաթիվ $A=(a_{ij})$ մատրիցի և $b=(b_1,...,b_m)$ վեկտորի երկարություններից: Առանց ապացույցի նշենք, որ ապացուցված է հետևյալ փաստը. եթե ԱՄԲՈՂՋԱԹԻՎ ԳԾ խնդիրն ունի լուծում ապա այն ունի լուծում, որի երկարությունը բազմանդորեն է սահմանափակված $A=(a_{ij})$ մատրիցի և $b=(b_1,...,b_m)$ վեկտորի երկարություններից: Այնպես որ ԱՄԲՈՂՋԱԹԻՎ ԳԾ պատկանում է $NP$ դասին:

ԱՄԲՈՂՋԱԹԻՎ ԳԾ խնդրի լրիվությունը մենք կապացուցենք երեք տարբեր ճանապարհով:

**Ճանապարհ 1**: Նախ անմիջապես նկատենք, որ 0-1 Ուսապարկ խնդիրը հանդիսանում է ԱՄԲՈՂՋԱԹԻՎ ԳԾ խնդրի մասնավոր դեպքը, այնպես որ 0-1 Ուսապարկ$\prec$ ԱՄԲՈՂՋԱԹԻՎ ԳԾ: Քանի որ 0-1 Ուսապարկ խնդիրն ինքը հանդիսանում է $NP$ –լրիվ խնդիր, ապա ԱՄԲՈՂՋԱԹԻՎ ԳԾ-ն նս $NP$ –լրիվ է:

**Ճանապարհ 2**: Ցույց տանք, որ ԲԱԶՄՈՒԹՅԱՆ ԾԱԾԿՈՒՅԹ$\prec$ ԱՄԲՈՂՋԱԹԻՎ ԳԾ:

Դիտարկենք ԲԱԶՄՈՒԹՅԱՆ խնդրի մի որևէ անհատ խնդիր. $k$ բնական թիվը, $A=\{a_1,...,a_n\}$ բազմությունը և նրա ենթաբազմությունների $\boldsymbol{A}=\{A_1,...,A_m\}$ ընտանիքը, որը ծածկում է $A=\{a_1,...,a_n\}$ բազմությունը, այսինքն՝

$$\bigcup_{i=1}^{m} A_i = A:$$

Դիտարկենք $H=(h_{ij})$ $m\times n$ մատրիցը (նկար 2), որտեղ

$$h_{ij}=\begin{cases}1 & \text{եթե} \quad a_j\in A_i,\\ 0 & \text{եթե} \quad a_j\notin A:\end{cases}$$

| | | | | | | | |
|---|---|---|---|---|---|---|---|
| | | $a_1$ | $a_2$ | . | $a_j$ | . | $a_n$ |
| $x_1$ | $A_1$ | $h_{11}$ | $h_{12}$ | . | $h_{1j}$ | . | $h_{1n}$ |
| $x_2$ | $A_2$ | $h_{21}$ | $h_{22}$ | . | $h_{2j}$ | . | $h_{2n}$ |
| . | . | . | | . | | . | |
| $x_i$ | $A_i$ | $h_{i1}$ | $h_{i2}$ | . | $h_{ij}$ | . | $h_{in}$ |
| . | . | . | | . | | . | |
| $x_m$ | $A_m$ | $h_{m1}$ | $h_{m2}$ | . | $h_{mj}$ | . | $h_{mn}$ |

նկար 2

Նկատենք, որ քանի որ $\boldsymbol{A}=\{A_1,...,A_m\}$ ընտանիքը ծածկում է $A=\{a_1,...,a_n\}$ բազմությունը, ապա $H=(h_{ij})$ մատրիցի յուրաքանչյուր սյան մեջ գոյություն ունի առնվազն մեկ հատ մեկ:

$\boldsymbol{A}=\{A_1,...,A_m\}$ ընտանիքի ցանկացած $A_{i_1},...,A_{i_k}$ ենթաընտանիքի համապատասխանեցնենք $x_1,...,x_m$ թվերը, որտեղ

$$x_i=\begin{cases}1 & \text{եթե} \quad A_i\in\{A_{i_1},...,A_{i_k}\},\\ 0 & \text{եթե} \quad A_i\notin\{A_{i_1},...,A_{i_k}\}:\end{cases}$$

Այս ձևով, մենք ցանկացած $A_{i_1},...,A_{i_k}$ ենթաընտանիքի համապատասխանեցրինք 0,1 թվերի ինչ-որ $m$-յակ: Նկատենք, որ տեղի ունի նաև հակառակը, 0,1 թվերի ցանկացած $m$-յակի համապատասխանում է $\boldsymbol{A}=\{A_1,...,A_m\}$ ընտանիքի ինչ-որ մի ենթաընտանիք: Ավելին, $A_{i_1},...,A_{i_k}$ ենթաընտանիքը կկազմի $\boldsymbol{A}=\{A_1,...,A_m\}$ ընտանիքի ենթածածկույթ այն և միայն այն դեպքում, երբ $H=(h_{ij})$ մատրիցի $i_1,...,i_k$ տողերով ծնված մատրիցի ցանկացած սյան մեջ կա առնվազն մեկ հատ մեկ, այլ կերպ ասած $A_{i_1},...,A_{i_k}$ ենթաընտանիքին համապատասխանող $x_1,...,x_m$ թվերը կբավարարեն

$$h_{11}x_1+...+h_{m1}x_m\geq 1$$
$$h_{12}x_1+...+h_{m2}x_m\geq 1$$
$$.$$
$$.$$
$$.$$
$$h_{1n}x_1+...+h_{mn}x_m\geq 1$$
$$x_1+...+x_m=k$$

համակարգին:

**Ճանապարհ 3**: Ցույց տանք, որ ՇՐՋԻԿ ԳՈՐԾԱԿԱԼ ≺ ԱՄԲՈՂՋԱԹԻՎ ԳԾ: Դիտարկենք ՇՐՋԻԿ ԳՈՐԾԱԿԱԼ խնդրի՝ $C=(c_{ij})$, $c_{ij}\in Z^+$ և $L\in Z^+$ անհատ խնդիրը: $i$-ից $j$ ճանապարհին համապատասխանեցնենք $x_{ij}\in\{0,1\}$ փոփոխականը: Եթե գործակալի երթուղում $i$-ից բնակավայրին անմիջապես հաջորդում է $j$-րդ բնակավայրը, ապա վերցնենք $x_{ij}=1$, հակառակ դեպքում՝ $x_{ij}=0$: Նկատենք, որ ցանկացած երթուղուն՝ այս ձևով համապատասխանացված $x_{ij}$ թվերը կբավարարեն

$$\sum_{i=1}^{n}x_{ij}=\sum_{j=1}^{n}x_{ij}=1$$

պայմաններին: Դժվար չէ տեսնել, որ հակառակը կարող է ճիժտ չլինել, այսինքն եթե ունենք վերը նշված պայմաններին բավարարող $x_{ij}$ թվեր, ապա նրանց կարող է ընդհանրապես չհամապատասխանել որևէ երթուղի: Օրինակ, վերցնենք 6 բնակավայր և դիցուք՝

$$x_{12}=x_{23}=x_{31}=x_{45}=x_{56}=x_{64}=1:$$

Նկատենք, որ այս ձևով սահմանված $x_{ij}$ թվերին համապատասխանում է երկու՝ միմյանց հետ չհատվող ցիկլեր:

ՇՐՋԻԿ ԳՈՐԾԱԿԱԼ խնդրի՝ $C=(c_{ij})$, $c_{ij}\in Z^+$ և $L\in Z^+$ անհատ խնդրին համապատասխանեցնենք ԱՄԲՈՂՋԱԹԻՎ ԳԾ հետևյալ խնդիրը

գոյություն ունեն արդյոք $x_{ij}\in\{0,1\}$, $u_i\in Z^+$ թվեր, որոնք բավարարում են հետևյալ համակարգին

$$\sum_{i=1}^{n}\sum_{j=1}^{n} c_{ij}x_{ij} \le L,$$

$$\sum_{i=1}^{n} x_{ij} = \sum_{j=1}^{n} x_{ij} = 1$$

$$u_i - u_j + nx_{ij} \le n-1,\ 2 \le i \ne j \le n:$$

Ցույց տանք, որ ՇՐՋԻԿ ԳՈՐԾԱԿԱԼ խնդրի՝ $C=(c_{ij})$, $c_{ij}\in Z^+$ և $L\in Z^+$ անհատ խնդրում պատասխանը դրական է այն և միայն այն դեպքում, երբ վերը նշված ԱՄԲՈՂՋԱԹԻՎ ԳԾ խնդրում պատասխանը դրական է:

Դիցուք ՇՐՋԻԿ ԳՈՐԾԱԿԱԼ խնդրին պատկանող անհատ խնդրում պատասխանը դրական է և $L$-ից ոչ մեծ երկարություն ունեցող երթուղին $1,i_1,...,i_{n-1},1$-ն է: Դիտարկենք հետևյալ ձևով սահմանված $x_{ij}\in\{0,1\}$, $u_i\in Z^+$ թվերը.

$x_{ij}=1$ այն և միայն այն դեպքում, երբ գործակալի $1,i_1,...,i_{n-1},1$ երթուղում $i$-րդ բնակավայրին անմիջապես հաջորդում է $j$-րդ բնակավայրը;

$$u_1=0, u_{i_1}=1, u_{i_2}=2, ..., u_{i_{n-1}}=n-1:$$

Նկատենք, որ այս ձևով սահմանված թվերը բավարարում են վերը նշված համակարգին:

Այժմ ենթադրենք, որ ԱՄԲՈՂՋԱԹԻՎ ԳԾ խնդրում է պատասխանը դրական: $i$-րդ բնակավայրից $j$-րդ բնակավայրը տանող ճանապարհը մտցնենք երթուղու մեջ այն և միայն այն դեպքում, երբ $x_{ij}=1$: Պարզ է, որ այս ձևով սահմանված երթուղին կպարունակի յուրաքանչյուր բնակավայր մտնող և դուրս եկող ճանապարհ: Ցույց տանք, որ սահմանված երթուղին բաղկացած է մեկ ցիկլից: Ենթադրենք հակառակը: Դիցուք գոյություն ունեն գոնե երկու ցիկլեր: Ընտրենք այդ ցիկլերից այն, որը չի անցնում 1 բնակավայրով: Դիցուք այն անցնում է $j_1,..., j_s$ բնակավայրերով, ընդ որում նշված հերթականությամբ: Ունենք՝ $x_{j_1j_2}=x_{j_2j_3}=...=x_{j_sj_1}=1$: Հետևաբար՝

$$u_{j_1}-u_{j_2}+n\le n-1,$$
$$u_{j_2}-u_{j_3}+n\le n-1,$$
$$\dots\dots\dots$$
$$u_{j_s}-u_{j_1}+n\le n-1:$$

Գումարելով՝ կստատանք, որ $ns\le(n-1)s$: Հակասություն:

Թեորեմն ապացուցված է:

# Գրականություն

Գտնված սխալների, առաջարկությունների, ինչպես նաև դասախոսություն-ներն e-mail-ով ստանալու համար կարող էք դիմել vahanmkrtchyan2002@yahoo.com հասցեով:

Կոմբինատորային ալգորիթմներ և ալգորիթմների վերլուծություն
Վահան Վ. Մկրտչյան

Դասախոսություն 21: 2-ԻՐԱԳՈՐԾԵԼԻՈՒԹՅՈՒՆ խնդրի $P$ դասին պատկանելը: Մոտարկում: Մոտավոր ալգորիթմներ Գագաթային ծածկույթ, Շրջիկ Գործակալ և Բազմության ծածկույթ խնդիրների համար:

Ստորև ցույց կտանք, որ գոյություն ունի 2-ԻՐԱԳՈՐԾԵԼԻՈՒԹՅՈՒՆ խնդիրը լուծող բազմանդամային ալգորիթմ:

$f(x_1,...,x_n)=D_1\&...\&D_r$ կոնյունկտիվ նորմալ ձևի համար դիտարկենք $G_f$ օրգրաֆը, որտեղ $V(G_f)\equiv\{x_1,...,x_n,\overline{x_1},...,\overline{x_n}\}$ և ցանկացած $\alpha,\beta\in V(G_f)$ համար $(\alpha,\beta)\in E(G_f)$ այն և միայն այն դեպքում, երբ $f(x_1,...,x_n)=D_1\&...\&D_r$ կոնյունկտիվ նորմալ ձևում գոյություն ունի $\overline{\alpha}\vee\beta$ կամ $\beta\vee\overline{\alpha}$ դիզյունկցիա: Այլ կերպ ասած $\alpha,\beta$ լիտերալները կազմում են աղեղ $G_f$ օրգրաֆում այն և միայն այն դեպքում, երբ $f(x_1,...,x_n)=D_1\&...\&D_r$-ում գոյություն ունի $\alpha\to\beta=\overline{\alpha}\vee\beta$ դիզյունկցիա:

Նկատենք, որ եթե $(\alpha,\beta)\in E(G_f)$ ապա $(\overline{\beta},\overline{\alpha})\in E(G_f)$: Ապացուցենք հետևյալ թեորեմը

**Թեորեմ**: $f(x_1,...,x_n)=D_1\&...\&D_r$ կոնյունկտիվ նորմալ ձևն իրագործելի չէ այն և միայն այն դեպքում, երբ գոյություն ունի $x\in\{x_1,...,x_n\}$ փոփոխական այնպես, որ $x$-ը և $\overline{x}$-ը պատկանում են $G_f$ օրգրաֆի միևնույն ուժեղ կապակցվածության բաղադրիչին:

**Ապացույց**: Ենթադրենք, որ ինչ-որ մի $x\in\{x_1,...,x_n\}$ փոփոխականի երկու լիտերալներն էլ պատկանում են $G_f$ օրգրաֆի միևնույն ուժեղ կապակցվածության բաղադրիչին, և հետևաբար՝ գոյություն ունեն $x$-ը $\overline{x}$-ին և $\overline{x}$-ը $x$-ին միացնող ուղիներ: Վերցնենք ցանկացած $\alpha \equiv (\alpha_1,...,\alpha_n)$ հավաքածու: Ցույց տանք, որ $f(\alpha_1,...,\alpha_n)=0$: Քննարկենք երկու դեպք

**Դեպք 1**: $\alpha \equiv (\alpha_1,...,\alpha_n)$ հավաքածուում $x$ փոփոխականի արժեքը 1-է: Նկատենք, որ այդ դեպքում $\overline{x}$-ի արժեքը հավասար է 0-ի, և հետևաբար $x$-ը $\overline{x}$-ին միացնող ուղու վրա գոյություն ունի $(\gamma,\delta)\in E(G_f)$ աղեղ այնպես, որ $\gamma$-ն ընդունում է մեկ արժեք, իսկ $\delta$-ն ընդունում է զրո արժեք: $G_f$ օրգրաֆի սահմանումից հետևում է, որ $\overline{\gamma}\vee\delta$ դիզյունկցիան հանդիսանում է դիզյունկցիա $f(x_1,...,x_n)=D_1\,\&\ldots\&\,D_r$-ում: Նկատենք, որ այս դիզյունկցիան ընդունում է զրո արժեք $\alpha \equiv (\alpha_1,...,\alpha_n)$ հավաքածուի վրա և հետևաբար $f(\alpha_1,...,\alpha_n)=0$:

**Դեպք 2**: $\alpha \equiv (\alpha_1,...,\alpha_n)$ հավաքածուում $x$ փոփոխականի արժեքը 0-է: Նկատենք, որ այդ դեպքում $\overline{x}$-ի արժեքը հավասար է 1-ի, և հետևաբար $\overline{x}$-ը $x$-ին միացնող ուղու վրա գոյություն ունի $(\gamma,\delta)\in E(G_f)$ աղեղ այնպես, որ $\gamma$-ն ընդունում է մեկ արժեք, իսկ $\delta$-ն ընդունում է զրո արժեք: $G_f$ օրգրաֆի սահմանումից հետևում է, որ $\overline{\gamma}\vee\delta$ դիզյունկցիան հանդիսանում է դիզյունկցիա $f(x_1,...,x_n)=D_1\,\&\ldots\&\,D_r$-ում: Նկատենք, որ այս դիզյունկցիան ընդունում է զրո արժեք $\alpha \equiv (\alpha_1,...,\alpha_n)$ հավաքածուի վրա և հետևաբար $f(\alpha_1,...,\alpha_n)=0$:

Երկու դեպքերի քննարկման արդյունքում ունենք, որ $f(x_1,...,x_n)=D_1\,\&\ldots\&\,D_r$ կոնյունկտիվ նորմալ ձևն իրագործելի չէ:

Հակառակը, ենթադրենք, որ ոչ մի $x\in\{x_1,...,x_n\}$ փոփոխականի համար $x$-ը և $\overline{x}$-ը չեն պատկանում $G_f$ օրգրաֆի միևնույն ուժեղ կապակցվածության բաղադրիչին:

Կառուցենք հավաքածու, որի վրա $f(x_1,...,x_n)=D_1\,\&\ldots\&\,D_r$ կոնյունկտիվ նորմալ ձևը կընդունի մեկ արժեք:

Քանի դեռ գոյություն ունի $x_1,...,x_n$ փոփոխականների լիտերալ, որի արժեքը դեռ որոշված չէ, կրկնենք հետևյալ քայլերը.

**Քայլ 1**: Վերցնենք $G_f$ օրգրաֆի $\alpha$ գագաթ, որի արժեքը դեռ որոշված չէ և որից հասանելի չէ $\overline{\alpha}$ գագաթը (նկատենք, որ այդպիսին գոյություն ունի);

**Քայլ 2**: $x_1,...,x_n$ փոփոխականների արժեքներն ընտրենք այնպես, որ $\alpha$ գագաթից հասանելի բոլոր գագաթներն (այդ թվում հենց ինքը $\alpha$ գագաթը) ընդունեն ճիշտ արժեք, և բոլոր այն գագաթները, որոնցից հասանելի է $\overline{\alpha}$ գագաթը (այդ թվում հենց ինքը $\overline{\alpha}$ գագաթը) ընդունեն սխալ արժեք:

Նախ նկատենք, որ **Քայլ 1**-ում ընտրված $\alpha$ գագաթից չեն կարող հասանելի լինեն $\beta,\overline{\beta}$ լիտերալներ: Իրոք, եթե այդպես լիներ, ապա ըստ $G_f$ օրգրաֆի սիմետրիկության, $\overline{\beta}$ զագաթից հասանելի կլիներ $\overline{\alpha}$ զագաթը, և հետևաբար՝ $\alpha$ զագաթից հասանելի կլիներ $\overline{\alpha}$ զագաթը, ինչը հակասում է $\alpha$ զագաթի ընտրությանը: Երկրորդ, նկատենք նաև, որ գոյություն չունի $\gamma$ լիտերալ, որին նախորդ քայլերում վերագրված լինեինք սխալ արժեք և որը հասանելի լինի $\alpha$ զագաթից: Իրոք, եթե այդպես չլիներ, ապա $\overline{\gamma}$ լիտերալից հասանելի կլիներ $\overline{\alpha}$ զագաթը, և հետևաբար, համաձայն **Քայլ 2**-ի, $\alpha$ լիտերալի համար մենք արդեն ընտրած կլինեինք որոշակի արժեք, ինչը կհակասեր այն բանին, որ $\alpha$ լիտերալի արժեքը դեռ անորոշ էր: Այս երկու հատկություններից հետևում է, որ նկարագրված ալգորիթմը կոռեկտ է:

Արդյունքում կառուցված հավաքածուի համար ստանում ենք, որ $G_f$ օրգրաֆի աղեղ, որը ճիշտ արժեքից անցնում է սխալ արժեքի (միակ դեպքը, երբ "$\rightarrow$" իմպլիկացիա ֆունկցիան ընդունում է սխալ արժեք): Հետևաբար, կառուցած հավաքածուն իրագործում է $f(x_1,...,x_n) = D_1 \& ... \& D_r$ կոնյունկտիվ նորմալ ձևը:

Վերջում նշենք, որ ապացուցված թեորեմից հետևում է, որ գոյություն ունի 2-ԻՐԱԳՈՐԾԵԼԻՈՒԹՅՈՒՆ խնդիրը լուծող բազմանդամային ալգորիթմ: Իրոք, բավական է կառուցել $f(x_1,...,x_n) = D_1 \& ... \& D_r$ կոնյունկտիվ նորմալ ձևին համապատասխանող $G_f$ օրգրաֆի ուժեղ կապակցվածության բաղադրիչները և ստուգել, թե տեղի ունեն արդյոք **Թեորեմ** –ի պայմանները:

**Մոտարկում**: Մինչև հիմա մենք դիտարկել ենք խնդիրներ, որոնք ունեն կամ էլ հեշտությամբ վերաձևակերպվում են հետևյալ տեսքով:

Տրված են $E = \{e_1,...,e_n\}$ էլեմենտների բազմությունը և $c: E \rightarrow R$ գնային ֆունկցիան: Պահանջվում է գտնել $e \in E$ տարր այնպես, որ $c(e) \rightarrow \min$ կամ $c(e) \rightarrow \max$:

Ինչպես արդեն նշել ենք, այն խնդիրները, որոնք ունեն քիչ թե շատ գործնական նշանակություն հիմնականում $NP$–լրիվ են: Սա նշան է այն բանի, որ չարժե փորձել մշակել այդպիսի խնդիրները լուծող էֆֆեկտիվ (բազմանդամային) ալգորիթմ: Մյուս կողմից, հաշվի առնելով, որ այդպիսի խնդիրները ունեն գործնական նշանակություն, մենք շահագրգռված ենք նրանց լուծելու մեջ՝ անկախ այն բանից, թե նրանք $NP$–լրիվ են, թե ոչ:

Նմանատիպ իրադրության մեջ, կարելի է առաջարկել երկու տարբերակ: Առաջինի էությունը կայանում է նրանում, որ մենք կարող ենք նախագծել ալգորիթմներ, որոնք ***չեն աշխատում*** բազմանդամային ժամանակում, բայց ճշգրիտ լուծում են դիտարկվող խնդիրը: Երկրորդ մոտեցումը կայանում է “ախորժակը” սահմանափակելու մեջ, այլ կերպ ասած, մենք կարող ենք որոնել բազմանդամային ալգորթիմներ, որոնք կառուցում են այնպիսի լուծումներ, որոնք լավագույնը չեն, բայց որոնց համար գնային ֆունկցիայի արժեքն այնքան էլ “հեռու չէ” լավագույնից: Ընդհանրապես, ընդունված է նմանատիպ ալգորիթմներին անվանել “մոտավոր” ալգորիթմներ, իսկ նրանց կառուցած լուծման շեղումը լավագույն լուծումից՝ ալգորիթմի մոտարկման աստիճան: Ավելի հստակ՝ եթե դիտարկում ենք օպտիմիզացիոն $\Pi$ խնդիրը, որտեղ $\Pi$ –ն հետևյալն է

Տրված են $E=\{e_1,...,e_n\}$ էլեմենտների բազմությունը և $c:E\to R$ գնային ֆունկցիան: Պահանջվում է գտնել $e\in E$ տարր այնպես, որ $c(e)\to\min$ կամ $c(e)\to\max$ ;

ապա բազմանդամային բարդություն ունեցող $A$ ալգորիթմին կանվանենք $\Pi$ խնդիրը լուծող $1+\varepsilon$ , $\varepsilon\geq 0$ մոտավոր ալգորիթմ, եթե ցանկացած $I\in\Pi$ անհատ խնդրի համար, $A$-ն կառուցում է $A(I)\in E=\{e_1,...,e_n\}$ տարր այնպես, որ $\frac{c(A(I))}{OPT(I)}\leq 1+\varepsilon$ , եթե $\Pi$-ն մինիմիզացիոն խնդիր է, և $\frac{OPT(I)}{c(A(I))}\leq 1+\varepsilon$ եթե $\Pi$-ն մաքսիմիզացիոն խնդիր է, որտեղ $OPT(I)$-ն $c:E\to R$ գնային ֆունկցիայի օպտիմալ արժեքն է $E=\{e_1,...,e_n\}$ էլեմենտների բազմության վրա:

Նկատենք, որ իրականում 1 մոտավոր ալգորիթմն իրենից ներկայացնում է $\Pi$ խնդիրը լուծող բազմանդամային ալգորիթմ: Փաստորեն, եթե $P\neq NP$ և $\Pi$-ն $NP$–լրիվ է, ապա գոյություն չունի $\Pi$ խնդիրը լուծող 1 մոտավոր ալգորիթմ:

Ստորև կդիտարկվեն որոշ $NP$–լրիվ խնդիրներ, որոնց լուծման համար համար կնկարագրվեն մոտավոր ալգորիթմներ:

**Գագաթային ծածկույթ**: Նախ հիշենք ԳԱԳԱԹԱՅԻՆ ԾԱԾԿՈՒՅԹ խնդիրը

Տրված է $G$ գրաֆը և $k$ բնական թիվը:
Պահանջվում է պարզել $\beta(G)\leq k$ թե ոչ, այլ կերպ ասած, պահանջվում է պարզել, թե գոյություն ունի արդյոք $G$ գրաֆի՝ ոչ ավել, քան $k$ գագաթ պարունակող ծածկույթ:
Ինչպես արդեն նշել ենք, այս խնդիրը $NP$–լրիվ է: Դիտարկենք ԳԱԳԱԹԱՅԻՆ ԾԱԾԿՈՒՅԹ խնդրի կառուցման տարբերակը, այսինքն, հետևյալ խնդիրը

Տրված է $G$ գրաֆը: Պահանջվում է կառուցել $G$ գրաֆի նվազագույն թվով գագաթներ պարունակող ծածկույթ:

Նկատենք, որ այս խնդիրն իրենից ներկայացնում է վերևում նկարագրված տիպի խնդիր: Իրոք, այստեղ, որպես $E=\{e_1,...,e_n\}$ էլեմենտների բազմություն վերցնենք $G$ գրաֆի գագաթային ծածկույթների բազմությունը, իսկ $c:E\to R$ գնային ֆունկցիան սահմանենք որպես ծածկույթին պատկանող գագաթների քանակ:

Դիտարկենք հետևյալ ալգորիթմը

Ալգորիթմ 1

**ՔԱՅԼ 1**: $V':=\varnothing$; $E':=E(G)$

**ՔԱՅԼ 2**:

Քանի դեռ $E'\neq\varnothing$

Վերցնել $G$ գրաֆի կամայական $e=(u,v)\in E'$ կող;

$V':=V'\cup\{u,v\}$;

$E'$-ից հեռացնել $u$ և $v$ գագաթներին ինցիդենտ կողերը:

**ՔԱՅԼ 3**: Վերադարձնել $V'$-ը:

Նախ նկատենք, որ ցանկացած $G$ գրաֆի համար Ալգորիթմ 1-ը կառուցում է $G$ գրաֆի գագաթային ծածկույթ: Ցույց տանք, որ այն կառուցում է այնպիսի ծածկույթ, որը լավագույնից տարբերվում է ոչ ավել քան երկու անգամ:

Իրոք, դիցու ք $V'$-ը $G$ գրաֆի այն ծածկույթն է, որը կառուցում է Ալգորիթմ 1-ը: Նկատենք, որ համաձայն **ՔԱՅԼ 2**-ի $V'$-ն իրենից ներկայացնում է $G$ գրաֆի՝ զույգ առ զույգ ընդհանուր գագաթ չունեցող $e_1,\ldots,e_k$ կողերի ծայրակետերը, և հետևաբար՝ $|V'|=2k$: Մյուս կողմից, ենթադրենք, որ $V^*$-ն իրենից ներկայացնում է $G$ գրաֆի՝ նվազագույն թվով գագաթներ պարունակող ծածկույթ: Քանի որ $e_1,\ldots,e_k$ կողերը չունեն ընդհանուր գագաթ, ապա $V^*$-ը, այս կողերը ծածկելու համար, պետք է պարունակի կողերի ծայրակետերից գոնե մեկը, և հետևաբար՝ $|V^*|\geq k$: Արդյունքում՝

$$\frac{c(A(I))}{OPT(I)}=\frac{|V'|}{|V^*|}\leq\frac{2k}{k}=2:$$

Նկատենք, որ Ալգորիթմ 1-ն ունի բազմանդամային, նունիսկ գծային բարդություն, և հետևաբար՝ այն հանդիսանում է 2-մոտավոր ալգորիթմ ԳԱԳԱԹԱՅԻՆ ԾԱԾԿՈՒՅԹ խնդրի համար:

Առաջին հայացքից, թվում է, թե ստորև նկազրված ալգորիթմը պետք է լինի ավելի փոքր գործակցով մոտավոր ալգորիթմ,

Ալգորիթմ 2

**ՔԱՅԼ 1**: $V':=\varnothing$;

**ՔԱՅԼ 2**: Քանի դեռ $E(G)\neq\varnothing$

Վերցնել $G$ գրաֆի $v\in V(G)$ գագաթ, որի աստիճանը ամենամեծն է;

$V':=V'\cup\{v\}$;

$G$ գրաֆից հեռացնել $v$ գագաթը:

**ՔԱՅԼ 3**: Վերադարձնել $V'$-ը:

սակայն, ինչպես ցույց է տալիս նկար 1-ում բերված գրաֆը, այս ալգորիթմը ***նույնիսկ մոտավոր ալգորիթմ չէ***, քանի որ այն կարող է կամայապես սխալվել:

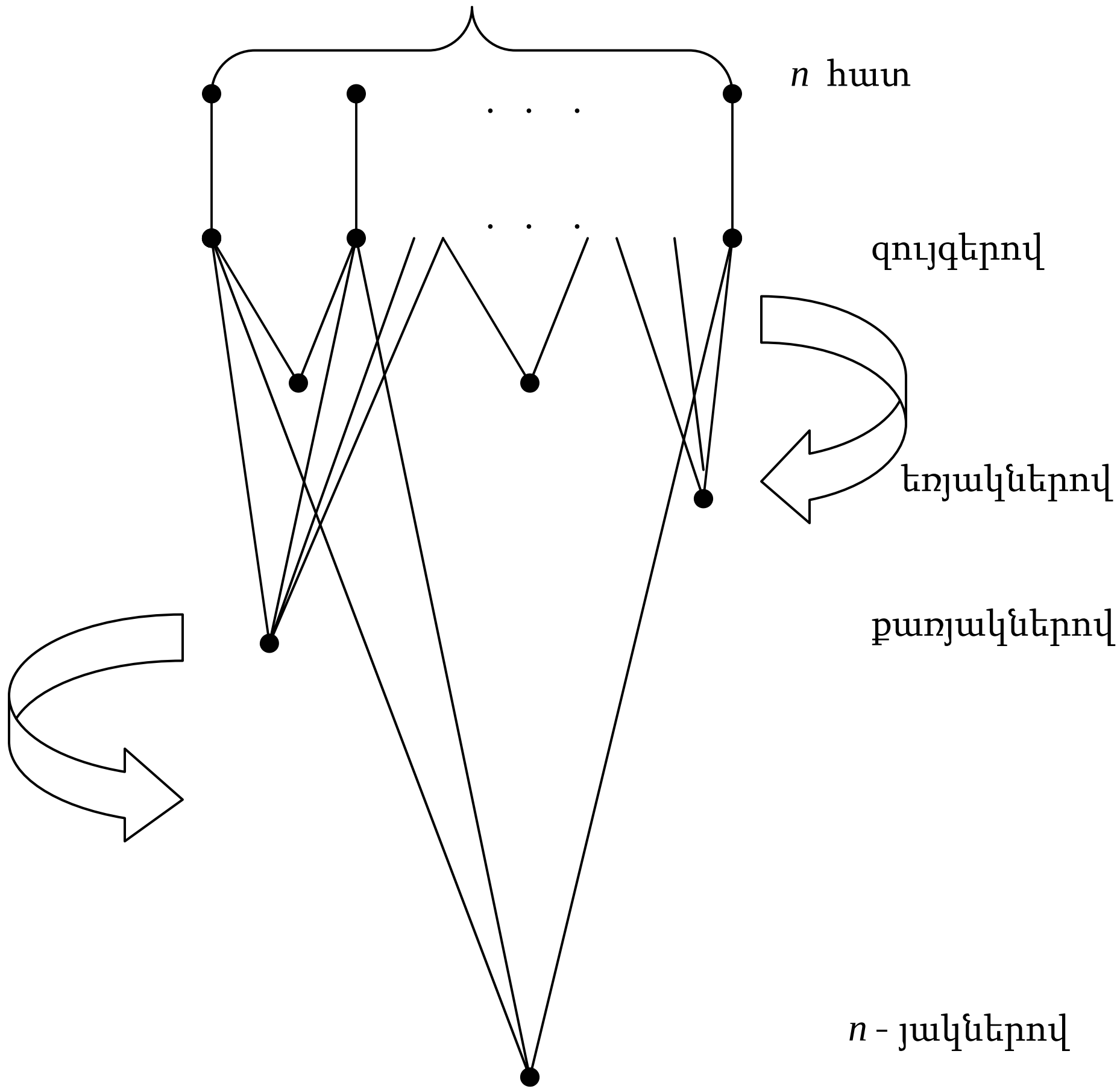

նկար 1

Այս գրաֆում նվազագույն գագաթային ծածկույթի հզորությունը $n$-է, իսկ Ալգորիթմ 2-ի շնորհիվ կառուցված ծածկույթի հզորությունը՝

$$1+1+...\left[\frac{n}{2}\right]+n=\sum_{k=1}^{n}\left[\frac{n}{k}\right]\geq\sum_{k=1}^{n}\left(\frac{n}{k}-1\right)=n\sum_{k=1}^{n}\frac{1}{k}-n$$

և հետևաբար՝

$$\frac{c(A(I))}{OPT(I)}\geq\sum_{k=1}^{n}\frac{1}{k}-1\rightarrow+\infty \text{ երբ՝ } n\rightarrow+\infty:$$

**Բազմության ծածկույթ**: Նախ հիշենք ԲԱԶՄՈՒԹՅԱՆ ԾԱԾԿՈՒՅԹ խնդիրը

Տրված է $k$ բնական թիվը, $A=\{a_1,...,a_n\}$ բազմությունը և նրա ենթաբազմությունների $\boldsymbol{A}=\{A_1,...,A_m\}$ ընտանիքը, որը ծածկում է $A=\{a_1,...,a_n\}$ բազմությունը, այսինքն՝

$$\bigcup_{i=1}^{m}A_i=A:$$

Պահանջվում է պարզել, թե գոյություն ունի արդյոք $\boldsymbol{A}=\{A_1,...,A_m\}$ բազմության $k$ ենթածածկույթ, այսինքն՝ $A_{i_1},...,A_{i_k}$ տարրեր, որոնք ծածկում են $A=\{a_1,...,a_n\}$ բազմությունը

$$\bigcup_{j=1}^{k} A_{i_j}=A:$$

Դիտարկենք ԲԱԶՄՈՒԹՅԱՆ ԾԱԾԿՈՒՅԹ խնդրի կառուցման տարբերակը, այսինքն, հետևյալ խնդիրը

Տրված է $A=\{a_1,...,a_n\}$ բազմությունը և նրա ենթաբազմությունների $\boldsymbol{A}=\{A_1,...,A_m\}$ ընտանիքը: Պահանջվում է կառուցել $\boldsymbol{A}=\{A_1,...,A_m\}$ ծածկույթի նվազագույն թվով բազմություններ պարունակող ենթածածկույթ:

Նկատենք, որ այս խնդիրն իրենից ներկայացնում է վերևում նկարագրված տիպի խնդիր: Իրոք, այստեղ որպես $E=\{e_1,...,e_n\}$ էլեմենտների բազմություն վերցնենք $\boldsymbol{A}=\{A_1,...,A_m\}$ բազմության ենթածածկույթների բազմությունը, իսկ $c:E\to R$ գնային ֆունկցիան սահմանենք որպես ենթածածկույթին պատկանող բազմությունների քանակ:

Ինչպես արդեն նշել ենք, այս խնդիրը $NP$–լրիվ է: Դիտարկենք հետևյալ “ագահ” ալգորիթմը

Ալգորիթմ 3

ՔԱՅԼ 1: $U:=A$; $\boldsymbol{A'}:=\varnothing$;

ՔԱՅԼ 2: Քանի դեռ $U\neq\varnothing$

Վերցնել $\boldsymbol{A}=\{A_1,...,A_m\}$ ծածկույթի այնպիսի $A_i\in\boldsymbol{A}$, որի համար $|A_i\cap U|\to\max$;

$U:=U\setminus A_i$; $\boldsymbol{A'}:=\boldsymbol{A'}\cup\{A_i\}$

ՔԱՅԼ 3: Վերադարձնել $\boldsymbol{A'}$-ը:

Նախ նկատենք, որ Ալգորիթմ 3-ը բազմանդամային է: Իրոք, ՔԱՅԼ 2-ում առկա ցիկլը կկատարվի ոչ ավել, քան $\min\{m,n\}$ անգամ, իսկ ցիկլի յուրաքանչյուր իտերացիա կարելի է իրականացնել $O(m\cdot n)$ ժամանակում: Հետևաբար, Ալգորիթմ 3-ի բարդությունը կլինի $O(m\cdot n)\cdot\min\{m,n\}$:

$n$ բնական թվի համար $H(n)$–ով նշանակենք հարմոնիկ շարքի առաջին $n$ անդամների գումարը, այսինքն՝

$$H(n)=1+\frac{1}{2}+...+\frac{1}{n}:$$

**Լեմմա 1**: $H(n) \le 1 + \ln n$:

**Ապացույց**: Նախ նկատենք, որ

$$H(n) = 1 + \frac{1}{2} + \ldots + \frac{1}{n} = 1 + \sum_{k=1}^{n-1} \frac{1}{k+1}:$$

Մյուս կողմից, եթե $x \in [k, k+1]$, ապա

$$\frac{1}{k+1} \le \frac{1}{x} \le \frac{1}{k},$$

և հետևաբար՝

$$\frac{1}{k+1} = \int_k^{k+1} \frac{1}{k+1} dx \le \int_k^{k+1} \frac{1}{x} dx,$$

որտեղից՝

$$H(n) = 1 + \sum_{k=1}^{n-1} \frac{1}{k+1} \le 1 + \sum_{k=1}^{n-1} \int_k^{k+1} \frac{1}{x} dx = 1 + \int_1^n \frac{1}{x} dx = 1 + \ln n:$$

Ստորև փորձելու ենք պարզել, թե ինչքան է կազմում Ալգորիթմ 3-ի կառուցած ենթածածկույթի շեղումը $\boldsymbol{A} = \{A_1, \ldots, A_m\}$ ծածկույթի լավագույն ենթածածկույթից:

Դիցուք ունենք $|\boldsymbol{A}'|$ դրամ: Նկատենք, որ Ալգորիթմ 3-ը $\boldsymbol{A}'$ ծածկույթը կառուցում է $|\boldsymbol{A}'|$ քայլում, ընդ որում յուրաքանչյուր քայլում $A$ բազմության չծածկված տարրերի $U$ բազմությունից ոմանք ծածկվում են: $|\boldsymbol{A}'|$ դրամը բաժանենք 1-ական դրամների, և յուրաքանչյուր քայլում ընտրված $A_i \in \boldsymbol{A}$ բազմության շնորհիվ ծածկված տարրերին, որոնք մինչ այդ պահը ծածկված չէին, տանք $\frac{1}{|A_i|}$ դրամ:

Քանի որ $\boldsymbol{A}'$-ը հանդիսանում է $A$ բազմության ծածկույթ, ապա պարզ է, որ ցանկացած $x \in A$ տարր կստանա որոշակի $c_x$ դրամ:

**Լեմմա 2**: Ցանկացած $X \in \boldsymbol{A} = \{A_1, \ldots, A_m\}$ բազմության համար $\sum_{x \in X} c_x \le H(|X|)$:

**Ապացույց**: Դիցուք $|X| = k$: Դիտարկենք $X$ բազմության այն տարրերը, որոնք առաջին անգամ ստանում են որոշակի գումար: Նկատենք, որ Ալգորիթմ 3-ի քայլ 2-ից հետևում է, որ այն բազմությունը, որի շնորհիվ այս տարրերը ստանում են գումար, պարունակում է առնվազն $|X| = k$ տարր: Հետևաբար, այս տարրերի ստացած գումարը չի գերազանցում $\frac{1}{k}$–ն: $k_1$-ով նշանակենք $X$ բազմության մնացած տարրերի քանակը, այսինքն այն տարրերի, որոնք առաջինը գումար չեն ստանում: Ունենք՝

$$\sum_{x\in X} c_x = \frac{1}{k}+...+\frac{1}{k}+(\text{մնացած } k_1 \text{ տարրերի ստացած գումարը}) \le$$

$$\le \frac{1}{k}+\frac{1}{k-1}+...+\frac{1}{k_1+1}+(\text{մնացած } k_1 \text{ տարրերի ստացած գումարը})$$

Դիտարկենք $X$ բազմության $k_1$ տարրերից նրանք, որոնք առաջին անգամ ստանում են որոշակի գումար: Կրկին Ալգորիթմ 3-ի քայլ 2-ից հետևում է, որ այն բազմությունը, որի շնորհիվ այս տարրերը ստանում են գումար, պարունակում է առնվազն $k_1$ տարր: Հետևաբար, այս տարրերի ստացած գումարը չի գերազանցում $\frac{1}{k_1}$–ն: $k_2$-ով նշանակենք $X$ բազմության $k_1$ տարրերից մնացած տարրերի քանակը, այսինքն այն տարրերը, որոնք առաջինը գումար չեն ստանում: Ունենք՝

$$\sum_{x\in X} c_x \le \frac{1}{k}+\frac{1}{k-1}+...+\frac{1}{k_1+1}+(\text{մնացած } k_1 \text{ տարրերի ստացած գումարը}) =$$

$$=\frac{1}{k}+\frac{1}{k-1}+...+\frac{1}{k_1+1}+(\frac{1}{k_1}+...+\frac{1}{k_1})+(\text{մնացած } k_2 \text{ տարրերի ստացած գումարը}) \le$$

$$\le \frac{1}{k}+\frac{1}{k-1}+...+\frac{1}{k_1+1}+\frac{1}{k_1}+...+\frac{1}{k_2+1}+(\text{մնացած } k_2 \text{ տարրերի ստացած գումարը})$$

Շարունակելով՝ կստանանք լեմմայի ապացույցը:

Օգտվելով **Լեմմա 2**-ից, գնահատենք Ալգորիթմ 3-ի կառուցած $\boldsymbol{A}'$ ենթածածկույթի շեղումը $\boldsymbol{A}=\{A_1,...,A_m\}$ ծածկույթի լավագույն $\boldsymbol{A}^*$ ենթածածկույթից: Ունենք՝

$$|\boldsymbol{A}'| = \sum_{x\in X} c_x \le \sum_{X\in \boldsymbol{A}^*}\sum_{x\in X} c_x \le \sum_{X\in \boldsymbol{A}^*} H(|X|) \le |\boldsymbol{A}^*| H(\max\{|X| : X\in \boldsymbol{A}^*\}) \le$$

$$\le |\boldsymbol{A}^*| H(\max\{|X| : X\in \boldsymbol{A}\})$$

Հաշվի առնելով **Լեմմա** 1-ը, վերջնականապես կստանանք՝

$$|\boldsymbol{A}'| \le |\boldsymbol{A}^*| H(n) \le |\boldsymbol{A}^*|(1+\ln n):$$

Նկատենք, որ իրականում Ալգորիթմ 3-ն ընդհանրացնում է Ալգորիթմ 2-ը: Իրոք, հիշենք, որ ԳԱԳԱԹՆԵՐՈՎ ԾԱԾԿՈՒՅԹ-≺ ԲԱԶՄՈՒԹՅԱՆ ԾԱԾԿՈՒՅԹ բերման ժամանակ, $G$ գրաֆին և $k$ բնական թվին՝ ԳԱԳԱԹՆԵՐՈՎ ԾԱԾԿՈՒՅԹ խնդրի անհատ խնդրին, մենք համապատասխանեցրինք

$$X_i = \{x\in E(G): x\text{-ը և } v\text{-ն ինցիդենտ են}\},\ 1\le i\le p,$$

$$E(G) = X_1 \cup ... \cup X_p$$

ԲԱԶՄՈՒԹՅԱՆ ԾԱԾԿՈՒՅԹ խնդրի անհատ խնդիրը, ընդ որում ցույց տվեցինք, որ $G$ գրաֆում $\beta(G)\le k$ այն և միայն այն դեպքում, երբ $E(G)$ բազմության $\{X_1,...,X_p\}$ ծածկույթը պարունակում է $k$ ենթածածկույթ:

Պարզ է, որ Ալգորիթմ 2-ում ամենամեծ աստիճան ունեցող գագաթի ընտրությունը նույնն է, ինչ-որ Ալգորիթմ 3-ում այնպիսի $A_i \in \boldsymbol{A}$ տարրի ընտրությունը, որի համար $|A_i \cap U| \to \max$: Հետևաբար, կարող ենք ասել, որ Ալգորիթմ 2-ի համար

$$\frac{c(A(I))}{OPT(I)} \le H(\max\{|X| : X \in \{X_1,...,X_p\}\}) \le 1+\ln\Delta(G) \le 1+\ln(|V(G)|-1),$$

որտեղ $\Delta(G)$–ով նշանակված է $G$ գրաֆում ամենամեծ աստիճան ունեցող գագաթի աստիճանը: Նկատենք, որ $\Delta(G) \le 3$ պայմանին բավարարող գրաֆների համար ունենք

$$\frac{c(A(I))}{OPT(I)} \le H(\max\{|X| : X \in \{X_1,...,X_p\}\}) \le H(3) \le 1+\frac{1}{2}+\frac{1}{3}=\frac{11}{6}<2$$

և հետևաբար, այս գրաֆների դասում Ալգորիթմ 2-ն ապահովում է 2-ից փոքր գործակցով մոտարկում:

Վերջում նշենք, որ $P \ne NP$ ենթադրության դեպքում ապացուցված է, որ ԳԱԳԱԹԱՅԻՆ ԾԱԾԿՈՒՅԹ խնդրի համար գոյություն չունի $1+\varepsilon \le 10\sqrt{5}-21 \approx 1,360$ մոտավոր ալգորիթմ (տես [4]-ը): Նշենք նաև, որ մինչև օրս չլուծված խնդիր է հանդիսանում ԳԱԳԱԹԱՅԻՆ ԾԱԾԿՈՒՅԹ խնդրի համար $10\sqrt{5}-21<1+\varepsilon<2$ մոտավոր ալգորիթմի գոյության հարցը:

**Շրջիկ գործակալ**: Պարզվում է, որ գոյություն ունեն խնդիրներ, որոնք ընդհանրապես թույլ չեն տալիս մոտարկում, այլ կերպ ասած, որոնց համար $P \ne NP$ ենթադրության դեպքում կարելի է ապացուցել, որ նրանք ընդհանրապես չունեն $1+\varepsilon$ մոտավոր ալգորիթմ ցանկացած $\varepsilon \ge 0$ համար:

Այդպիսի խնդրի օրինակ է հանդիսանում ՇՐՋԻԿ ԳՈՐԾԱԿԱԼ խնդիրը, որի ձևակերպումը հետևյալն էր

Ունենք $n \ge 3$ բնակավայրեր, որոնցից մեկում գտնվում է գործակալը: Նա պետք է շրջագայի այդ բնակավայրերը և վերադառնա մեկնավայրը՝ յուրաքանչյուր բնակավայրում գտնվելով ճիշտ մեկ անգամ: Հայտնի է նաև բնակավայրերի միջև հեռավորությունը: Խնդիրը կայանում է հետևյալում. ինչ հերթականությամբ պետք է գործակալը շրջանցի բնակավայրերը, որպեսզի նրա անցած ճանապարհի երկարությունը լինի նվազագույնը:

Բնակավայրերը համարակալենք $1,2,...,n$ թվերով և ենթադրենք, որ գործակալը գտնվում է $1$ բնակավայրում: $c_{ij}$-ով նշանակենք $i$–րդ բնակավայրից $j$–րդ բնակավայր տանող ճանապարհի երկարությունը:

Եթե գործակալն ընտրել է բնակավայրերի շրջանցման $1,i_1,...,i_{n-1},1$ հերթականությունը, ապա նրա անցած ճանապարհի երկարությունը կլինի՝

$$c_{1,i_1}+c_{i_1,i_2}+...+c_{i_{n-1},1}:$$

ՇՐՋԻԿ ԳՈՐԾԱԿԱԼ խնդրի ճանաչման տարբերակը, ձևակերպվում է հետևյալ կերպ.

Տրված է $C=(c_{ij})$, $c_{ij}\in Z^{+}$ ,$c_{ii}=0$ մատրիցը և $L\in Z^{+}$ բնական թիվը:
Գոյություն ունի $2,...,n$ թվերի այնպիսի $i_1,...,i_{n-1}$ տեղափոխություն, որ

$$c_{1,i_1}+c_{i_1,i_2}+...+c_{i_{n-1},1}\le L:$$

Հատուկ նշենք, որ պարտադիր չէ, որ $c_{ij}$ թվերը բավարար են $c_{ij}=c_{ji}$ հավասարությանը կամ եռանկյան անհավասարությանը՝ $c_{ij}\le c_{ik}+c_{kj}$, չնայած մենք ապացուցել ենք, որ նույնիսկ այս առնչություններին բավարարող ՇՐՋԻԿ ԳՈՐԾԱԿԱԼ խնդիրը մնում է $NP$–լրիվ: Ստորև կդիտարկենք ՇՐՋԻԿ ԳՈՐԾԱԿԱԼ խնդրի կառուցման տարբերակը, որտեղ պահանջվում է գտնել ամենակարճ երթուղին:

**Թեորեմ**: Եթե $P\ne NP$, ապա ցանկացած $\varepsilon\ge 0$ համար ՇՐՋԻԿ ԳՈՐԾԱԿԱԼ խնդրի կառուցման տարբերակի համար գոյություն չունի $1+\varepsilon$ մոտավոր ալգորիթմ:
**Ապացույց**: Ենթադրենք, որ ինչ-որ մի $\varepsilon\ge 0$ համար գոյություն ունի ՇՐՋԻԿ ԳՈՐԾԱԿԱԼ խնդիրը լուծող Ա $1+\varepsilon$ մոտավոր ալգորիթմ: Ցույց տանք, որ այդ դեպքում $P=NP$:

$P=NP$ ցույց տալու համար, ցույց տանք, որ օգտագործելով Ա ալգորիթմը, մենք կարող ենք բազմանդամային ժամանակում լուծել $NP$–լրիվ խնդիր հանդիսացող ՀԱՄԻԼՏՈՆՅԱՆ ՑԻԿԼ խնդիրը: Ցանկացած $G$ գրաֆի, որի գագաթների բազմությունը՝ $V(G)=\{v_1,...,v_n\}$-ն է, համապատասխանեցնենք $v_1,...,v_n$ բնակավայրերը, որոնց միջև հեռավորությունները սահմանված են հետևյալ կերպ

$$c(v_i,v_j)=\begin{cases}1 & \text{եթե}\quad (v_i,v_j)\in E(G),\\ (1+\varepsilon)|V(G)|+1 & \text{եթե}\quad (v_i,v_j)\notin E(G):\end{cases}$$

Դիտարկենք հետևյալ ալգորիթմը
**Քայլ** 1: Ա ալգորիթմի միջոցով կառուցել $v_1,...,v_n$ բնակավայրերը շրջանցող երթուղի:
**Քայլ** 2: Եթե այդ երթուղու երկարությունը չի գերազանցում $(1+\varepsilon)|V(G)|$-ն, ապա $G$ գրաֆը համիլտոնյան է, հակառակ դեպքում՝ $G$ գրաֆը համիլտոնյան չէ:

Նախ նկատենք, որ այս ալգորիթմը բազմանդամային է: Իրոք, $c(v_i,v_j)$-թվերը հաշվվում են ըստ $G$ գրաֆի բազմանդամային ժամանակում, իսկ Ա ալգորիթմը բազմանդամային էր ըստ ենթադրության:

Ցույց տանք, որ այս ալգորիթմը լուծում է ՀԱՄԻԼՏՈՆՅԱՆ ՑԻԿԼ խնդիրը: Նախ նկատենք, որ քանի որ $c(v_i, v_j) \geq 1$, ապա $v_1, ..., v_n$ բնակավայրերը շրջանցող ցանկացած երթուղու երկարությունն առնվազն $|V(G)|$-է:

Ենթադրենք $G$ գրաֆը համիլտոնյան է: Նկատենք, որ այդ դեպքում $G$ գրաֆի համիլտոնյան ցիկլին համապատասխանող երթուղու երկարությունը $|V(G)|$-է: Հետևաբար, ամենակարճ երթուղու երկարությունը $|V(G)|$-է: Քանի որ Ա $1+\varepsilon$ մոտավոր ալգորիթմ է, ապա այն կկառուցի մի երթուղի, որի երկարությունը չի գերազանցում $(1+\varepsilon)|V(G)|$-ն, և հետևաբար Քայլ 2-ում մեր ալգորիթմը կպատասխանի, որ $G$ գրաֆը համիլտոնյան է:

Ենթադրենք $G$ գրաֆը համիլտոնյան չէ: Նկատենք, որ այդ դեպքում $v_1, ..., v_n$ բնակավայրերը շրջանցող ցանկացած երթուղի պարունակում է գոնե մեկ կող, որի երկարությունը $(1+\varepsilon)|V(G)|+1$ է: Հետևաբար, Ա ալգորիթմի կառուցած երթուղու երկարությունը կլինի առնվազն

$$(1+\varepsilon)|V(G)|+1+(|V(G)|-1) > (1+\varepsilon)|V(G)|,$$

և հետևաբար՝ Քայլ 2-ում մեր ալգորիթմը կպատասխանի, որ $G$ գրաֆը համիլտոնյան չէ: Այսպիսով նկարագրված ալգորիթմը լուծում է ՀԱՄԻԼՏՈՆՅԱՆ ՑԻԿԼ խնդիրը բազմանդամային ժամանակում, և հետևաբար՝ $P = NP$: Թեորեմն ապացուցված է:

Պարզվում է, որ եթե դիտարկենք ՇՐՋԻԿ ԳՈՐԾԱԿԱԼ$_\Delta$ խնդիրը, որը բաղկացած է ՇՐՋԻԿ ԳՈՐԾԱԿԱԼ խնդրի այն անհատ խնդիրներից, որոնցում առկա $c_{ij}$ թվերը բավարարում են $c_{ij} \leq c_{ik} + c_{kj}$՝ եռանկյան անհավասարությանը, ապա այդ խնդիրն ունի մոտավոր ալգորիթմ: Ստորև կդիտարկվեն երկու այդպիսի ալգորիթմներ:

Ալգորիթմ 4

**ՔԱՅԼ 1**: Դիտարկել $G$ գրաֆը, որի գագաթների բազմությունը ՇՐՋԻԿ ԳՈՐԾԱԿԱԼ$_\Delta$ խնդրում առկա բնակավայրերն են, և որի կողերի երկարությունները հենց բնակավայրերի միջև հեռավորություններն են: $G$ գրաֆում կառուցել $T$ - մինիմալ կմախքային ծառ:

**ՔԱՅԼ 2**: Դիտարկել $2T$ գրաֆը, որը ստացվում է $T$ ծառի կողերի կրկնապատկումից: Նկատենք, որ այդ $2T$ գրաֆում ցանկացած գագաթի աստիճան զույգ է, հետևաբար այն էյլերյան է: Կառուցել $2T$ գրաֆի մի որևէ $W$ էյլերյան ցիկլ:

**ՔԱՅԼ 3**: $W$ ցիկլից կառուցել բնակավայրերի շրջանցման երթուղի՝ հեռացնելով կրկնությունները, այլ կերպ ասած շարժվել $W$ ցիկլով, և եթե հանդիպում ենք գագաթ, որի հաջորդ գագաթն արդեն այցելել ենք, ապա այցելել $W$ ցիկլի վրա գտնվող հաջորդ չայցելված գագաթը:

Նկատենք, որ Ալգորիթմ 4-ը բազմանդամային է:Իրոք, դա հետևում է մինիմալ կմախքային ծառ և էյլերյան ցիկլ խնդիրների համար բազմանդամային ալգորիթմների գոյության փաստից:

**Թեորեմ**: Ալգորիթմ 4-ը հանդիսանում է 2-մոտավոր ալգորիթմ ՇՐՋԻԿ ԳՈՐԾԱԿԱԼ$_{\Delta}$ խնդրի համար:

**Ապացույց**: Դիցուք $H^*$–ը բնակավայրերի շրջանցման ամենակարճ երթուղին է, իսկ $H$–ը՝ բնակավայրերի շրջանցման այն երթուղին է, որը կառուցում է Ալգորիթմ 4-ը: Նկատենք, որ քանի որ բավարարվում է եռանկյան անհավասարությունները, ապա

$$c(H) \le c(W):$$

Մյուս կողմից, քանի որ $T$-ն մինիմալ կմախքային ծառ է, ապա $H^*$ երթուղու ցանկացած $e$ կողի համար

$$c(T) \le c(H^* - e) \le c(H^*):$$

Ունենք՝

$$c(H) \le c(W) = 2c(T) \le 2c(H^*)$$

և հետևաբար՝

$$\frac{c(A(I))}{OPT(I)} = \frac{c(H)}{c(H^*)} \le 2:$$

Թեորեմն ապացուցված է:

Դիտարկենք հետևյալ խնդիրը
ՄԻՆԻՄԱԼ ԿԱՏԱՐՅԱԼ ԶՈՒԳԱԿՑՈՒՄ
Տրված է զույգ թվով գագաթներ պարունակող լրիվ գրաֆ, որի կողերին վերագրված են ինչ-որ թվեր:
Պահանջվում է գտնել գրաֆի այնպիսի կատարյալ զուգակցում, որին պատկանող կողերին համապատասխանող թվերի գումարը նվազագույնն է:

Առանց ապացույցի նշենք, որ ապացուցված է, որ գոյություն ունի ՄԻՆԻՄԱԼ ԿԱՏԱՐՅԱԼ ԶՈՒԳԱԿՑՈՒՄ խնդիրը լուծող բազմանդամային ալգորիթմ: Օգտվելով այս փաստից, մենք կնկարագրենք մի ալգորիթմ, որը ՇՐՋԻԿ ԳՈՐԾԱԿԱԼ$_{\Delta}$ խնդիրը լուծում է 3/2-մոտավորությամբ:

Ալգորիթմ 5
**ՔԱՅԼ** 1: Դիտարկել $G$ գրաֆը, որի գագաթների բազմությունը ՇՐՋԻԿ ԳՈՐԾԱԿԱԼ$_{\Delta}$ խնդրում առկա բնակավայրերն են, և որի կողերի երկարությունները հենց բնակավայրերի միջև հեռավորություններն են: $G$ գրաֆում կառուցել $T$ - մինիմալ կմախքային ծառ:

**ՔԱՅԼ 2**: Դիտարկել $T$ գրաֆի կենտ աստիճան ունեցող $v_1,...,v_{2k}$ գագաթները, և կառուցել $v_1,...,v_{2k}$ գագաթներով ծնված լրիվ գրաֆի $F$ մինիմալ կատարյալ զուգակցում:
**ՔԱՅԼ 3**: Դիտարկել $T+F$ գրաֆը, որը ստացվում է $T$ ծառի կողերին $F$ մինիմալ կատարյալ զուգակցման կողերի ավելացումից: Նկատենք, որ այդ $T+F$ գրաֆում ցանկացած գագաթի աստիճան զույգ է, հետևաբար այն էյլերյան է: Կառուցել $T+F$ գրաֆի մի որևէ $W$ էյլերյան ցիկլ:
**ՔԱՅԼ 4**: $W$ ցիկլից կառուցել բնակավայրերի շրջանցման երթուղի՝ հեռացնելով կրկնությունները, այլ կերպ ասած շարժվել $W$ ցիկլով, և եթե հանդիպում ենք գագաթ, որի հաջորդ գագաթն արդեն այցելել ենք, ապա այցելել $W$ ցիկլի վրա գտնվող հաջորդ չայցելված գագաթը:

Նկատենք, որ Ալգորիթմ 5-ը բազմանդամային է:
**Թեորեմ**: Ալգորիթմ 5-ը հանդիսանում է 3/2-մոտավոր ալգորիթմ ՇՐՋԻԿ ԳՈՐԾԱԿԱԼ$_{\Delta}$ խնդրի համար:

**Ապացույց**: Դիցուք $H^*$−ը բնակավայրերի շրջանցման ամենակարճ երթուղին է, իսկ $H$−ը՝ բնակավայրերի շրջանցման այն երթուղին է, որը կառուցում է Ալգորիթմ 5-ը: Նկատենք, որ քանի որ բավարարվում է եռանկյան անհավասարությունները, ապա

$$c(H) \le c(F) + c(T):$$

Մյուս կողմից, քանի որ $T$-ն մինիմալ կմախքային ծառ է, ապա $H^*$ երթուղու ցանկացած $e$ կողի համար

$$c(T) \le c(H^* - e) \le c(H^*):$$

Մյուս կողմից, նկատենք, որ $H^*$ երթուղուց կարելի կառուցել միայն $v_1,...,v_{2k}$ գագաթները շրջանցող $H'$ երթուղի, ընդ որում քանի որ բավարարվում է եռանկյան անհավասարությունները, ապա կարելի է հասնել այն բանին, որ

$$c(H') \le c(H^*):$$

Նկատենք, որ $H'$ երթուղին բաղկացած է երկու կատարյալ զուգակցումներից, և քանի որ $F$-ը $v_1,...,v_{2k}$ գագաթներով ծնված լրիվ գրաֆի մինիմալ կատարյալ զուգակցում էր, ապա

$$2c(F) \le c(H') \le c(H^*) \text{ կամ } c(F) \le c(H^*)/2$$

և հետևաբար՝

$$\frac{c(A(I))}{OPT(I)} = \frac{c(H)}{c(H^*)} \le \frac{c(F)+c(T)}{c(H^*)} \le 3/2:$$

Թեորեմն ապացուցված է:

# Գրականություն

Գտնված սխալների, առաջարկությունների, ինչպես նաև դասախոսություն-ներն e-mail-ով ստանալու համար կարող էք դիմել vahanmkrtchyan2002@yahoo.com հասցեով:

Կոմբինատորային ալգորիթմներ և ալգորիթմների վերլուծություն
Վահան Վ. Մկրտչյան

Դասախոսություն 22: Մոտարկում: Մոտավոր ալգորիթմներ Մաքսիմալ կտրվածք, Ուսապարկ, Տեղավորում խնդիրների համար:

**Մաքսիմալ կտրվածք**: Նախ ձևակերպենք ՄԱՔՍԻՄԱԼ ԿՏՐՎԱԾՔ խնդիրը

Տրված է $G$ գրաֆը և $k$ բնական թիվը:
Պահանջվում է պարզել, թե գոյություն ունի արդյոք $V^{'} \subseteq V(G)$ այնպես, որ $\left|(V^{'},V(G)\backslash V^{'})\right| \geq k$, որտեղ $(V^{'},V(G)\backslash V^{'})$ -ով նշանակված է $G$ գրաֆի այն կողերի բազմությունը, որոնց ինցիդենտ գագաթներից մեկը պատկանում է $V^{'}$–ին, իսկ մյուսը՝ $V(G)\backslash V^{'}$-ին:

Առանց ապացույցի նշենք, որ ՄԱՔՍԻՄԱԼ ԿՏՐՎԱԾՔ խնդիրը $NP$–լրիվ է: Հատկանշական է, որ այս խնդրին հարակից՝ ՄԻՆԻՄԱԼ ԿՏՐՎԱԾՔ խնդրի համար գոյություն ունի այն լուծող բազմանդամային ալգորիթմ:
Դիտարկենք ՄԱՔՍԻՄԱԼ ԿՏՐՎԱԾՔ խնդրի կառուցման տարբերակը, այսինքն, հետևյալ խնդիրը

Տրված է $G$ գրաֆը: Պահանջվում է կառուցել $G$ գրաֆի առավելագույն թվով կողեր պարունակող կտրվածք, այսինքն՝ այնպիսի $V^{'} \subseteq V(G)$, որ $\left|(V^{'},V(G)\backslash V^{'})\right| \to \max$:

Դիտարկենք հետևյալ ալգորիթմը

Ալգորիթմ 1
**ՔԱՅԼ 1**: Վերցնել $G$ գրաֆի մի որևէ սկզբնական $(V^{'},V(G)\backslash V^{'})$ կտրվածք (կարելի է սկսել օրինակ $(\varnothing,V(G))$-ից):
**ՔԱՅԼ 2**: Եթե հնարավոր է $V^{'}$ $(V(G)\backslash V^{'})$-ից հանել մի գագաթ և ավելացնել այն $V(G)\backslash V^{'}$ $(V^{'})$-ին, դրանով իսկ ավելացնելով $(V^{'},V(G)\backslash V^{'})$ կտրվածքի կողերի քանակը, ապա վարվել այդ ձևով, հակառակ դեպքում՝ ավարտել Ալգորիթմ 1–ի աշխատանքը՝ վերադարձնելով $(V^{'}, V(G)\setminus V^{'})$ կտրվածքը:

Նախ նկատենք, որ ցանկացած $G$ գրաֆի համար Ալգորիթմ 1-ը կառուցում է $G$ գրաֆի կտրվածք, և քանի որ ցանկացած $V^{'} \subseteq V(G)$ համար $\left|(V^{'}, V(G)\setminus V^{'})\right| \leq \left|E(G)\right|$, ապա պարզ է, որ Ալգորիթմ 1-ն աշխատում է բազմանդամային ժամանակում:

Ցույց տանք, որ այն կառուցում է այնպիսի կտրվածք, որը լավագույնից տարբերվում է ոչ ավել քան երկու անգամ:

Իրոք, դիցուք $(V^{'}, V(G)\setminus V^{'})$-ն այն կտրվածքն է, որ կառուցում է Ալգորիթմ 1-ը, իսկ՝ $(V_0, V(G)\setminus V_0)$-ն $G$ գրաֆի մի որևէ մաքսիմալ կտրվածք է: Նշանակենք՝

$$V_1 = V_0 \cap V^{'};\ V_2 = V_0 \setminus V^{'};$$

$$V_3 = V^{'} \setminus V_0;\ V_4 = V(G)\setminus (V^{'} \cup V_0):$$

Նկատենք, որ

$$(V^{'}, V(G)\setminus V^{'}) = (V_1 \cup V_3, V_2 \cup V_4);$$

$$(V_0, V(G)\setminus V_0) = (V_1 \cup V_2, V_3 \cup V_4):$$

$1 \leq i \leq j \leq 4$ համար $e_{ij}$–ով նշանակենք այն կողերի քանակը, որոնց ինցիդենտ գագաթներից մեկը պատկանում է $V_i$-ին, իսկ մյուսը՝ $V_j$-ին:

Համաձայն Ալգորիթմ 1–ի Քայլ 2-ի

- ցանկացած $v \in V_1$ համար $v$–ն $V_1 \cup V_3$-ին միացնող կողերի քանակը չի գերազանցում $v$–ն $V_2 \cup V_4$-ին միացնող կողերի քանակը, հետևաբար՝
$$e_{13} \leq 2e_{11} + e_{13} \leq e_{12} + e_{14};$$
- ցանկացած $v \in V_2$ համար $v$–ն $V_2 \cup V_4$-ին միացնող կողերի քանակը չի գերազանցում $v$–ն $V_1 \cup V_3$-ին միացնող կողերի քանակը, հետևաբար՝
$$e_{24} \leq 2e_{22} + e_{24} \leq e_{12} + e_{23};$$
- ցանկացած $v \in V_3$ համար $v$–ն $V_1 \cup V_3$-ին միացնող կողերի քանակը չի գերազանցում $v$–ն $V_2 \cup V_4$-ին միացնող կողերի քանակը, հետևաբար՝
$$e_{13} \leq 2e_{33} + e_{13} \leq e_{23} + e_{34};$$
- ցանկացած $v \in V_4$ համար $v$–ն $V_2 \cup V_4$-ին միացնող կողերի քանակը չի գերազանցում $v$–ն $V_1 \cup V_3$-ին միացնող կողերի քանակը, հետևաբար՝
$$e_{24} \leq 2e_{44} + e_{24} \leq e_{14} + e_{34}:$$

Գումարելով ստացված անհավասարությունները, կստանանք՝

$$2(e_{13} + e_{24}) \leq 2(e_{12} + e_{14} + e_{23} + e_{34})$$

կամ՝

$$e_{13} + e_{24} \leq e_{12} + e_{14} + e_{23} + e_{34}:$$

Ունենք՝

$$OPT(I) = \left|(V_0, V(G) \setminus V_0)\right| = e_{13} + e_{14} + e_{23} + e_{24} \le (e_{12} + e_{14} + e_{23} + e_{34}) +$$
$$+ (e_{12} + e_{14} + e_{23} + e_{34}) = 2(e_{12} + e_{14} + e_{23} + e_{34}) = 2\left|(V', V(G) \setminus V')\right| = 2c(A(I)),$$

հետևաբար՝

$$\frac{OPT(I)}{c(A(I))} \le 2 :$$

Արդյունքում ունենք, որ Ալգորիթմ 1-ը հանդիսանում է 2-մոտավոր ալգորիթմ ՄԱՔՍԻՄԱԼ ԿՏՐՎԱԾՔ խնդրի համար:

**Ուսապարկ**: Վերևում տեսանք, որ Գագաթային ծածկույթ խնդիրը թույլ է տալիս 2-մոտավոր ալգորիթմ, և նշեցինք, որ $P \ne NP$ ենթադրության դեպքում այս խնդրի համար գոյություն չունի $1+\varepsilon \le 10\sqrt{5} - 21 \approx 1,360$ մոտավոր ալգորիթմ: Այնուհետև նշեցինք, որ Շրջիկ գործակալ խնդիրը $P \ne NP$ ենթադրության դեպքում ընդհանրապես չունի մոտավոր ալգորիթմ ցանկացած ճշտությամբ: Ստորև կդիտարկենք մեկ այլ “ծայրահեղություն”, ավելի կոնկրետ, ցույց կտանք, որ Ուսապարկ խնդրի համար գոյություն ունի $1+\varepsilon$-մոտավոր ալգորիթմ ***ցանկացած*** $\varepsilon > 0$ :

Նախ հիշենք Ուսապարկ խնդիրը.

Ունենք որոշ թվով առարկաներ: Հայտնի է նրանցից յուրաքանչյուրի գինն ու ծավալը: Անհրաժեշտ է որոշակի տարողություն ունեցող ուսապարկով տեղափոխել այս առարկաներից այնպիսիները, որոնց ծավալների գումարը չգերազանցի ուսապարկի ծավալը և որոնց գումարային գինը լինի հնարավորին չափ մեծ: Խնդրի մաթեմատիկական ձևակերպումը կայանում էր հետևյալում. տրված են $c_1,...,c_n$, $v_1,...,v_n$, և $V$ ոչ բացասական թվերը: Անհրաժեշտ է $x_1,...,x_n$ փոփոխականների համար ընտրել 0 կամ 1 արժեքներ, որ բավարարվի $x_1v_1 + ... + x_nv_n \le V$ պայմանը և $(x_1c_1 + ... + x_nc_n)$ արտահայտությունը ստանա իր առավելագույն հնարավոր արժեքը: Այս խնդիրը կարճ կգրենք $(c_1,...,c_n,V,v_1,...,v_n)$:

Ինչպես հիշում ենք, Ուսապարկ խնդրի համար մենք նկարագրել էինք ալգորիթմ, որը այն լուծում էր $O(n^3c_0^{\,2})$ ժամանակում, որտեղ $c_0 = \max_{1\le i\le n} c_i$: Դիցուք ունենք $\varepsilon > 0$ թիվ: Նկարագրենք Ուսապարկ խնդիրը լուծող $1+\varepsilon$-մոտավոր ալգորիթմ:

Դիտարկենք Ուսապարկ խնդրի՝ $(c_1,...,c_n,V,v_1,...,v_n)$ անհատ խնդիրը: Վերցնենք $b$ բնական թիվ (հետո կճշտենք, թե ինչպես ընտրել $b$-ն): $(c_1,...,c_n,V,v_1,...,v_n)$ խնդրից անցնենք $(c'_1,...,c'_n,V,v_1,...,v_n)$ խնդրին, որտեղ $c'_1,...,c'_n$ թվերը ստացվում են $c_1,...,c_n$ թվերից վերջին $b$ բիթերի զրոյացումից, այսինքն՝

$$i=1,...,n \text{ համար } c_i^{'}=2^b\left[\frac{c_i}{2^b}\right]:$$

$(c_1^{'},...,c_n^{'},V,v_1,...,v_n)$ խնդրի լուծումը, այսինքն՝ առարկաների այն $I'\subseteq\{1,...,n\}$ բազմությունը, որի գումարային ծավալը չի գերազանցում $V$-ն և որոնց գումարային արժեքը մաքսիմալն է, համարենք $(c_1,...,c_n,V,v_1,...,v_n)$ խնդրի մոտավոր լուծում (այն կարելի է կառուցել մեր նկարագրած ալգորիթմի միջոցով): Ենթադրենք նաև, որ $I^*\subseteq\{1,...,n\}$ բազմությունը $(c_1,...,c_n,V,v_1,...,v_n)$ խնդրի ճշգրիտ լուծումն է: Նկատենք, որ

$$\sum_{i\in I^*}c_i\geq\sum_{i\in I'}c_i\geq\sum_{i\in I'}c_i^{'}\geq\sum_{i\in I^*}c_i^{'}\geq\sum_{i\in I^*}(c_i-2^b)\geq\sum_{i\in I^*}c_i-n2^b:$$

Եթե մենք ցանկանում ենք, որ նկարագրված ալգորիթմը $1+\varepsilon$ ճշտությամբ մոտարկի լավագույն լուծումը, ապա պետք է տեղի ունենա

$$\frac{OPT(I)}{c(A(I))}=\frac{\sum_{i\in I^*}c_i}{\sum_{i\in I'}c_i}\leq 1+\varepsilon$$

անհավասարությունը: Քանի որ $\sum_{i\in I^*}c_i\geq c_0$ ($I^*\subseteq\{1,...,n\}$ բազմությունը $(c_1,...,c_n,V,v_1,...,v_n)$ խնդրի լավագույն լուծումն է), ապա

$$\frac{OPT(I)}{c(A(I))}=\frac{\sum_{i\in I^*}c_i}{\sum_{i\in I^*}c_i-n2^b}\leq\frac{1}{1-\frac{n2^b}{\sum_{i\in I^*}c_i}}\leq\frac{1}{1-\frac{n2^b}{c_0}}:$$

Ընտրենք $b$-ն այնպես, որ

$$\frac{1}{1-\frac{n2^b}{c_0}}\leq 1+\varepsilon:$$

Ունենք

$$1\leq(1+\varepsilon)(1-\frac{n2^b}{c_0})=1-\frac{n2^b}{c_0}+\varepsilon-\varepsilon\frac{n2^b}{c_0},$$

որտեղից՝

$$(1+\varepsilon)\frac{n2^b}{c_0}\leq\varepsilon \text{ կամ } 2^b\leq\frac{c_0\varepsilon}{n(1+\varepsilon)}:$$

Վերցնենք $b=\left[\log_2\frac{c_0\varepsilon}{n(1+\varepsilon)}\right]$: Պարզ է, որ այդ դեպքում

$$\frac{OPT(I)}{c(A(I))}\leq 1+\varepsilon:$$

Ցույց տանք, որ նկարագրված ալգորիթմը բազմանդամային է: Քանի որ բոլոր $c_1^{'},...,c_n^{'}$ թվերը պատիկ են $2^b$-ին, ապա պարզ է, որ $I'\subseteq\{1,...,n\}$ բազմությունը կարելի է գտնել լուծելով $(\left\lceil \frac{c_1}{2^b} \right\rceil,...,\left\lceil \frac{c_n}{2^b} \right\rceil, V, v_1,...,v_n)$ խնդիրը (բոլորից ընդհանուր հանել $2^b$-ն): Արդյունքում $I' \subseteq \{1,...,n\}$ բազմությունը կգտնենք $O(n^3(\frac{c_0}{2^b})^2)$ ժամանակում: Նկատենք, որ համաձայն $b$ թվի ընտրության՝

$$O(n^3(\frac{c_0}{2^b})^2) = O(n^3(\frac{n(1+\varepsilon)}{\varepsilon})^2) = O(n^5\,\frac{(1+\varepsilon)^2}{\varepsilon^2}),$$

ինչն իրենից ներկայացնում է ***բազմանդամ յուրաքանչյուր, նախապես տրված*** $\varepsilon > 0$ ***թվի համար***: Հետևաբար, մեր նկարագրած ալգորիթմն ունի բազմանդամային բարդություն և ապահովում է $1+\varepsilon$ ճշտություն:

Տեղավորում: Նախ ձևակերպենք ՏԵՂԱՎՈՐՈՒՄ խնդիրը

Տրված են $u_1,...,u_n$ առարկաները, նրանց $s(u_1),...,s(u_n) \in [0,1]$ չափերը (առարկայի չափն իրենից ներկայացնում է ռացիոնալ թիվ), և $k$ բնական թիվը:
Պահանջվում է պարզել, թե հնարավոր է արդյոք $u_1,...,u_n$ առարկաները տեղավորել միավոր տարողություն ունեցող $k$ արկղերում:

Առանց ապացույցի նշենք, որ ՏԵՂԱՎՈՐՈՒՄ խնդիրը $NP$-լրիվ է:
Դիտարկենք ՏԵՂԱՎՈՐՈՒՄ խնդրի կառուցման տարբերակը, այսինքն, հետևյալ խնդիրը

Տրված են $u_1,...,u_n$ առարկաները և նրանց $s(u_1),...,s(u_n) \in [0,1]$ չափերը: Պահանջվում է $u_1,...,u_n$ առարկաները տեղավորել հնարավորին չափ քիչ միավոր տարողություն ունեցող արկղերում:

Պատկերացնենք ունենք անվերջ թվով միավոր տարողություն ունեցող արկղեր, որոնք համարակալված են 1,2,3,…. թվերով: Դիտարկենք հետևյալ պարզ ալգորիթմը.
հերթական առարկան տեղավորել ամենափոքր համար ունեցող արկղում, որտեղ այն կարելի է տեղավորել:
Չնայած իր պարզ ձևակերպմանը, այս ալգորիթմը լուծում է ՏԵՂԱՎՈՐՈՒՄ խնդիրը 2-մոտավորությամբ: Իրոք, նկատենք, որ $u_1,...,u_n$ առարկաների տեղավորման համար անհրաժեշտ են առնվազն $\left\lceil \sum_{i=1}^{n} s(u_i) \right\rceil$ միավոր տարողությամբ արկղ, հետևաբար՝ $OPT(I) \geq \left\lceil \sum_{i=1}^{n} s(u_i) \right\rceil$: Մյուս կողմից, նկատենք, որ մեր նկարագրած ալգորիթմի աշխատանքի ընթացքում ժամանակի ցանկացած պահին կարող է լինել ամենաշատը մեկ արկղ, որը լցված լինի կեսից քիչ, հետևաբար՝

$$c(A(I)) \le \left\lceil 2\sum_{i=1}^{n} s(u_i) \right\rceil,$$

և հետևաբար՝

$$\frac{c(A(I))}{OPT(I)} \le \frac{\left\lceil 2\sum_{i=1}^{n} s(u_i) \right\rceil}{\left\lceil \sum_{i=1}^{n} s(u_i) \right\rceil} \le 2 :$$

# Գրականություն

Գտնված սխալների, առաջարկությունների, ինչպես նաև դասախոսություն-ներն e-mail-ով ստանալու համար կարող էք դիմել vahanmkrtchyan2002@yahoo.com հասցեով:

# Քննական Հարցաշար

1. Որոնման ալգորիթմներ և նրանց ներկայացումը ծառերի միջոցով: Ալգորիթմի բարդություն: Կարգավոր բազմության տարրի որոնում:
2. Կեղծ մետաղադրամի որոնում:
3. Բազմությունների հավասարության ստուգում:
4. Երկընթաց հաջորդականության ամենամեծ տարրի որոնում:
5. Մրցաշարային խնդիրներ. Հաղթողի, հաղթողի և պարտվողի որոշումը:
6. Մրցաշարային խնդիրներ. Առաջին և երկրորդ տեղերի որոշումը:
7. Մրցաշարային խնդիրներ. Առաջին, երկրորդ և երրորդ տեղերի որոշումը:
8. Ով ով է:
9. Տեսակավորման խնդիրներ: Ներքնից զնահատական: Տեսակավորում տեղավորման եղանակով: Տեսակավորում շարքերի ձուլման եղանակով:
10. Տեսակավորման խնդիրներ: Տեսակավորում ձուլման և տեղավորման եղանակով:
11. Հաջորդականության միջնակետի որոնումը:
12. Գրաֆի լայնությամբ շրջանցում: Էյլերյան ցիկլ: Գրաֆի կապակցվածության բաղադրիչներ:
13. Գրաֆի խորությամբ շրջանցում: Կողմնորոշված գրաֆի ուժեղ կապակցվածության բաղադրիչների որոնում:
14. Օրգրաֆում կարճագույն ուղու և գրաֆում կարճագույն ճանապարհի գտնելու խնդիրներ: Դեյկստրայի ալգորիթմի նկարագիրը:
15. Օրգրաֆում կարճագույն ուղու և գրաֆում կարճագույն ճանապարհի գտնելու խնդիրներ: Ֆլոյդի ալգորիթմի նկարագիրը: Օրգրաֆի տրանզիտիվ փակում:
16. Կապակցված գրաֆի մինիմալ կմախքային ծառը գտնելու Կրասկալի և Պրիմի ալգորիթմների նկարագիրը և վերլուծությունը:
17. Դինամիկ ծրագրման մեթոդ: Ռեսուրսների բաշխման խնդիր, ուսապարկի խնդիր, երկու հաջորդականությունների ամենաերկար ընդհանուր ենթահաջորդականության գտնելու խնդիր:
18. Դինամիկ ծրագրման մեթոդ: Մի քանի մատրիցների բազմապատկման խնդիր: Բազմանկյան տրիանգուլյացիայի խնդիր:
19. Ալգորիթմական խնդիրների $P, NP$ դասերը: $NP$-լրիվություն: Բերումներ:
20. 3-Իրագործելիություն և Գրաֆի ներկում խնդիրների $NP$-լրիվությունը:
21. Խմբավորում, Անկախ բազմություն, Գագաթներով ծածկույթ խնդիրների $NP$-լրիվությունը:

22. Կատարյալ ծածկույթ և Ներկայացուցիչների համակարգ խնդիրների $NP$ -լրիվությունը:
23. Ուսապարկ, 0-1 Ուսապարկ և Տրոհում խնդիրների $NP$ -լրիվությունը:
24. Բազմության ծածկույթ և Ստուգող թեստ խնդիրների $NP$ -լրիվությունը:
25. Համիլտոնյան կոնտուր խնդրի $NP$ -լրիվությունը:
26. Համիլտոնյան ցիկլ և շրջիկ գործակալ խնդիրների $NP$ -լրիվությունը:
27. Ամբողջաթիվ Գծային Ծրագրավորման խնդրի $NP$ -լրիվությունը: Երեք ապացույց:
28. 2-Իրագործելիություն խնդրի $P$ –դասին պատկանելը:
29. Մոտարկում: Գագաթային և բազմության ծածկույթ:
30. Մոտարկում: Շրջիկ գործակալ:
31. Մոտարկում: Մաքսիմալ կտրվածք:
32. Մոտարկում: Ուսապարկ և Տեղավորում: